\newlength{\extralineskip}
  \documentstyle[12pt]{article}
\newcounter{equnum}[section]
\def\theequnum{\thesection.\arabic{equnum}}
\newcommand{\beq}{$$ \refstepcounter{equnum}}
\newcommand{\eeq}{\eqno (\theequnum) $$}

\newcommand{\bd}{\begin{displaymath}}
\newcommand{\ed}{\end{displaymath}}

\def\hu{{\hat u}}
\def\proj{{\cal P}}
\def\lat{{{\cal L}^D}}
\def\Winf{{\cal W}}
\def\gbar{{\bar{g}}}
\def\Gee{{\cal G}}
\def\Hoo{{\cal H}}
\def\Koo{{\cal K}}
\def\ch2{{\bar\chi\chi}}
\def\e{\, {\rm e}}
\def\IR{{\rm I}\!{\rm R}}
\def\inbar{\,\vrule height1.5ex width.4pt depth0pt}
\def\IC{\relax\hbox{$\inbar\kern-.3em{\rm C}$}}
\def\IR{\relax{\rm I\kern-.18em R}}
\def\IZ{\relax\ifmmode\mathchoice
{\hbox{\kern-.4em Z}}{\hbox{Z\kern-.4em Z}}
{\lower.9pt\hbox{Z\kern-.4em Z}}
{\lower1.2pt\hbox{Z\kern-.4em Z}}\else{Z\kern-.4em Z}\fi}
\def\pvint{{\int\!\!\!\!\!\!-}}

\def\tr{~{\rm tr}~}
\def\ps2{{\bar{\psi}\psi}}
\makeatletter
\newdimen\normalarrayskip              
\newdimen\minarrayskip                 
\normalarrayskip\baselineskip
\minarrayskip\jot
\newif\ifold             \oldtrue            \def\new{\oldfalse}
\def\arraymode{\ifold\relax\else\displaystyle\fi} 
\def\@arrayskip{\ifold\baselineskip\z@\lineskip\z@
     \else
     \baselineskip\minarrayskip\lineskip2\minarrayskip\fi}
\def\@arrayclassz{\ifcase \@lastchclass \@acolampacol \or
\@ampacol \or \or \or \@addamp \or
   \@acolampacol \or \@firstampfalse \@acol \fi
\edef\@preamble{\@preamble
  \ifcase \@chnum
     \hfil$\relax\arraymode\@sharp$\hfil
     \or $\relax\arraymode\@sharp$\hfil
     \or \hfil$\relax\arraymode\@sharp$\fi}}
\def\@array[#1]#2{\setbox\@arstrutbox=\hbox{\vrule
     height\arraystretch \ht\strutbox
     depth\arraystretch \dp\strutbox
     width\z@}\@mkpream{#2}\edef\@preamble{\halign \noexpand\@halignto
\bgroup \tabskip\z@ \@arstrut \@preamble \tabskip\z@ \cr}%
\let\@startpbox\@@startpbox \let\@endpbox\@@endpbox
  \if #1t\vtop \else \if#1b\vbox \else \vcenter \fi\fi
  \bgroup \let\par\relax
  \let\@sharp##\let\protect\relax
  \@arrayskip\@preamble}
\makeatother

\begin{document}

\begin{titlepage}

\baselineskip=12pt

\rightline{UBC/S-96/2}
\rightline{OUTP-96-22P}
\rightline{hep-th/9605140}
\rightline{   }
\rightline{Revised Version}
\rightline{   }
\rightline{\today}

\vskip 0.5truein
\begin{center}

{\LARGE\bf Fermionic Matrix Models}\\
\vskip 0.4truein
{\large\bf Gordon W. Semenoff}\\
\medskip
{\it Department of Physics and Astronomy, University of British Columbia\\
Vancouver, British Columbia, Canada V6T 1Z1}\\

\bigskip
\medskip

{\large\bf Richard J. Szabo}\\
\medskip
{\it Department of Theoretical Physics, University of Oxford\\ 1 Keble
Road, Oxford OX1 3NP, U.K.}\\

\vskip 1.0 truein

\end{center}

\begin{abstract}

\baselineskip=12pt

We review a class of matrix models whose degrees of freedom are
matrices with anticommuting elements. We discuss the properties of the
adjoint fermion one-, two- and gauge invariant $D$-dimensional matrix
models at large-$N$ and compare them with their bosonic counterparts
which are the more familiar Hermitian matrix models. We derive and
solve the complete sets of loop equations for the correlators of these
models and use these equations to examine critical behaviour.  The
topological large-$N$ expansions are also constructed and their
relation to other aspects of modern string theory such as integrable
hierarchies is discussed. We use these connections to discuss the
applications of these matrix models to string theory and induced gauge
theories.  We argue that as such the fermionic matrix models may
provide a novel generalization of the discretized random surface
representation of quantum gravity in which the genus sum alternates
and the sums over genera for correlators have better convergence
properties than their Hermitian counterparts.  We discuss the use of
adjoint fermions instead of adjoint scalars to study induced gauge
theories. We also discuss two classes of dimensionally reduced models,
a fermionic vector model and a supersymmetric matrix model, and
discuss their applications to the branched polymer phase of string
theories in target space dimensions $D>1$ and also to the meander
problem.

\end{abstract}

\end{titlepage}

\clearpage\newpage

\baselineskip=14pt

\tableofcontents

\newpage

\setcounter{footnote}0

\baselineskip=14pt

\section{Introduction: From Hermitian to Fermionic Random Matrix Models}

In this Paper we shall review the theory and applications of random
matrix models whose matrices have anticommuting elements.  Typically,
these models possess a $U(N)$ symmetry and the matrices transform under
the adjoint representation of the symmetry group.  They can be
analyzed, and often solved, in the limit of large $N$ where they have
an interesting and non-trivial structure.  Uses of these models ranges
from attempts to formulate dynamical random surface theories with
alternating genus sums to the solution of combinatorial problems such
as the meander problem which we shall review, to lattice gauge
theories and the formulation of models of induced gauge theory.

Fermionic matrix models have several features which distinguish them
from the more familiar bosonic Hermitian matrix models.  They are not
eigenvalue models in the sense that matrices with anticommuting
elements cannot be diagonalized by unitary transformations.  They can
nevertheless often be analyzed by methods similar to those used for
Hermitian matrix models, such as the method of loop equations.
Furthermore, the integration over anticummuting numbers is
straightforward and is often more convergent than ordinary integrals
over commuting variables.  In perturbation theory, this improved
convergence is seen as the result of an alternating perturbation
series.

To help motivate the study of fermionic matrix models, we first
present a brief overview of the techniques which are used to study the
large $N$ limit of the standard Hermitian matrix models.

\subsection{Hermitian Matrix Models, String Theory and
Induced Gauge Theories}

For completeness, and so that we can later compare techniques for
solving Hermitian and fermionic matrix models, we begin this Section
with a brief review of some of the technical aspects of solving random
Hermitian matrix models in the large-$N$ limit.  We also discuss some
of their uses, particularly in lower dimensional quantum gravity,
string theory and lattice gauge theories. A more detailed discussion
of the material can also be found in many other reviews that will be cited as
we proceed. For a particularly lucid discussion, see \cite{fgz}.

\subsubsection{One-matrix Models}

A Hermitian one-matrix model is defined as a statistical theory of
Hermitian matrices with partition function
\beq
Z=\int d\phi~\e^{-N\tr V(\phi)}
\label{hermpart}
\eeq
where $V(\lambda)$ is a smooth potential and
\beq
d\phi\equiv\prod_{k=1}^Nd\phi_{kk}~
{}~\prod_{i<j}d~{\rm Re}~\phi_{ij}~d~{\rm Im}~\phi_{ij}
\label{hermmeas}
\eeq
is the integration measure on the space of $N\times N$ Hermitian
matrices ($d\phi_{ij}$ is always understood as an ordinary
Riemann-Lebesgue measure), i.e.  the Haar measure on the Lie algebra
${\rm Herm}(N)$.  This model is invariant under the adjoint action of
the unitary group $U(N)$,
\beq
\phi\to U\phi U^\dagger~~~~~,~~~~~U\in U(N)
\label{unitarytransf}\eeq
which restricts the observables to those which are invariant functions of
$\phi$. The free energy $\log Z$ can be expanded in a power series in the
parameter $\frac{1}{N^2}$.  The leading term is of order $N^2$ and
occurs in the infinite-$N$ limit.  Each order in this expansion can be
represented as an infinite series of ``fat-graphs" with the
topological property that all graphs of a given order can be drawn
without crossing lines of the graphs on a two-dimensional surface of
particular genus - the graphs with genus $g$ contribute the term of
order $N^{2-2g}$.  The leading term is the sum of genus zero (or
planar) graphs, i.e. those which can be drawn on a plane or the
surface of a sphere.

In this and several more elaborate Hermitian matrix models, the planar
and higher genus graphs can be summed explicitly.  These Hermitian
matrix models can be thought of as examples of a $D=0$ quantum field
theory where the topological large-$N$ expansion which was originally
proposed by 'tHooft for quantum chromodynamics (QCD) \cite{thooft} is
explicitly solvable \cite{biz,bipz}. In QCD and in Yang-Mills theory,
this expansion is intractable, except in (1 + 1) dimensions where
Yang-Mills theory and QCD with fundamental representation matter are
solvable.

The main applications of Hermitian matrix models are as
non-perturbative approaches to low-dimensional string theory where the
large-$N$ expansion of the matrix model coincides with the genus
expansion \cite{brekaz,dougshenk,grossmig} of the string partition
function.  For instance, for a polynomial potential of the form
\beq
V(\phi)=\frac{1}{2}\phi^2+\frac{\bar g}{K}\phi^K
\label{potdiscr}\eeq
the perturbative expansion of (\ref{hermpart}) in the coupling $\bar
g$ in terms of fat-graphs coincides with the formal sum over discretizations by
regular $K$-gons of 2-dimensional compact
Riemann surfaces. The fattening of lines in the Feynman graphs
represents the 2 indices that a matrix has. In this interpretation,
the Gaussian $\phi^2$ term in (\ref{potdiscr}) represents the free
boson term (with unit mass $m^2=1$) for the non-kinematical quantum field
theory
(\ref{hermpart}) which gives the propagator of the (fat) Feynman
graphs,
\beq
(2\pi)^{-N^2/2}\int d\phi~\phi_{ij}\phi_{k\ell}\e^{-N\tr\phi^2/2}=\frac{1}{N}
\delta_{i\ell}\delta_{kj}
\label{hermprop}\eeq
The $\phi^K$ interaction term produces $K$-valence vertices.

Diagrammatically, the partition function (\ref{hermpart}) is the sum
over all possible connected and disconnected Feynman diagrams
constructed by linking the propagators (\ref{hermprop}) and $K$-point
vertices together in an orientation-preserving way. This is just the
usual Wick expansion in quantum field theory. The Wick contractions
between matrices, as in (\ref{hermprop}), assign a product of 2
delta-functions, one for the inner 2 indices of a matrix pair and one
for the outer ones. The connected diagrams, divided by the appropriate
symmetry factors for topologically equivalent graphs, are obtained by
expanding the free energy $\log Z$. This gives the fat-graph expansion
\cite{fgz}
\beq
\log Z=\sum_{\cal F}\frac{(-\bar g)^vN^{v-e+l}}{|G({\cal F})|}
\label{hermfatgraph}\eeq
where the sum is over all fat-graphs $\cal F$ with $v$ vertices, $e$
propagators and $l$ index loops, and $|G({\cal F})|$ is the order of
the symmetry group of $\cal F$ (the group of permutations of the
vertices of $\cal F$).  The connected diagrams generate a
2-dimensional lattice whose dual lattice is a discretization of a
Riemann surface by regular polygons. This dual lattice of regular
polygons is constructed by associating polygon faces with $K$-point
vertices, sides with propagators, and polygon vertices with closed
loops. The number $v-e+l$ appearing in (\ref{hermfatgraph}) is then
the Euler characteristic of the Riemann surface.  A smooth Riemann
surface is well approximated by its polygonization when the number of
polygons is large and the area of each polygon is infinitesimal.
General polynomial potentials would allow a variety of polygons in the
polygonization of the Riemann surface.  If the area of each polygon is
taken as one, the statistical sum in (\ref{hermfatgraph}) is a
statistical sum over connected Riemann surfaces where each term in the
sum is dual to a Feynman graph.

Note that the sign in front of the factor of $N$ in (\ref{hermpart})
is taken to be negative to ensure the convergence of the Gaussian
statistical model ($\bar g=0$ in (\ref{potdiscr})) and hence all
Feynman graphs.  However, the sum over Riemann surfaces in (\ref{hermfatgraph})
has positive weights only when $\bar g<0$.  For these values of $\bar
g$, the integration over Hermitian matrices in (\ref{hermpart})
diverges. It turns out that one can make sense of it only when $N$ is
infinite.

The corresponding discretized surface model (\ref{hermfatgraph}) is
associated with the (Euclidean) statistical ensemble of random
surfaces with partition function
\beq
Z_{\rm str}
=\sum_{h=0}^\infty\int Dg~\e^{-\Lambda A(\Sigma^h;g)+G^{-1}\chi(h)}
\label{qugravd0}\eeq
where the action is the Einstein-Hilbert action for pure gravity with
cosmological term. Here the functional integration is over all metrics
$g_{\mu\nu}(x)$ on the 2-surface $\Sigma^h$ with $h$ `handles',
$\Lambda$ is the cosmological constant (or string tension) which
multiplies the area $A(\Sigma^h;g)=\int_{\Sigma^h}
\sqrt{g}$ of $\Sigma^h$, and
$G$ is the gravitational constant which multiplies the topological scalar
curvature term
\beq
\chi(h)=\int_{\Sigma^h}\sqrt{g}R(g)/4\pi=2-2h
\label{euler}\eeq
with $\chi(h)$ the Euler characteristic of $\Sigma^h$. This
statistical model can be regarded as a $D=0$ dimensional string
theory, i.e. a pure theory of surfaces with no coupling to additional
matter degrees of freedom on the string world-sheet or the propagation
of strings in a non-existent embedding space.

The fat-graph expansion (\ref{hermfatgraph}) of the Hermitian matrix
model (\ref{hermpart}) represents a discretized version of the
continuum quantum gravity model (\ref{qugravd0}). This latticized
model is the dynamically triangulated random surface theory
\cite{amdur,david1,kaz1,kazkost}
\beq
Z=\sum_{h=0}^\infty\e^{G^{-1}\chi(h)}\sum_{T_h}\e^{-\Lambda{\cal N}(T_h)}
\label{dyntriang}\eeq
where the second sum in (\ref{dyntriang}) is over all possible
triangulations $T_h$ at fixed genus $h$ and ${\cal N}(T_h)$ is the
dynamical variable which counts the number of triangles in $T_h$ (so
that $\Lambda$ plays the role here of a chemical potential). The dual
lattice formed by the fat graphs of the matrix model represents the
triangulation of the surface.  The propagator (\ref{hermprop})
produces a factor of $N^{2-2h}$ in the free energy
(\ref{hermfatgraph}) associated with a graph of genus $h$
\cite{thooft}.  The coupling constants of (\ref{dyntriang}) are
therefore related to those in (\ref{hermpart}),(\ref{potdiscr}) by
\cite{thooft}
\beq
N=\e^{G^{-1}}~~~~~,~~~~~\bar g=-\e^{-\Lambda}
\label{Ngrel}\eeq

It is also possible to couple matter in (\ref{dyntriang}) (represented by more
complicated potentials in (\ref{hermpart})) by including matter fields $X_i$ at
each vertex $i$ of the dual triangulation lattice $\cal F$ (of fixed
coordination number), adding appropriate interaction terms between nearest
neighbour vertices to the action in (\ref{dyntriang}) and summing over all of
the $X_i$. For instance, to couple the Ising model to 2-dimensional quantum
gravity we can replace the weight in (\ref{dyntriang}) by the Ising spin
partition function
\beq
Z_I(\beta;T_h)=\sum_{S_i=\pm1}\e^{\beta\sum_{\langle i,j\rangle\in{\cal
F}}S_iS_j}
\label{Ising}\eeq
The large-$N$ limit of the Hermitian matrix model exhibits phase
transitions which correspond to the continuum limits of the
discretized random surface theories
\cite{brekaz,dougshenk,fgz,grossmig}.

Normally, Hermitian 1-matrix models are treated as statistical
theories of the eigenvalues of the Hermitian matrices \cite{bipz} and
their partition function and correlators are computable in the
large-$N$ limit using techniques such as loop equations
\cite{ackm,david3,kaz} or orthogonal polynomials \cite{biz}.
This is accomplished by diagonalizing the Hermitian matrices in
(\ref{hermpart}).  The measure can be written as \cite{fgz}
\beq
d\phi~=~[dU]~\prod_{i=1}^Nd\lambda_i~\Delta^2(\lambda)
\eeq
where $[dU]$ is the Haar measure for integration over the unitary
transformations (\ref{unitarytransf}) which diagonalize $\phi$,
$\lambda_i$ are the eigenvalues of $\phi$ and the Vandermonde
determinant,
\beq
\Delta(\lambda)=\det_{i,j}[\lambda_i^{j-1}]=\prod_{i<j}
\left(\lambda_i-\lambda_j\right)=\e^{\sum_{i<j}\log(\lambda_i-\lambda_j)}
\label{vandermonde}
\eeq
is the Jacobian.  The partition function (up to an
irrelevant overall numerical factor) is
\beq
Z=\int[dU]~\int_{-\infty}^{+\infty}\prod_{i=1}^Nd\lambda_i~\Delta^2(\lambda)
\e^{-N\sum_{j=1}^NV(\lambda_j)}
\label{eigenvaluemodel}\eeq

The Haar measure for integration on the group $U(N)$ of unitary matrices
\beq
[dU]=\frac{\prod_{k=1}^Nk!}{(2\pi)^{N(N+1)/2}}\prod_{i,j}dU_{ij}~\delta
\left(\sum_{k=1}^NU_{ik}U_{jk}^\dagger-\delta_{ij}\right)
\eeq
is normalized ($\int[dU]=1$) and has the symmetries
\beq
\int[dU]~F(U^\dagger)=\int[dU]~F(U)~~~,~~~\int[dU]~F(VU)=\int[dU]~F(U)=\int[dU]
{}~F(UV)
\eeq
for any $V\in U(N)$.  These imply that some of its lower order moments
are
\beq
\int[dU]~U_{ij}=0~~~~~,~~~~~\int[dU]~U_{ij}U_{k\ell}^\dagger=
\delta_{i\ell}\delta_{jk}
\eeq
\beq\new{\begin{array}{c}
\int[dU]~U_{ij}U_{mn}U_{k\ell}^\dagger
U_{pq}^\dagger=\frac{1}{N^2-1}\left(\delta_{i\ell}
\delta_{jk}\delta_{mq}\delta_{np}
+\delta_{iq}\delta_{jp}\delta_{m\ell}\delta
_{nk}\right)
\\  +\frac{1}{N(N^2-1)}\left(\delta_{jk}\delta_{\ell
m}\delta_{np}\delta_{qi}
+\delta_{jp}\delta_{qm}\delta_{nk}\delta_{\ell i}\right)\end{array}}
\eeq
If the group were $SU(N)$ rather than $U(N)$, there is an aditional
identity which follows from invariance of the determinant,
\beq
\int[dU]~U_{i_1j_1}U_{i_2j_2}\cdots U_{i_Nj_N}=c\cdot\epsilon_{i_1i_2\cdots
i_N}\epsilon_{j_1j_2\cdots j_N}
\eeq
where $c$ is a constant. Similar expressions can be obtained for the
higher-order correlators of the Haar measure.

Since the integration over unitary matrices in (\ref{eigenvaluemodel})
decouples, the effective statistical theory of eigenvalues involves an
action of order $N^2$ describing $N$ degrees of freedom. This makes
the model (formally) an exactly-solvable one in the $N\to\infty$
limit. In particular, in this limit it can be evaluated by the
saddle-point approximation. The Vandermonde determinant in
(\ref{eigenvaluemodel}) acts as a hard-core repulsive term in the
action which prevents its minimum from being simply the minimum of the
potential. The stationary condition for the effective action in
(\ref{eigenvaluemodel}) leads to the saddle-point equation
\beq
V'(\lambda_i)=\frac{1}{N}\sum_{j<i}\frac{1}{\lambda_i-\lambda_j}~~~~~
,~~~~~ i=1,\dots,N
\label{saddlepteq}\eeq
and the lowest-order contribution to (\ref{hermpart}) (i.e. the number
of planar fat-graphs), which dominates for $N\to\infty$, is obtained
by substituting into (\ref{eigenvaluemodel}) the saddle-point value
determined by (\ref{saddlepteq}). The common way to study the
large-$N$ limit is to introduce the normalized spectral density of
eigenvalues
\beq
\rho(\lambda)=\frac{1}{N}\sum_{i=1}^N\delta(\lambda-\lambda_i)~~~,~~~\lambda\in
\IR
\label{rhofinN}\eeq
which for finite-$N$ is supported at the discrete points
$\lambda_1,\dots,\lambda_N$. We can then represent all summations
above as integrations over $\lambda\in\IR$ using (\ref{rhofinN}). To
treat the large-$N$ limit, we order the eigenvalues so that
$\lambda_1<\lambda_2<\dots<\lambda_N$ and introduce a non-decreasing
differentiable function $\lambda(x)$ of $x\in[0,1]$ with
$\lambda_i=N\lambda(i/N)$ for $i=1,\dots,N$. We then regard the continuous
function $\rho(\lambda)\equiv\frac{dx}{d\lambda (x)}$ as the formal
large-$N$ limit of (\ref{rhofinN}). The positive quantity
$N\rho(\lambda)d\lambda$ is the number of eigenvalues in a range
$d\lambda$ about $\lambda$, and it is supported on some finite
intervals $[a_i,b_i]\subset
\IR$. The saddle-point equation (\ref{saddlepteq}) then becomes
\beq
V'(\lambda)=2\pvint ~d\alpha~\frac{\rho(\alpha)}{\lambda-\alpha}
\label{saddlelargeN}\eeq
and the large-$N$ saddle-point free energy is
\beq
\lim_{N\to\infty}\frac{1}{N^2}\log Z=-\int
d\alpha~\rho(\alpha)V(\alpha)+\int\!\!\pvint~d\alpha~d\beta~\rho
(\alpha)\rho(\beta)\log|\alpha-\beta|
\label{saddlefreeherm}\eeq
where $\rho(\alpha)$ is the solution of (\ref{saddlelargeN}). Note
that (\ref{saddlelargeN}) now also follows from the functional
variation of (\ref{saddlefreeherm}) with respect to the spectral
density $\rho(\alpha)$.

To solve (\ref{saddlelargeN}) for the eigenvalue distribution
$\rho(\alpha)$, we introduce its Hilbert transform
\beq
\omega(z)\equiv\left\langle\frac{\tr}{N}\frac{1}{z-\phi}\right\rangle=\frac{1}
{N}\sum_{i=1}^N\frac{1}{z-\lambda_i}=\int d
\alpha~\frac{\rho(\alpha)}{z-\alpha}~~~,~~~z\in\IC
\label{omherm}\eeq
and note that the saddle-point equation (\ref{saddlelargeN}) is the
real part of the discontinuity of the function $\omega(z)$ across the
support of $\rho$, i.e.
\beq
\omega(\lambda\pm i0)=V'(\lambda)/2\mp
i\pi\rho(\lambda)~~~~~,~~~~~\lambda\in~{\rm supp}~\rho
\eeq
In (\ref{omherm}), $\langle\cdot\rangle$ denotes the normalized average with
respect to the matrix ensemble (\ref{hermpart}),
\beq
\langle Q(\phi)\rangle\equiv\frac{1}{Z}\int d\phi~\e^{-N\tr V(\phi)}Q(\phi)
\label{normavg}\eeq
and the function (\ref{omherm}) is the generating function, via its
$\frac{1}{z}$-expansion, for the correlators $\langle\frac
{\tr}{N}\phi^k\rangle$ of the Hermitian matrix model. For $s\in\IR^+$,
the quantity $\frac{\tr}{N}\phi^s$ can be thought of as an operator
which creates a loop of length $s$ on the surface \cite{kaz}, and the
potential $V(z)$ is a source for the inverse Laplace transform of
(\ref{omherm}),
\beq
\frac{\tr}{N}V(\phi)=\int_{0-i\infty}^{0+i\infty}\frac{dz}{2\pi
i}~V(z)\omega(z)
\label{laplaceom}\eeq
This method of determining the solutions to the saddle-point equation
by solving an equation via the analytic properties of a function such
as (\ref{omherm}) is known as the Riemann-Hilbert method.
Alternatively, one can derive explicitly an equation for $\omega(z)$
by multiplying (\ref{saddlepteq}) by $\frac{1}{N}\frac{1}{z-
\lambda_i}$ and summing over $i=1,\dots,N$. This leads to
\beq
\omega(z)^2+\omega'(z)/N-V'(z)\omega(z)=\int d\alpha~\rho(\alpha)\frac{V'(z)
-V'(\alpha)}{\alpha-z}
\label{hermloopeq}\eeq
where the right-hand side of (\ref{hermloopeq}) is a polynomial of degree
$K-2$ when the potential $V(\lambda)$ is a polynomial of degree $K$.

The saddle-point technique relies heavily on the facts that (a) the
matrix model has a finite large-$N$ limit, or equivalently a finite
contribution from planar (genus 0) diagrams to the free energy, and
(b) the matrices $\phi$ are diagonalizable. Another way to arrive at
the equation (\ref{hermloopeq}) is to exploit the invariance of the
partition function (\ref{hermpart}) under arbitrary changes $\phi\to
f(\phi)$ of the Hermitian matrix variables. The resulting equations
represent the full set of Schwinger-Dyson equations of motion for the
non-kinematical field theory (\ref{hermpart}). They are obtained by
performing a shift of the matrix variables by an infinitesimal loop of
length $k\in\IZ^+$, i.e.  the change of variables
$\phi\to\phi+\varepsilon\phi^k$ in the matrix integral
(\ref{hermpart}), where $\varepsilon$ is an infinitesimal real number.
This leads to the identity
\beq
\sum_{\ell=0}^{k-1}\left\langle\frac{\tr}{N}\phi^\ell\frac{\tr}{N}\phi^{k-
\ell-1}\right\rangle=\left\langle\frac{\tr}{N}\phi^kV'(\phi)\right\rangle
\label{loopfin}\eeq
Multiplying (\ref{loopfin}) by $1/z^{k+1}$ and summing over all
$k\in\IZ^+$ then leads to the loop equation (\ref{hermloopeq}), which
represents the full set of equations of motion of the model because
(\ref{omherm}) contains a complete set of operators. In fact, this
equation could also have been obtained from the shift
\beq
\phi\to\phi+\frac{\varepsilon}{z-\phi}
\label{hermshift}\eeq
under which the invariance of (\ref{hermpart}) leads to the matrix
identity
\beq
0=\int d\phi~\frac{\partial}{\partial\phi_{ij}}\left\{\left(\frac{1}{z-\phi}
\right)_{k\ell}\e^{-N\tr V(\phi)}\right\}
\label{matrixid}\eeq
After expanding the terms in (\ref{matrixid}) into averages and
summing over all $i=k$ and $j=\ell$, we arrive at the loop equation
(\ref{hermloopeq}). However, if the potential $V$ in (\ref{matrixid})
is not bounded from below, then the integration by parts on the
right-hand side of (\ref{matrixid}) gives infinite contributions from
boundary terms at infinity and is not formally zero. This derivation
of the loop equations is appealing because it does not explicitly
involve the eigenvalue distribution of the model.

As an example, consider the simplest case of the Gaussian model where
$V(\phi)=m^2\phi^2/2$. The loop equation (\ref{hermloopeq}) at $N=\infty$ then
leads to a quadratic equation for $\omega(z)$ whose solution is
\beq
\omega(z)=\frac{1}{2}\left(m^2z-\sqrt{m^4z^2-4m^2}\right)
\label{wignerherm}\eeq
where the branch of the square root is chosen so that
$\omega(z)\sim1/z$ at $|z|\to\infty$ which follows from the definition
(\ref{omherm}) and the normalization condition $\int
d\lambda~\rho(\lambda)=1$. Computing the discontinuities of
(\ref{wignerherm}) across its branch cuts leads to the well-known
Wigner semi-circle distribution for the eigenvalues of $\phi$,
\beq
\rho(\lambda)=\frac{m}{\pi}\sqrt{1-\frac{m^2\lambda^2}{4}}
\eeq
which has support on the interval $(-\frac{2}{m},\frac{2}{m})$.

Generally, the Wilson loop average $\langle\frac{\tr}{N}\e^{s\phi}\rangle$
yields a weighted superposition of loops of different length on the surface and
it has the same behaviour in the continuum as the correlator
 $\langle\frac{\tr}{N}\phi^s\rangle$.
The resolvent function $\omega(z)$ is the inverse Laplace transform of the
Wilson loop (see (\ref{omherm}))
\beq
W(s)\equiv\left\langle\frac{\tr}{N}\e^{s\phi}\right\rangle=-
\oint_{\cal C}\frac{dz}{2\pi i}~\e^{sz}\omega(z)
\label{loopops}\eeq
where the contour $\cal C$ encircles the singularities of $\omega(z)$ with
 counterclockwise orientation. The loop equation (\ref{hermloopeq}) at
$N=\infty$ can then also be written as \cite{kaz}
\beq
V'(\partial_s)W(s)=\frac{1}{N}\int_0^sdt~W(t)W(s-t)
\label{loopglue}\eeq
where the right-hand side of (\ref{loopglue}) can be interpreted as the
operation of gluing 2 boundary loops together \cite{kaz}. (\ref{loopglue})
is very similar to a simplified (zero-dimensional) version of the
Makeenko-Migdal equations for the Wilson loop functionals $W[C]$ in
multi-colour QCD \cite{mig1}.

In the large-$N$ limit the branch cut singularities of the solution $\omega(z)$
to the quadratic equation (\ref{hermloopeq}) determine the spectral density
$\rho(\lambda)$. There will be regions in general in coupling constant space
where the analytic features of the continuous function $\rho$ change (e.g.
it becomes negative and acquires multiple branch cuts) \cite{bipz,fgz}. For
a discretization potential (\ref{potdiscr}), the area, or number of $K$-point
vertices $v$, appears in the free energy $F$ as the power $\bar
g^v$, and so the average area of each diagram is $\langle
v\rangle\sim\partial_{\bar g}\log F$. These singular
points of the free energy therefore describe critical behaviours in the
random matrix models at which the areas of the individual discretization
polygons become infinite \cite{kopnevnus,fgz}. As $\bar g$ then approaches
its critical value $\bar g_c$, the areas of the individual polygons may be
rescaled to zero giving a continuum surface with finite area. This is precisely
what is needed to define continuum surfaces, in that one tunes the coupling
$\bar g$ to the point where the perturbation series for (\ref{hermpart})
diverges and the integral becomes dominated by 'tHooft diagrams with infinite
numbers of vertices.  A phase transition for $\Lambda\to\Lambda_c$ at $D=0$ is
indeed possible -- it is a third order Gross-Witten type phase transition
\cite{gw} and it occurs for $N\to\infty$ when the number of degrees of freedom
becomes infinite. The continuum limit is therefore reached as $N\to\infty$ and
$\Lambda\to\Lambda_c$ where the discrete partition function (\ref{dyntriang})
approaches the physical continuum one (\ref{qugravd0}).
The singularities in the coupling constants in this case occur at points where
some of the zeroes of the singular parts of $\omega(z)$ (i.e. the zeroes of
$\rho(\alpha)$) coalesce with an endpoint of its branch cuts \cite{fgz}.
These divergences arise because of the entropy factor (number of graphs) in
(\ref{dyntriang}), whereas for $\Lambda$ large enough ($\bar g$ small) the
sum over triangulations $T_h$ in (\ref{dyntriang}) converges.

One prediction of the continuum Liouville field theory approach to
non-critical string theory and quantum gravity
\cite{pol1,kniz,david1,diska} (see \cite{fgz,mir} for reviews) is
the occurence of a critical string exponent $\gamma_{\rm str}$, which is
defined in terms of the area dependence of the free energy for
surfaces of fixed large area $A$ as
\beq
\log Z_{\rm str}(A)\sim A^{(\gamma_{\rm str}-2)\chi(h)/2-1}
\label{largeAdep}\eeq
If we couple 2-dimensional gravity to conformal field theory, in
particular the unitary discrete series which are labelled by an integer
$m\geq2$ (see \cite{ginsparg} for a review) and the central charge
\beq
D=1-\frac{6}{m(m+1)}~~~~~,
\label{centralcharge}\eeq
then the continuum Liouville theory prediction for the critical exponent
$\gamma_{\rm str}$ is
\beq
\gamma_{\rm str}=\frac{1}{12}\left(D-1-\sqrt{(D-1)(D-25)}\right)=-\frac{1}{m}
\label{stringexpherm}\eeq
The constant $\gamma_{\rm str}$ is universal and it represents the nature of
the geometry. If we think of the gravity-coupled conformal field theory as a
string theory where the conformal matter fields are identified as the string
embedding functions and the 2-dimensional space-time is thought of as the
string world-sheet \cite{pol1}, then the central charge (\ref{centralcharge})
represents the (fractal) dimension of the embedding space.
The case $m=2$ corresponds to $D=0$, i.e. $\gamma_{\rm str}=-\frac{1}{2}$
for pure gravity. The case $m=3$ corresponds to $D=\frac{1}{2}$, i.e.
$\gamma_{\rm str}=-\frac{1}{3}$ for a half-boson or fermion, or equivalently
for the conformal field theory of the 2-dimensional critical Ising model
coupled to gravity (for which the continuum limit is also associated with the
magnetization transition of the ordinary Ising model). As (\ref{stringexpherm})
ceases to make sense for $D>1$, there is a
``conformal barrier" in the model at $D=1$ (which corresponds to the
conformal field theory of a single free boson).

These features of string theory are all reproduced by the Hermitian one-matrix
model with polynomial potentials which therefore serve as tools for extracting
non-perturbative information about string theories, i.e. the one-matrix models
can be explicitly solved non-perturbatively at each order of the
$\frac{1}{N}$-expansion. For instance, for $K=3$
in (\ref{potdiscr}), we can consider the ``double-scaling limit"
\cite{brekaz,dougshenk,grossmig,kosmet} where we take the limits
$N\to\infty$ and $\bar g\to \bar g_c$ in a correlated fashion so that the
renormalized string coupling $N(\bar g-\bar g_c)^{(2-\gamma_{\rm str})/2}$
remains finite. Notice that the area dependence in (\ref{largeAdep}) is
realized in terms of the coupling $\bar g$ in the matrix model partition
function (c.f. (\ref{Ngrel})), except that in the latter case the overall
exponent is down by 1 power (c.f. (\ref{loopops})) because of the Laplace
transformation from the area dependence. The limit $N=\infty$ by itself
corresponds to planar diagrams
or genus 0 (the spherical approximation), while the higher-genus terms
($h\geq1,\chi(h)\leq0$) in the $\frac{1}{N}$-expansion are suppressed by
$\frac{1}{N^{2h}}$ for $N\to\infty$ but are enhanced as $\bar g\to\bar g_c$.
The double scaling limit of Hermitian 1-matrix
models therefore results in a coherent contribution from all genus surfaces
and allows the construction of the genus expansion of 2-dimensional quantum
gravity \cite{brekaz,dougshenk,grossmig,fgz}. In terms of the renormalized
cosmological constant
\beq
x\equiv\Lambda_R=(1-\bar g/\bar g_c)N^{4/5}
\label{scalingz}\eeq
it can be shown that the leading singular part, as $\bar g\to\bar g_c$, of the
specific heat (or string susceptibility)
\beq
\chi(x)\equiv-\frac{\partial^2\log Z}{\partial x^2}
\label{hermsus}\eeq
obeys the transcendental Painlev\'e I equation
\cite{brekaz,dougshenk,grossmig,fgz}
\beq
x=\chi^2(x)-\frac{1}{3}\chi''(x)
\label{painleveeq}\eeq
which is also known as the string equation \cite{banks,grossmig2}.

The characteristic property of this second order differential equation is
that its only removeable singularities in the complex plane are double poles
which have residue 2 and correspond to double zeroes of the free energy
$\log Z$. Its solutions are determined by 2 boundary conditions. The
 solutions which
have an asymptotic expansion for $x$ large (the topological expansion) that
begins with the leading spherical result $\chi(x)\sim\sqrt{x}$ are determined
as
\beq
\chi(x)=\sqrt{x}\left(1-\sum_{k=1}^\infty\chi_k~x^{-5k/2}\right)
\label{painsol}\eeq
where $\chi_k>0$ for large-$k$ grow asymptotically as $(2k)!$. The
solution for $\chi(x)$ (or the topological expansion of the free energy)
is therefore not Borel summable and thus does not define a unique function
 \cite{fgz}.
Thus the Hermitian matrix models reproduce the well-known fact that quantum
gravity in 2-dimensions is ill-defined as a statistical theory of random
surfaces because the topological genus expansion is not Borel summable (and
therefore has terrible convergence properties). This is reflected directly in
the Hermitian matrix integrals (\ref{hermpart}) in which the divergence of the
large-$N$ expansion is a consequence of the fact that the integration over
Hermitian matrices diverges in the region of interest. The coefficients of the
genus expansion (\ref{painsol}) can be explicitly determined by the Bessis
method of orthogonal polynomials for the eigenvalue model
(\ref{eigenvaluemodel}) \cite{biz} or alternatively by examining the
$\frac{1}{N}$-expansions of the loop equations \cite{ackm}. In this case, the
string exponent $\gamma_{\rm str}=-\frac{1}{2}$ appears for the discretization
potentials (\ref{potdiscr}), while the values $\gamma_{\rm str}=-\frac{1}{m}$
are associated with more complicated polynomial potentials for which similar
Painlev\'e expansions can be constructed \cite{fgz}.

\subsubsection{Multi-matrix Models}

Although the simplest Hermitian one-matrix models above describe pure
2-dimensional quantum gravity, the simplest $\IZ_2$-symmetric multi-matrix
models describe gravity interacting with matter in a $D\leq1$ dimensional
embedding space \cite{alb,bredougkaz,crn,daul,doug1,fgz,ginzinn,grossmig1}. The
typical multi-matrix models are defined by the partition functions
\beq
Z_n=\int\prod_{\ell=1}^nd\phi_\ell~\e^{-N\tr S[\phi]}
\label{hermmulti}\eeq
where $\phi_\ell$ are $N\times N$ Hermitian matrices and the action is
\beq
S[\phi]=\sum_{\ell=1}^nV_\ell(\phi_\ell)-\sum_{\ell=1}^{n-1}\phi_\ell\phi
_{\ell+1}
\label{actionmulti}\eeq
The generalization of the one-matrix eigenvalue distribution follows from the
Itzykson-Zuber formula \cite{iz}
\beq
I[\phi_m,\phi_\ell]\equiv\int[dU]~\e^{\phi_mU\phi_\ell
U^\dagger}=\frac{\det_{i,j}\left[\e^{\lambda_i^{(m)}
\lambda_j^{(\ell)}}\right]}{\Delta(\lambda^{(m)})\Delta(\lambda^{(\ell)})}
\label{itzykherm}\eeq
where $\lambda_i^{(\ell)}$ are the eigenvalues of the Hermitian matrix
$\phi_\ell$. This formula enables one to write (\ref{hermmulti}) as
the multi-eigenvalue
model
\beq
Z_n=\int_{-\infty}^{+\infty}
\prod_{\ell=1}^n\prod_{i=1}^Nd\lambda_i^{(\ell)}~\Delta(\lambda^{(1)})
\Delta(\lambda^{(n)})\e^{-N\sum_{j=1}^NS[\lambda_j]}
\label{2mateigen}\eeq

The diagrammatic expansion of these multi-matrix models generates a sum over
discretized surfaces, where the different matrices $\phi_\ell$ represent
$n$ different matter states that can exist at the vertices. The cross-terms
between the $\phi_\ell$'s in (\ref{actionmulti}) links the graphs generated by
each $\phi_\ell$ together. The quantity
(\ref{hermmulti}) thereby admits an interpretation as the partition function
of 2-dimensional gravity coupled to matter, or of string theory in a $D\leq1$
dimensional embedding space. In particular, by taking $n\to\infty$ (a matrix
chain) one can represent a $D=1$ model (i.e. a single free boson) coupled to
gravity. When $V_\ell=V_{n+1-\ell}$, the matrix problem has a $\IZ_2$-symmetry
corresponding to the mapping $\phi_\ell\to\phi_{n+1-\ell}$ of Hermitian
matrices, which manifests itself as an Ising-like reflection symmetry in
the discretized random surface model. In these cases the statistical theory
can be solved for using techniques such as loop
 equations \cite{alf,gava,marsh,staud,mak2}, orthogonal
polynomials or canonical commutation relations \cite{daul,fgz}. The
continuum limits of the
discretized random surface models in the case of the $\IZ_2$-symmetric
2-matrix model ($n=2$ in (\ref{hermmulti})) represent the $(p,q)$-minimal
models of conformal field theory \cite{daul,fgz,ginsparg,mir}, with
the relatively prime integers $p$ and $q$ associated with
the scaling behaviours generated by the matrices $\phi_1$ and $\phi_2$. The
central charge of these models is
\beq
D=1-\frac{6(p-q)^2}{pq}
\eeq
and the associated string susceptibility exponent is
\beq
\gamma_{\rm str}=-\frac{2|p-q|}{p+q-|p-q|}
\eeq
The unitary discrete series which is generated by the Hermitian one-matrix
 models is then recovered for $(p,q)=(m+1,m)$, and the $\IZ_2$ symmetry of
 the two-matrix model corresponds to the well-known $p$-$q$ duality of
conformal field theory. The large-$N$ phase transitions in these cases can
 also be studied using saddle
point methods analogous to those in Hermitian one-matrix models \cite{fgz}.

\subsubsection{The Kazakov-Migdal Model}

Unitary matrix models play a role in 2-dimensional QCD \cite{bg,gw}, mean-field
computations in lattice gauge theory \cite{dz} and various other approaches
to higher-dimensional continuum gauge theories \cite{mms}. Recently, a
unitary matrix model for induced QCD has been proposed by Kazakov and
 Migdal \cite{km}. Their model is the bosonic lattice field theory
\beq\new{\begin{array}{ll}
Z_{KM}=&\int\prod_{x\in\lat}d\phi(x)~\prod_{\langle x,y\rangle\in\lat}[dU(x,y)]
\\&\times\exp\left\{-N\tr\left(\sum_{x\in\lat}V(\phi(x))-\sum_{\langle x,y
\rangle\in\lat}\phi(x)U(x,y)\phi(y)U^\dagger(x,y)\right)\right\}\end{array}}
\label{kazmigmodel}\eeq
where $\lat$ is a $D$-dimensional oriented hypercubic lattice, $\phi(x)$ is a
scalar
field which, for each site $x\in\lat$, is an $N\times N$ Hermitian matrix and
$U(x,y)$ is a gauge field which, for each link $\langle x,y\rangle\in\lat$
connecting nearest neighbour sites $x$ and $y$, is an $N\times N$ unitary
matrix. This model is invariant under the gauge transformation
\beq
\phi(x)\to\Xi(x)\phi(x)\Xi^\dagger(x)~~~,~~~U(x,y)\to\Xi(x)U(x,y)
\Xi^\dagger(y)
\label{scalargaugetransf}\eeq
where $\Xi(x)$ is an arbitrary $U(N)$-valued function of $x\in\lat$. The
second term in the action in (\ref{kazmigmodel}) is the usual gauge invariant
kinetic term for a scalar field in the adjoint representation of the colour
gauge group \cite{dz,wilson}. The absence of the usual Wilson kinetic term for
 the gauge field
makes this model exactly solvable in the large-$N$ limit. The model
(\ref{kazmigmodel}) is the natural $D>1$ dimensional extension of the
Hermitian one-matrix model (\ref{hermpart}). It reduces to the standard
$D\leq1$ matrix chains discussed above if the lattice $\lat$ is just a
1-dimensional sequence of points in which case the gauge field can be
absorbed by a unitary transformation (\ref{scalargaugetransf}) of $\phi(x)$.

For the Gaussian potential $V(\phi)=m^2\phi^2/2$, it is straightforward
to calculate the Gaussian integrals over the Hermitian matrix fields to
 rewrite (\ref{kazmigmodel}) as the lattice gauge theory
\beq
Z_{KM}=\int\prod_{\langle x,y\rangle\in\lat}[dU(x,y)]\e^{-S_{\rm ind}[U(x,y)]}
\eeq
where the induced action is given by the large mass expansion \cite{km}
\beq
S_{\rm ind}[U]=-\frac{1}{2}\sum_{\Gamma\in\lat}\frac{|\tr U(\Gamma)|^2}{l(
\Gamma)m^{2l(\Gamma)}}
\label{indscalar}\eeq
The sum in (\ref{indscalar}) is over all closed loops $\Gamma$ with $l(\Gamma)$
links and $U(\Gamma)$ denotes the path-ordered product of the gauge fields
along $\Gamma$ with counterclockwise orientation. Alternatively, we can treat
(\ref{kazmigmodel}) using standard matrix model techniques.
As usual, because of the gauge invariance (\ref{scalargaugetransf}) the
integral (\ref{kazmigmodel}) depends only on the eigenvalues $\phi_i(x)$ of the
Hermitian matrices $\phi(x)$. Using the Itzykson-Zuber formula
(\ref{itzykherm}) it can be written as the eigenvalue model
\bd
Z_{KM}=\int_{-\infty}^{+\infty}\prod_{x\in\lat}\prod_{i=1}^Nd\phi_i(x)~\prod
_{x\in\lat}\Delta^2(\phi(x))\exp\left(
-N\sum_{x\in\lat}\sum_{i=1}^NV(\phi_i(x))\right)
\ed
\beq
\times\prod_{\langle x,y\rangle\in\lat}\frac{\det_{i,j}\left[\e^{N\phi_i(x)
\phi_j(y)}\right]}{\Delta(\phi(x))\Delta(\phi(y))}
\label{kmeigenmodel}\eeq
When $N$ is large the integral (\ref{kmeigenmodel}) is dominated by the
saddle-point of the effective eigenvalue action. The saddle-point equation is
\beq
\frac{2D}{N}\frac{\partial}{\partial\phi_i}\log I[\phi,\chi]\biggm|_{\chi=\phi}
=V'(\phi_i)-\frac{1}{N}\sum_{j<i}\frac{1}{\phi_i-\phi_j}~~~,~~~i=1,\dots,N
\label{kmsaddle}\eeq
for each site $x\in\lat$.
The solutions $\Phi_i(x)$ to (\ref{kmsaddle}) are called the master fields
of the theory and the value of the integral (\ref{kmeigenmodel})
when $N\to\infty$ is equal to the
integrand evaluated when $\phi_i(x)$ is set equal to the master field
$\Phi_i(x)$. The mean-field approximation to the partition function consists
of assuming that the master field $\Phi_i(x)$ is frozen at some constant
value $\Phi_0$ at each site $x$ of the lattice $\lat$.

The saddle-point equation (\ref{kmsaddle}) can be written in
terms of the one-link pair correlator for the gauge fields,
\beq
\frac{1}{N}C_{ij}\equiv\frac{\int[dU]~|U_{ij}|^2\e^{N\tr\phi(x)U\phi(y)U^
\dagger}}{\int[dU]~\e^{N\tr\phi(x)U\phi(y)U^\dagger}}
\label{scalarpaircor}\eeq
as
\beq
\frac{1}{N}\sum_{k=1}^NC_{ik}\phi_k=\frac{1}{2D}\left(V'(\phi_i)-\frac{1}{N}
\sum_{j<i}\frac{1}{\phi_i-\phi_j}\right)
\label{saddlecor}\eeq
In the large-$N$ limit, as usual we replace the index of the eigenvalues of
$\phi$ with a continuous label and introduce the eigenvalue density $\rho(
\lambda)$ for the master field $\lambda=\Phi$. Then the saddle-point
equation (\ref{saddlecor}) is
\beq
\int d\alpha~\rho(\alpha)C(\lambda,\alpha)\alpha=\frac{1}{2D}\left(V'(\lambda)
-2\pvint~d\alpha~\frac{\rho(\alpha)}{\lambda-\alpha}\right)
\label{kmsaddlecont}\eeq
The singular non-linear integral equation (\ref{kmsaddlecont}) is rather
complicated \cite{mig4} and its explicit solution even for the simplest
Gaussian potential $V(\phi)=m^2\phi^2/2$ is non-trivial \cite{gross}. In this
latter case, the quantities appearing in (\ref{kmsaddlecont}) can be solved for
explicitly to yield the usual semi-circle eigenvalue distribution \cite{dms}
\beq\new{\begin{array}{c}
F(\lambda)\equiv\int
d\alpha~\rho(\alpha)C(\lambda,\alpha)\alpha=\Pi\lambda=\frac{2\lambda}{\mu+\sqrt{\mu^2+4}}\\C(\alpha,\beta)=\frac{\Pi^{-1}}{\alpha^2-(\Pi+\Pi^{-1})\alpha\beta
+\beta^2+\mu}~~~,~~~\rho(\alpha)=\frac{1}{\pi}\sqrt{\mu-\frac{\mu^2\alpha^2}{4}}
\end{array}}
\eeq
where
\beq
m^2=D\sqrt{\mu^2+4}-(D-1)\mu~~~~~,~~~~~\mu=\Pi^{-1}-\Pi
\eeq
 An alternative way to solve for the
eigenvalue distribution, which is equivalent to this large-$N$ Riemann-Hilbert
 method, is again by the method of loop equations, which
in this case provides as well the full set of one-link correlators of the gauge
and scalar fields \cite{dms,mak1} (see \cite{mak2} for a review).

It was originally hoped that the Kazakov-Migdal model would be a description
of QCD in
the limit where the number $N$ of colours is large \cite{km}. It was thought
 that the
model would have a second order order phase transition and that the critical
behaviour should be represented by continuum QCD \cite{km,ksw,mig3}, the only
 known non-trivial
4-dimensional quantum field theory with non-abelian gauge symmetry. It was
shown, however, that even in the simplest Gaussian model, for which the
phase transition was thought to occur at $m^2=2D$, there is no such critical
behaviour \cite{gross,kmak1}. This problem, combined with other problems such
 as the ``hidden"
local $\IZ_N$-symmetry which forces the Wilson loops to vanish \cite{kmsw},
 has led to the
concensus that the Kazakov-Migdal model does {\it not} induce QCD.
Nevertheless, the model (\ref{kazmigmodel}) is interesting in its own right
both as a higher-dimensional gauge theory which is exactly solvable at
large-$N$ and as an interesting example of a matrix model in dimensions
greater than 1 for which a solution in the large-$N$ limit may be attainable.
In the former interpretation the model can be studied using the standard
techniques of lattice gauge theory (see \cite{dz} for a review), while in the
 latter case one can employ the
basic methods of matrix models discussed above.

Furthermore, it has been recently suggested that,
as a matrix model, the partition function (\ref{kazmigmodel}) may serve as
some sort of statistical random surface model for strings in dimension $D>1$
\cite{kmsw,mak5,makmean}. For instance, some remarkable self-consistent scaling
solutions with non-trivial scaling indices have been found for a quartic
potential $V$ in (\ref{kazmigmodel}) at large-$N$ and for any $D$
\cite{mig3,mig4}. However, it is not clear what physical system these scaling
solutions are associated with. The relation of the Kazakov-Migdal model to
discretized random surfaces and strings has also been recently suggested in its
equivalence at large-$N$ with the gauged Potts matrix model \cite{mak5}.
 The latter model has a natural connection with triangulated random surface
 theories. The mechanism which governs the discretized random surface approach
to string theory in dimensions $D>1$, where the susceptibility formula
(\ref{stringexpherm}) breaks down, has been a subject of much discussion over
the years. The random surface models for $D\geq1$ are perfectly well-defined
and there is no obvious pathology at $D=1$. It has been suggested \cite{das}
that the conformal barrier is due to some change in the geometry at $D=1$.
There is some evidence that this change is a transition to a tree-like or
branched polymer phase (rather than a stringy phase)
\cite{amb1,amb2,amdur2,amdurjon,ambthor}. The polymer trees are thought of as
connecting 2-dimensional baby universes together \cite{makmean} so that the
string constant is modified to \cite{amdurjon,durhuus}
\beq
\gamma=\frac{\gamma_{\rm str}}{\gamma_{\rm str}-1}
\label{babystr}\eeq
with $\gamma_{\rm str}=-1/m<0$ the critical exponent of each 2-dimensional
surface in the $D$-dimensional embedding space.

In this picture, the case $\gamma_{\rm str}=-1$, when there is no critical
behaviour at all on the 2-dimensional surfaces, leads to an exponent
$\gamma=1/2$, which is the typical mean field value for pure branched polymers.
The exponents $\gamma_{\rm str}<0$ are generically associated with a
2-dimensional, or stringy, random surface summation, while $\gamma>0$ signifies
a fragmentation of the surface or a ``crumpling" of the strings
\cite{amdur,amdur1,amdur3}. The multicritical string exponents $\gamma_{\rm
str}=-1/m$ lead to $\gamma=1/(m+1)$ due to the polymerization. This differs
from the mean field value $\gamma=1/2$ due to the effects of the 2-dimensional
gravitational dressing, just like the critical indices of the Ising model on a
random surface differ from the usual mean field exponents of the Ising model on
a fixed regular lattice. The polymer sum is the 1-dimensional reduction of the
dynamically triangulated random surface model (\ref{dyntriang}),
\beq
Z_{\rm poly}=\sum_{h=0}^\infty\kappa^{2h-2}\sum_{{\cal
P}_h}\e^{-\Lambda L({\cal P}_h)}\int\prod_{i\in{\cal
P}_h}d^DX_i~\e^{-\frac{1}{2}\sum_{\langle i,j\rangle\in{\cal P}_h}|X_i-X_j|^2}
\label{polymersum}\eeq
where the sum is over all polymers ${\cal P}_h$ with $h$ loops, $L({\cal P}_h)$
is the intrinisic length of the polymer chains, the constant $\kappa$ is
related to the fugacity of the branching chains, and $X_i$ are matter degrees
of freedom placed at the vertices $i$ of the polymer graphs. These models can
be
described by vector field theories \cite{vec1,vec2,nish,nishyon,zinn1}, where
the reduction from matrix to vector degrees of freedom is equivalent to the
reduction above from random surfaces to randomly branched chains. The
Kazakov-Migdal model yields an explicit realization of the crumpled surfaces
above \cite{mak5,makmean}, and the above picture of the branched polymer phase
is expected for any matrix model describing discretized random surfaces in
target space dimensions $D>1$.

\subsection{Adjoint Fermion Matrix Models}

Although the scalar matrix models discussed above reproduce some nice
features of non-critical string theory and 2-dimensional quantum gravity, and
they provide non-perturbative approaches to these theories which serve as
useful
 tools for string theory (beyond the realm of the usual perturbative
 approaches), they have several undesirable characteristics. For
instance, as mentioned above, there is the problem that the observables
 of the Hermitian matrix models are not well-defined because the
integrations over Hermitian matrices diverge in the
region of interest for a description of discretized quantum gravity (reflecting
the divergence of the genus sum in the random surface model). Furthermore, in
Hermitian matrix models the stringy phase does not exist in dimension $D>1$
because of the conformal barrier, and even their $D>1$ generalizations
have instability problems associated with the unbounded scalar actions
\cite{mak2}. It is therefore desirable to look for alternative matrix models
which describe more general types of string theories and which have
well-defined convergent observables. In particular, it would be interesting
to find some alternative candidate to the Kazakov-Migdal model which has
the same solvability features and might really induce QCD in the continuum
 limit.

One such possibility has been recently introduced by Makeenko and Zarembo
 \cite{mz} who
considered a class of matrix models where the degrees of freedom are
matrices whose elements are anticommuting Grassmann numbers. The simplest
example is the fermionic one-matrix model \cite{akm,mz,mss,semsz} which is
defined by the partition function
\beq
Z_1=\int d\psi~d\bar\psi~\e^{N^2\tr V(\bar\psi\psi)}
\label{part1}
\eeq
where $V$ is some potential and $\psi$ and $\bar\psi$ are independent
complex Grassmann-valued $N\times N$ matrices, i.e. matrices with
anti-commuting
nilpotent elements,
\beq\new{\begin{array}{c}
\psi_{ij}\psi_{kl}=-\psi_{kl}\psi_{ij}~~~,~~~\psi_{ij}\bar\psi_{kl}=-\bar\psi_
{kl}\psi_{ij}~~~,~~~\bar\psi_{ij}\bar\psi_{kl}=-\bar\psi_{kl}\bar\psi_{ij}\\
(\psi_{ij})^2=(\bar\psi_{ij})^2=0\end{array}}
\eeq
The rules for left differentiation are given by the anticommutators
\beq
\left\{\frac{\partial}{\partial\psi_{ij}},\psi_{k\ell}\right\}=
\left\{\frac{\partial}{\partial\bar\psi_{ij}},\bar\psi_{k\ell}\right\}=
\delta_{ik}\delta_{j\ell}~~~,~~~\left\{\frac{\partial}{\partial\psi_{ij}}
,\bar\psi_{k\ell}\right\}=\left\{\frac{\partial}{\partial\bar\psi_{ij}},
\psi_{k\ell}\right\}=0
\eeq
and the usual rules for complex conjugation of these Grassmann numbers are
\beq
\bar\psi^*_{ij}=-\psi_{ji}~~~,~~~\psi_{ij}^*=-\bar\psi_{ji}~~~,~~~
(\bar\psi_{ij}\psi_{k\ell})^*=\psi_{k\ell}^*\bar\psi_{ij}^*
\eeq
The integration measure in (\ref{part1}) (the Haar measure on the Grassmann
 algebra ${\rm Grass}(N)$),
\beq
d\psi~d\bar{\psi}\equiv\prod_{i,j}d\psi_{ij}~d\bar{\psi}_{ij}
\nonumber
\eeq
is defined using the usual Berezin rules for integrating Grassmann variables,
\beq
\int d\psi_{ij}~\psi_{ij}=1~~~~~,~~~~~\int d\psi_{ij}~1=0
\label{grassint}\eeq
which is equivalent to left differentiation. We normalize all traces here and
in the following as
\beq\tr A\equiv\frac{1}{N}\sum_{i=1}^NA_{ii}
\eeq

Matrix models of this kind were originally motivated by the studies of induced
gauge theories using adjoint matter where the Yang-Mills interactions of gluons
are induced by loops with heavy adjoint scalar fields \cite{km,mak2} as in the
Kazakov-Migdal model or other kinds of matter such as heavy adjoint or
fundamental representation fermions \cite{kmak2,mak2,mz,mig2,mig5,semsz}. In
these latter types of adjoint theories one can think of the quark fields of the
model as being represented by the fermion matrices in the adjoint
representation of the gauge group $U(N)$. This differs from previous approaches
to lattice gauge theories which used fermion fields that transform in the
fundamental representation of the gauge group \cite{ban,ham}. In
particular, since there is no asymptotic freedom in the gauge coupling
constant in the adjoint fermion model (if the number of fermion species is
large enough), the kinetic term for the gauge field
is not essential in this case (just like in quantum electrodynamics) \cite{mz}.

To see the connection between the fermionic matrix model (\ref{part1}) and a
 statistical theory of discretized random surfaces as is described by
the Hermitian matrix models, consider the simple Grassmann matrix field
 theory defined by
\beq
Z_1^{(S)}=\int d\psi~d\bar\psi~\e^{N^2\tr(\ps2+g(\ps2)^K/K)}
\label{part1S}\eeq
The perturbative expansion of (\ref{part1S}) is the series
\beq
Z_1^{(S)}=\sum_{k\geq0}\frac{(N^2g)^k}{K^kk!}
\int d\psi~d\bar\psi~\left[\tr(\ps2)^K\right]^k\e^{N^2\tr\ps2}
\label{part1Spert}\eeq
The perturbative solution of (\ref{part1S}) therefore requires the evaluation
of the normalized fermionic Gaussian moments
\beq
\left\langle\!\!\left\langle
\left[\tr(\ps2)^K\right]^k\right\rangle\!\!\right\rangle\equiv\frac{
\int d\psi~d\bar\psi~[\tr(\ps2)^K]^k\e^{N^2\tr\ps2}}{\int
 d\psi~d\bar\psi~\e^{N^2\tr\ps2}}
\label{gaussfermmoms}\eeq
These moments can be obtained from the generating functional
\beq
Z_1(\eta,\bar\eta)
=\frac{\int d\psi~d\bar\psi~\e^{N\tr(N\ps2+\bar\eta\psi+\bar\psi\eta)}
}{\int d\psi~d\bar\psi~\e^{N^2\tr\ps2}}~~~~~,
\label{momgenfn}\eeq
where $\eta$ and $\bar\eta$ are also independent $N\times N$
 Grassmann-valued matrices, through the identity
\beq\new{\begin{array}{ll}
\left\langle\!\!\left\langle
\bar\psi_{i_1j_1}\psi_{k_1\ell_1}\cdots\bar\psi_{i_nj_n}
\psi_{k_n\ell_n}\right\rangle\!\!\right\rangle&=\frac{\partial}
{\partial\eta_{j_1i_1}}\frac{
\partial}{\partial\bar\eta_{\ell_1k_1}}\cdots\frac{\partial}{\partial
\eta_{j_ni_n}}\frac{\partial}{\partial\bar\eta_{\ell_n
k_n}}Z_1(\eta,\bar\eta)\biggm
|_{\eta=\bar\eta=0}\\&=\frac{\partial}{\partial\eta_{j_1i_1}}\frac{
\partial}{\partial\bar\eta_{\ell_1k_1}}\cdots\frac{\partial}{\partial
\eta_{j_ni_n}}\frac{\partial}{\partial\bar\eta_{\ell_n
k_n}}\e^{-\tr\bar\eta\eta}\biggm
|_{\eta=\bar\eta=0}\end{array}}
\label{fermgaussmomgenfn}\eeq

In particular, the fermion propagator is
\beq
\left\langle\!\!\left\langle
\bar\psi_{ij}\psi_{k\ell}\right\rangle\!\!\right\rangle
=\frac{1}{N}\delta_{i\ell}\delta_{kj}
\label{fermprop}\eeq
and it coincides with the scalar propagator (\ref{hermprop}). Thus the
fermionic
 propagator has the same diagrammatic representation as in the Hermitian case.
 The difference in the two models lies in the vertices that each produces. The
 fat graphs in the fermionic case have $2K$-point vertices. However,
a $2K$-point vertex is topologically equivalent to the contraction of
 two $(K+1)$-point vertices along an external leg. Thus the dual lattice
 of the fat graph discretization of surfaces in the fermionic one-matrix
 model (\ref{part1S}) with degree $K$ polynomial potential
 $V(\ps2)=\ps2+g(\ps2)^K/K$ is the same as the triangulations
 of 2-surfaces produced in the Hermitian one-matrix model
(\ref{hermpart}) with a degree $K+1$ polynomial potential
 $V(\phi)=\phi^2/2+g\phi^{K+1}/(K+1)$. This correspondence shows that the
combinatorics of the surface discretizations in the fermionic case are modified
because of a doubling of degrees of freedom at each vertex.

However, an important difference in the statistical theory generated by
(\ref{part1S}) now occurs. The Wick contraction rules given by
(\ref{fermgaussmomgenfn}) yield a product of 2 delta-functions (as in
(\ref{fermprop})) in each pairing $\ps2$, one for each contraction of the inner
and outer indices of $\bar\psi$ with $\psi$. A contraction of the form
$\psi\bar\psi$ gives the same result but with the opposite sign. This leads to
the well-known Feynman rule from perturbative quantum field theory that
Feynman diagrams for fields which obey Fermi statistics have
 an extra factor of $(-1)^L$ compared to the bosonic case,
where $L$ is the number of closed fermion loops. Since for the fat graphs of
the fermionic matrix model the number $L$ is the area or number of vertices of
the triangulated random surface \cite{fgz,thooft}, this indicates that the
genus expansion of the free energy in the fermionic case may be an alternating
series.

\subsubsection{Penner Matrix Models}

In the case of fermionic matrix models, since it is not possible to diagonalize
matrices which have anticommuting elements using a unitary transformation, they
are not natural eigenvalue models. It has been argued though that many of the
analytical tools which are used to analyze Hermitian matrix models, such as the
concept of ``eigenvalue distribution'' and a ``master field'', are also useful
in the fermionic case \cite{mz}. In fact, Makeenko and Zarembo showed that
the adjoint fermion matrix model (\ref{part1}) has many of the features of the
more familiar Hermitian one-matrix model and that its loop equations are
identical to those for the Hermitian one-matrix model with generalized Penner
potential \cite{akm,mz}
\beq
Z_P=\int d\phi~\e^{-N^2\tr(V(\phi)-2\log\phi)}
\label{penner}\eeq
When $N$ is infinte, the models (\ref{part1}) and (\ref{penner}) therefore have
the same solution. However, beyond the leading order in the large-$N$
expansion, the loop equations for the 2 models should be solved with different
boundary conditions and the solution is different in the 2 cases.

The formal equivalence between the models defined by (\ref{part1}) and
(\ref{penner}) at large-$N$ can be seen by inserting the matrix-valued
delta function
\beq
1=\int d\phi~\delta(\phi-\ps2)~~~~~,
\label{matdelta}\eeq
where $\phi$ is a Hermitian matrix, to write (\ref{part1}) as
\beq
Z_1=\int d\psi~d\bar\psi~\int d\phi~\e^{N^2\tr V(\phi)}\delta(\phi-\ps2)
\eeq
and using the identity
\beq
\int\frac{d\lambda}{(2\pi)^{N^2}}~\int d\psi~d\bar\psi~\e^{i\tr\lambda(\phi-
\ps2)}=\int\frac{d\lambda}{(2\pi)^{N^2}}~{\det}[-i(I\otimes\lambda)]
\e^{i\tr\lambda\phi}
\label{id}\eeq
where $I$ is the $N\times N$ identity matrix. In the infinite-$N$ limit,
the integral over the Hermitian matrices $\lambda$
is evaluated on a saddle point.  In that case, when $\phi$ has no zero
eigenvalues, the coordinate transform $\lambda=\lambda'\phi^{-1}$ can be
used to show
\beq
\int\frac{d\lambda}{(2\pi)^{N^2}}~{\det}^N(-i\lambda)\e^{i\tr\lambda\phi}
\propto {\det}^{-2N}\phi
\eeq
so that
\beq
Z_1\sim\int d\phi~\e^{N^2\tr(V(\phi)-2\log\phi)}
\label{pennerferm}\eeq
However, the integral over Hermitian matrices in (\ref{penner}) is ill-defined
for finite $N$ because of the logarithmic divergence at $\phi=0$. This
representation of the fermionic matrix model partition function in terms
of a Hermitian model with effective action involving a hard-core logarithmic
 interaction can be
thought of as the analog of the effective eigenvalue representation involving
the Vandermonde determinant in Hermitian matrix models.

The Penner matrix models (\ref{penner}) were originally used to calculate
the virtual Euler characteristics of the moduli spaces of compact Riemann
surfaces \cite{pen}. This latter quantity is defined as follows \cite{dv}.
 The  moduli space ${\cal M}_{h,n}$ of a Riemann surface $\Sigma_n^h$ of
genus $h$ with $n$ distinguished punctures can be discretized by a simplicial
 decomposition. The simplices can be represented by dual fat graphs $\cal F$
 such that the dimension $\dim{\cal F}$ of each simplex is the number of
lines in the graph minus the number $n$ of punctures in the Riemann surface.
 Then the virtual Euler characteristic of ${\cal M}_{h,n}$ is
\beq
\chi_V({\cal M}_{h,n})=\sum_{{\cal F}\in{\cal M}_{h,n}}\frac{(-1)^{\dim{\cal
 F}}}
{|G({\cal F})|}
\label{virtualeuler}\eeq
where $|G({\cal F})|$ is the order of a stabalizer of the subgroup of the
 mapping class group of $\Sigma^h_n$ which fixes the topological class
of the fat graph $\cal F$ (this is precisely the order of the symmetry group of
the fat graph itself).

A Hermitian matrix model which reproduces the topological invariant
 (\ref{virtualeuler}) should therefore have very special features.
 Since arbitrary valence vertices appear in the simplicial discretization
of ${\cal M}_{h,n}$, the matrix model should include arbitrary
powers of interaction terms $\tr\phi^m$, $m\geq3$. The weighting
 factor for the order of the symmetry group in (\ref{virtualeuler})
 appears in the perturbative expansion of the free energy if we
include a coupling factor of $1/m$ in front of the $m$-th interaction
term, as is standard in perturbative quantum field theory. Finally,
to produce the correct sign in (\ref{virtualeuler}) for each fat
graph $\cal F$ the potential must associate a factor of $-1$ for
each vertex \cite{pen,dv}, which we recall was precisely the
case for the fermionic diagrammatics. Thus the matrix model potential
 which reproduces these Feynman rules and whose free energy perturbative
 expansion therefore coincides with (\ref{virtualeuler}) is
\beq
V_P(\phi)=-\sum_{m=2}^\infty\frac{\phi^m}{m}=\phi+\log(1-\phi)
\label{orpennerpot}\eeq

It has been shown that the Hermitian matrix model with Penner potential
 (\ref{orpennerpot}) has a continuum limit which is described by a
 logarithmic susceptibility behaviour at criticality \cite{cdl,dv}.
 Such a critical behaviour is associated with a string constant $\gamma_{
\rm str}=0$, or equivalently central charge $D=1$, i.e. a free boson
conformal field theory \cite{ginsparg}. Thus the Penner matrix models are
 also deeply connected to $D=1$ string theory. More precisely, they describe
 matter compactified on a self-dual radius
circle interacting in $D=1$ dimension with 2-dimensional quantum gravity
\cite{cdl,dv,tan}. By adding more complicated polynomial interactions as
in (\ref{penner}), it is possible to localize multi-critical points with
 $\gamma_{\rm str}=-1/m$ which are associated with the $D=1$ string
 compactified to the critical radius of the Kosterlitz-Thouless phase
transition \cite{akm,cdl,tan,mak5}. These models therefore capture
various intriguing features of 2-dimensional quantum gravity and certain
 topological aspects of moduli space \cite{wit}. One important technical
feature
of Penner matrix models is that in the derivation of their loop equations, the
field transformation (\ref{hermshift}) must be slightly
 modified because it produces a singular pole term $\langle\tr(1/\phi)
\rangle$ from the variation of $\log\phi$ in the action. This term is
 easily removed by the considering instead the field redefinition \cite{akm}
\beq
\phi\to\phi+\varepsilon\sum_{k=1}^\infty\frac{\phi^k}{z^{k+1}}=\phi\left(1+
\frac{\varepsilon}{z(z-\phi)}\right)
\label{pennershift}\eeq
However, the resulting loop equations lead to an algebraic equation of
order $2K$ when the potential in (\ref{penner}) is a polynomial of degree $K$.

The connection between the Penner models and matrix
models involving fermionic degrees of freedom were originally suggested in
the work of Gilbert and Perry \cite{gilper} (see also \cite{alvarez}) who
considered a $D=0$ random matrix model with $N=2$ supersymmetry using a
manifestly Hermitian supermatrix formalism. They showed that the integration of
the fermionic matrix degrees of freedom out of the matrix integral reduces the
model exactly to the Penner matrix model, and that in the double-scaling
limit the solutions of the string equation in the case of the cubic
superpotential exhibit critical behaviour that lies in the same universality
class as the Penner model. These `mixed' types of fermionic matrix models
are therefore deeply related to $D=1$ bosonic string theory, but differ
from the bosonic matrix models in the type of universal behaviour that they
exhibit (e.g. there are 2 distinct universality classes displayed)
\cite{gilper}.

It has been conjectured by Makeenko and Zarembo \cite{mz} and Ambj\o rn,
 Kristjansen and
Makeenko \cite{akm} that the fermionic one-matrix model also corresponds to
a similar sort
of statistical theory of triangulated random surfaces, except that now the
genus expansion has alternating signs \cite{akm}. The resulting convergence of
the sum over genera is reflected in the feature of the fermionic matrix model
that its partition function and observables are all well-defined since the
integrals over Grassmann variables always converge, in contrast to Hermitian
matrix models \cite{fgz,kaz}. In fact, one can formally perform the integral
(\ref{part1}) at finite-$N$ by inserting the matrix-valued delta function
(\ref{matdelta}) and using (\ref{id}) as
\beq
\int\frac{d\lambda}{(2\pi)^{N^2}}~\int d\psi~d\bar\psi~\e^{i\tr\lambda(\phi-
\ps2)}={\det}^N\left(-\frac{\partial}{\partial\phi}\right)\delta(\phi)
\eeq
to obtain
\beq
Z_1={\det}^N\left(\frac{\partial}{\partial\phi}\right)\e^{N^2\tr V(\phi)}
\biggm\vert_{\phi=0}
\label{part1exact}
\eeq
Similarly, any correlator at finite-$N$ is given by
\beq\new{\begin{array}{ll}
\left\langle\prod_j\tr(\bar\psi\psi)^{p_j}\right\rangle&\equiv\frac{\int
d\psi~d\bar\psi~\e^{N^2\tr V(\ps2)}\prod_j\tr(\ps2)^{p_j}}{\int
d\psi~d\bar\psi~\e^{N^2\tr V(\ps2)}}\\&=\frac{{\det}^N\left(
\frac{\partial}{\partial\phi}\right)\left(\prod_j\tr\phi^{p_j}\right)\e^{N^2\tr
V(\phi)}}{{\det}^N\left(\frac{\partial}{\partial\phi}\right)\e^{N^2\tr V(\phi)
}}\biggm\vert_{\phi=0}\end{array}}
\eeq

For a given potential $V(z)$, the expression (\ref{part1exact}) can be expanded
to give a finite polynomial in the parameters of the potential which coincides
with the usual perturbative expansion (which is a finite sum because of the
nilpotency of the Grassmann variables). To generate a full statistical random
surface model with arbitrary numbers of vertices, one needs to take the
large-$N$ limit. This is in contrast to the bosonic cases of the last
Subsection where the large-$N$ limit was required to make the perturbative
expansion a well-defined quantity at each order in $1/N$. However, in spite of
this good convergence of the partition function (\ref{part1}), it has
been shown that this model has a non-trivial phase structure in the infinite
$N$ limit \cite{akm,mz,mss,semsz}. In particular, the adjoint fermion
one-matrix
model possesses the usual multi-critical behaviour \cite{fgz,kaz} with third
order phase transition and string susceptibility with critical exponent
$\gamma_{\rm str}=-1/m$, $m\geq2$. It may also have a first order phase
transition \cite{mss}. Although a first order phase transition is unusual for a
matrix model, the phase structure of this model resembles the critical
behaviour that occurs near the triple point of a liquid-vapour-solid system.
There the vapour phase, for example, can be supercooled to the boundary between
the liquid and the solid phase where the phase transition must then occur and
be of first order. The existence of 2 critical points in the adjoint fermion
1-matrix model is similar to the occurence of 2 distinct universality classes
in the Gilbert-Perry supermatrix model.

\subsubsection{Complex Matrix Models}

{}From an analytic point of view, the loop equations of the adjoint fermion
matrix model are similar in structure to those of the Hermitian one-matrix
model with generalized Penner potential. However, form a combinatorical point
of view, there is another bosonic matrix field theory which is similar in
structure to the fermionic ones above. Consider the {\it complex} one-matrix
model which is defined by the partition function
\beq
Z_C=\int\prod_{i,j}d\phi_{ij}~d\phi^\dagger_{ij}~\e^{-N^2\tr
V(\phi^\dagger\phi)}
\label{complexpart}\eeq
where the integration is over the space of all $N\times N$ complex-valued
matrices $\phi$. The complex matrix model is equivalent to the Hermitian matrix
model in the double scaling limit \cite{ackm,mak3} and its complete set of
correlators can be written down in closed form expressions \cite{amjurk}. It
can also be written as an eigenvalue model since the action in
(\ref{complexpart}) depends only on the radial part $\phi^\dagger\phi$ of
$\phi$ which is a Hermitian matrix. As a
statistical theory of random surfaces, it has been argued that the fat-graphs
of the complex matrix model (\ref{complexpart}) represent `checkered' surfaces
which are formed from the doubling of vertex degrees of freedom as discussed
above for the fermionic matrix model.

Except for the appropriate minus signs and convergence properties, the
perturbative expansion of the model (\ref{complexpart}) yields the same
generating function for triangulations of random surfaces as in the fermionic
case because the non-zero Wick contractions occur only among the pairings
$\phi^\dagger\phi$. This is a consequence of the much larger unitary and charge
conjugation invariances of (\ref{complexpart})),
\beq
\phi\to V\phi U^\dagger~~~{\rm with}~~~\{U,V\}\in U(N)\otimes U(N)~~~~~;~~~~~
\phi\to\phi^\dagger
\label{complexuninv}\eeq
which restricts the non-zero correlators to those of invariant functions of
$\phi^\dagger\phi$.
Thus the combinatorical factors associated with the counting of Feynman
diagrams in both cases are the same, and in the case of the complex matrix
model it is known that, because of the doubling of degrees of freedom in
(\ref{complexpart}), the free energy of the Hermitian matrix model with an even
polynomial potential is twice the free energy $\log Z_C$ defined with the same
potential. Furthermore, any given correlator of the $N\times N$ complex matrix
model can be obtained from that of the $2N\times 2N$ Hermitian matrix model
with an even potential and dividing by 2 \cite{amjurk,mak3}. As we shall see in
the following, the loop equations for the adjoint fermion one-matrix model are
derived in a manner similar to those of the complex matrix model. In the latter
case the loop equations follow from the same shifts (\ref{pennershift}) as for
the Penner matrix models. These correspondences lead to many other similarities
between fermionic matrix models and these classes of bosonic matrix models.

\subsubsection{Fermionic Matrix Models at $D>0$}

Makeenko and Zarembo also introduced a class of fermionic two-matrix models
 \cite{mz}. The equations of
motion for these generalizations are high-degree polynomial equations whose
solutions are not as immediate as for the case of the one-matrix
models \cite{semsz}. In
these models, as well as the one-matrix models, the $\frac{1}{N}$-expansion of
the Schwinger-Dyson equations can be related to discrete Virasoro and
$W$-constraints which have recently been used to discuss the integrable
structure of matrix models \cite{kharch1,kharch2,marsh} and the
relations between matrix models and
certain topological quantum field theories such as topological gravity in
2-dimensions \cite{dijk1,dijk2,makmarsh,wit}. These constraints also provide
an alternative way of examining the continuum
limit relevant to string theory (see \cite{mir} and \cite{mor} for
recent reviews).

Originally, fermionic matrix models were introduced by Khokhlachev and
Makeenko for adjoint fermions \cite{kmak2}, and by Migdal for fundamental
fermions \cite{mig2}, as another proposal of a model which could be
considered as inducing Yang-Mills theory in the continuum limit. This model
is the most general fermionic matrix model which is the
natural $D$-dimensional generalization of (\ref{part1}) minimally coupled to a
unitary matrix field \cite{kmak2,mak2,mz,semsz}. It has been shown that
there are several advantages of this fermionic matrix model over its Hermitian
counterpart which is the Kazakov-Migdal model of induced QCD \cite{km}. It is
by now well-known that if the large-$N$ phase transition in the Kazakov-Migdal
model occurs before the one separating the local confinement and perturbative
Higgs phases, then one gets normal area law confinement in the intermediate
region bounded by these 2 phase transitions \cite{mak2}. Moreover, the
solvability features of the Kazakov-Migdal model are masked by the fact
that the extra local $U(1)$ gauge symmetry of the theory must be spontaneously
broken before the continuum limit is reached \cite{kmsw,kmak1,mak2}.
Furthermore, the Gaussian model in this case is unstable due to an unlimited
Bose condensation of scalar particles for an action which is unbounded from
below (and if a stabalizing self-interaction is added then it leads to the
Higgs phenomenon) \cite{kmak1}. In the fermionic case, however, the
perturbative expansion for weakly fluctuating gauge fields resembles that
of ordinary QCD and a large-$N$ first order phase transition occurs with
decreasing fermion mass which separates the perturbative regime, associated
with the restoration of the area law, from the strong coupling phase with
unbroken $U(1)$ gauge symmetry and associated local confinement \cite{kmak2}.
Moreover, since fermions can never condense, the Gaussian fermionic action has
no instability (or Higgs phase) for any value of the fermion mass (but this
does not exclude the existence of a composite Higgs phase associated with a
fermion chiral condensate $\langle\ps2\rangle\neq0$).

These higher-dimensional adjoint fermion matrix models may also provide
important information concerning the $D>1$ phase of string theory. Some recent
works have appeared describing polymer models using fermionic degrees of
freedom. Fermionic vector models \cite{semsz1} share many of the random
geometry interpretations of the usual scalar vector models which describe the
statistical mechanics of randomly branching polymers. However, they generate
better defined statistical models and moreover they provide interesting toy
examples of many of the features of fermionic matrix models. Supersymmetric
matrix models \cite{ammakz},\cite{makmean}--\cite{makpe} which combine the
adjoint fermion and
complex matrix models above contain a dimensional reduction due to the
supersymmetry and also therefore describe a statistical theory of discrete
filamentary surfaces. Thus the use of fermionic degrees of freedom also enables
one to construct more exotic types of matrix models and to exploit their
convergence features to describe more complicated combinatorical problems.

\subsection{Outline}

This Review will focus on the above aspects of adjoint fermion matrix models.
We shall discuss in detail the technical features involved in solving these
models in the large-$N$ limit. In particular, we shall discuss how to
explicitly solve for the observables of these models and examine their phase
structures. Although not yet completely understood, we shall at various places
indicate what these matrix models may have to do with string theories and other
continuum field theories such as Yang-Mills theory. Throughout we shall compare
these models to the bosonic matrix models, highlighting the essential
differences and similarities.

The structure of this Paper is as follows. We begin in Section 2 by
considering, for the sake of illustration, a vector version of (\ref{part1})
which is the fermionic analog of the scalar $O(N)$ vector
models \cite{schel,vec1,vec2,zinn1} which have
recently been studied as `test models' of concepts for the more complicated
Hermitian matrix models. There we
show how the fermionic nature of the degrees of freedom can be used to
explicitly solve for the observables of such random models, and we show how the
construction of the topological expansion carries through explicitly in these
cases. The nicest feature of these vector models is that everything can be
solved for {\it exactly} with relative ease, and the critical behaviour
of the more complicated matrix models becomes rather transparent in these
simplified versions. We also discuss the relevance of these fermionic vector
models for random polymers and establish exactly the alternating nature of
their genus expansions.

We then discuss the adjoint fermion one-matrix model (\ref{part1}) in
more detail in Section 3, and in particular we present the full explicit
analysis of its critical behaviour in the case of a symmetric potential. We
discuss the physical and mathematical bearings the third and possibly first
order phase transitions have \cite{mss}, and we examine the
$\frac{1}{N}$-expansion about the third order multi-critical point and present
the argument \cite{akm} which indicates that the genus expansion may be a
convergent, alternating series version of the Painlev\'e expansion which
otherwise coincides with the genus expansion of the Hermitian one-matrix models
(this is exemplified by the foregoing vector model results). We also show that
the large-$N$ expansion of the loop equations in this case can be interpreted
as Virasoro constraints which allows one to interpret the fermionic one-matrix
model as an integrable system \cite{mor}.

We then consider the fermionic two-matrix model
in Section 4 as the natural next higher-dimensional gauge invariant
generalization of (\ref{part1}). There we derive the complete sets of loop
equations for these models, and hence illustrate that even in the simple cases
of symmetric potentials the equations of motion are far too complex for any
explicit solution \cite{semsz}. However, the loop equations for these
two-matrix models are quite similar to those of the Hermitian two-matrix
model \cite{alf,gava,marsh,staud}, so that one expects the same
values of $\gamma_{\rm str}$ to appear and hence a similar string
 theoretical interpretation of these higher-dimensional fermionic matrix
models. We also show how the $\frac{1}{N}$-expansion of the
two-matrix model is related to integrable hierarchies \cite{mor}.

In Sections 5
and 6 we generalize the adjoint fermion two-matrix model to $D$ dimensions as
a lattice gauge theory. We compare this class of lattice gauge
theories to both the Kazakov-Migdal model and previous attempts at using
fermions in the fundamental representation of the gauge group to model the
quark fields. We review the present status of the quark
confinement problem for these theories, and in particular the progress thus far
in solving the loop equations of the models at strong coupling. As of yet, this
is still a highly non-trivial and unsolved problem, and in particular it is not
known how to extrapolate the strong coupling solutions to solutions of the loop
equations in other phases of the model, specifically where the area law
behaviour of Wilson loops holds \cite{kmak2,mak2,mz,semsz}. Here we shall also
analyse in detail the critical behaviour of the Itzykson-Zuber integral
(\ref{itzykherm}) in both the bosonic and fermionic cases. We shall show that
even in the bosonic case it may have a non-trivial phase structure, as has been
suggested recently by Makeenko in \cite{mak5} where it was shown that
extended Wilson loops in the Kazakov-Migdal model exhibit a continuum
limit due to a singular behaviour of the Itzykson-Zuber correlator of
the gauge fields.

Finally, in Section 7 we combine the adjoint fermion one-matrix model with the
complex bosonic matrix model of Subsection 1.2.2 above. In certain instances,
the model has a supersymmetry between the bosonic and fermionic degrees of
freedom which can be exploited to solve complicated combinatorical problems
that may be relevant for the dimensionally reduced phases of string theory in
target space dimensions $D>1$. We illustrate this on the particular
combinatorical problem of evaluating meander numbers (i.e. the number of
foldings of a closed polymer chain). We show how to formulate the loop
equations in these models in terms of random variables, a technique which has
been exploited recently for the solution of Hermitian matrix models. We also
demonstrate how the dimensional reduction of the supersymmetric matrix model
reproduces characteristics of a branched polymer theory, and comment on the
potential uses of super-matrices for theories of supergravity and superstrings.
Generally, throughout the Review we single out the numerous unsolved problems
in this
field, all of which are interesting problems which certainly deserve future
investigation. Although for the most part this Paper is a review, a good
portion of the analyses presented in the other Sections is original work.

\section{Fermionic Vector Models}

Scalar $O(N)$-symmetric vector models have been studied over the past few years
as an interesting testing ground of ideas for the more complex Hermitian matrix
models \cite{vec1,vec2,eyal,nishyon,schel,zinn1}. Besides this, as mentioned in
the previous Section, they describe statistical models of discrete filamentary
surfaces relevant for the branched polymer phase of $D>1$ string theories
\cite{amdur1,amdur3,anderson,das,dupl,nish}. We would expect the same to be
true of some simplified vector version of the fermionic matrix model
(\ref{part1}). Therefore, as a preliminary discussion and to
illustrate some of the features of statistical ensembles with
fermionic degrees of freedom in a context simpler than the matrix
theories, we consider a toy model with $N$ Grassmann variables
$\psi_i$, $i=1,\ldots,N$, and $N$ independent conjugate variables
$\bar\psi_i$. The model is specified by the partition function
\cite{semsz,semsz1}
\beq
Z_0=\int d\psi~d\bar\psi~\e^{NV(\bar\psi\psi)}
\label{part0}
\eeq
where
\beq
\ps2\equiv\sum_{i=1}^N\bar{\psi}_i\psi_i~~~~~,
\label{psisum}\eeq
$V(z)$ is some ``potential'' function of $z$ and the integration
measure is
\beq
d\psi~d\bar\psi\equiv\prod_{i=1}^Nd\psi_i~d\bar\psi_i
\label{vecmeas}\eeq

The integration over Grassmann variables in (\ref{part0}) is
well-defined and finite if $\e^{NV(z)}$ has a well-defined Taylor
expansion to order $N$ in the variable $z$.  As we shall show in the
following, this simplicity is what makes this model explicitly solvable.  The
model possesses a continuous symmetry,
\beq
\psi_i\to\sum_{j=1}^NU_{ij}\psi_j~~,~~\bar\psi_i\to\sum_{j=1}^N\bar\psi_j
(U^{-1})_{ji}~~~{\rm with}~~~U\in GL(N,\IC)
\label{ctsvecsymm}
\eeq
parametrized by the Lie group $GL(N,\IC)$ of invertible complex-valued $N\times
N$ matrices. There is a further discrete symmetry under the ``chiral"
transformation
\beq
\psi_i\rightarrow\bar\psi_i~~,~~\bar\psi_i\rightarrow-\psi_i~~~{\rm for}~~{\rm
any}~~i
\label{chiralsymvec}\eeq
Together, these two symmetries restrict the observables
of the model to those which are functions only of
$\ps2$. The statistical model is then completely determined once the
moments
\beq
M^k\equiv\left\langle(\bar\psi\psi)^k\right\rangle\equiv\frac{\int
d\psi~d\bar\psi~(\ps2)^k\e^{NV(\ps2)}}{\int d\psi~d\bar\psi~\e^{NV(\ps2)}}
\eeq
are known. Since nilpotency of the components $\psi_i$ and $\bar\psi_i$
imply that
\beq
(\bar\psi\psi)^{N+1}=0
\eeq
there are $N$ independent moments $M^1,\ldots,M^N$.  Also, symmetry
and nilpotency of the vector components $\psi_i$ and $\bar\psi_i$ imply that
\beq
M^k=\frac{N!}{(N-k)!}\left\langle\prod_{j=1}^k\bar\psi_j\psi_j\right\rangle
{}~~~,~~~0<k\leq N
\label{Mkcorrs}\eeq
and so the correlators $\langle f(\ps2)\rangle$ for finite-$N$ are polynomials
in the parameters of the potential in (\ref{part0}).

\subsection{Finite and Infinite $N$ Solutions}

The first feature we will stress of the fermionic vector model (\ref{part0})
is the extent to which it can be solved at both finite and infinite-$N$. The
explicit form of the solution at infinite-$N$ is obtainable even in the scalar
models because of the linearity of the vector theories in contrast to the
non-linearity of the equations of motion involving matrix fields. The
solvability here at finite-$N$ is a consequence of the Grassmann integrations
in (\ref{part0}). More precisely, the partition function (\ref{part0}) at
finite $N$ can be formally evaluated by inserting the delta function
\beq
1=\int_{-\infty}^{+\infty}dx~\delta(x-\bar\psi\psi)
\eeq
and using the identity
\beq
\int_{-\infty}^{+\infty}\frac{dy}{2\pi}~\int d\psi~d\bar\psi~\e^{iy(x-\ps2)}=
\int_{-\infty}^{+\infty}\frac{dy}{2\pi}~N!(-iy)^N\e^{iyx}
=N!\left(-\frac{\partial}{\partial x}\right)^N\delta(x)
\label{part0derivid}\eeq
to obtain
\beq
Z_0=N!\cdot\left(\frac{\partial}{\partial
x}\right)^N\e^{NV(x)}\biggm\vert_{x=0}
\label{z0}\eeq
Similarly, the correlators for $N$ finite are given by
\beq
M^n=\frac{\left(\frac{\partial}{\partial x}\right)^Nx^n
\e^{NV(x)}}{\left(\frac{\partial}{\partial x}\right)^N\e^{NV(x)}}\biggm|_{x=0}
\eeq

These moments can always be obtained
from a (not unique) distribution function $\rho$ defined by
\beq
M^n=\int d\alpha~\rho(\alpha)\alpha^n
\label{rho0}
\eeq
where the integral is over the support of $\rho$ in the complex plane. When
$N$ and therefore the number of moments is finite the support of the
spectral function $\rho(\alpha)$ is
concentrated near the origin in the complex $\alpha$-plane
\beq\new{\begin{array}{ll}
\rho(\alpha)&\equiv\left\langle\delta(\alpha-\ps2)\right\rangle=\frac{1}{2\pi}
\int dw~\e^{-iw\alpha}\left\langle\e^{iw\ps2}\right\rangle
\\&=\frac{1}{2\pi}\int dw~\e^{-iw
\alpha}~\frac{\left(\frac{\partial}{\partial z}\right)^N\e^{iwz+NV(z)}}{\left(
\frac{\partial}{\partial z}\right)^N\e^{NV(z)}}\biggm|_{z=0}\\&=\frac{1}{Z_0}
\left(-\frac{\partial}{\partial\alpha}+NV'(\alpha)\right)^N\delta(\alpha)
\end{array}}
\label{rhovec}\eeq
The spectral density (\ref{rhovec}) completely determines the solution of the
random vector model (\ref{part0}). The most general potential which we shall
consider in this Section is a combination of
a logarithm and a polynomial of degree $K\leq\infty$,
\beq
V(z)=\kappa\log z+\sum_{n=1}^K\frac{g_n}{n}z^n
\label{pot}
\eeq

When $N$ is infinite, the distribution has an infinite number of
moments and the support of the spectral function need no longer be concentrated
at the origin. In this case the solution of the vector model simplifies because
the correlators factorize,
\beq
\langle f(\ps2)\rangle= f(\langle\ps2\rangle)+{\cal O}(1/N)
\eeq
This factorization property follows from the fact that for an arbitrary
potential of the form (\ref{pot}) the connected correlators are given by
\beq
\langle(\ps2)^{n}\rangle_{\rm conn}=\frac{1}{N^n}\left(\frac{\partial}{\partial
g_1}\right)^n\log Z_0
\eeq
and
\beq
\log Z_0= NF_0
\label{freevec}\eeq
so that the free energy $F_0$ is of order one at $N\rightarrow\infty$. It
follows that
\beq
M^n=(M^1)^n+{\cal O}(1/N)
\eeq
and the statistical model in the large-$N$ limit is completely determined by
the first moment
\beq
M^1=\langle\ps2\rangle
\eeq

The spectral density in the large-$N$ limit
\beq
\rho(\alpha)=\lim_{N\to\infty}\sum_{k=0}^N\frac{M^k}{k!}\left(-\frac
{\partial}{\partial\alpha}\right)^k\delta(\alpha)=\e^{-M^1\frac{\partial}{
\partial\alpha}}~\delta(\alpha)=\delta(\alpha-M^1)
\label{rholargeN}\eeq
is a point mass localized at the point $\alpha=M^1$. The expectation value of
any observable of the vector model (\ref{part0}) at large-$N$ is therefore
\beq
\langle f(\ps2)\rangle=f(M^1)
\label{veclargeNavgs}\eeq
Thus the fermionic vector model (\ref{part0}) provides a rare example of a
statistical theory which is exactly solvable at both finite and infinite $N$.
We shall discuss the applications of the fermionic vector model as a random
geometry theory \cite{semsz1} in the last Subsection of this Section.

\subsection{Loop Equations}

An alternative way to solve for the moments in a random distribution of
variables, which will generalize to the matrix theories we are ultimately
interested in, is by the Makeenko-Migdal method of loop equations
\cite{mig1}. Consider the ``propagator''
\beq
\omega(z)\equiv \left\langle\frac{1}{z-\bar\psi\psi}
\right\rangle\eeq
The asymptotic expansion of this function is a series in the moments
\beq
\omega(z)=\frac{1}{z}+\sum_
{n=1}^N\frac{\langle(\ps2)^n\rangle}{z^{n+1}}=\frac{1}{z}+\sum_{k=1}^N
\frac{M^k}{z^{k+1}}
\label{vecom}\eeq
When $N$ is finite, since this series is finite,
$\omega(z)$ has singularities only at the origin in the complex
$z$-plane.  It can be expressed as an integral over the spectral density
\beq
\omega(z)=\int d\alpha~\frac{\rho(\alpha)}{z-\alpha}
\label{omrho}\eeq
where $\rho(\alpha) $ is the spectral function whose formal solution
is given in (\ref{rhovec}) and the integration in (\ref{omrho}) can be taken
over any contour in the complex $\alpha$-plane which goes through the origin.

When $N$ is infinite, the support of the spectral function can be
deduced from the analytic structure of $\omega(z)$.  If the support is
in a compact region of the complex plane, the propagator has the
asymptotic form
\beq
\lim_{\vert z\vert\rightarrow\infty}\omega(z) = 1/z
\label{bcvec}\eeq
and the integral over $\alpha$ in (\ref{omrho}) is on a contour or set of
contours in the complex plane which contain the support of
$\rho(\alpha)$.  The resulting integral is analytic in the region
outside of the support of $\rho$. The positions of branch and pole
singularities of $\omega$ determines the support of $\rho$ which in
the large-$N$ limit is generally some contour in the complex plane.
The continuous function $\rho$ itself can be found by computing the
discontinuity of $\omega(z)$ across its support, where
\beq
\omega(\beta\pm\epsilon_\perp)=\pvint~d\alpha~\frac{\rho(\alpha)}{\beta-
\alpha}\mp\frac{\epsilon_\perp}{|\epsilon_\perp|}\pi\rho(\beta)~~~{\rm
for}~~~\beta\in~{\rm supp}~\rho
\label{disc}
\eeq
and $\epsilon_\perp(\beta)$ is a complex number with infinitesimal
amplitude and direction perpendicular to the integration contour in
(\ref{disc}) at the point $\beta$.

The loop equation for $\omega(z) $ follows from the identity
\beq
0=\int d\psi~d\bar\psi~\frac{\partial}{\partial\psi_i}
\left(\psi_j\frac{1}{z-\bar\psi\psi}\e^{NV(\bar\psi\psi)}\right)
\eeq
which is a consequence of the Grassmann integration rules (\ref{grassint}).
Dividing by $Z_0$ and expanding out the averages in this equation gives
\beq
0=\delta_{ij}\left\langle\frac{1}{z-\ps2}\right\rangle+\left\langle\bar\psi_i
\psi_j\left(\frac{1}{z-\ps2}\right)^2\right\rangle+N\left\langle\bar\psi_i
\psi_j\frac{1}{z-\ps2}V'(\ps2)\right\rangle
\eeq
and then summing over $i=j=1,\dots,N$ leads to
\beq
\frac{1}{N}\left(z\frac{\partial}{\partial z}+1\right)\omega(z)+\omega(z)+
\left\langle \frac{1}{z-\bar\psi\psi}\ps2 V'(\bar\psi\psi)\right\rangle=0
\label{loopvec}
\eeq
The correlator in (\ref{loopvec}) involving the potential $V(\ps2)$ can be
expressed, using (\ref{omrho}), as an integral over a contour $\cal C$ in the
complex plane which encircles the singularities of $\omega(z)$ with
counterclockwise orientation. The loop equation (\ref{loopvec}) can then be
written as the integro-differential equation
\beq
\frac{1}{N}\left(z\frac{\partial}{\partial
z}+1\right)\omega(z)+\omega(z)-\oint_{\cal C}\frac{d\lambda}{2\pi
i}~\frac{V'(\lambda)\lambda}{z-\lambda}\omega(\lambda)=0
\label{loopveccont}\eeq

When the potential is of the form (\ref{pot}), we have
\beq
zV'(z)=\sum_{n=0}^K g_n z^n
\label{potderiv}\eeq
where $g_0=\kappa$. To simplify (\ref{loopvec}) we use the identity
\beq\new{\begin{array}{ll}
\bar\psi\psi V'(\bar\psi\psi)&=zV'(z)+\left(\bar\psi\psi
V'(\bar\psi\psi)-zV'(z)\right)\\&=zV'(z)+(\bar\psi\psi-z)\sum_{n=1}^Kg_n
\sum_{m=0}^{n-1}z^m(\bar\psi\psi)^{n-m-1}\end{array}}
\eeq
and obtain the loop equation
\beq
\frac{1}{N}\left(z\frac{\partial}{\partial z}+1\right)\omega(z)+\omega(z)+
zV'(z)\omega(z)=P(z)
\label{loopvec1}
\eeq
where $P(z)$ is a polynomial of degree $K-1$ which is determined by
the first $K-1$ moments $M^1,\ldots, M^{K-1}$
\beq
P(z)=\sum_{m=1}^Kg_m\sum_{p=0}^{m-1}\left\langle(\bar\psi\psi)^{m-p-1}\right
\rangle z^p~=~\sum_{m=1}^K g_m \sum_{p=0}^{m-1} M^{m-p-1}z^p
\label{polyvec}\eeq
This equation can also be obtained from the integro-differential equation
(\ref{loopveccont}). When the potential is given by (\ref{potderiv}), the
contour integral in (\ref{loopveccont}) can be evaluated by computing the
residues at $\lambda=z$ and $\lambda=\infty$ to get
\beq
\oint_{\cal C}\frac{d\lambda}{2\pi
i}~\frac{V'(\lambda)\lambda}{z-\lambda}\omega(\lambda)=-V'(z)z\omega(z)+P(z)
\label{contour}\eeq

Generally, when the loop equations are solved, the moments in (\ref{polyvec}),
which appear as constants in the equation, must be found self-consistently.
Substituting the asymptotic expansion (\ref{vecom}) into the loop equation
(\ref{loopvec1}) and equating the coefficients of $z^p$, we find that the
moments are determined in terms of the coupling constants of the potential
(\ref{pot}) by the set of recursive equations
\beq
\sum_{k\geq1}g_kM^k=1-\kappa
\label{momalg}\eeq
\beq
\kappa M^p+\sum_{k\geq1}g_kM^{k+p}
=\left(1-\frac{p}{N}\right)M^p~~~{\rm for}~~~1\leq p\leq N
\label{momrec}\eeq
The loop equation must be solved with the asymptotic boundary condition
(\ref{bcvec}). Its solution is readily found for arbitrary $N$ by integrating
the first order linear ordinary differential equation (\ref{loopvec1}) as
\beq
\omega(z)=\frac{N}{z}z^{-N}\e^{-NV(z)}\int dz~P(z)z^N\e^{NV(z)}
\label{omvecN}
\eeq
The overall constant $N$ is fixed by the asymptotic behaviour (\ref{bcvec})
which can be found by expanding (\ref{omvecN}) in $1/z$. The leading term is
found by
\beq\new{\begin{array}{ll}
\lim_{|z|\to\infty}\omega(z)&\sim\frac{1}{z}\lim_{\vert
z\vert\rightarrow\infty}\frac{N\int_{z_0}^z dw~P(w)w^N\e^{NV(w)} }{z^N
\e^{NV(z)} }\\&=\frac{1}{z}\lim_{\vert
z\vert\rightarrow\infty}\frac{NP(z)z^N\e^{NV(z)} } {( N/z+NV'(z))z^N\e^{NV(z)}
}=\frac{1}{z}\end{array}}
\eeq
as expected.

Thus, the linearity of the vector model equations of motion lead to a rare
example of a field theory whose loop equations can be solved exactly and
explicitly for any $N$. For the adjoint fermion matrix models that we shall
soon consider, the loop equations will be non-linear and such a solution at
arbitrary $N$ will not be possible, even though the partition function is
explicitly computable for finite $N$ as in the vector case. Notice also that
the moments appearing in the recursion relations (\ref{momalg}) and
(\ref{momrec}) can be expressed as
\beq
M^k=k\frac{\partial}{\partial g_k}F_0~~~,~~~k\geq1
\label{MkdF0}\eeq
and so the loop equations are equivalent to a system of differential equations
in coupling constant space which determine the partition function $Z_0$ of the
model. In the matrix generalizations we shall exploit this feature to relate
the solutions of the loop equations to integrable hierarchies of differential
equations. In the bosonic cases, these appear as Virasoro constraints on a
$\tau$-function which is characterized as a symmetry of the string equation.
The fact that this is readily observed in the vector model counterparts owes to
the simplicity and linearity of these models.

As expected, the loop equations simplify at large-$N$ because of factorization.
The propagator (\ref{vecom}) at $N=\infty$ becomes
\beq
\omega(z)=\frac{1}{z}+\sum_{k=1}^\infty\frac{(M^1)^k}{z^{k+1}}=\frac{1}{z-M^1}
\label{propinf}\eeq
and the loop equation (\ref{loopvec}) at $N=\infty$ is
\beq
\omega(z)=\frac{P(M^1,z)}{zV'(z)-1}
\label{omegavecsol}\eeq
where
\beq
P(M^1,z)=\sum_{m=1}^Kg_m\sum_{p=0}^{m-1}(M^1)^{m-p-1}z^p
\eeq
The solution (\ref{propinf}) agrees with that found in the last Subsection. The
coincidence of the 2 solutions (\ref{propinf}) and (\ref{omegavecsol}) is
equivalent to the equations (\ref{momalg}),(\ref{momrec}) which determine the
single independent moment $M^1$. For a potential of the form (\ref{pot}), at
large-$N$ the correlator $M^1$ is determined as the solution of the $K$-th
order algebraic equation (\ref{momalg}),
\beq
M^1V'(M^1)=1
\label{M1algeq}\eeq

\subsection{The Gaussian Model}

The simplest model is the Gaussian model which is defined by the potential
\beq
V(z)=tz
\label{gausspot}\eeq
The Gaussian potential (\ref{gausspot}) in all of the models we present in this
Review will always serve as an interesting consistency check of the formalisms
and will moreover present partial solutions for more complicated examples.
In this case, the distribution function (\ref{rhovec}) becomes
\beq
\rho(\alpha)=\left(1+\frac{1}{Nt}\frac{\partial}{\partial\alpha}\right)^N\delta
(\alpha)
\label{rhovfin}\eeq
The $N\to\infty$ limit of (\ref{rhovfin}) is
\beq
\rho(\alpha)=\e^{\frac{1}{t}\frac{\partial}{\partial\alpha}}\delta(\alpha)
=\delta(\alpha-1/t)
\label{rhovinf}\eeq
which identifies the first moment of the distribution as
\beq
M^1=\langle\ps2\rangle=1/t
\label{gaussmom1}\eeq

The finite-$N$ loop equation (\ref{loopvec}) in the case of the Gaussian
potential (\ref{gausspot}) is the first order ordinary differential equation
\beq
\frac{z}{N}\omega'(z)+\left(1+\frac{1}{N}-tz\right)\omega(z)+t=0
\label{finNgauss}\eeq
whose solution is
\beq
\omega(z)=\frac{1}{z}\left(1-\frac{1}{Nz}\frac{\partial}{\partial
t}\right)^N\frac{1}{t}
\label{omgaussN}\eeq
where we have used the boundary condition (\ref{bcvec}). In the infinite-$N$
limit (\ref{omgaussN}) becomes
\beq
\omega(z)=\frac{t}{z}\e^{-\frac{1}{z}\frac{\partial}{\partial t}}\left(\frac{1}
{t}\right)=\frac{t}{z(t-1/z)}
\eeq
which coincides with the large-$N$ solution (\ref{omegavecsol}) for the
Gaussian potential (\ref{gausspot}). The propagator $\omega(z)$ at both finite
and infinite $N$ agrees with the spectral density obtained directly above.
Alternatively, we obtain from (\ref{omvecN}) the general solution
\beq
\omega(z)=\frac{N}{z}\frac{\int_{z_0}^zdw~w^Nt\e^{Ntw}}{z^N\e^{Ntz}}=\frac{1}
{z}\sum_{k=0}^N\frac{N!}{(N-k)!}\frac{1}{t^kN^kz^k}
\eeq
which identifies the Gaussian moments
\beq
M^k=\frac{N!}{(N-k)!}\frac{1}{(Nt)^k}
\label{gaussmomk}\eeq
as expected from (\ref{Mkcorrs}). In the large-$N$ limit, it is readily seen
that $M^k=(M^1)^k$ where the moment $M^1$ is given by (\ref{gaussmom1}).

\subsection{Four-Fermi Vector Model}

In the simple Gaussian example above, the free energy is analytic in the
coupling constant $t$ and there are thus no critical points. In the general
case, however, the moment $M^1$, which completely specifies the solution of the
vector model at large-$N$, is determined as the solution of a $K$-th order
polynomial equation and, from (\ref{MkdF0}), the free energy $F_0$ will have a
non-analytic cut structure leading to some sort of critical behaviour. For
example, the simplest non-Gaussian model is defined by the quadratic potential
\beq
V(z)=tz+\frac{g}{2}z^2
\label{4fermipot}\eeq
The propagator $\omega(z)$ can be computed formally at finite $N$ by evaluating
the integral
\beq
\omega(z)=\frac{N}{z}z^{-N}\e^{-N(tz+gz^2/2)}\int_{z_0}^zdw~w^N(t+gw+gM^1)
\e^{N(tw+gw^2/2)}
\label{4fermiomN}\eeq
which, after shifting the integration variable $w$, can be expressed in terms
of error functions.

For $p=1$, the 2 relations (\ref{momalg}) and (\ref{momrec}) become
\beq
tM^1+gM^2=1~~~~~,~~~~~tM^2+gM^3-(1-1/N)M^1=0
\label{4fermirecrelns}\eeq
In the large-$N$ limit, (\ref{4fermirecrelns}) yields a
quadratic equation for the moment $M^1$,
\beq
tM^1+g(M^1)^2-1=0
\label{quadM1}\eeq
Note that this quadratic equation also follows from the large-$N$ limit of the
generating function (\ref{4fermiomN}),
\beq
\omega(z)=\frac{1}{z-M^1}=\frac{t+g(M^1+z)}{tz+gz^2-1}
\label{4fermiomlargeN}\eeq
Expanding both sides of (\ref{4fermiomlargeN}) to order $1/z^3$ and equating
coefficients, we recover (\ref{quadM1}). The solution of (\ref{quadM1}) which
is regular at $g=0$ (and thus consistently reproduces the Gaussian solution
(\ref{gaussmom1}) when $g\to0$) is
\beq
M^1=\frac{t}{2g}\left(1-\sqrt{1+\frac{4g}{t^2}}\right)
\label{4fermiM1}\eeq
The $N=\infty$ free energy can be obtained by integrating up the identity
$M^1=\frac{\partial F_0^{(0)}}{\partial t}$. The second set of relations in
(\ref{4fermirecrelns}) at finite-$N$ then yield a relation among the various
partial derivatives of $F_0$ which can be used to recursively construct the
large-$N$ expansion of the vector model order by order in $1/N$. This approach
was advocated for the scalar version of the four-Fermi model above in
\cite{schel}. This construction and the physical relevance of the ensuing
critical behaviour is the topic of the next Subsection.

\subsection{Critical Behaviour and the $\frac{1}{N}$-expansion}

One of the nicest features of vector models in general is that one can
straightforwardly construct the $\frac{1}{N}$-expansion of the free energy
explicitly. In the fermionic case, this is an especially useful probe of the
characteristics of the more complicated fermionic matrix theories which are
hoped to describe 2-dimensional quantum gravity coupled to some sort of
fermionic matter placed at the vertices of the discretization. We shall
conclude our discussion of fermionic vector models in this Subsection by
examining the explicit critical behaviour and double scaling limit of the
vector theories described above \cite{semsz1}. This will provide much
insight into some general ideas that will be encountered in the matrix models
later on where such explicit constructions have yet to be carried out.

\subsubsection{Random Polymer Models}

We start by examining what sort of random geometry theory the fermionic vector
model describes. For completeness, let us begin by briefly reviewing the status
of the $O(N)$ vector model with partition function \cite{vec2,nishyon,schel}
\beq
Z_S(t,g;N)=\int_{-\infty}^{+\infty}\prod_{i=1}^Nd\phi_i~\exp\left
\{-t\sum_{i=1}^N\phi_i^2+\frac{g}
{N}\left(\sum_{i=1}^N\phi_i^2\right)^2\right\}
\label{bosevec}\eeq
The model (\ref{bosevec}) isinvariant under the orthogonal transformation
\beq
\phi_i\to \sum_{j=1}^NS_{ij}\phi_j~~~{\rm with}~~~S\in O(N)
\label{orthtransf}\eeq
The continuous (connected) $SO(N)$ part of the full orthogonal group symmetry
in (\ref{orthtransf}) is the analog of the $GL(N,\IC)$ symmetry
(\ref{ctsvecsymm}), while the discrete reflection symmetry $\phi_i\to-\phi_i$,
for any $i$, is the bosonic analog of the chiral symmetry (\ref{chiralsymvec}).
The
symmetry (\ref{orthtransf}) restricts the observables of the theory to those
which are functions of
\beq
\phi^2\equiv\sum_{i=1}^N\phi_i^2
\label{phisquared}\eeq

When the coupling constants $t$ and $g$ are positive, it is straightforward to
show that the formal expression (\ref{bosevec}) counts the number of randomly
branching polymers, both those with a tree-like structure and those
with arbitrarily many loops. The Feynman graphs produced in this case do not
have enough structure to specify a Riemann surface \cite{fgz} and instead they
describe `discrete filamentary surfaces'. As we shall see below, this is a
consequence of the behaviour of the susceptibility of these models which
exhibit a positive string constant indicative of a fragmentation of the
surface (into a branched polymer phase in this case) \cite{amdur}. The fact
that the integral (\ref{bosevec}) is divergent is a reflection of the
divergence of the statistical sum.

This statistical sum over polymers coincides with the expansion of the
free energy
\beq
F_S\equiv -\frac{1}{N}\log\left( \frac{ Z_S(t,g;N)}{ Z_S(t,0;N)}\right)
\label{freeS}\eeq
in Feynman diagrams. The propagator is
\beq
\left\langle\!\left\langle\phi_i\phi_j\right\rangle\!\right\rangle
\equiv\frac{\int\prod_kd\phi_k~
\phi_i\phi_j\e^{-t\phi^2}}{\int\prod_kd\phi_k~\e^{-t\phi^2}}=\delta_{ij}~~~~~,
\label{scalarpropvec}
\eeq
the vertex is
\beq
\left\langle\!\left\langle
\phi_i\phi_j\phi_k\phi_l\right\rangle\!\right\rangle=
\frac{1}{(2t)^2}\left(\delta_{ij}\delta_{kl}+
\delta_{il}\delta_{jk}+\delta_{ik}\delta_{jl}\right)
\label{4ptvec}\eeq
and the free energy is the sum of all connected diagrams with four-point
couplings. The standard rules for evaluating the Gaussian integrals as in
(\ref{scalarpropvec}) and (\ref{4ptvec}) are given by the Wick contraction
rules in which each pairing of vector components is assigned a delta-function
in the index contraction. The dual graphs to these Feynman diagrams, defined by
associating a vertex (molecule) in the center
of each of the scalar loops and lines (bonds) connecting vertices by crossing
each of the Feynman 4-point couplings, are random walk diagrams
\cite{anderson,dupl,nishyon,semsz1}. The number of molecules $n$ appears as the
power in $N^{n-1}$ and corresponds to the number of index loops of the Feynman
graphs. The number of bonds $b$ is given by the power in $(g/Nt^2)^b$ and
corresponds to the number of vertices, or equivalently contacts of these index
loops.

The random polymer model generated by (\ref{freeS}) is therefore of the form
\beq
F_S=-\sum_{{\cal P}_{b,\ell}}N^{-\ell}\left(\frac{g}{t^2}\right)^b
\label{statsumfreeS}\eeq
where the sum is over all polymer graphs ${\cal P}_{b,\ell}$ with $b$ bonds and
\beq
\ell=b-n+1
\label{polymerloops}\eeq
loops. Thus the vector model partition function is the generating function for
the number of polymer configurations with $b$ bonds and $\ell$ loops.
(\ref{statsumfreeS}) identifies the parameters of the vector model with the
discretized action terms for a random walk model as
\cite{amdur1,amdur2,amdur3,nish}
\beq
\frac{1}{N}=\e^\mu~~~~~,~~~~~\frac{g}{t^2}=\e^{-L}
\label{fuglength}\eeq
where $\mu$ is the fugacity and $L$ the length of the branched polymer chain.
Note that the sum (\ref{statsumfreeS}) includes the self-bonding polymers which
are generated by Wick contracting several propagators into single loops (as
opposed to multi-loops) and occur in the expansion only for $\ell\geq1$
\cite{nishyon,semsz1}.

{}From an analytic point of view, the perturbative expansion of the
partition function
\beq \frac{Z_S(t,g;N)}{Z_S(t,0;N)}=
\sum_{n\geq0}\frac{1}{n!}\left(\frac{g}{N}\right)
^n\left\langle\!\!\left\langle
\left(\phi^2\right)^{2n}\right\rangle\!\!\right\rangle
\eeq
is completely determined by the Gaussian moments
\beq
\left\langle\!\!\left\langle
\left(\phi^2\right)^{2n}\right\rangle\!\!\right\rangle
=\frac{1}{Z_S(t,0;N)}\left(\frac{\partial^2}{\partial
t^2}\right)^nZ_S(t,0;N)=t^{N/2}\left(\frac{
\partial^2}{\partial t^2}\right)^nt^{-N/2}
\label{gaussmomO(N)}\eeq
as
\beq
\frac{Z_S(t,g;N)}{Z_S(t,0;N)}=\sum_{n=0}^{N_\Lambda}
\frac{(N+4n-2)!!}{2^{2n} n!(N-2)!!}\left(\frac{g}{Nt^2}\right)^n
\label{zscutoff}\eeq
Here we have introduced an ``ultraviolet" cutoff $N_\Lambda\in\IZ^+$
to make the partition function well-defined. In the limit
$N_\Lambda\to\infty$, the series (\ref{zscutoff}) is a non-Borel
summable asymptotic series reflecting the divergence of the original
integral and also the divergence of the statistical sum.
As in the case of random surfaces, even though the
series is divergent, if arranged as a power series in $1/N$, rather
than $g$, the terms in this series are individually convergent and it
is the sum over genera $\ell$ which is asymptotic. With the cutoff $N_\Lambda$
in (\ref{zscutoff}), the partition function is an analytic function of $N$, but
it is only well-defined at $N=\infty$ when this ultraviolet cutoff is removed.

The random polymer model above can also be generated by the fermonic vector
model with partition function
\beq Z_F(t,g;N)=\int
d\psi~d\bar\psi~\e^{t\ps2-\frac{g}{2N}(\ps2)^2}
\label{part4fermivec}\eeq
As we shall now see, the model (\ref{part4fermivec}) possesses a random
geometry interpretation similar to that of the $O(N)$
vector model. A main difference with the $O(N)$ vector model is that the
integration over Grassmann variables in the generating function
(\ref{part4fermivec}) for the polymers is a well-defined finite polynomial in
the coupling constants
$g$ and $t$. The dimension $N$ itself provides a cutoff on the number
of terms (and polymers) in the Feynman diagram expansion of
(\ref{part4fermivec}). Here, rather than making the partition function
integration well-defined as in the bosonic case, the large-$N$ limit is needed
to generate the full ensemble of randomly-branched chains.

The Feynman diagrams for the fermion vector theory (\ref{part4fermivec}) have
propagator
\beq
\left\langle\!\!\left\langle\bar\psi_i\psi_j\right\rangle\!\!\right\rangle
\equiv\frac{\int
d\psi~d\bar\psi~\bar\psi_i\psi_j\e^{t\ps2}}{\int
d\psi~d\bar\psi~\e^{t\ps2}}=\delta_{ij}
\label{4fermiprop}\eeq
and the four-Fermi interaction vertex is
\beq
\left\langle\!\!\left\langle\bar\psi_i\psi_j\bar\psi_k\psi_l
\right\rangle\!\!\right\rangle
=\frac{1}{t^2}\left(\delta_{ij}\delta_{kl}-\delta_{il}\delta_{jk}\right)
\label{4fvertex}\eeq
These Feynman rules have the same graphical representation as for the $O(N)$
vector model above with a left-handed orientation for the lines (ingoing lines
into a vertex for $\bar\psi_i$ components and outgoing lines for $\psi_j$
components). Now the graphs are formed from all connected 4-point diagrams
which preserve this orientation. The Feynman rules also associate a factor of
$-1$ to each fermion loop in a Feynman graph. The Wick rules for the evaluation
of the Gaussian fermionic integrals such as (\ref{4fermiprop}) and
(\ref{4fvertex}) associate delta-functions in the indices for each contraction
of the form $\ps2$ and a minus sign for an interchange in the order of
$\psi\bar\psi$. Notice that these Gaussian moments can likewise be obtained
from a generating functional
\beq
Z_0(\eta,\bar\eta)=\frac{\int
d\psi~d\bar\psi~\e^{Nt\ps2+\bar\eta\psi+\bar\psi\eta}}{\int
d\psi~d\bar\psi~\e^{Nt\ps2}}~~~~~,
\label{vecmomgenfn}\eeq
where $\eta$ and $\bar\eta$ are also independent Grassmann-valued vectors, as
\beq
\left\langle\!\!\left\langle
\prod_k\bar\psi_{i_k}\psi_{j_k}\right\rangle\!\!\right\rangle=\prod_k\frac{
\partial}{\partial\eta_{i_k}}\frac{\partial}{\partial\bar\eta_{j_k}}Z_0(\eta,
\bar\eta)\biggm|_{\eta=\bar\eta=0}=\prod_k\frac{
\partial}{\partial\eta_{i_k}}\frac{\partial}{\partial\bar\eta_{j_k}}\e^{-\bar
\eta\eta/Nt}\biggm|_{\eta=\bar\eta=0}
\label{fermavggenfn}\eeq
which also leads to the fermionic Feynman-Wick rules with the appropriate minus
signs.

The perturbative expansion of (\ref{part4fermivec})
\beq
\frac{Z_F(t,g;N)}{Z_F(t,0;N)}=\sum_{n\geq0}\frac{(-1)^n}{n!}\left(\frac{g}{2N}
\right)^n\left\langle\!\!\left\langle(\ps2)^{2n}\right\rangle\!\!\right\rangle
\eeq
is completely determined by the normalized Gaussian moments
\beq
\left\langle\!\!\left\langle(\ps2)^{2k}\right\rangle\!\!\right\rangle
=\frac{1}{Z_F(t,0;N)}\left(\frac{\partial^2}{\partial
t^2}\right)^kZ_F(t,0;N)=t^{-N}\left(\frac{\partial^2}{\partial t^2}\right)^kt^N
\label{gaussvecmom}\eeq
to be \beq \frac{Z_F(t,g;N)}{Z_F(t,0;N)}=\sum_{n=0}^{N_2}(-1)^n
\frac{N!}{n!(N-2n)!}\left(\frac {g}{2Nt^2}\right)^n
\label{fermpert}\eeq
where $N_2=N/2$ (respectively $(N-1)/2$) when $N$ is even (odd). It is
straightforward to check that (\ref{fermpert}) coincides with (\ref{z0}) for
the four-Fermi vector potential. The
perturbation series is a finite sum which represents the same sort of random
walk distribution as in the scalar theory above except that it only includes
polymers with up to $N_2$ bonds. From a diagrammatic point of
view, the alternating nature of the series arises from the minus signs
associated with fermion loops.  Term by term, this series can be made
identical with the terms of same order in $g$ in (\ref{zscutoff}) by
the analytical continuation $N\rightarrow -N/2$ in (\ref{fermpert}) (so that
$N_\Lambda=N_2$ in (\ref{zscutoff})). The factor of 2 is associated with the
doubling of degrees of freedom in the fermionic case. Thus, after the
substitution $N\rightarrow N/2$, the large $N$ expansion of the fermionic
vector model is identical to that of the $O(N)$ vector model {\it except} that
it is an alternating series in $1/N$. The coefficient of $1/N^\ell$ in
the former and $1/(-N)^\ell$ in the latter are identical. Now, however, the
alternating nature of the fermionic vector series makes its $N\to\infty$ limit
Borel summable, and as such it defines a better behaved statistical theory.

Notice that the combinatorical factors occuring in the Feynman diagram
expansion of this fermionic vector model are identical to those obtained in the
{\it complex} vector model
\beq
Z_C(t,g;N)=\int\prod_{i=1}^Nd\phi_i~d\phi_i^*~\exp\left\{-t\sum_{i=1}^N\phi_i^*
\phi_i+\frac{g}{2N}\left(\sum_{i=1}^N\phi_i^*\phi_i\right)^2\right\}
\label{complexvecpart}\eeq
where the integration is over the whole of $\IC^N\simeq\IR^{2N}$. The model
(\ref{complexvecpart}) is invariant under the unitary transformations
\beq
\phi_i\to\sum_{j=1}^NU_{ij}\phi_j~~~{\rm with}~~~U\in U(N)
\eeq
This symmetry, along with the discrete charge conjugation symmetry
$\phi_i\to\phi_i^*$ for any $i$, restricts the observables of the model to
those which are functions only of $\sum_i\phi_i^*\phi_i$. Although the complex
vector model (\ref{complexvecpart}) exhibits the same doubling of degrees of
freedom as in the fermionic case as well the same Wick contraction
(non-vanishing only among $\phi_i^*$ and $\phi_j$) and Feynman rules without
the minus signs, it still leads to an asymptotic series expansion as in the
case of the $O(N)$ vector model above. In analogy with the matrix case, it can
be thought of as describing the statistical mechanics of `checkered'
filamentary surfaces.

\subsubsection{Double Scaling Limit}

The large $N$ expansion of the $O(N)$ vector model above is a saddle point
computation of the integral (\ref{bosevec}). After integration over angular
variables, the partition function (\ref{bosevec}) can be written in terms of
the
radial coordinate in Euclidean $N$-space as
\beq
Z_S(t,g;N)=\frac{2\pi^{N/2}}{\Gamma(N/2)}\int_0^\infty
d\phi~\phi^{N-1}\e^{-t\phi^2+\frac{g}{N}\phi^4}
\label{zsradial}\eeq
In the infinite-$N$ limit, which counts the tree-graphs ($\ell=0$),
the integral (\ref{zsradial}) can be evaluated using the saddle-point
approximation. Rescaling $\phi\to\phi/\sqrt{N}$ the stationary condition for
the effective action $-Nt\phi^2+gN\phi^4+N\log\phi$ in (\ref{zsradial})
is \beq 2t\phi^2-4g\phi^4=1
\label{statcond}\eeq
The solution of (\ref{statcond}) which is regular at $g=0$ and which minimizes
the effective action is
\beq
\phi_0^2=\frac{t}{4g}\left(1-\sqrt{1-\frac{4g}{t^2}}\right)
\label{saddlepoint}\eeq
The tree-level free energy is then the value of the effective action
in (\ref{zsradial}) evaluated at the saddle-point (\ref{saddlepoint}),
\beq\new{\begin{array}{ll}
F_S^{(0)}&\equiv\lim_{N\to\infty}-\frac{1}{N}\log Z_S(t,g;N)\\&=\frac{1}{2}
\left\{\frac{1}{2}+\frac{t}{4g}\left(t-\sqrt{t^2-4g}\right)-\log
\left(\frac{t}{4g}-\frac{1}{4g}\sqrt{t^2-4g}\right)\right\}\end{array}}
\label{fstree}\eeq

The free energy (\ref{fstree}) becomes non-analytic at the critical point
\beq
g=g_c\equiv t^2/4
\label{critptvec}\eeq
where the 2 solutions of the quadratic equation (\ref{statcond}) coalesce.
There the minimum of the effective action in (\ref{zsradial}) disappears and it
becomes unbounded \cite{vec1,vec2,eyal,zinn1}, so that the saddle point
solution is no longer valid. There is a second order phase transition at
the coupling $g=g_c$ with susceptibility exponent
\cite{amdur1,amdur2,amdur3,anderson,dupl,nishyon}
\beq
\gamma_{\rm str}^{(0)}=1/2
\label{polymerstr1/2}\eeq
This critical point is identified as the ``continuum" limit of the random
polymer theory where the number of branches, and hence the lengths, of the
tree-graphs becomes infinite. The higher-loop contributions (molecular
networks) can be found in \cite{schel} and their ``continuum" limit is
associated with
an infinite number of molecules, thus tracing out a continuum filamentary
surface \cite{amdur1,amdur2,amdur3,anderson,dupl}. Notice that since a negative
string susceptibility constant is indicative of a locally 2-dimensional random
geometry, the critical exponent (\ref{polymerstr1/2}) is inherently related to
a dimensionally reduced discretization.

We shall now see that the fermionic vector model
(\ref{part4fermivec}) possesses a similar critical behaviour. To explicitly
carry out the $\frac{1}{N}$-expansion of
the fermionic vector model, we introduce a scalar Hubbard-Stratonovich field
$\varphi$ defined by the identity
\beq
1=\int d\varphi~\e^{-\frac{g}{2N}(\varphi+i\ps2)^2}
\label{hubdef}\eeq
into the partition function integral (\ref{part4fermivec}) to write it as
\beq
Z_F(t,g;N)=\int d\varphi~\e^{-g\varphi^2/2N}\int
d\psi~d\bar\psi~\e^{(t-ig\varphi/N)\ps2}=N!\int
d\varphi~(t-ig\varphi/N)^N\e^{-g
\varphi^2/2N}
\label{zfhub}\eeq
When $N\to\infty$ the integral in (\ref{zfhub}) is determined by the
saddle-point value of $\varphi$. Rescaling $\varphi\to\varphi/N$, this can be
found from the stationary condition for the effective action
\beq
S_F(\varphi)\equiv-Ng\varphi^2/2+N\log(t-ig\varphi)
\label{effaction}\eeq
appearing in (\ref{zfhub}) which is \beq tg\varphi-ig^2\varphi^2+ig=0
\label{fermsaddle}\eeq
The solution of (\ref{fermsaddle}) which is regular at $g=0$ is \beq
\varphi_0=\frac{t}{2ig}\left(1-\sqrt{1-\frac{4g}{t^2}}\right)
\label{hub0}\eeq
Substituting (\ref{hub0}) into (\ref{effaction}) we get the tree-level
fermionic free energy
\beq
F_F^{(0)}=\frac{1}{2}-\frac{t}{4g}\left(t-\sqrt{t^2-4g}\right)-\log\left
(\frac{t}{2}+\sqrt{t^2-4g}\right)
\label{fermfree0}\eeq
Note that the saddle points in (\ref{saddlepoint}) and (\ref{hub0}) are related
by the correspondence
\beq
\phi^2=\frac{i}{2}\varphi
\label{saddlerelns}\eeq
which makes more precise the analytical continuation between the bosonic and
fermionic models discussed above. Notice also that the free energy
(\ref{fermfree0}) is related to the free energy (\ref{fstree}) of the scalar
model as
\beq
F_F^{(0)}=1-2F_S^{(0)}
\label{freeencorr}\eeq
as anticipated since the large-$N$ limit of the 2 models represents the same
combinatorical problem of enumerating tree-graphs.

The higher-loop contributions (which count the polymer networks with a given
number of molecules) can be found by carrying out the saddle point calculation
of the integral (\ref{zfhub}) to higher orders. For this, we decompose the
Hubbard-Stratonovich field as
\beq
\varphi=\varphi_0+\varphi_q
\eeq
and expand the action (\ref{effaction}) in a Taylor series about the
saddle-point value (\ref{hub0}) in terms of the fluctuation fields $\varphi_q$.
Using the saddle-point equation (\ref{fermsaddle}) when evaluating the
higher-order derivatives $S_F^{(n)}(\varphi_0)$, this Taylor series is found to
be
\beq
S_F(\varphi)=S_F(\varphi_0)-\frac{N}{2}\left(g+g^2\varphi_0^2\right)
\varphi_q^2-N\sum_{n=3}^\infty\frac{(-g\varphi_0)^n}{n}\varphi_q^n
\label{actiontaylor}\eeq
The genus 1 free energy is then obtained from the fluctuation determinant that
arises from Gaussian integration over the quadratic part in $\varphi_q$ of
(\ref{actiontaylor}),
\beq
F_F^{(1)}=\frac{1}{2N}\log\left(g+g^2\varphi_0^2\right)=\frac{1}{2N}\log\left(
2g-\frac{t^2}{2}+\frac{t}{2}\sqrt{t^2-4g}\right)
\label{fermfree1}\eeq
which also agrees with the 1-loop free energy of the $O(N)$ vector model
\cite{schel}.

The $\frac{1}{N}$-expansion of the fermionic free energy also becomes
non-analytic at the critical point $g=g_c=t^2/4$. It exhibits the same critical
behaviour as the $\phi^4$ theory above and it therefore lies in the same
universality class as this statistical model. Notice that this critical
behaviour, and also the free energies above, could have been obtained as well
from the results of the previous Subsection with the identification
$\phi^2\leftrightarrow M^1/2$ there. It is straightforward to carry
out the double-scaling limit of the fermionic vector model in much the same way
as in the bosonic case \cite{vec2,schel,nishyon}. This limit is associated with
the continuum limit of the polymer network at $N\to\infty$, $g\to g_c$ in such
a way that a coherent contribution from all orders of the perturbative and
$1/N$ expansions is obtained. We approach the critical point $g_c$ by defining
a dimensionless ``lattice spacing'' $a$ by
\beq
at^2=g_c-g
\eeq
and taking the continuum limit $a\to0$. With the rescalings mentioned above,
the contribution to an arbitrary $\ell$-loop vacuum diagram with $\ell\geq1$ is
$N^{-b}(\sqrt{a})^{-b}N^n=N^{1-\ell}(\sqrt{a})^{1-\ell-n}$. As shown in
\cite{vec2}, the maximum number of vertices that a 4-point polymer diagram
with $\ell$ loops can have is $n=2(\ell-1)$, so that the most singular
behaviour of an $\ell$-loop diagram in the continuum limit $a\to0$ is
$(Na^{3/2})^{1-\ell}$. The proper continuum limit wherein a finite contribution
from arbitrary genus polymers is obtained is thus the ``double-scaling" limit
where $N\to\infty$ and $a\to0$ in a correlated fashion so that the renormalized
``cosmological constant" (or ``linear string tension")
\beq
\Lambda_R\equiv Na^{3/2}
\eeq
remains finite. The double scaling limit enables an explicit construction of
the genus expansion of the continuum polymer theory from the vector model.

We can now take the double scaling limit of the partition function
(\ref{part4fermivec}) and write it as a loop expansion in the linear string
tension $\Lambda_R$. However, as noted for the $O(N)$ vector model \cite{vec2},
the tree and 1-loop contributions are singular in this limit. The saddle-point
value (\ref{hub0}) can be written in terms of the lattice spacing as
\beq
g\varphi_0=2(1-2\sqrt{a})/it
\eeq
from which it follows that the genus 0 and 1
free energies (\ref{fermfree0}) and (\ref{fermfree1}) are given by
\beq\new{\begin{array}{c}
NF_F^{(0)}\sim-\frac{N}{2}-N\log\left(\frac{t}{2}\right)+6N^{1/3}\Lambda_R
^{2/3}-\frac{32}{3}\Lambda_R\\
NF_F^{(1)}\sim\frac{1}{2}\log\left(\frac{t^2}{2}\right)+\frac{1}{2}\log
\left(\frac{2\Lambda_R^{1/3}}{N^{1/3}}\right)
\end{array}}
\label{nonuniv}\eeq
in the continuum limit $a\to0$. The $\Lambda_R$-dependent terms in
(\ref{nonuniv}) diverge in the double scaling limit and represent a
non-universal behaviour of the statistical polymer system. The tree-level and
one-loop order Feynman diagrams should therefore be subtracted in the
definition of the double-scaling limit leading to a renormalized partition
function $Z_R(\Lambda_R,t)$ that only contains contributions from the
$\ell$-loop diagrams with $\ell\geq2$.

This renormalized partition function is obtained by integrating over that part
of the action involving $n\geq3$ vertices in the fluctuation field $\varphi_q$
weighted against the Gaussian form in (\ref{actiontaylor}). To pick out the
finite contribution in the double-scaling limit, we rescale the fluctuation
field as $\varphi_q\to2N^{1/3}\Lambda_R^{1/6}g_c^{-1/2}\varphi_q$ and note that
with this rescaling we have
\beq
-\frac{N}{2}\left(g+g^2\varphi_0^2\right)\varphi_q^2\to-\frac{1}{2}\varphi_q^2
{}~~~,~~~-N(-g\varphi_0)^n\varphi_q^n\to-\frac{N(g_c)^{n/2}}{2^n}
\left(\frac{it}{2}\right)^nN^{-n/3}\Lambda_R^{-n/6}\varphi_q^n
\label{rescaling}\eeq
in the continuum limit $a\to0$. The $n\geq4$ vertex terms in
(\ref{rescaling}) vanish in the double scaling limit, and therefore the exact
renormalized partition function in the double scaling limit is (up to
irrelevant normalization factors)
\beq
Z_R(\Lambda_R,t)=\frac{\int
d\varphi_q~\e^{-\frac{1}{2}\varphi_q^2+\frac{it^6}{512}
\Lambda_R^{-1/2}\varphi_q
^3}}{\int
d\varphi_q~\e^{-\frac{1}{2}\varphi_q^2}}=\sum_{k=0}^\infty\frac{i^kt^{6k}}
{k!(512)^k}~\Lambda_R^{-k/2}\left(\left\langle\!\!\left\langle
\phi^{3k}\right\rangle\!\!\right\rangle\biggm|_{N=1,t=\frac{1}{2}}\right)
\label{zren}\eeq
The Gaussian moments in (\ref{zren}) can be evaluated as in
(\ref{gaussmomO(N)}). The odd moments vanish, while the even moments yield a
factor $(3k-1)!!$. Thus the double-scaled renormalized partition function
admits the exact genus expansion
\beq
Z_R(\Lambda_R,t)=\sum_{\ell=0}^\infty(-1)^\ell
{}~t^{12\ell}~\frac{(6\ell-1)!!}{(2\ell)!(512)^{2\ell}}~\Lambda_R^{-\ell}
\label{renloopexp}\eeq

The partition function (\ref{renloopexp}) has a similar structure as that in
the $O(N)$ vector model where the genus expansion is an asymptotic series with
zero radius of convergence \cite{vec2}. In the fermionic case, however, the
genus expansion is an alternating sum, and is therefore Borel summable. The
convergence of the sum over genera is easily seen in the integral expression
(\ref{zren}) where the unbounded cubic term contains a factor of $i$ which
makes the overall integration there finite. The Borel summability of the genus
expansion is a feature unique to the fermionic models that does not usually
occur for random geometry theories. In this sense, the fermionic vector model
represents some novel discretized surface theory in which the topological
expansion uniquely specifies the generating function of the statistical theory.
The identification $-\frac{1}{N}=\e^\mu$ in the fermionic case suggests that
the associated random polymer theory has a complex-valued ``fugacity"
$\mu=i\pi+\mu_0$, $\mu_0\in\IR$, with doubly-degenerate degrees of freedom at
each vertex. It would be
interesting to give these fermionic properties of the theory a direct
interpretation in terms of a random geometry model. From an analytic point of
view, the genus sum alternates relative to that of the $O(N)$ vector model
because the saddle-point (\ref{hub0}) is imaginary in the fermionic vector
model (\ref{part4fermivec}) so that its saddle-point expansion is the
analytical
continuation $\phi_0^2=\frac{i}{2}\varphi_0$ of that for the scalar model.

The above analysis can be straightforwardly generalized to an interaction of
the form $g(\ps2)^K$, which will then represent a random polymer model with up
to $2K$-valence vertices. The critical behaviour is the same as that in a
$\phi^{2K}$ scalar vector model and leads to the same susceptibility exponent
$\gamma_{\rm str}^{(0)}=\frac{1}{2}$, i.e. such a theory of random polymers is
universal. To generate more complicated polymer models, for instance those with
matter degrees of freedom at the vertices of the discretization
\cite{amdur1,amdur2,amdur3,nish}, one must study vector models with more
complicated interactions, such as those which were considered quite generally
at the beginning of this Section. To treat such models defined as in
(\ref{part0}) at $N=\infty$, we could use the first part of the identity
(\ref{part0derivid}) to write the partition function as
\beq
Z_0=\frac{(-i)^NN!}{2\pi}\int dz~dw~\e^{NV(z)+iwz+N\log w}
\label{partlargeN}\eeq
If the potential is a polynomial of degree $m$ ($\kappa=0,K=m$ in (\ref{pot})),
then we can rescale $z\to z/N$ and the coupling constants $g_k\to N^k\cdot g_k$
simultaneously so that the effective action in (\ref{partlargeN}) is
$NV(z)+iNwz+N\log w$. The integral (\ref{partlargeN}) at large-$N$ is
determined by the saddle-point value of this effective action. In the
2-dimensional complex space of the variables $w$ and $z$, the stationary
conditions are
\beq
V'(z)+iw=0~~~~~,~~~~~iz+1/w=0
\label{2saddles}\eeq
which can be combined into the single equation
\beq
zV'(z)=1
\label{ONstatgen}\eeq

The equation (\ref{ONstatgen}) is identical to the stationary condition for the
$O(N)$ vector model defined with potential $V(\phi^2)$ \cite{vec2,eyal,zinn1}.
Thus the critical behaviour of the fermionic vector model (\ref{part0}) is the
same as that for the $O(N)$ vector model with the same polynomial potential
(\ref{pot}). The genus expansion is generated by the 2-dimensional saddle-point
evaluation of the integral (\ref{partlargeN}). The imaginary saddle-point
values given by (\ref{2saddles}) will lead to an alternating genus expansion in
the double-scaling limit for the fermionic theory, leading to a Borel summable
polymer model, in contrast to the scalar case. In this case the critical point
is again that point in coupling constant space where the function $zV'(z)$
vanishes. This point coincides with the simple pole of residue 1 in the
propagator $\omega(z)$ in (\ref{omegavecsol}) and the solutions of the loop
equations in (\ref{M1algeq}) cease to exist. For a potential of the form
(\ref{pot}) with $\kappa=0,K=m$, we can adjust the coupling constants in such a
way that the critical point is a zero of $zV'(z)$ of order $m$. The leading
singularity of the free energy will then be $a^{(m+1)/m}$ \cite{vec2,nishyon}
which leads to the critical susceptibility exponent (c.f. (\ref{largeAdep}))
\beq
\gamma_{\rm str}^{(0)}=1-1/m~~~~~,~~~~~m=2,3,\dots
\eeq
This is the multicritical series for generalized random polymer systems in
dimension $D\geq0$ \cite{amdur1,amdur2,amdur3,nish} which interpolates between
the Cayley tree at $m=\infty$ with $\gamma_{\rm str}^{(0)}=1$ and the ordinary
random walk we discussed earlier at $m=2$ with $\gamma_{\rm
str}^{(0)}=\frac{1}{2}$. In the
$O(N)$ vector models, the former case would represent a phase of bosonic string
theory (as in the Penner model) in target space dimension $D\geq1$ while the
latter case would represent a phase
of pure 2-dimensional quantum gravity. In the general case, the potential
(\ref{pot}) leads to discrete filamentary surfaces which have vertices of even
valence up to $2m$. The $N^0$-component of the vector model free energy
represents the self-avoiding random walk (i.e. no loops, $\ell=0$), and it can
be computed by dividing the statistical sum by $N$ and then taking the $N\to0$
limit. This method of isolating the constant configurations in a random surface
model is known as the `replica trick' and it will be encountered again in
Section 7. It would be interesting to determine precisely what physical systems
the fermionic vector models represent in the continuum limit.

Given the convergence properties of the fermionic vector models, they can be
combined with bosonic models to obtain supersymmetric-type vector theories
representing new sorts of generating functions for random geometry theories
\cite{semsz1}. Some supersymmetric generalizations of the $O(N)$ vector
model have been studied in \cite{dadda,mckane,ohta,schnitzer} and it would be
interesting to find physical applications of these models. We shall discuss
some of these supersymmetric theories in Section 7. The main lessons we wish to
draw here concerning our ``toy model" analysis of this Section is that random
geometry models involving fermionic degrees of freedom admit solutions
analogous to those of the more conventional bosonic theories, except that the
overall models have better convergence properties and lead to better defined
statistical theories. In particular, the Borel summability will be argued later
on to hold as well in the adjoint fermion one-matrix models. As we shall see,
this is expected to be only true for odd polynomial potentials \cite{akm} as it
is only in that case that the matrix model possesses a chiral symmetry and
imaginary endpoints for the support of the spectral distribution (i.e. an
analytical continuation of a Hermitian spectral density). In the case of the
simpler fermionic vector models the partition function is always invariant
under chiral transformation of the fermionic vector components. Furthermore, in
the case of fermionic matrix models the correspondence with a scalar theory is
more complicated -- a polynomial fermion model can be analytically continued to
a Hermitian matrix model with a generalized Penner potential. It should
therefore represent a Borel summable generating function for the virtual Euler
characteristics of the discretized moduli spaces of Riemann surfaces (rather
than just the generating function for a random surface triangulation itself).
In the vector case, the fermionic model represents the same type of random
surface theory as the corresponding $O(N)$ vector field theory. The results of
the vector model analysis above therefore clarify and confirm many of the
matrix model arguments that are presented.

\section{Adjoint Fermion One-matrix Models}

In this Section we will analyse in detail the adjoint fermion one-matrix
model (\ref{part1}). It can be viewed as a $D=0$ dimensional quantum field
theory of a self-interacting Dirac fermion which transforms under the adjoint
action of a ``colour'' gauge group. The model possesses the symmetry
\beq
\psi\rightarrow U\psi V^{-1}~~,~~\bar\psi\rightarrow V\bar\psi U^{-1}
{}~~~{\rm with}~~~\{U,V\}\in GL(N,\IC)\otimes GL(N,\IC)
\eeq
In spite of
this large degree of symmetry, it is not possible to diagonalize a matrix with
anticommuting elements. Thus, unlike the more familiar Hermitian one-matrix
models \cite{biz,bipz}, the model (\ref{part1}) cannot be written as a
statistical theory of eigenvalues. Nevertheless, in the large-$N$ limit it
shares many of the properties of such a theory. Furthermore, the large degree
of symmetry restricts the observables to those which are essentially
invariant functions of $\bar\psi\psi$.

The chiral transformation
\beq
\psi\to\bar\psi~~~,~~~\bar\psi\to-\psi
\label{ch}\eeq
is the analog of the reflection symmetry $\phi\to-\phi$ in a Hermitian 1-matrix
model \cite{bipz}. The invariant traces transform under (\ref{ch}) as
\beq
\tr(\bar\psi\psi)^k\to(-1)^{k+1}\tr(\bar\psi\psi)^k
\eeq
Thus, (\ref{ch}) is a symmetry of the model when the potential is an {\it odd}
polynomial. It is interesting that the analog in Hermitian matrix models is the
reflection symmetry when the potential there is an even polynomial and in that
case one expects the eigenvalue distribution to be symmetric and all odd
moments of the distribution vanish, $\langle\tr\phi^{2k+1}\rangle=0$
\cite{bipz} \footnote{\baselineskip=12pt Such symmetric matrix models are
usually referred to as `reduced' matrix models.}. In the
present model, this symmetry introduces the feature that all {\it even} moments
vanish,
\beq
\langle\tr(\bar\psi\psi)^{2k}\rangle=0
\eeq

When $N$ is finite, because of the anticommuting property of the elements of
$\psi$ and $\bar\psi$, there is an integer $k_0\leq N^2$ such that
\beq
\langle\tr(\bar\psi\psi)^{k_0}\rangle=0
\eeq
There are also a finite number of non-zero correlators of the form $\langle
\prod_i\tr(\ps2)^{k_i}\rangle$.
In the large-$N$ limit, correlators of the matrix model factorize
\beq
\langle{\tr}f(\ps2){\tr}g(\ps2)\rangle=\langle\tr f(\ps2)\rangle\langle\tr g(
\ps2)\rangle+{\cal O}(1/N^2)
\eeq
This factorization property follows from the existence of a finite
large-$N$ limit for the correlators $\langle\tr(\bar\psi\psi)^k\rangle$ for
arbitrary polynomial potential $V(\ps2)=\sum_{n\geq1}\frac{g_n}{n}(\ps2)^n$,
since then the connected correlators are given by
\beq\new{\begin{array}{ll}
\langle\tr(\bar\psi\psi)^p\tr(\bar\psi\psi)^k\rangle_{\rm
conn}&\equiv\langle\tr(\ps2)^p\tr(\ps2)^k\rangle-\langle\tr(\ps2)^p\rangle
\langle\tr(\ps2)^k\rangle\\&=
\frac{1}{N^2}p\frac{\partial}{\partial g_p}\left\langle\tr(\bar\psi\psi)^k
\right\rangle\sim~\frac{1}{N^2}\end{array}}
\label{conncorrsN2}\eeq
and
\beq
\left\langle\tr(\ps2)^k\right\rangle=\frac{1}{N^2}k\frac{\partial}{\partial
g_k}\log Z_1
\label{1fermcorrs}\eeq
Factorization and symmetry imply that the large-$N$ limit of the model
is completely characterized by the set of correlators $\langle\tr(\bar\psi\psi)
^k\rangle$.

When $N$ is finite, the moment generating function
\beq
\omega(z)=\left\langle\tr\frac{1}{z-\bar\psi\psi}\right\rangle
=\sum_{k=0}^{N^2}\left\langle\tr(\bar\psi\psi)^k\right\rangle\frac{1}{z^{k+1}}
\label{gen}\eeq
has singularities only at the origin in the complex $z$-plane. The potential
$V(z)$ is now a source for the inverse Laplace transformation of the Wilson
loop
$\tr\frac{z}{z^2-\ps2}$,
\beq
\tr V(\ps2)=\int_{0-i\infty}^{0+i\infty}\frac{dz}{2\pi
i}~V(z)\tr\frac{z}{z^2-\ps2}
\label{Vres}\eeq
The set of moments can always be obtained from a (not unique)
distribution function $\rho$ with support in the complex plane
\beq
\langle\tr(\bar\psi\psi)^k\rangle=\int d\alpha~\rho(\alpha)\alpha^k~~~{\rm
with}~~~\int d\alpha~\rho(\alpha)=1
\label{mom}
\eeq
The support of $\rho$ can be deduced from the position of the
singularities of $\omega$ in (\ref{gen}).  When $N$ and therefore the
number of moments is finite the support of $\rho$ is concentrated near
the origin in the complex plane just as in (\ref{rhovec})
\beq
\rho(\alpha)=\langle\tr\delta(\alpha-\ps2)\rangle=
\sum_{k=0}^{N^2}\frac{1}{k!}\left\langle\tr(\bar\psi\psi)^k\right\rangle\left(-
\frac {\partial}{\partial\alpha}\right)^k \delta(\alpha)
\eeq
In the large-$N$ limit, the spectral function $\rho(\alpha)$ can be a
function with support on some contour in the complex plane.  The
distribution function $\rho$ is the analog in the fermionic matrix
model of the density of eigenvalues in Hermitian one-matrix models
as the quantity which completely specifies the solution of the model in the
infinite $N$ limit \cite{bipz}.

The generating functions for the connected correlators are
\beq
\omega_n(z_1,\ldots,z_n)=\left\langle\tr\frac{1}{z_1-\ps2}\cdots
\tr\frac{1}{z_n-\bar\psi\psi}\right\rangle_{\rm conn}
\label{multiloop}\eeq
and in the Hermitian case they are associated with the (inverse Laplace
transforms of) the sum over discretized open surfaces with $n$ boundaries
\cite{kaz}. When the potential is a polynomial
$V(\ps2)=\sum_{k\geq0}g_k(\ps2)^k$, the expansion of the multi-loop correlators
(\ref{multiloop}) in $\frac{1}{z_1},\dots,\frac{1}{z_n}$ can be obtained, using
(\ref{conncorrsN2}),(\ref{1fermcorrs}), from the free energy as
\beq
\omega_n(z_1,\ldots,z_n)={\cal L}(z_1)\cdots{\cal L}(z_n)\log Z_1
\label{multins}\eeq
by successive applications of the loop insertion
operator\footnote{\baselineskip=12pt Note that with this definition we have
${\cal L}(z)V(w)=\frac{1}{z-w}$ which acts as a delta-function when integrated
along the imaginary axis as in (\ref{Vres}).}
\beq
{\cal L}(z)=\frac{1}{N^2}\sum_{k\geq0}\frac{1}{z^{k+1}}\frac{\partial}
{\partial g_k}
\label{loopins}\eeq
The single-loop correlator $\omega_1(z)\equiv\omega(z)$ as in (\ref{omrho})
is analytic in $z$ away from the support of $\rho$ in the complex
plane. The distribution function can be determined as before by computing the
discontinuity (\ref{disc}) of $\omega(z)$ across its support. Notice that since
the signs of the actions in (\ref{part1}) and (\ref{penner}) are opposite (see
(\ref{pennerferm})), the connected correlators of the fermionic matrix model
alternate in sign relative to those of the generalized Hermitian Penner model
(\ref{penner}). This indicates that the large-$N$ genus expansion of the
fermionic matrix model (\ref{part1}) is an alternating series. As we shall see,
this feature will follow from the different boundary conditions that must be
used to define (\ref{part1}) and (\ref{penner}).

\subsection{Loop Equations}

The loop equation for the single-loop correlator $\omega(z)$ follows from
the identity
\beq
\int d\psi~d\bar\psi~\frac{\partial}{\partial\psi_{ij}}\left[\left(\psi\frac{
1}{z-\ps2}\right)_{k\ell}\e^{N^2\tr V(\ps2)}\right]=0
\label{loop1}\eeq
In contrast to Hermitian matrix models \cite{kaz}, the identity (\ref{loop1})
is exact for fermionic matrices. Dividing by $Z_1$ in (\ref{loop1}) and
expanding out the expectation values gives
\beq
0=\delta_{ik}\left\langle\left(\frac{1}{z-\ps2}\right)_{j\ell}\right\rangle-
\left\langle\psi_{k\ell}\left(\frac{1}{(z-\ps2)^2}\bar\psi\right)_{ji}\right
\rangle-N\left\langle\left(\psi\frac{1}{z-\ps2}\right)_{k\ell}\left(\bar\psi
V'(\ps2)\right)_{ji}\right\rangle
\label{loop2}\eeq
Setting $i=k,j=\ell$ and summing over $i,j=1\dots,N$ then leads to
\beq
-z\left(\omega(z)^2+\omega_2(z,z)\right)+2\omega(z)+\oint_{\cal C}
\frac{d\lambda}{2\pi i}~\frac{V'(\lambda)\lambda}{z-\lambda}\omega(\lambda)=0
\label{loopcont}\eeq
where the contour $\cal C$ encircles the cut (and possibly pole)
singularities of $\omega(z)$ with counterclockwise orientation and
(\ref{loopcont}) should again be solved with the boundary condition
(\ref{bcvec}). When the
potential is a polynomial of degree $K$ as in (\ref{pot}) (with $\kappa=0$),
then the contour integral in (\ref{loopcont}) can be obtained as in
(\ref{contour}) and the loop equation (\ref{loopcont}) becomes
\beq
-z\omega(z)^2+\left(2-z V'(z)\right)\omega(z)+V'(z)+P(z)=z\omega_2(z,z)
\label{loop}\eeq
where $P(z)$ is a polynomial of degree $K-2$
\beq
P(z)=\sum_{k=2}^K g_k\sum_{p=0}^{k-2}\left\langle\tr(\bar\psi\psi)^{k-1-p}
\right\rangle z^p
\label{polynomial}\eeq

Note that the loop equation (\ref{loopcont}) can also be derived from the
Schwinger-Dyson equations expressing the invariance of the partition function
(\ref{part1}) under arbitrary changes of variables. Under the field
transformations
\beq
\psi\to\psi\left(1+\epsilon\frac{1}{z(z-\ps2)}\right)~~~~,~~~~\bar\psi
\to\bar\psi~~~~~,
\label{transf}\eeq
where $\epsilon$ is an infinitesimal parameter, the integration measure in
(\ref{part1}) changes by
\beq
d\psi~d\bar\psi\to d\psi~d\bar\psi~\left(1-\epsilon\left[N^2\left(\tr\frac{1}
{z-\ps2}\right)^2-2N\tr\frac{1}{z(z-\ps2)}\right]\right)
\eeq
and then the invariance of (\ref{part1}) to first order in $\epsilon$ under the
transformations (\ref{transf}) leads directly to (\ref{loopcont}). Notice also
that the field transformation (\ref{transf}) is similar to the shift
(\ref{pennershift}) used to derive the loop equations of the Penner matrix
model \cite{akm}.

Factorization implies that the connected correlators are all suppressed by
factors of $1/N^2$ and the term on the right-hand side of the loop equation
(\ref{loop}) vanishes in the large-$N$ limit. Then the loop equation has
the solution
\beq
\omega(z)=\frac{1}{z}-\frac{V'(z)}{2}+\frac{1}{z}\sqrt{1+\left(\frac{zV'(z)}
{2}\right)^2+zP(z)}
\label{omcut}\eeq
where the sign of the square root is chosen to satisfy the asymptotic boundary
condition (\ref{bcvec}). The branches of the square root must be placed so that
it is negative near the origin in order to cancel the pole at $z=0$. If the
potential is a polynomial of order $K$, then the solution (\ref{omcut}) in
general will possess a square root singularity with $K$ branch cuts and the
spectral density $\rho$ will have $K$ contours in its support.

The simplest solution of the model is the
one-cut solution which assumes that the singularities of $\omega(z)$ consist
of only a single square root branch cut, so that the distribution function
$\rho$ has support only on one arc in the complex plane with endpoints at some
complex values $a_1$ and $a_2$. The simplest one-cut solution of the
homogeneous part of the equation
\beq
\omega(z+\epsilon_\perp)+\omega(z-\epsilon_\perp)=2/z-V'(z)
\label{homoeqn}\eeq
is $\sqrt{(z-a_1)(z-a_2)}$. Dividing (\ref{homoeqn}) through by this function
gives the discontinuity equation
\beq
\frac{\omega(z+\epsilon_\perp)}{\sqrt{(a_1-z)(a_2-z)}}-\frac{\omega(z-\epsilon_
\perp)}{\sqrt{(a_2-z)(z-a_1)}}=\frac{V'(z)-2/z}{\sqrt{(a_2-z)(z-a_1)}}
\eeq
from which it follows that the one-cut
solution for $\omega(z)$ can be represented in the form
\beq
\omega(z)=\oint_{{\cal C}_z}\frac{dw}{4\pi i}~\frac{V'(w)-2/w}{z-w}
\sqrt{\frac{(z-a_1)(z-a_2)}{(w-a_1)(w-a_2)}}
\label{omcont}\eeq
where the closed contour ${\cal C}_z$ encloses the support of the spectral
function but not the point $w=z$. The absence of any terms regular in $z$ in
(\ref{omcont}) follows from the large-$|z|$ behaviour of the 1-loop correlator.

The endpoints of the cut can then be found by
expanding (\ref{omcont}) in $\frac{1}{z}$ and imposing the asymptotic boundary
condition (\ref{bcvec}) on the solution
(\ref{omcont}), which leads to the two equations
\beq
0=\oint_{{\cal C}}\frac{dw}{2\pi i}~\frac{V'(w)-2/w}{\sqrt{(w-a_1)(w-a_2)}}
=\oint_{{\cal C}}\frac{dw}{2\pi i}~\frac{V'(w)}{\sqrt{(w-a_1)(w-a_2)}}
+\frac{2}{\sqrt{a_1a_2}}
\label{1cut1}\eeq
\beq
2=\oint_{{\cal C}}\frac{dw}{2\pi i}~\frac{wV'(w)-2}{\sqrt{(w-a_1)(w-a_2)}}
=\oint_{{\cal C}}\frac{dw}{2\pi i}~\frac{wV'(w)}{\sqrt{(w-a_1)(w-a_2)}}
\label{1cut2}\eeq
In particular these equations show that the points $a_1$ and $a_2$ cannot
lie on the real axis. If they did, the solution (\ref{omcont}) would have a
pole at $z=0$ and the contour $\cal C$ would encircle the origin. But then
(\ref{1cut1}) and (\ref{1cut2}) would have no real solutions. Notice also that
in this case the degree-$2K$ polynomial that appears under the square root
in (\ref{omcut}) must have $K-1$ double roots. This yields $K-1$ conditions
that fully determine the polynomial $P(z)$ in (\ref{polynomial}).

To determine the precise location of the support contour of $\rho$ in the
complex plane, we first use the observation \cite{mz} that the large-$N$
equation (\ref{loop}) for the single-loop correlator is identical to the
loop equation for the generalized Penner model (\ref{penner}). It follows that
the matrix models (\ref{part1}) and (\ref{penner}) are equivalent at
any order of the $\frac{1}{N}$-expansion and therefore belong to the same
universality class \cite{akm,mz}. Notice, however, that this does not imply
that all observables in the 2 models are the same. In particular, the endpoints
$a_1$ and $a_2$ of the one-cut ansatz in the fermionic case are complex-valued.
The same is true of the spectral distribution function where in addition
the requirement of positivity of $\rho$ is lost in the fermionic case.
Nevertheless, the solution at $N=\infty$ is the same in both models and this
fact can be used to derive some important properties of the one-cut solution
for the fermionic one-matrix model.

In particular, in the large-$N$ limit the spectral density therefore obeys
the saddle-point equation \cite{bipz}
\beq
\frac{2/\beta-V'(\beta)}{2}=\pvint~d\alpha~\frac{\rho(\alpha)}{\beta-
\alpha}~~~~,~~~~\beta\in~{\rm supp}~\rho
\label{saddle}\eeq
Note that this equation can be obtained from the discontinuity
(\ref{disc}),(\ref{homoeqn})
of the loop correlator (\ref{omcut}), and it also follows from the local
minimization condition $\frac{\delta F_0}{\delta\rho}=0$ for the free energy
\beq
F_0=\lim_{N\to\infty}-\frac{1}{N^2}\log Z_P=\int d\alpha~\rho(\alpha)\left(
V(\alpha)-2\log\alpha\right)+\int\!\!\pvint~d\alpha~d\beta~\rho(\alpha)
\rho(\beta)\log(\alpha-\beta)
\label{free}\eeq
with respect to the distribution function $\rho$. Note the change in sign of
the fermionic free energy relative to the Hermitian one (compare (\ref{penner})
and (\ref{pennerferm})). The double
integral in (\ref{free}) is evaluated by integrating up the saddle-point
equation (\ref{saddle}). This introduces a logarithmic divergence at
$\beta=0$ arising from the Penner potential in (\ref{penner}) which we
remove by subtracting from (\ref{free}) the Gaussian free energy $F_G$
defined by setting $V(\ps2)=g_1\ps2$ in (\ref{part1})
\beq
F_0-F_G=\frac{1}{2}\int d\alpha~\rho(\alpha)\left(V(\alpha)-2\log\alpha\right)
+\pvint~d\alpha~\rho(\alpha)\log\alpha
\label{freeg}\eeq
where we have ignored terms independent of the general potential couplings
$g_k$, $k>1$, in (\ref{free}).

The support contour
of $\rho$ can now be determined from the David primitive function \cite{david}
\beq
G(w)=\int_{a_1}^wdz~\left(\frac{2}{z}-V'(z)-2\omega(z)\right)
\label{david}\eeq
The branch points of $\omega(z)$ in (\ref{omcut}) (i.e. the solutions to
(\ref{1cut1}) and (\ref{1cut2})) fix the endpoints of the support of $\rho$,
but not its position in the complex plane. From (\ref{omcut}) and the analogy
above with the Hermitian Penner matrix model (for which $\rho$ is positive and
real-valued) it follows that the support of $\rho$ is an arc connecting $a_1$
to $a_2$ in the complex
plane along which $G(w)$ is purely imaginary and which can be embedded in a
region where ${\rm Re}~G(w)<0$ \footnote{\baselineskip=12pt This first property
of $G(w)$ follows from the Hermitian matrix model definition
$\rho(\lambda)\equiv(\frac{d\lambda(x)}{dx})^{-1}>0$. The second property
follows from the fact that global variations along ${\rm supp}~\rho$ of the
planar free energy $F_0$ in (\ref{free}) are proportional to $G$, $\delta
F_0\propto\delta G$, so that positivity of the real part of $\delta F_0$
ensures global stability of the ground state solution (\ref{free}) determined
by $\rho$.}. This feature, however, depends
strongly on the boundary conditions used in (\ref{penner}) \cite{david}.

The general $n$-loop correlators (\ref{multiloop}) in the spherical
 approximation can also be found
by applying the loop insertion operators (\ref{loopins}) $n-1$ times to
$\omega(z)$ as prescribed by (\ref{multins}). For example, applying the
 differential operator ${\cal L}(z)$ to the boundary conditions (\ref{1cut1})
 and (\ref{1cut2}) we can evaluate ${\cal L}(z)a_i$ as
\beq
\left({\cal L}(z)a_1\right)\cdot\oint_{\cal C}\frac{dw}{2\pi
 i}~\frac{V'(w)-2/w}{(w-a_1)^{3/2}(w-a_2)^{1/2}}=\frac{1}{(z-a_1)^{3/2}(z-a_2)
^{1/2}}
\label{lzai}\eeq
where we have used the identity
 \beq
{\cal L}(w)V'(z)=\frac{\partial}{\partial z}\frac{1}{w-z}
\label{loopinsvprime}\eeq
and ${\cal L}(z)a_2$ is obtained from (\ref{lzai}) by interchanging $a_1$
and $a_2$. Then applying ${\cal L}(w)$ to the one-cut solution (\ref{omcont})
we arrive at the two-loop correlator
\beq\new{\begin{array}{ll}
\omega_2(z,w)&=\frac{1}{N^2}\oint_{{\cal C}_z}\frac{dw'}
{4\pi
i}~\frac{1}{(z-w')(w-w')^2}\sqrt{\frac{(w'-a_1)(w'-a_2)}{(z-a_1)(z-a_2)}}\\&
=\frac{1}{N^2}\frac{1}{4(w-z)^2}\left(\frac{(w-a_1)(z-a_2)
+(w-a_2)(z-a_1)}{\sqrt{(w-a_1)(w-a_2)(z-a_1)(z-a_2)}}-2\right)\end{array}}
\label{2loop1cut}\eeq

(\ref{2loop1cut}) is identical to the 2-loop correlator of the
Hermitian one-matrix model \cite{amjurk}. Therefore all the
 multi-loop correlators (\ref{multiloop}) for $n\geq2$ are the same as those in
the Hermitian 1-matrix model with the same polynomial potential $V$ (and
 when $V$ is odd in the fermionic case these correlators are the same as
 those for a Hermitian model with a symmetric potential). As we shall discuss
in Subsection 3.4, this indicates a certain equivalence between the
 genus expansions of the fermionic and Hermitian one-matrix models. As
(\ref{multiloop}) represents the complete set of operators for the
fermionic matrix model, the loop equation (\ref{loopcont}) therefore
determines the complete set of equations of motion of the model. Notice
that the 2-loop correlator (\ref{2loop1cut}) depends on the potential $V$
 in (\ref{part1}) only implicitly through the endpoints $a_1$ and $a_2$
(but not explicitly). This is not so for the higher-order multi-loop
 correlators of the matrix model \cite{amjurk}. The system of standard
 Schwinger-Dyson equations for the connected correlators of the model
 $\langle\prod_{j=1}^n\tr(\ps2)^{p_j}\rangle_{\rm conn}$ can now be obtained by
expanding the multiloop correlators (\ref{multiloop}) in powers of
 $\frac{1}{z_1},\dots,\frac{1}{z_n}$ and using the loop equation
 (\ref{loopcont}).

\subsection{The Gaussian Model}

The solution of the model for the Gaussian potential (\ref{gausspot}) is
\beq
\omega(z)=\frac{1}{z}-\frac{t}{2}+\frac{1}{2z}\sqrt{4+t^2z^2}
\label{omgauss}\eeq
and the distribution function is
\beq
\rho(\alpha)=\frac{t}{2\pi i}\sqrt{1+\frac{4}{t^2\alpha^2}}~~~;~~~\alpha\in
{}~{\rm supp}~\rho
\label{rhogauss}\eeq
This is the analog in the fermionic case of the Wigner semi-circle law for a
Gaussian distribution of Hermitian random matrices \cite{bipz}. As mentioned
already, one of the crucial features of the fermionic models, as compared to
Hermitian models, is that the endpoints of the support region of the spectral
distribution function lie off of the real axis and the support contour is in
general embedded in some region of the complex plane. In the present case the
endpoints are situated on the imaginary axis at $\pm2i/t$, and to satisfy
(\ref{omrho}) and the normalization condition $\int d\alpha~\rho(\alpha)=1$
the support contour connecting the points $\pm2i/t$ must be chosen so that it
avoids the origin in the complex $\alpha$-plane.

To determine the precise support contour for $\rho$ which
connects these points, we evaluate the David function (\ref{david})
\beq\new{\begin{array}{ll}
G(z)&=-\int_{-\frac{2i}{|t|}}^z\frac{dw}{w}~\sqrt{4+t^2w^2}
\\&=-\sqrt{4+t^2z^2}-{}~{\rm
sgn}(t)\log\left(\frac{\sqrt{4+t^2z^2}-2}{\sqrt{4+t^2z^2}+2}\right)-i\pi
{}~{\rm sgn}~t\end{array}}
\label{davidgauss}\eeq
where the branch cut of the square roots in (\ref{davidgauss}) is taken to be
the straight line joining the points $\pm2i/t$. Notice that ${\rm Re}~G(z)
\to\mp\infty$ as $z\to\pm\infty$ and ${\rm Re}~G(z)\to+\infty$ as $z\to0$.
A careful study of the equation ${\rm Re}~G(z)=0$ and of the region where
${\rm Re}~G(z)<0$ shows that the support contour of $\rho$ cannot cross the
imaginary axis for $|{\rm Im}~z|>2/|t|$ and that it crosses the real axis
at some non-zero values of order $\pm1/t$. The regions ${\rm Re}~G(z)<0$
are to the right of these crossing points (but note that ${\rm Re}~G(z)$
changes sign across ${\rm supp}~\rho$). Thus the support contour of
(\ref{rhogauss}) can be taken to be the counterclockwise oriented half circle
of radius $2/|t|$ in the first and fourth quadrants of the complex
$\alpha$-plane. It is easy to verify that with this definition of $\rho$
the equations (\ref{mom}) and (\ref{omrho}) are indeed satisfied. Again, since
the free energy is analytic in the couplings for this simple Gaussian case
(i.e. a free fermion field theory), there is no critical behaviour in this
model.

\subsection{Critical Behaviour of a Non-Gaussian Model}

We now analyse some non-trivial polynomial potentials at large-$N$
and discuss the ensuing phase structure of the model.

\subsubsection{The Cubic Potential}

The simplest symmetric case is the cubic potential
\beq
V(z)=tz+\frac{g}{3}z^3
\label{potc}\eeq
for which
\beq
\omega(z)=\frac{1}{z}-\frac{t}{2}-\frac{g z^2}{2}+\frac{1}{2 z}
\sqrt{g^2 z^6+2tg z^4+(t^2+4g\xi ) z^2+4}
\label{ocubic}\eeq
where $\xi $ is the as yet unknown correlator
\beq
\xi=\langle\tr\ps2\rangle
\eeq
In this case the vanishing of all even moments, $\int d\alpha~\rho(\alpha)
\alpha^{2k}=0$, implies that the endpoints of the support contour of the
continuous function $\rho$ lie in the complex plane and are symmetric on
reflection through the origin. Furthermore, an application of Wick's
theorem shows that the series (\ref{gen}) in the odd moments is alternating.

Generically the square root in $\omega(z)$ has three branch cuts, so
that in the general case the distribution function $\rho$ will have three
disjoint and symmetric (about the origin) support contours.
The one-cut solution for (\ref{ocubic}) takes the form
\beq
\omega(z)=\frac{1}{z}-\frac{t}{2}-\frac{g z^2}{2}+\frac{gz^2\pm b}{2z}
\sqrt{ z^2+4/b^2}
\label{om1cut}\eeq
where comparing the polynomial coefficients in (\ref{om1cut}) with
those of (\ref{ocubic}) shows that the parameter $b$ and the
correlator $\xi $ are determined by the two equations
\beq
\pm b^3-tb^2+2g=0
\label{b}\eeq
\beq
b^3-(t^2+4g\xi )b\pm8g=0
\label{C}\eeq
The sign ambiguity here can be eliminated by requiring that at $g=0$ the
correct Gaussian value $b(g=0,t)=t$ for $b$ be attainable. This is the
boundary condition that is relevant for an interpretation of this matrix model
as a discretized random surface theory, i.e. for a consistent perturbative
expansion of the model in the coupling constant $g$. It means that
we take the positive sign in the above equations. The choice of negative
sign yields solutions with boundary conditions at $g=0$ appropriate to
generalized Penner models \cite{akm} (e.g. they yield real-valued endpoints
for ${\rm supp}~\rho$). This can also be seen directly from
the contour integrals (\ref{1cut1}) and (\ref{1cut2}) by computing the residues
at $\infty$. For any odd polynomial
potential, (\ref{1cut2}) is an identity since there is no residue at
infinity, while for the potential (\ref{potc}) the equation (\ref{1cut1})
yields exactly (\ref{b}) with the sign ambiguity arising from the possible
choices of sign of the square root $\sqrt{a_1a_2}$.

We assume henceforth that $t$ is a positive constant. The 3
solutions of (\ref{b}) are
\beq
b_0(x,t)=\frac{t}{3}\left(\beta^{1/3}(x)+\beta^{-1/3}(x)+1\right)
\label{bb}\eeq
\beq
b_\pm(x,t)=\frac{t-b_0(x,t)}{2}\pm\frac{i\sqrt{3}t}{6}\left(\beta^{1/3}(x)-
\beta^{-1/3}(x)\right)
\label{bim}\eeq
where
\beq
\beta(x)=2x-1+2\sqrt{x(x-1)}
\label{beta}\eeq
and we have introduced the dimensionless scaling parameter
\beq
x=1-27g/2t^3\equiv1-g/g_c
\label{x}\eeq
When $x\leq0$ ($g\geq g_c\equiv\frac{
2t^3}{27}$) or $x\geq1$ ($g\leq0$), $\beta(x)$ is a monotone real-valued
function with $\beta(x)\geq1$ for $x\geq1$ and $\beta(x)\leq-1$ for $x\leq0$.
In the region $0<x<1$ ($0<g<g_c$), $\beta(x)$ is a unimodular complex-valued
function. The function (\ref{bb}) is always real-valued and the
region $0<x<1$ is the region wherein all 3 roots (\ref{bb}), (\ref{bim})
of the cubic equation (\ref{b}) are real. These 3 roots can all be obtained
from (\ref{bb}) by choosing the 3 inequivalent cube roots of $\beta(x)$.
For $x\notin(0,1)$ the solutions (\ref{bim}) are complex.

For the fermionic matrix model, where the distribution function $\rho$ can be
complex-valued, there is no immediate reason to disregard generic
complex-valued endpoints for the support of $\rho$. However, the free energy
(\ref{freeg}) for the cubic potential (\ref{potc}) up to terms independent of
$b$ and $g$ is
\beq
F_0(x,t)-F_G(t)=\frac{t(3b(x,t)-t)}{6b^2(x,t)}+\log|b(x,t)|
\label{freec}\eeq
where we have used the spectral density determined by (\ref{disc}) and
(\ref{om1cut}) as
\beq
\rho(\alpha)=\frac{1}{2\pi i}\left(b+g\alpha^2\right)\sqrt{1+\frac{4}
{b^2\alpha^2}}~~~~;~~~~\alpha\in~{\rm supp}~\rho
\label{rho}\eeq
with $b(g,t)$ given by (\ref{bb}) and (\ref{bim}) (or equivalently the moments
$\xi$ and $\langle\tr(\ps2)^3\rangle$ determined by expanding (\ref{om1cut}) to
order $\frac{1}{z^4}$). In arriving at (\ref{freec}) we have used the boundary
conditions (\ref{b}) and (\ref{C}), and the fact that in the difference between
the two logarithmic integrations in (\ref{freeg}) only the residue at
$\alpha=0$ survives. The free energy (\ref{freec}) is complex-valued for the
complex values (\ref{bim}) of $b(x,t)$ for $x\notin(0,1)$.
Such a free energy corresponds to an unstable state and we therefore consider
only the real-valued solutions to (\ref{b}).
The support contour on which (\ref{rho}) is defined is again found
from the David function ({\ref{david}) which for the cubic potential
(\ref{potc}) has the same qualitative properties as (\ref{davidgauss})
\cite{mss}. Thus the support contour in
(\ref{rho}) can be taken as the counterclockwise oriented half-circle of
radius $2/|b|$ in the first and fourth quadrants of the complex
$\alpha$-plane. The boundary condition (\ref{C}) now follows from evaluating
the correlator
$\xi=\int d\alpha~\rho(\alpha)\alpha$ with this distribution function.

There are 2 critical points in this large-$N$ matrix model, at $g=0$ and
$g=g_c$, which separate 3 phases determined by the analytic structure of the
function (\ref{beta}), i.e. the one-cut solution is a non-analytic function
of $x$ about $x=0$ and $x=1$ where it acquires a square root branch cut.
For $x\geq1$ the solution
\beq
b_0^>(x,t)=\frac{t}{3}\left[1+\left(2x-1+2\sqrt{x(x-1)}\right)^{1/3}+
\left(2x-1+2\sqrt{x(x-1)}\right)^{-1/3}\right]
\label{x>1}\eeq
of (\ref{b}) satisfies the Gaussian boundary condition $b_0(x=1,t)=t$. When
$x\leq 0$ the real solution for $b$ is
\beq
b_0^<(x,t)=\frac{t}{3}\left[1-\left|2x-1+2\sqrt{x(x-1)}\right|^{1/3}-
\left|2x-1+\sqrt{x(x-1)}\right|^{-1/3}\right]
\label{x<0}\eeq

As $x$ is varied between 0 and 1, $\beta(x)$ has modulus one and phase which
varies from $\pi$ to $0$, i.e. $\beta(x)=\e^{i\phi(x)}$ where
\beq
\phi(x)=\arctan\left(\frac{2\sqrt{x(1-x)}}{2x-1}\right)\in[0,\pi]
\label{phi}\eeq
The arctangent function in (\ref{phi}) is well-defined only up to an
integral multiple of $2\pi$, and the three real solutions for $b$ are
\beq
b^{(n)}(x,t)=\frac{t}{3}\left[1+2\cos\left\{\frac{1}{3}\arctan\left(\frac
{2\sqrt{x(1-x)}}{2x-1}\right)+\frac{2n\pi}{3}\right\}\right]
\label{0<x<1}\eeq
where $0<x<1$ and $n=0,1,2$. The branch which matches (\ref{x<0}) is the one
with $n=1$, whereas the branch which matches (\ref{x>1}) is the one with $n=0$.
The branch with $n=2$ does not connect with either solution. Any of these 3
branches can be used to define the one-cut solution (\ref{om1cut}). The
free energy (\ref{freec}) is positive for all $x\in(0,1)$ for the $n=0$
branch, negative for all $x\in(0,1)$ for the $n=1$ branch, and for the
$n=2$ branch it is positive for $0<x<\frac{1}{2}$ and flips sign for the
rest of the interval at $x=\frac{1}{2}$. Thus the $n=1$ branch in (\ref{0<x<1})
is the ground state solution in the region $0<x<1$.

The free energy associated with this stable one-cut solution is
discontinuous across $g=0$, and thus with this choice of branch in the regime
$0<x<1$ the Gaussian point of this matrix model is a critical point of a
first order phase transition. The other one-cut solution which is a
perturbation of the Gaussian solution is metastable but can still be
thought of as a valid solution of the model since the energy barrier between
the stable and metastable one-cut solutions is infinite
at $N=\infty$ (the height of the barrier is of order $N^2$). Ordinarily, the
infinite energy barrier prevents tunneling and also a phase transition from
occuring \cite{david} \footnote{\baselineskip=12pt In the Hermitian one-matrix
model, instanton effects, which correspond to a single eigenvalue of the matrix
model climbing to the top of the barrier, are responsible for the divergence of
the perturbation series at large orders. These instanton effects are suppressed
as $\e^{-K(g)N}$ in the large-$N$ limit for fixed $g>g_c$, but remain finite in
the double-scaling limit. Thus the scaling behaviour of the $m=2$ Hermitian
one-matrix model corresponds to a {\it complex} solution of the Painlev\'e I
equation, and in this context the large-order behaviour can be understood in
terms of an instanton calculation of barrier penetration effects. For details,
see \cite{david}.}. However, if we restrict attention to one-cut solutions and
follow them over the range of $g$, we must encounter a discontinuity of
the free energy somewhere, i.e. a first order phase transition.

This is similar to the situation in the Hermitian
one-matrix model with symmetric polynomial potential of degree 6 \cite{jurk}.
There a phase transition occurs due to an infinite volume effect, as opposed
to a large-$N$ effect where the only possibilities could be second or third
order phase transitions. There is also the possibility that the loop
correlator (\ref{ocubic}) evolves into a three-cut phase at $g=0$,
corresponding to a third order phase transition (see below), but
there is no immediate indication of this since in the fermionic case the
spectral measure $\rho(\alpha)d\alpha$ need not be positive. This possibility
is also suggested by the exact form (\ref{ocubic}) of the loop correlator.
Although the one-cut ansatz (\ref{om1cut}) is insensitive to a change in
sign of $g$, the analytic properties of (\ref{ocubic}) are affected
by the passage through $g=0$ (i.e. the sign of the square root flips in order
to satisfy the boundary condition (\ref{bcvec})). The model therefore cannot be
analytically
continued to negative values of $g$, and the resulting peculiarities in the
$\frac{1}{N}$-expansion of the model are related to the occurence
of complex-valued endpoints for the support of the distribution function
$\rho$. The existence of 3 phases in this matrix model
and the possibility of a first order phase transition at the Gaussian point
$g=0$ are completely unlike what occurs in the conventional polynomial
Hermitian matrix models \cite{fgz,kaz} or in Penner models \cite{akm,cdl,tan}
\footnote{\baselineskip=12pt In polynomial Hermitian matrix models criticality
is the result of $m$ zeroes of $\rho(\alpha)$ coalescing with one of the
endpoints of ${\rm supp}~\rho$ with string constant $\gamma_{\rm str}=-1/m$.
For generalized Penner models, in addition to this multi-critical behaviour
there are critical points with $\gamma_{\rm str}=0$ for which logarithmic
scaling violations occur \cite{dv} and which are the result of the coalescence
of two endpoints of ${\rm supp}~\rho$ \cite{akm,cdl,tan}. In these cases, the
multi-critical coupling constants are negative and separate the unique one-cut
phase from a multi-cut phase \cite{bipz,fgz}. In the simplest symmetric case of
a quartic plus quadratic interaction, the endpoint of the one-cut solution
obeys a quadratic equation whose 2 solutions coalesce at criticality (as in the
vector model of Section 2 above).}.
Notice, however, that the free energy (\ref{freec}) with the choice of
stable branch for $x\in(0,1)$ is continuous across the critical point $g=g_c$.

The scaling behaviour of the matrix model in the vicinity of its
critical points is determined by the string susceptibility
\beq
\chi(g,t)=-\frac{1}{N^2}\frac{\partial^2\log Z_1}{\partial g^2}=-\frac{1}{3}
\frac{\partial}{\partial g}\left\langle\tr(\ps2)^3\right\rangle
\label{chi}\eeq
where the correlator $\langle\tr(\ps2)^3\rangle$ can be read off from the
$\frac{1}{z^4}$ coefficient of the large-$z$ expansion of (\ref{om1cut}). The
critical exponent $\gamma_{\rm str}^{(i)}$ at each critical point
$g_c^{(i)}$ is defined by the leading non-analytic behaviour of (\ref{chi})
\cite{fgz}
\beq
\chi(g,t)\sim_S(g-g_c^{(i)})^{-\gamma_{\rm str}^{(i)}}~~~{\rm as}~~~g
\to g_c^{(i)}
\eeq
where $\sim_S$ denotes the most singular part of the function in a
neighbourhood of the critical point. In terms of the scaling variable
(\ref{x}), the susceptibility (\ref{chi}) is
\beq\new{\begin{array}{ll}
\chi(x,t)&=\frac{1}{3}\frac{\partial}{\partial
x}\left(\frac{10(x-1)}{9b^6(x,t)}-\frac{4}{g_cb^3(x,t)}\right)\\&
=\frac{972}{t^6(\beta^{2/3}+\beta^{1/3}+1)^6}\left[1-8x+8x^2+4\sqrt{
\frac{x}{x-1}}\left(1-3x+2x^2\right)\right]\end{array}}
\label{chi1}\eeq
{}From (\ref{chi1}) we find that the leading singular parts of the
susceptibility
near each of the two critical points $g=g_c$ and $g=0$ are respectively
\beq
\chi(x,t)\sim_S-\frac{15552}{t^6}\sqrt{x}~~~{\rm as}~~~x\to0
\label{sc1}\eeq
\beq
\chi(x,t)\sim_S\frac{11648}{3t^6}\sqrt{x-1}~~~{\rm as}~~~x\to1
\label{sc2}\eeq
Both critical points therefore have string constant
\beq
\gamma_{\rm str}=-1/2
\eeq
which coincide with those of the usual $m=2$ quantum gravity models
\cite{kaz}.

In particular, this shows that with the choice of stable branch for
$x\in(0,1)$ the phase transition at the non-zero critical coupling $g=g_c$
is of third order. Notice that the spectral density at this critical point is
given by
\beq
\rho_c(\alpha)=\frac{g_c}{2\pi
i\alpha}\left(\alpha^2+\frac{27b_c}{2t^3}\right)\sqrt{\alpha^2+\frac{4}{b_c^2}}
\eeq
where $b_c(t)=b(g_c,t)$. For either the $n=0$ or $n=2$ branches in
(\ref{0<x<1}) we find $b_c(t)=2t/3$, and therefore a zero of $\rho(\alpha)$ at
criticality coalesces with each of the symmetrical endpoints of its support.
The critical point $g=g_c$ therefore
enjoys all of the properties of a conventional $m=2$ multi-critical
point \cite{fgz,kaz}. It represents a third order phase transition with
string susceptibility exponent $\gamma_{\rm str}=-1/2$, at criticality the
zeroes of the spectral distribution function coalesce with the endpoints of
the cut, and the scaling behaviour of functions near this critical point
coincides with that of the Hermitian one-matrix model with symmetric quartic
potential. The 2 critical points of the fermionic one-matrix model, which arise
as those points in parameter space where the cubic equation (\ref{b}) which
determines the one-cut solution acquires a double real root, provide
information relevant to the perturbative and topological expansions of the
theory. In the 2 phases outside of the region $0<x<1$ a unique one-cut solution
with real free energy exists\footnote{\baselineskip=12pt The solution of a
Riemann-Hilbert problem is always unique \cite{bipz}.}, while in the phase
$x\in(0,1)$ a multi-cut
solution as well as the multi-branch one-cut solutions can in addition
exist. The scaling behaviours (\ref{sc1}) and (\ref{sc2}) indicate that the
2 transitions into the multi-cut phase would both be of third order, while
the transitions into the stable one-cut phase are of first and third order.
The existence of a single-cut or multi-cut phase in the region $0<x<1$ is
determined by which one of these 2 possibilities is in fact the vacuum state.
It would be interesting to investigate this point further, although there is
no immediate way to determine the various parameters of the three-cut ansatz
due to the appearence of the unknown correlator $\xi$ in (\ref{ocubic}).
The appearence of this unknown variable is another one of the distinguishing
analytic features of the fermionic matrix models.

\subsubsection{General Polynomial Potentials}

We now discuss the critical behaviour associated
with higher order potentials. For simplicity we consider the
chirally symmetric case where the potential (\ref{pot}) is a generic
odd polynomial, i.e. $\kappa=g_{2k}=0$ for all $k$ in (\ref{pot}), with
$K=\deg V>3$ an odd integer. The solution for the loop correlator is then
\beq\new{\begin{array}{l}
\omega(z)=\frac{1}{z}-\sum_{k=1}^{\frac{K+1}{2}}\frac{g_{2k-1}}{2}z^{2(k-1)}
\\+\frac{1}{2z}\left(4+\sum_{k,m=1}^{\frac{K+1}{2}}g_{2k-1}g_{2m-1}z^{2(k+m-1)}
+\sum_{k,m=1}^{m+k\leq\frac{K+1}{2}}g_{2(k+m)-1}\xi_{2m}z^{2k}\right)^{1/2}
\end{array}}
\label{omK}\eeq
where $\xi_{2m}$ are the as yet unknown moments $\xi_{2m}=\langle\tr(\ps2)^
{2m-1}\rangle$. The one-cut solution for (\ref{omK}) takes the form
\beq
\omega(z)=\frac{1}{z}-\sum_{k=1}^{\frac{K+1}{2}}\frac{g_{2k-1}}{2}z^{2(k-1)}+
\frac{1}{2z}\left(g_Kz^{K-1}+\sum_{k=1}^{\frac{K-3}{2}}a_{2k}z^{2k}+b\right)
\sqrt{z^2+4/b^2}
\label{K1cut}\eeq
where we have fixed the sign in front of the endpoint parameter $b$ by the same
convention as before. The one-cut ansatz (\ref{K1cut}) along with the
general solution (\ref{omK}) together involve $K-1$ unknown parameters
-- $b$, the $(K-3)/2$ polynomial coefficients $a_{2k}$, and the
$(K-1)/2$ correlators $\xi_{2k}$.

These parameters can be found by comparing the various polynomial coefficients
of (\ref{omK}) with those of (\ref{K1cut}), which leads to the set of equations
\beq
b^3+8a_2-b\left(g_1^2+\sum_{k=1}^{\frac{K-1}{2}}g_{2k+1}\xi_{2k}\right)=0
\label{parstart}\eeq
\beq
2b^3+8ba_4+4a_2^2-b^2\left(2g_1g_3+\sum_{k=1}^{\frac{K-3}{2}}g_{2k+3}
\xi_{2k}\right)=0
\eeq
\beq
2bg_Ka_{K-5}+8a_{K-3}-b\sum_{k=1}^{K-2}g_{2k-1}g_{2K-2k-3}=0
\eeq
\beq
2b^2g_Ka_{K-3}+4g_K^2-b^2\sum_{k=1}^{K-1}g_{2k-1}g_{2K-2k-1}=0
\eeq
\beq
2b^2a_{K-3}+8g_K-b\left(g_K\xi+\sum_{k=1}^{\frac{K-1}{2}}g_{2k-1}g_{K-2k}
\right)=0
\eeq
\beq
2b^3g_K+8bg_Ka_2-b^2\sum_{k=1}^{\frac{K+1}{2}}g_{2k-1}g_{K-2k+2}=0
\eeq
When $K>7$ we have in addition the sets of equations
\bd
2b^3a_{2(m-1)}+8ba_{2m}+4a_2a_{2(m-1)}+\sum_{k=1}^{m-2}\left(b^2a_{2k}a_{2
(m-k-1)}+4a_{2k}a_{2(m-k)}\right)
\ed
\beq
-b^2\left(\sum_{k=1}^mg_{2k-1}
g_{2(m-k)+1}+\sum_{k=1}^{\frac{K+1-2m}{2}}g_{2(m+k)-1}\xi_{2k}\right)=0
\eeq
for $3\leq m\leq\frac{K-3}{2}$, and
\bd
2b^2g_Ka_{2m-K-1}+8ba_{2m-K+1}+b^2\sum_{k=1}^{m-2}a_{2k}a_{2(m-k-1)}
\ed
\beq
+4\sum_{k=1}^{m-1}a_{2k}a_{2(m-k)}-b^2\sum_{k=1}^mg_{2k-1}g_{2(m-k)+1}=0
\label{parend}\eeq
for $\frac{K+3}{2}\leq m\leq K-3$.

(\ref{parstart})--(\ref{parend}) yield a complete set of equations for
the $K-1$ unknown coefficients of the one-cut solution (\ref{K1cut})
in terms of the coupling constants of the potential (\ref{pot}). The
parameter $b$ can alternatively be found from the contour
integral (\ref{1cut1}) which leads to a $K$-th order equation for $b$
\beq
b^K+\sum_{k=1}^{\frac{K+1}{2}}\frac{(-1)^k2^{k-1}(2k-3)!!}{(k-1)!}g_{2k-1}
b^{K-2k+1}=0
\label{bK}\eeq
Since $K$ is odd this equation always has a real solution, and as
before the one-cut solution can always be constructed. The spectral
density is
\beq
\rho(\alpha)=\frac{1}{2\pi i}\left(g_K\alpha^{K-1}+\sum_{k=1}^{\frac{K-3}{2}}
a_{2k}\alpha^{2k}\right)\sqrt{1+\frac{4}{b^2\alpha^2}}~~~;~~~\alpha\in~{\rm
supp}~\rho
\label{distrK}\eeq
The first $(K-1)/2$ moments of this distribution function are given by the
solutions to (\ref{parstart})--(\ref{parend}).

In general (\ref{bK}) will acquire multiple real roots at some
coupling $g_{2k-1}^c$ which will be a critical point of a third order
phase transition with string constant $\gamma_{\rm str}=-1/2$. We
expect that the phase with multiple real roots will be bounded by
other coupling constant values so that the model will contain several
critical points corresponding possibly to different order phase
transitions. Notice that since the potential now depends on more parameters,
we can adjust them in such a way that $m-1$ regular zeroes of (\ref{distrK})
coalesce with a cut end-point $\pm2i/b$ at criticality for the same critical
coupling $g_{2k-1}^c$, i.e. so that
\beq
\rho_c(\alpha)\sim(\alpha^2+4/b_c^2)^{m-1/2}
\eeq
where we neglect possibly other irrelevant zeroes. $g_{2k-1}^c$ will then be an
$m$-th order multi-critical point \cite{fgz,kaz} with susceptibility exponent
\beq
\gamma_{\rm str}=-1/m
\eeq

\subsection{The Topological Expansion}

The fermionic one-matrix model possesses a novel critical
behaviour which includes the usual multi-critical behaviour that occurs in
Hermitian one-matrix models, and furthermore, in the case of the
simplest symmetric potential, there may also be a first order phase transition
at zero coupling. This would imply that the perturbative expansion of the
theory in $g$ has a preferred direction through values of the coupling with
a definite sign (corresponding to $-~{\rm sgn}~t$). It means that
perturbation theory near the Gaussian point $g=0$ does not correctly
reflect the properties of the theory when $g$ is small and negative and this
fact is important for the interpretation of the fermionic one-matrix model
as a statistical theory of discretized random surfaces (because then the model
cannot be continued to values of couplings with ${\rm sgn}~g=-~{\rm sgn}~t$).
As mentioned before, this effect seems to be merely an artifact of the
fermionic nature of the matrix degrees of freedom here. The other critical
point, which is the usual $m$-th order multi-critical point
with third order phase transition and string susceptibility with critical
exponent $\gamma_{\rm str}=-1/m$, gives the continuum limit of the topological
genus expansion relevant to string theory. In the Hermitian case, this
continuum limit corresponds to $D\leq1$ modes of pure 2-dimensional quantum
gravity \cite{fgz,kaz}. We shall now examine the topological
$\frac{1}{N}$-expansion of the fermionic one-matrix model, which determines
explicitly the number of 'tHooft diagrams of a given genus, and present the
argument \cite{akm} indicating why one expects that this results in a genus
expansion
which is an alternating series but otherwise coincides with the usual
Painlev\'e expansion \cite{fgz} (but otherwise no conclusive argument is
available as of yet).

The crucial point of the argument is the identification of the fermionic
matrix model (\ref{part1}) with the generalized Penner model (\ref{penner})
and the fact that the multi-critical points of the adjoint fermion model
belong to the same universality class as those of the Hermitian one-matrix
model with the same potential $V$ \cite{akm,fgz,mz}. In particular, the value
of $\gamma_{\rm str}$ for the $m$-th multi-critical points is the same to all
genera (i.e. powers of $1/N^2$) in the 2 models \cite{akm}. However, we expect
that the fermionic genus series alternates relative to the Hermitian case
because the branch points of the one-loop correlator (\ref{omcut}) are
complex-valued in the fermionic case. Using the usual identification with the
Penner matrix model, it follows that the moment functions
\beq
M_k=\oint_{{\cal C}}\frac{dw}{2\pi i}~\frac{V'(w)-2/w}{(w-a_2)^{k+1/2}
(w-a_1)^{1/2}}~~~~,~~~~J_k=\oint_{{\cal C}}\frac{dw}{2\pi i}~\frac
{V'(w)-2/w}{(w-a_2)^{1/2}(w-a_1)^{k+1/2}}
\label{momcrit}\eeq
defined for $k\geq0$ can be used to determine the critical points of the
matrix model \cite{ackm}. We restrict attention as above to a generic chirally
symmetric potential, so that $a_2=-a_1=iy$ where $y=2/|b|\in\IR^+$.
{}From (\ref{omcut}) and (\ref{omcont}) it follows that the 1-cut ansatz for
the loop correlator is
\beq
\omega(z)=\frac{1}{z}-V'(z)+M(z)\sqrt{z^2+y^2}
\eeq
where
\beq
M(z)=\oint_{\cal C}\frac{dw}{2\pi
i}~\frac{1}{z-w}\frac{V'(z)-V'(w)+2/w-2/z}{\sqrt{w^2+y^2}}
\label{mz}\eeq
and $zM(z)$ is a polynomial of degree $K-1$. The algebraic coefficients of
(\ref{mz}) can be found by calculating the residue of the contour integral
there at infinity (see the last Subsection). At an $m$-th multi-critical point,
$m-1$ zeroes of $M(z)$ coalesce with the branch point $y$. From (\ref{omcont})
we then have
\beq
M_k\propto\frac{\partial^{k-1}M(z)}{\partial
z^{k-1}}\biggm|_{z=iy}~~~~~,~~~~~k\geq1
\eeq
and so the condition for the $m$-th multi-critical point is equivalent to the
requirement that \cite{ackm}
\beq
M_1(iy_c)=M_2(iy_c)=\ldots=M_{m-1}(iy_c)=0~~~~,~~~~M_m(iy_c)\neq0
\label{momcond}\eeq
where $y_c$ denotes the endpoint parameter $y$ at criticality.

In the case at hand, the moment functions in (\ref{momcrit}) are related by
\beq
M_k=(-1)^kJ_k
\label{mj}\eeq
and moreover we have
\beq
M_k=(-i)^k\tilde{M}_k
\label{mreal}\eeq
where
\beq
\tilde{M}_k=\sum_{n\geq0}(-1)^ng_{2n+1}y^{2n}\oint_{{\cal C}}\frac{dw}
{2\pi i}~\frac{1}{(w-y)^{k+1/2}(w+y)^{1/2}}+\frac{2}{y(-y)^k}
\label{mtild}\eeq
is real-valued. In particular, the 0-th order moment function is
\beq
M_0=\sum_{n\geq0}(-1)^{n+1}g_{2n+1}\frac{(2n)!}{2^{2n}(n!)^2}y^{2n}+\frac{2}{y}
\label{m0}\eeq
and the higher order moments are related to it by
\beq
M_k=\frac{2^k}{(2k-1)!}\frac{\partial^k}{\partial
a_2^k}M_0\biggm|_{a_2=-a_1=iy}~~~~~,~~~~~k\geq1
\label{m1}\eeq
We also introduce the cut parameter
\beq
d=a_2-a_1
\label{cutpar}\eeq
which in the present case is purely imaginary
\beq
d=2iy\equiv i\tilde{d}
\label{d}\eeq

The free energy $F_P=\frac{1}{N^2}\log Z_P$ admits the genus expansion
\beq
F_P=\sum_{h=0}^\infty\frac{1}{N^{2h}}F_h
\label{freegenus}\eeq
where the genus 0 free energy $F_0$ is given by (\ref{freeg}). From
(\ref{multins}) it follows that the one-loop correlator $\omega(z)$
has the genus expansion
\beq
\omega(z)=\sum_{h=0}^\infty\frac{1}{N^{2h}}\omega^{(h)}(z)
\label{omgenus}\eeq
where
\beq
\omega^{(h)}(z)=N^2{\cal L}(z)F_h
\label{omh}\eeq
The higher genus contributions to the single-loop correlator are obtained
by iterating the genus zero contribution $\omega^{(0)}(z)$ (i.e. the $N=\infty$
solution to
(\ref{loopcont})). Substituting the genus expansion (\ref{omgenus}) into
(\ref{loopcont}) and equating the $\frac{1}{N^{2h}}$ coefficients, we obtain
an iterative equation for $h\geq1$
\beq\new{\begin{array}{c}
-z\left(2\omega^{(0)}(z)\omega^{(h)}(z)+\sum_{h'=1}^{h-1}\omega^{(h')}(z)
\omega^{(h-h')}(z)+N^2{\cal L}(z)\omega^{(h-1)}(z)\right)\\
+2\omega^{(h)}(z)+\oint_{\cal C}\frac{d\lambda}{2\pi
i}~\frac{V'(\lambda)\lambda}{z-\lambda}\omega^{(h)}(\lambda)=0\end{array}}
\label{omiter}\eeq
which determines $\omega^{(h)}(z)$ entirely in terms of $\omega^{(h')}(z)$
with $h'<h$. The solution $\omega^{(h)}(z)$ to (\ref{omiter}) can be expressed
in terms of $2\cdot(3h-1)$ lower moment functions (\ref{momcrit}) \cite{ackm}.

For instance, to find $\omega^{(1)}(z)$ in the one-cut phase, from
(\ref{2loop1cut}) we have
\beq
{\cal
L}(z)\omega^{(0)}(z)=\omega_2(z,z)=\frac{1}{N^2}\frac{(a_1-a_2)^2}{16(z-a_1)^2
(z-a_2)^2}
\label{2loopzz}\eeq
and so (\ref{omiter}) for $h=1$ yields
\beq
\omega^{(1)}(z)=\frac{1}{\sqrt{(z-a_1)(z-a_2)}}\oint_{\cal C}\frac{dw}{2\pi
i}~\frac{1}{(w-z)M(w)}\frac{(a_1-a_2)^2}{16(w-a_1)^2(w-a_2)^2}
\label{omgenus1}\eeq
which is unambiguous provided that it is analytic at the zeroes of $M(z)$.
After some algebra it can be written as \cite{ackm}
\beq\new{\begin{array}{ll}
\omega^{(1)}(z)=&\frac{1}{8dM_1}\frac{1}{\sqrt{(z-a_1)(z-a_2)^3}}
-\frac{1}{8dJ_1}\frac{1}{\sqrt{(z-a_1)^3(z-a_2)}}\\&+\frac{1}
{16M_1}\left(\frac{1}{\sqrt{(z-a_1)(z-a_2)^5}}-\frac{M_2}{8dM_1
\sqrt{(z-a_1)(z-a_2)^3}}\right)\\&+\frac{1}{16J_1}
\left(\frac{1}{\sqrt{(z-a_1)^5(z-a_2)}}+\frac{J_2}{8dJ_1
\sqrt{(z-a_1)^3(z-a_2)}}\right)\end{array}}
\label{om1mom}\eeq
In the case of the cubic potential (\ref{potc}) we have
\beq
M(z)=(gz^2+b)/2z
\label{mzcubic}\eeq
so that (\ref{omgenus1}) is
\beq
\omega^{(1)}(z)=\frac{y^2z}{8g(z^2+p^2)\sqrt{z^2+y^2}}\left(\frac{y}{(z^2+y^2)^
2}-\frac{p}{(p^2+y^2)^2}\right)
\label{om1cubic}\eeq
which is manifestly analytic at $z^2=-p^2\equiv-b/g$. After some rewriting, it
can be checked that the terms in (\ref{om1mom}) can all
be expressed in terms of the loop insertion operator ${\cal L}(z)$ acting on
some quantities, and that the genus 1 free energy determined from (\ref{omh})
is \cite{ackm}
\beq
F_1=\frac{1}{24}\log M_1+\frac{1}{24}\log J_1+\frac{1}{6}\log d
\label{freegenus1}\eeq
This method of iterative solution can be carried out order by order in the
$\frac{1}{N}$-expansion.

The general structure of the terms in the topological expansion
(\ref{freegenus}) as found from (\ref{omh}) has been studied in detail by
Ambj\o rn-Chekhov-Kristjansen-Makeenko \cite{ackm} and they have shown that the
higher-genus terms can be written symbolically in the form
\beq
F_h=-{\sum_{\alpha_i,\beta_j>1}}^{\prime}\langle\alpha_1,\ldots,\alpha_s;
\beta_1,\ldots,\beta_\ell|\alpha,\beta,\gamma\rangle_h~
f_h(\alpha_1,\ldots,\alpha_s;\beta_1,
\ldots,\beta_\ell;\alpha,\beta,\gamma)
\label{freeatg}\eeq
where $h\geq1$ and
\beq
f_h(\alpha_1,\ldots,\alpha_s;\beta_1,\ldots,\beta_\ell;
\alpha,\beta,\gamma)=\frac{M_{\alpha_1}\cdots M_{\alpha_s}
J_{\beta_1}\cdots J_{\beta_\ell}}{M_1^\alpha J_1^\beta d^\gamma}
\label{f}\eeq
The brackets in (\ref{freeatg}) are rational functions of the non-negative
integers $\alpha$, $\beta$ and $\gamma$, the indices $\alpha_i$ and
$\beta_j$ lie in the interval $[2,3h-2]$, and $h-1\leq\gamma\leq4h-4$.
The prime on the sum in (\ref{freeatg})
means that the summation is over the sets of indices obeying the restrictions
\beq
s-\alpha\leq0~~~,~~~s=\alpha\iff s=\alpha=0~~~~~;~~~~~
\ell-\beta\leq0~~~,~~~\ell=\beta\iff\ell=\beta=0
\eeq
\beq
\alpha-s+\beta-\ell=2h-2~~~~~;~~~~~\sum_{i=1}^s\left(\alpha_i-1\right)
+\sum_{j=1}^\ell\left(\beta_j-1\right)+\gamma=4h-4
\label{ind}\eeq
The first relation in (\ref{ind}) follows from the invariance of the partition
function $Z_P=\e^{\sum_hN^{2-2h}F_h}$ under simultaneous rescalings of $N$ and
the spectral density $\rho$, $N\to\lambda\cdot N$,
$\rho\to\frac{1}{\lambda}\rho$. The second relation in (\ref{ind}) follows from
the invariance of $Z_P$ under the rescalings $N\to\lambda^2\cdot N$,
$g_j\to\lambda^{j-2}\cdot g_j$. The restriction of the integer $\gamma$ to the
range $h-1\leq\gamma\leq4h-4$ is a consequence of the double-scaling limit of
the Hermitian one-matrix model \cite{ackm} in which
\beq
F_h\sim\Lambda_R^{(2-\gamma_{\rm str})(1-h)}
\eeq
where $\Lambda_R\sim x$ is the renormalized cosmological constant.

In the Hermitian case, the higher genus coefficients can be related to
intersection indices on the moduli space and the virtual Euler characteristics
of the discretized moduli space of compact Riemann surfaces \cite{ackm}. This
follows from the identification of the Hermitian one-matrix model with
the Kontsevich and Kontsevich-Penner matrix models which all have the same
double-scaling limits. In this way, the topological expansion can be related to
the intersection indices
represented by Kontsevich matrix models which relate the 1-matrix models
to topological gravity \cite{kont} (see the next Subsection) in terms of a
discretization of moduli space similar to that used to define the virtual Euler
characteristic for Penner matrix models (c.f. Subsection 1.2.1). Given the
genus $h$ contribution to the free energy (\ref{freegenus}) the genus $h$
contribution to any connected correlator can be found from (\ref{multins}). We
refer to \cite{ackm} for the technical details of this iterative determination
of the free energy (\ref{freeatg}) and its relation to the Kontsevich matrix
model.

Substituting the relations (\ref{mj}), (\ref{mreal}) and (\ref{d}) into
(\ref{f}) we find
\beq
f_h=i^{(\alpha+\beta-\sum_{i=1}^s\alpha_i-\sum_{j=1}^\ell\beta_j
-\gamma)}\tilde{f}_h
\eeq
where $\tilde{f}_h$ is real-valued and is defined from $f_h$ by replacing
$M_k$, $J_k$ and $d$ by $\tilde{M}_k$, $\tilde{J}_k=(-1)^k\tilde{M}_k$ and
$\tilde{d}$ in (\ref{f}). Using the restrictions (\ref{ind}) we then find that
\beq
F_h=(-1)^{h-1}\tilde{F}_h
\label{freealt}\eeq
where $\tilde{F}_h$ is real-valued and is defined from $F_h$ by replacing
$f_h$ by $\tilde{f}_h$ in (\ref{freeatg}). (\ref{freealt}) shows that the
fermionic free energy alternates in sign according to genus in some sense.
To show that it alternates relative to the Hermitian case, we need to
determine the scaling behaviour of $\tilde{M}_k$ near the critical point.

For example, for the cubic potential (\ref{potc}), consider the moment
functions (\ref{momcrit}) near the $m=2$ multi-critical point $g_c=2t^3/27$
\footnote{\baselineskip=12pt This critical point and the value $y_c$ can also
be found from the moment condition (\ref{momcond}).}.
Expanding the boundary condition (\ref{1cut1}) to leading order in
$y-y_c$, after some algebra we find that it  can be expressed as
\beq
3M_2(iy_c)y_c(y-y_c)^2=-4x
\eeq
which is similar to the scaling behaviour in the Hermitian
one-matrix model with the same odd polynomial potential (i.e. (\ref{penner})
without the logarithm term) \cite{akm,fgz}. Moreover, the moment functions of
interest have the leading order scaling behaviours
\beq
\tilde{M}_2=\frac{2}{y_c^3}-g_c=-2g_c<0~~~,~~~\tilde{M}_1=\tilde{M}_2(iy_c)
(y-y_c)=-2g_c(y-y_c)>0
\label{momsc}\eeq
and
\beq
\tilde{d}_c=2y_c>0
\label{dsc}\eeq

Since (\ref{momsc}) and (\ref{dsc}) have the same signs as $M_1$, $M_2$ and
$d$ for the corresponding asymmetric Hermitian model \cite{akm,fgz}, it
follows from (\ref{freealt}) that the topological expansion (\ref{freegenus})
about the $m=2$ multi-critical point is an alternating series relative to that
of the Hermitian one-matrix model with the same potential. Aside from this
alternating nature, the genus expansion (\ref{freegenus}) resembles the usual
Painlev\'e expansion \cite{fgz} (for suitable normalization of the cosmological
constant). It is expected that this is also true for general higher-order
multi-critical points, and thus it is conjectured that the genus expansion
(\ref{freegenus}) about an $m$-th order multi-critical point in the scaling
limit alternates according to
\beq
F_h=(-1)^{h-1}F_h^H
\eeq
where $F_h^H$ is the genus $h$ contribution to the free energy of an
$m$-th multi-critical Hermitian model obtained from a symmetric potential.
Being an alternating series the topological expansion of chirally symmetric
fermionic one-matrix models may be Borel summable and thus these matrix
models provide some novel worldsheet discretization of the string theory.
However, it should be expected that the double-scaling limit of the
fermionic one-matrix model differs from that of the Hermitian one-matrix
model by more than just signs. For instance, the alternating nature of the
genus expansion above is no longer true for generic (non-chirally symmetric)
fermionic potentials $V(\ps2)$ \cite{akm} (essentially because the cut
endpoints do not lie on the imaginary axis in these cases), whereas in the
Hermitian case it is well-known that the generic matrix model free energy
coincides (modulo a factor of 2) in the continuum limit with that of a reduced
matrix model
\cite{ackm,fgz,mak3}. The topological expansion of the fermionic one-matrix
model therefore represents a rather unusual random surface theory which
deserves future investigation.

\subsection{Virasoro Algebra Constraints and Integrable Hierarchies}

The loop equation (\ref{loopcont}) for any polynomial potential $V(\ps2)
=\sum_{k\geq0}g_k(\ps2)^k$ can be represented as a set of discrete Virasoro
constraints imposed on the partition function (\ref{part1}). From
(\ref{multins}) and (\ref{loopins}) it follows that the
expansion of (\ref{loopcont}) in $1/z$ can be written as
\beq
\frac{1}{Z_1}\sum_{n=0}^\infty\frac{1}{z^{n+1}}L_nZ_1=0
\label{virconstrexp}\eeq
where the differential operators
\beq
L_n=\sum_{k\geq1}kg_k\frac{\partial}{\partial g_{k+n}}+\frac{1}{N^2}\sum_{k=0}
^n\frac{\partial^2}{\partial g_k\partial g_{n-k}}+2\frac{\partial}{\partial
g_n}
\label{virgen}\eeq
generate the discrete $c=0$ Virasoro algebra
\beq
[L_n,L_m]=(n-m)L_{n+m}
\label{viralg}\eeq
The loop equation (\ref{loopcont}) is therefore represented by the Virasoro
constraints
\beq
L_nZ_1=0~~~~~,~~~~~n\geq0
\label{virconstr}\eeq
These Virasoro constraints resemble those of the complex one-matrix
model (\ref{complexpart}) \cite{mak3}. Notice also that by definition we have
the additional constraint
\beq
\frac{\partial}{\partial g_0}Z_1=N^2Z_1
\label{g0constr}\eeq

It is instructive to see precisely what symmetry of the fermionic matrix model
the operators (\ref{virgen}) represent for each $n$. They are associated with
the invariance of the partition function (\ref{part1}) under the infinitesimal
shifts
\beq
\psi\to\psi+\epsilon\psi(\ps2)^n~~,~~n\geq1~~~~~~;~~~~~\bar\psi\to\bar\psi
\label{infshift}\eeq
of the fermionic variables, under which the potential $V$ in (\ref{part1})
changes by
\beq\new{\begin{array}{ll}
V(\ps2)=\sum_{k\geq0}g_k(\ps2)^k\to&\sum_{k\geq0}g_k\left(\ps2+\epsilon(\ps2)
^{n+1}\right)^k\\&~~~~=V(\ps2)+\epsilon\sum_{k>n}\left(g_k+(k-n)g_{k-n}\right)
(\ps2)^k\end{array}}
\label{potinfchange}\eeq
This variation in the partition function (\ref{part1}) is represented by the
action of the operators
\beq
L_n^{\rm cl}=\sum_{k\geq1}kg_k\frac{\partial}{\partial g_{k+n}}
\label{classvirgen}\eeq
on the partition function $Z_1$. The operators $L_n^{\rm cl}$ represent the
``classical" invariance of $Z_1$ and are the usual classical generators of the
Virasoro algebra (\ref{viralg}) (or, more precisely, of the Borel subalgebra of
the full Virasoro algebra). The integration measure in (\ref{part1}) under the
shifts (\ref{infshift}) changes by
\beq\new{\begin{array}{ll}
d\psi~d\bar\psi&\to
d\psi~d\bar\psi\cdot\det\left[\frac{\partial(\psi+\epsilon\psi
(\ps2)^n)}{\partial
\psi}\right]\\&\sim d\psi~d\bar\psi\cdot\left(1+\epsilon\tr\frac{\partial(\psi(
\ps2)^n)}{\partial\psi}\right)\\&=d\psi~d\bar\psi\cdot
\left(1+2\epsilon\tr(\ps2)
^n+\epsilon\sum_{k=0}^n\tr(\ps2)^k\tr(\ps2)^{n-k}\right)\end{array}}
\label{intmeaschange}\eeq
which is represented by the action of the last 2 terms in (\ref{virgen}) on
$Z_1$ and which can be thought of as encoding the ``quantum corrections" to the
classical Virasoro generators (\ref{classvirgen}). The fermionic Virasoro
operators (\ref{virgen}) differ from the standard bosonic ones in the last
derivative operator $2\frac{\partial}{\partial g_n}$ which is absent in the
scalar cases \cite{mak3}.

The loop equations of the fermionic matrix model represent the full set of
Ward identities (or equations of motion) of the model. The completeness of
these sets of equations is reflected in the fact that the Virasoro operators
(\ref{virgen}) form a closed algebra (\ref{viralg}). The advantage of
representing the loop equations of the matrix model in terms of Virasoro
constraints is that it represents an invariant formulation of the partition
function $Z_1$, i.e. $Z_1$ is determined as a solution of this set of
compatible differential equations. From this point of view one can now compare
these with
other solutions to the Virasoro constraints, for example those which arise
naturally from free fermion or free boson conformal field theory
\cite{ginsparg}, or
Kontsevich integrals \cite{fgz,mir,mor}. In the former case this implies a
certain string
theoretical duality, between 2-dimensional world-sheets and the spectral
surfaces which are associated to the configuration space of the string theory.
The analysis of the uniqueness of such representations of the Virasoro
constraints partitions the solutions into universality classes which can be
used to relate these models to the multicomponent KP and Toda-chain integrable
hierarchies \cite{kharch1}. In this case they also yield an alternative way of
studying
the double-scaling and continuum limits of the discretized random surface
theories \cite{dijk1,makmarsh}. Furthermore, the intersection numbers on the
(compactified) moduli
space of compact Riemann surfaces in 2-dimensional topological gravity are
known to be related to a number of recursion relations which are equivalent to
the Virasoro constraints of Hermitian 1-matrix models in the continuum limit
\cite{dijk1,kont,wit}.
The similarities between the Virasoro constraints in the fermionic case and the
bosonic ones are
another indication of the relation between the adjoint fermion 1-matrix
models and 2-dimensional quantum gravity. It still remains an unsolved
problem, however, as to what this precise connection really is.

Nonetheless, the loop equations are precisely just the above set of Virasoro
constraints imposed on the partition function $Z_1$. In the continuum limit,
this Virasoro symmetry would then represent the underlying conformal invariance
of the associated random surface theory, and it allows one to identify the
proper continuum (double-scaling) limit partition function with the correct
continuum Virasoro algebra invariance \cite{mak3}. It would be interesting to
relate the continuum loop equations of the fermionic one-matrix model more
precisely to the techniques of exactly solvable integrable systems (e.g. the
KdV or KP hierarchies) \cite{banks,doug2,grossmig2}, and also to Witten's
approach \cite{dijk2,wit} which is based on the interpretation of the double
scaling limit of the matrix model as a topological quantum field theory so that
the problem is reduced to the calculation of intersection indices on moduli
space.
It would also be very interesting to find a relation, based on the Virasoro
algebra constraints, between the adjoint fermion one-matrix model and conformal
field theory (e.g. in the Hermitian case the one-matrix model can be
represented in terms of correlation functions of a $D=1$ conformal field theory
\cite{mir,mor}, i.e. a Gaussian field theory of a free scalar field).

\section{Adjoint Fermion Two-matrix Models}

The simplest higher-dimensional generalization of the $D=0$ dimensional
model (\ref{part1}) is a non-dynamical gauge theory minimally coupled to a
fermionic two-matrix ensemble. The partition function of the two-matrix model
is
\beq
Z_2=\int[dU]~\int d\psi~d\bar\psi~d\chi~d\bar\chi~\e^{S[\psi,\bar\psi,\chi,
\bar\chi;U]}
\label{part2}\eeq
with action
\beq
S[\psi,\bar\psi,\chi,\bar\chi;U]=N^2\tr\left(\bar\psi U\chi U^\dagger+\bar\chi
U^\dagger\psi U+V(\ps2)+\tilde{V}(\ch2)\right)
\label{action2}\eeq
where $[dU]$ is Haar measure on the unitary group $U(N)$ and $V$ and
$\tilde V$ are independent potentials. The fermion fields
$\psi$, $\bar\psi$, $\chi$ and $\bar\chi$ are independent $N\times N$
Grassmann-valued matrices and the partition function (\ref{part2})
describes staggered self-interacting Dirac fermions which interact with
a gauge field on a $D=\frac{1}{2}$ dimensional lattice (i.e. a single link
shared by 2 fermions) and which transform under the adjoint representation of
the gauge group $U(N)$. Using a gauge transformation (see below) the unitary
matrices in (\ref{part2}) can be eliminated and one is left with the natural
fermionic analog of a Hermitian two-matrix model \cite{fgz,mz}. Just as the
fermionic one-matrix model provides some novel random theory of discretized
quantum gravity, the adjoint fermion two-matrix model (\ref{part2}) will
yield some novel random theory of discretized gravity interacting with some
type of matter. It may be that these matter fields are not restricted by the
$D=1$ conformal barrier as they are in the Hermitian cases, and one might
therefore obtain a matrix model representation of strings in $D>1$
dimensional target spaces. Alternatively, the Grassmann integrals over the
fermion matrix fields can be performed leaving an induced gauge theory. This
will be discussed in the next Section where we consider the generalizations
of (\ref{part2}) to arbitrary dimensions, i.e. the analog of the Kazakov-Migdal
model \cite{km} using (adjoint) fermions \cite{kmak2,mak2,mz,mig2,mig5,semsz}.

The model defined by (\ref{part2}) possesses a number of
symmetries which follow from the invariance properties of the Haar measure.
It is invariant under the gauge transformation
\beq
U\to V^\dagger UW~~~,~~~(\psi,\bar\psi)\to(V^\dagger\psi V,V^\dagger\bar\psi
V)~~~,~~~(\chi,\bar\chi)\to(W^\dagger\chi W,W^\dagger\bar\chi W)
\label{gaugetrans2}\eeq
where $\{V,W\}\in U(N)\otimes U(N)$. The model also has the
$U(1)$ gauge-symmetry
\beq
U\to {\cal Z}U
\label{znsym}\eeq
where $\cal Z$ is an element of the center of $U(N)$, i.e. a unimodular
complex number. This symmetry implies that any correlator of the theory must
contain the same number of $U$ and $U^\dagger$ matrices.
The $U(1)$ phase invariance of (\ref{part2})
\beq
(\psi,\bar\psi)\to(\e^{i\theta}\psi,\e^{-i\theta}\bar\psi)~~~~~,~~~~~
(\chi,\bar\chi)\to(\e^{i\theta}\chi,\e^{-i\theta}\bar\chi)
\label{charge}\eeq
leads to fermion number conservation in the gauge theory, i.e. any
correlator of the matrix model must contain the same number of fermion
and conjugate matrices.

In the symmetric case where $V=\tilde V$ the charge conjugation
\beq
U\to U^\dagger~~~~~,~~~~~(\psi,\bar\psi)\leftrightarrow(\chi,\bar\chi)
\label{sympot}\eeq
is a symmetry of the two-matrix model and it implies equality of a large
number of correlators of the model. When both potentials $V$ and $\tilde V$
in (\ref{action2}) are odd polynomials, the chiral transformation
\beq
(\psi,\bar\psi)\to(\bar\psi,-\psi)~~~~~,~~~~~(\chi,\bar\chi)\to(\bar\chi,
-\chi)
\label{chiral}\eeq
is a symmetry and as before it implies that all even $\ps2$ and $\ch2$
moments vanish,
\beq
\langle\tr(\ps2)^{2k}\rangle=\langle\tr(\ch2)^{2k}\rangle=0
\eeq
where the normalized averages are now with respect to the statistical ensemble
(\ref{part2}). In the $\IZ_2$-symmetric case when $V=-\tilde V$ is an odd
polynomial potential
the charge conjugation
\beq
U\to U^\dagger~~~,~~~(\psi,\bar\psi)\to(\chi,-\bar\chi)~~~,~~~(\chi,
\bar\chi)\to(-\psi,\bar\psi)
\label{chargeconj}\eeq
is an invariance of the model (\ref{part2}) and it relates the non-vanishing
$\ps2$ and $\ch2$ moments by
\beq
\langle\tr(\ps2)^{2k+1}\rangle=-\langle\tr(\ch2)^{2k+1}\rangle
\eeq
Another important symmetry in this case is the composition of the charge
conjugation invariance (\ref{chargeconj}) and the chiral symmetry
(\ref{chiral}) which leads to a mixed symmetry among the more general
correlators of the two-matrix model (\ref{part2}). As before, even though
the fermionic matrices in (\ref{part2}) cannot be diagonalized the
fermionic two-matrix model leads to solutions analogous to those of
Hermitian two-matrix models.

\subsection{Loop Equations}

The most general generating functions for the correlators of the two-matrix
model (\ref{part2}) are those which generate all possible observables
respecting
the symmetries (\ref{gaugetrans2})--(\ref{charge}). We therefore introduce
the even-even two-point correlator
\beq
\Gee(z,w)=\left\langle\tr\frac{1}{z-\ps2}U\frac{1}{w-\ch2}U^\dagger\right
\rangle
\label{even}\eeq
and the odd-odd two-point correlators
\beq
\Hoo(z,w)=\left\langle\tr\psi\frac{1}{z-\ps2}U\frac{1}{w-\ch2}\bar\chi U^
\dagger\right\rangle
\label{hodd}\eeq
\beq
\Koo(z,w)=\left\langle\tr\frac{1}{z-\ps2}\bar\psi U\chi\frac{1}{w-\ch2}U^
\dagger\right\rangle
\label{kodd}\eeq
We also have the usual generating functions for the $\ps2$ and $\ch2$ moments
\beq
\omega(z)=\left\langle\tr\frac{1}{z-\ps2}\right\rangle~~~~~,~~~~~\tilde\omega
(w)=\left\langle\tr\frac{1}{w-\ch2}\right\rangle
\label{omegas}\eeq
When $N$ is finite these functions are all analytic in the punctured complex
plane.

The even-even correlator has the asymptotic expansions
\beq
\Gee(z,w)=\frac{\omega(z)}{w}+\sum_{n=1}^{N^2}\frac{\Gee_n(z)}{w^{n+1}}
=\frac{\tilde\omega(w)}{z}+\sum_{n=1}^{N^2}\frac{\tilde\Gee_n(w)}{z^{n+1}}
\label{gasymp}\eeq
where
\beq
\Gee_n(z)=\left\langle\tr\frac{1}{z-\ps2}U(\ch2)^nU^\dagger\right\rangle~~~,~~~
\tilde\Gee_n(w)=\left\langle\tr(\ps2)^nU\frac{1}{w-\ch2}U^\dagger\right\rangle
\eeq
Similarly, the odd-odd correlators have the asymptotic expansions
\beq
\Hoo(z,w)=\sum_{n=0}^{N^2}\frac{\Hoo_n(z)}{w^{n+1}}=\sum_{n=0}^{N^2}
\frac{\tilde\Hoo_n(w)}{z^{n+1}}~~~,~~~\Koo(z,w)=\sum_{n=0}^{N^2}\frac{
\Koo_n(z)}{w^{n+1}}=\sum_{n=0}^{N^2}\frac{\tilde\Koo_n(w)}{z^{n+1}}
\label{hasymp}\eeq
where
\beq
\Hoo_n(z)=\left\langle\tr\psi\frac{1}{z-\ps2}U(\ch2)^n\bar\chi U^\dagger\right
\rangle~~~,~~~\tilde\Hoo_n(w)=\left\langle\tr\psi(\ps2)^nU\frac{1}{w-\ch2}\bar
\chi U^\dagger\right\rangle
\eeq
and
\beq
\Koo_n(z)=\left\langle\tr\frac{1}{z-\ps2}\bar\psi U\chi(\ch2)^nU^\dagger\right
\rangle~~~,~~~\tilde\Koo_n(w)=\left\langle\tr(\ps2)^n\bar\psi U\chi\frac{1}{w-
\ch2}U^\dagger\right\rangle
\eeq

The various symmetries (\ref{sympot})--(\ref{chargeconj}) imply some
noteworthy relations among these generating functions. For example,
in the symmetric case $V=\tilde V$ we have the symmetries
\beq
\Gee(z,w)=\Gee(w,z)~~~~,~~~~\Koo(z,w)=-\Hoo(-z,-w)~~~~,~~~~\omega(z)=\tilde
\omega(z)
\label{symmetric}\eeq
When the potential $V=-\tilde V$ is an odd polynomial, we have the symmetries
\beq
\Gee(z,w)=\Gee(-w,-z)~~~,~~~\Hoo(z,w)=\Hoo(w,z)~~~,~~~\Koo(z,w)=-\Hoo(-z,-w)
\label{asymmetric}\eeq
and the vanishing of the even moments in this case further implies that
\beq
\omega(z)+\tilde\omega(z)=2/z
\label{asymom}\eeq

An important observable when the two-matrix model (\ref{part2}) is viewed
as an induced gauge theory is the pair correlator of the gauge fields
\beq
\frac{1}{N}C_{ij}\delta_{i\ell}\delta_{jk}=\left\langle U_{ij}U^\dagger_{k
\ell}\right\rangle
\label{gaugecorr}\eeq
where the delta functions on the left-hand side of (\ref{gaugecorr}) arise
from the gauge invariance of (\ref{part2}). Unitarity implies that it obeys
the sum rule
\beq
\frac{1}{N}\sum_{i=1}^NC_{ij}=\frac{1}{N}\sum_{j=1}^NC_{ij}=1
\label{gaugenorm}\eeq
In the Hermitian case, when $V=\tilde V$ the pair correlator
$C_{ij}=N\langle|U_{ij}|^2\rangle$ can be computed from the double
discontinuity
of the scalar version of the generating function (\ref{even}) using
\beq
\Gee^H(z,w)\equiv\left\langle\tr\frac{1}{z-\phi}U\frac{1}{w-\phi}U^\dagger
\right\rangle=\int d\alpha~\rho(\alpha)\int
d\beta~\rho(\beta)\frac{C(\alpha,\beta)}{(z-\alpha)(w-\beta)}
\label{geeherm}\eeq
since there it depends only on the
moments of powers of the Hermitian fields \cite{dms,mak2}. This is {\it not}
the case for the adjoint fermion matrix ensemble (\ref{part2}) because it
involves a larger set of fermionic correlators than just those of the type
$\langle\tr(\ps2)^n\rangle$ and $\langle\tr(\ch2)^n\rangle$ (for instance one
needs to know the correlators of the form $\langle\tr\bar\psi^n\psi^n\rangle$).
There does not appear to be any direct way to generate these observables from
the loop equations. Moreover, there is no known analog of the Itzykson-Zuber
formula \cite{iz} for the integral
\beq
I[\psi,\chi]=\int[dU]~\e^{N^2\tr(\psi U\chi U^\dagger)}
\label{iz}\eeq
when $\psi$ and $\chi$ are Grassmann-valued matrices that transform under the
adjoint representation of $U(N)$. In the Hermitian
case the knowledge of the explicit form of (\ref{iz}) at least allows one
to formally determine $C_{ij}$ using saddle-point methods \cite{dms}.

For the fermionic matrix chain (\ref{part2}), factorization and symmetry
imply that the correlators (\ref{even})--(\ref{kodd}) and (\ref{gaugecorr})
generate the complete set of observables of the model at $N=\infty$.
The loop equations for the correlators (\ref{even})--(\ref{kodd}) can now
be derived as before and they will involve sets of mixed equations for the
even-even correlator with either of the odd-odd correlators. The loop
equations involving the generating function (\ref{hodd}) are as follows.
The first one follows from the identity
\beq
\int[dU]~\int d\psi~d\bar\psi~d\chi~d\bar\chi~\frac{\partial}{\partial
\psi_{ij}}\left[\left(\psi\frac{1}{z-\ps2}U\frac{1}{w-\ch2}U^\dagger\right
)_{k\ell}\e^{S[\psi,\bar\psi,\chi,\bar\chi;U]}\right]=0
\label{2matrixprelimid}\eeq
Expanding (\ref{2matrixprelimid}) into averages and summing over $i=k,j=\ell$
as before leads to
\beq\new{\begin{array}{c}
0=\left\langle\tr\frac{1}{z-\ps2}U\frac{1}{w-\ch2}U^\dagger\right\rangle
+\left\langle\tr\psi\frac{1}{z-\ps2}\bar\psi\tr\frac{1}{z-\ps2}U\frac{1}
{w-\ch2}U^\dagger\right\rangle\\+\left\langle\tr\psi\frac{1}{z-\ps2}U
\frac{1}{w-\ch2}\bar\chi
U^\dagger\right\rangle+\left\langle\tr\psi\frac{1}{z-\ps2}U\frac{1}{w-\ch2}
U^\dagger V'(\ps2)\bar\psi\right\rangle\end{array}}
\label{2loopexpl}\eeq
which at $N=\infty$, when factorization holds, gives
\beq
\left(2-z\omega(z)\right)\Gee(z,w)+\Hoo(z,w)+\oint_{\cal C}\frac{d\lambda}{
2\pi i}~\frac{V'(\lambda)\lambda}{z-\lambda}\Gee(\lambda,w)=0
\label{loop21}\eeq
where the contour $\cal C$ encircles the singularities of $\Gee(z,w)$
with counterclockwise orientation in the complex $z$-plane. Writing
the same sort of equation for $\bar\chi$ instead of $\psi$
\beq
\int[dU]~\int d\psi~d\bar\psi~d\chi~d\bar\chi~\frac{\partial}{\partial\bar\chi
_{ij}}\left[\left(U^\dagger\frac{1}{z-\ps2}U\frac{1}{w-\ch2}\bar\chi\right)_{
k\ell}\e^{S[\psi,\bar\psi,\chi,\bar\chi;U]}\right]=0
\eeq
leads to the $N=\infty$ loop equation
\beq
\left(2-w\tilde\omega(w)\right)\Gee(z,w)+\Hoo(z,w)+\oint_{\cal C}\frac{d
\lambda}{2\pi i}~\frac{\tilde V'(\lambda)\lambda}{w-\lambda}\Gee(z,\lambda)=0
\label{loop22}\eeq

Two more loop equations involving the two-point correlators (\ref{even})
and (\ref{hodd}) follow first from the identity
\beq
\int[dU]~\int d\psi~d\bar\psi~d\chi~d\bar\chi~\frac{\partial}{\partial\bar
\psi_{ij}}\left[\left(\frac{1}{z-\ps2}U\frac{1}{w-\ch2}\bar\chi U^\dagger
\right)_{k\ell}\e^{S[\psi,\bar\psi,\chi,\bar\chi;U]}\right]=0
\eeq
which leads to
\beq
\omega(z)\Hoo(z,w)-\omega(z)+w\Gee(z,w)-\oint_{\cal C}\frac{d\lambda}{2\pi i}~
\frac{V'(\lambda)\lambda}{z-\lambda}\Hoo(\lambda,w)=0
\label{loop23}\eeq
at large-$N$. Finally, we analogously have the identity
\beq
\int[dU]~\int d\psi~d\bar\psi~d\chi~d\bar\chi~\frac{\partial}{\partial \chi
_{ij}}\left[\left(U^\dagger\psi\frac{1}{z-\ps2}U\frac{1}{w-\ch2}\right
)_{k\ell}\e^{S[\psi,\bar\psi,\chi,\bar\chi;U]}\right]=0
\eeq
which yields the large-$N$ loop equation
\beq
\tilde{\omega}(w)\Hoo(z,w)-\tilde\omega(w)+z\Gee(z,w)-\oint_{\cal C}\frac
{d\lambda}{2\pi i}~\frac{\tilde V'(\lambda)\lambda}{w-\lambda}\Hoo(z,\lambda)=0
\label{loop24}\eeq
A similar set of loop equations involving the odd-odd two-point correlator
$\Koo(z,w)$ instead of $\Hoo(z,w)$ can also be derived as above with the
obvious modifications.

These loop equations could also have been obtained from the Schwinger-Dyson
equations expressing the invariance of the fermionic integration measure under
arbitrary changes of variables and the symmetries of the Haar measure $[dU]$.
For example, denote the generators of $U(N)$ by $T^a$, $a=1,\ldots,N^2$.
They obey the completeness and normalization relations
\beq
\sum_{a=1}^{N^2}(T^a)_{ij}(T^a)_{k\ell}=N\delta_{i\ell}\delta_{jk}~~~~~,~~~~~
\tr T^aT^b=\delta^{ab}
\label{ungen}\eeq
We can then represent the Grassmann-valued matrices $\psi$ as
\beq
\psi=\sum_{a=1}^{N^2}T^a\psi^a~~~{\rm where}~~~\psi^a=\tr T^a\psi
\eeq
The first loop equation (\ref{loop21}) then follows from the invariance of the
integration measure in the vanishing correlator
\beq
\left\langle\tr T^a\psi\frac{1}{z-\ps2}U\frac{1}{w-\ch2}U^\dagger\right\rangle
\equiv0
\label{sdc1}\eeq
under the field transformation
\beq
\psi\to\psi+\epsilon^aT^a
\label{sdtransf}\eeq
with $\epsilon^a$ infinitesimal parameters. The vanishing of (\ref{sdc1}) is a
consequence of the Grassmann integrations in (\ref{part2}) and gauge
invariance.  (\ref{loop21}) now follows explicitly by performing the shift
(\ref{sdtransf}), using the invariance of the integration measure to get
\beq
\frac{1}{Z_2}\int[dU]~\int d\psi~d\bar\psi~d\chi~d\bar\chi~T_{\ell
k}^a\frac{\partial}{\partial\psi_{ij}}\left\{\left(\psi
\frac{1}{z-\ps2}U\frac{1}
{w-\ch2}U^\dagger\right)_{k\ell}\e^S\right\}\epsilon^aT_{ij}^a=0
\eeq
and then summing over $a=1,\dots,N^2$ using (\ref{ungen}) and calculating the
derivatives $\frac{\partial}{\partial\psi^b}$. The other 3 loop equations
follow
from the invariance conditions analogous to (\ref{sdc1}), (\ref{sdtransf})
\cite{mz}.

The loop equations (\ref{loop21}) and (\ref{loop23}) can be combined to
give an equation which determines the one-loop correlator $\omega(z)$.
Substituting into these loop equations the asymptotic expansions
(\ref{gasymp}) and (\ref{hasymp}) in $1/w$ and equating the coefficients of
$\frac{1}{w^{n+1}}$ we get recursive relations determining
$\Gee_n(z)$ and $\Hoo_n(z)$ in terms of $\omega(z)$
\beq
\Hoo_n(z)=\left(z\omega(z)-2\right)\Gee_n(z)-\oint_{\cal C}\frac{d\lambda}{2
\pi i}~\frac{V'(\lambda)\lambda}{z-\lambda}\Gee_n(\lambda)
\label{asloop1}\eeq
\beq
\Gee_{n+1}(z)=\oint_{\cal C}\frac{d\lambda}{2\pi i}~\frac{V'(\lambda)\lambda}
{z-\lambda}\Hoo_n(\lambda)-\omega(z)\Hoo_n(z)~~~,~~~\Gee_0(z)\equiv\omega(z)
\label{asloop2}\eeq
The equation determining $\omega(z)$ now follows from expanding the loop
equation (\ref{loop24}) in $1/w$ using (\ref{gasymp}) and (\ref{hasymp})
and keeping only the leading-order term in $1/w$,
\beq
z\omega(z)=1+\oint_{\cal C}\frac{d\lambda}{2\pi i}~\tilde V'(\lambda)\lambda
\Hoo(z,\lambda)
\label{asloop3}\eeq

When the potential $\tilde V$ is a polynomial of degree $\tilde K$
\beq
\tilde V(w)=\sum_{n=1}^{\tilde K}\frac{\tilde g_n}{n}w^n
\eeq
the equation (\ref{asloop3}) becomes
\beq
z\omega(z)=1+\sum_{k=1}^{\tilde K}\tilde g_k\Hoo_{k-1}(z)
\label{poloop3}\eeq
and it involves $\tilde K$ unknown functions which are determined by the
recursive equations (\ref{asloop1}) and (\ref{asloop2}). The equations that
determine $\omega(z)$ in this case lead to a $(2\tilde K)$-th order polynomial
equation for the one-loop correlator $\omega(z)$. When in addition the
potential $V$ is a polynomial (\ref{pot}), the contour integrals in
(\ref{asloop1}) and (\ref{asloop2}) are
\beq
\oint_{\cal C}\frac{d\lambda}{2\pi i}~\frac{V'(\lambda)\lambda}{z-\lambda}
\Gee_n(\lambda)=-V'(z)z\Gee_n(z)+G_n(z)
\label{poloop11}\eeq
\beq
\oint_{\cal C}\frac{d\lambda}{2\pi i}~\frac{V'(\lambda)\lambda}{z-\lambda}
\Hoo_n(\lambda)=-V'(z)z\Hoo_n(z)+H_n(z)
\label{poloop12}\eeq
where $G_n(z)$ and $H_n(z)$ are polynomials of degree $K-1$
\beq
G_n(z)=\sum_{m=1}^Kg_m\sum_{p=0}^{m-1}\Gee_{m-p-1,n}z^p~~~,~~~H_n(z)=\sum_
{m=1}^Kg_m\sum_{p=0}^{m-1}\Hoo_{m-p-1,n}z^p
\label{ghpols}\eeq
and
\beq
\Gee_{m,n}=\langle\tr(\ps2)^mU(\ch2)^nU^\dagger\rangle~~~~~,~~~~~\Hoo_{m,n}=
\langle\tr\psi(\ps2)^mU(\ch2)^n\bar\chi U^\dagger\rangle
\label{ghcoeffs}\eeq
are the coefficients of the asymptotic expansions
\beq
\Gee_n(z)=\sum_{m=0}^\infty\frac{\Gee_{m,n}}{z^{m+1}}~~~~~,~~~~~\Hoo_n(z)=
\sum_{m=0}^\infty\frac{\Hoo_{m,n}}{z^{m+1}}
\eeq
An identical set of asymptotic equations determining $\tilde\omega(w)$ can
also be written down and they correspond to interchanging
tilde and un-tilde quantities in the above in the obvious way.

Once the single-loop correlators are known they can be substituted back into
the original loop equations and the even-even and odd-odd two-point
correlators can be found. In the case of the polynomial interactions above, the
contour integrals appearing in the loop equations can be determined as
\beq
\oint_{\cal C}\frac{d\lambda}{2\pi i}~\frac{V'(\lambda)\lambda}{z-\lambda}
\Gee(\lambda,w)=-V'(z)z\Gee(z,w)+\sum_{m=1}^Kg_m\sum_{p=0}^{m-1}
\tilde\Gee_{m-p-1}(w)z^p
\label{gcontz}\eeq
\beq
\oint_{\cal C}\frac{d\lambda}{2\pi i}~\frac{\tilde V'(\lambda)\lambda}{w-
\lambda}\Gee(z,\lambda)=-\tilde V'(w)w\Gee(z,w)+\sum_{m=1}^{\tilde K}\tilde
g_m\sum_{p=0}^{m-1}\Gee_{m-p-1}(z)w^p
\label{gcontw}\eeq
and similarly for the integrals involving $\Hoo(z,w)$.

\subsection{The Gaussian Model}

The asymmetric Gaussian potential
\beq
V(z)=-\tilde V(z)=mz
\label{gauss2}\eeq
describes a free Dirac fermion of mass $m$ on a lattice in $D=\frac{1}{2}$
dimensions. (\ref{poloop3}) is then
\beq
z\omega(z)=1-m\Hoo_0(z)
\eeq
where from (\ref{asloop1})
\beq
\Hoo_0(z)=(z\omega(z)-2)\omega(z)+mz\omega(z)-m
\eeq
The equation determining $\omega(z)$ is quadratic and it is the same as the
loop equation for the Gaussian one-matrix model (\ref{gausspot}) with
$t\equiv t_+$ where
\beq
t_\pm=m\pm1/m
\eeq
The one-loop correlator $\omega(z)$ is therefore given by (\ref{omgauss})
with this definition of $t$. The correlator $\tilde\omega(w)$ is then
determined by (\ref{asymom}).

The even-even correlator $\Gee(z,w)$ can now be determined by subtracting
the loop equation (\ref{loop22}) from (\ref{loop21}), using (\ref{gcontz}) and
(\ref{gcontw}), and substituting in
the one-loop correlators obtained above. We find
\beq
\Gee(z,w)=\frac{\left(t_++t_-\right)\left(\frac{1}{z}+\frac{1}{w}+\frac{1}{2z}
\sqrt{4+t_+^2z^2}-\frac{1}{2w}\sqrt{4+t_+^2w^2}\right)}{t_-(z+w)+
\sqrt{4+t_+^2z^2}+\sqrt{4+t_+^2w^2}}
\label{geegauss}\eeq
which when substituted back into (\ref{loop21}) gives the odd-odd two-point
correlator
\beq
\Hoo(z,w)=\frac{t_+(z+w)-\sqrt{4+t_+^2z^2}-\sqrt{4+t_+^2w^2}}{t_-(z+w)+
\sqrt{4+t_+^2z^2}+\sqrt{4+t_+^2w^2}}
\label{hoogauss}\eeq
The remaining generating function ${\cal K}(z,w)$ is given by
(\ref{asymmetric}). It is easy to see that these correlators are non-singular
for any $z$ and $w$ and obey the appropriate symmetries (\ref{asymmetric}).

\subsection{Cubic Interaction Model}

The simplest non-Gaussian model is associated with the asymmetric cubic
potential
\beq
V(z)=-\tilde V(z)=mz+\frac{g}{3}z^3
\label{cubic2}\eeq
In this case the asymptotic equation (\ref{poloop3}) is
\beq
z\omega(z)=1-m\Hoo_0(z)-g\Hoo_2(z)
\label{cubloop1}\eeq
where the unknown functions in (\ref{cubloop1}) are found by combining
(\ref{asloop1}) and (\ref{asloop2}) together using (\ref{poloop11}) and
(\ref{poloop12}) to generate the 3 equations
\beq
\Hoo_0(z)=-m-gz^2-gz\xi-\omega(z)\left[2-z(\omega(z)+m+gz^2)\right]
\label{cubloop2}\eeq
\bd
\Hoo_1(z)=-\Hoo_0(z)(2-z(\omega(z)+m+gz^2))(\omega(z)+m+gz^2)+(m+gz^2)\xi
\ed
\beq
+gz\Hoo_{0,0}(1-z(\omega(z)+m+gz^2))-g\Gee_{2,1}
\label{cubloop3}\eeq
\bd
\Hoo_2(z)=\Hoo_0(z)(2-z(\omega(z)+m+gz^2))^2(\omega(z)+m+gz^2)^2-\Hoo_1(z)
(2-z(\omega(z)+m+gz^2))
\ed
\beq
\times(\omega(z)+m+gz^2)-gz\Gee_{1,2}-g\Gee_{2,2}+g\Hoo_{1,1}(2-z(\omega(z)
+m+gz^2))
\label{cubloop4}\eeq
where as before $\xi=\langle\tr\ps2\rangle$. In arriving at
(\ref{cubloop2})--(\ref{cubloop4}) we have used the $\IZ_2$
charge-conjugation and chiral symmetries of the potential (\ref{cubic2}) to
deduce from (\ref{asymmetric}) that ${\cal H}_0(z)=-{\cal H}_0(-z)$ so that
${\cal H}_{1,0}={\cal
H}_{0,1}\equiv0$ here. These same symmetries also imply that
$\Gee_{2,1}=-\Gee_{1,2}$. Also, we have expanded the asymptotic loop equation
(\ref{asloop2}) for $n=1$ in powers of $1/z$ to find that
$\Hoo_{0,0}=\Gee_{1,1}$. Then
combining the 4 equations (\ref{cubloop1})--(\ref{cubloop4}) we find
that the one-loop correlator $\omega(z)$ is determined by
a complicated 6-th order equation
\beq\new{\begin{array}{ll}
0=&\Bigl[2g\{2-z(\omega(z)+m+gz^2)\}^2(\omega(z)+m+gz^2)^2+m\Bigr]\\&
\times\Bigl[\{z(\omega(z)+m+gz^2)-2\}\omega(z)-(m+gz^2)z-gz\xi\Bigr]\\&
-g(\omega(z)+m+gz^2)\{2-z(\omega(z)+m+gz^2)\}\Bigm[(m+gz^2)\xi\\&+gz\Hoo_{0,0}
\{1-z(\omega(z)+m+gz^2)\}-g\Gee_{2,1}\Bigm]+g^2z\Gee_{2,1}-g^2\Gee_{2,2}
\\&+g^2\{2-z(\omega(z)+m+gz^2)\}\Hoo_{1,1}+z\omega(z)-1\end{array}}
\label{cubiclong}\eeq

The solvability features of fermionic multi-matrix models are even worse
compared to Hermitian multi-matrix models where a degree $K$ polynomial
potential leads to a $K$-th order polynomial equation for the one-loop
correlator \cite{mak2}. Here the order of the equation doubles due to the
doubling of degrees of freedom in the fermionic case, just like in the
fermionic
one-matrix models, which can be understood from the result (\ref{part1exact})
which required 2 Hermitian matrices for its derivation. There does not
appear to be any way to directly solve the 6-th order equation
(\ref{cubiclong}), nor does there seem to be any way of reducing it to a
lower degree equation. Thus it is not possible to directly study the cut
structure of $\omega(z)$ and determine the ensuing phase structure of the
two-matrix model as before.

\subsection{A Penner-type Model}

In the Hermitian case there is still an exactly solvable non-Gaussian
multi-matrix model involving a non-polynomial interaction
\cite{mak4}--\cite{mak5},\cite{pan}. In
certain limiting cases it can be reduced to a polynomial interaction. The
analog of this in the fermionic case is the asymmetric logarithmic interaction
\beq
V(z)=-\tilde V(z)=mz+g\log(\gbar-z)
\label{penner2}\eeq
In this case there is an extra boundary condition imposed on the loop equations
by requiring that the residue of the simple pole at $z=\gbar$ vanish.

The asymptotic equation (\ref{asloop3}) for the potential (\ref{penner2}) reads
\beq
z\omega(z)=1-m\Hoo_0(z)-g\Hoo(z,\gbar)
\label{penloop1}\eeq
where the function $\Hoo_0(z)$ is determined from (\ref{asloop1}) as
\beq
\Hoo_0(z)=\left[2-z\left(\omega(z)-m-\frac{g}{z-\gbar}\right)\right]\omega(z)
-\left(m+\frac{g\gbar}{z-\gbar}\omega(\gbar)\right)
\label{penloop2}\eeq
and the function $\Hoo(z,\gbar)$ is determined by letting $w\to\gbar$ in
the loop equations (\ref{loop21}) and (\ref{loop22}) which lead to
\beq\new{\begin{array}{l}
\Hoo(z,\gbar)\\=\frac{\left(2-z\left(\omega(z)-m-\frac{g}{z-\gbar}\right)
\right)\left(\omega(z)-\frac{g}{z-\gbar}\Hoo(\gbar,\gbar)\right)-\gbar
\left(m\tilde\omega(\gbar)+\frac{g\gbar}{z-\gbar}\Gee(\gbar,\gbar)\right)}{
\left(\omega(z)-m-\frac{g}{z-\gbar}\right)\left(2-z\left(\omega(z)-m-
\frac{g}{z-\gbar}\right)\right)+\gbar}\end{array}}
\label{hoogbar}\eeq
Substituting (\ref{penloop2}) and (\ref{hoogbar}) into (\ref{penloop1})
leads to a complicated quartic equation for the one-loop correlator $\omega(z)$
\beq\new{\begin{array}{c}
0=m\left[\left(2-z\left(\omega(z)-m-\frac{g}{z-\gbar}\right)\right)\omega(z)
-\left(m+\frac{g\gbar}{z-\gbar}\omega(\gbar)\right)\right]+z\omega(z)-1\\
+g\left[\frac{\left(2-z\left(\omega(z)-m-\frac{g}{z-\gbar}\right)\right)
\left(\omega(z)-\frac{g}{z-\gbar}\Hoo(\gbar,\gbar)\right)-\gbar\left(
m\tilde\omega(\gbar)+\frac{g\gbar}{z-\gbar}\Gee(\gbar,\gbar)\right)}{\left(
\omega(z)-m-\frac{g}{z-\gbar}\right)\left(2-z\left(\omega(z)-m-\frac{g}
{z-\gbar}\right)\right)+\gbar}\right]\end{array}}
\label{penquartic}\eeq

In the Hermitian case the extra boundary condition at $z=\gbar$ leads to
a quadratic equation for $\omega(z)$ \cite{mak4,mak2}. Here, again because of
the doubling of the fermionic degrees of freedom, we obtain a
quartic equation which is not directly amenable to a multi-branch
solution but which is nonetheless explicitly solvable. It is not clear,
however, how to determine the precise cut structure or the existence of
phase transitions with these solutions due to the complicated structure
of their square root branches. Thus the only exactly solvable fermionic
multi-matrix models are the trivial Gaussian ones.

\subsection{The Genus Expansion and $W$-algebra Constraints}

Although the loop equations in Subsection 4.1 were derived at infinite $N$, it
is possible to examine the structure of the $\frac{1}{N}$-expansion of
the fermionic two-matrix model by including the irreducible correlators
(of order $1/N^2$) using the loop insertion operator (\ref{loopins}) in the
asymmetric case $V\neq\tilde V$. The extensions of the loop equations
(\ref{loop21}) and (\ref{loop23}) to finite $N$ are (see (\ref{2loopexpl}))
\beq
(2-z\omega(z))\Gee(z,w)+\Hoo(z,w)-{\cal L}(z)\Gee(z,w)+\oint_{\cal C}
\frac{d\lambda}{2\pi i}~\frac{V'(\lambda)\lambda}{z-\lambda}\Gee(\lambda,w)=0
\label{loop1n}\eeq
\beq
\omega(z)\Hoo(z,w)-\omega(z)+w\Gee(z,w)+{\cal L}(z)\Hoo(z,w)-\oint_{\cal C}
\frac{d\lambda}{2\pi i}~\frac{V'(\lambda)\lambda}{z-\lambda}\Hoo(\lambda,w)=0
\label{loop2n}\eeq
so that the recurrence relations (\ref{asloop1}) and (\ref{asloop2}) at
finite $N$ become
\beq
\Hoo_n(z)=(z\omega(z)-2)\Gee_n(z)-\oint_{\cal C}\frac{d\lambda}{2\pi i}~
\frac{V'(\lambda)\lambda}{z-\lambda}\Gee_n(\lambda)+{\cal L}(z)\Gee_n(z)
\label{asloop1n}\eeq
\beq
\Gee_{n+1}(z)=\oint_{\cal C}\frac{d\lambda}{2\pi i}~\frac{V'(\lambda)\lambda}
{z-\lambda}\Hoo_n(\lambda)-\omega(z)\Hoo_n(z)-{\cal L}(z)\Hoo_n(z)
\label{asloop2n}\eeq

We can introduce the analogs of the loop insertion operator (\ref{loopins})
for the correlators $\Gee_n(z)$ and $\Hoo_n(z)$ by
\beq
\Hoo_n(z)=\frac{1}{Z_2}{\cal L}_{2n+1}(z)Z_2
\label{hloopins}\eeq
where
\beq
{\cal L}_{2n+1}(z)\equiv\sum_{k\geq-n}\frac{1}{z^{k+n}}
\Winf_k^{(2n)}
\label{loopinseven}\eeq
and
\beq
\Gee_n(z)=\frac{1}{Z_2}{\cal L}_{2n}(z)Z_2
\label{gloopins}\eeq
where
\beq
{\cal L}_{2n}(z)\equiv\sum_{k\geq-n}\frac{1}{z^{k+n+1}}
\Winf_k^{(2n+1)}
\label{loopinsodd}\eeq
Here $\Winf_k^{(m)}$ are some differential operators determined by the
general potential $V(\ps2)=\sum_{k\geq0}g_k(\ps2)^k$. From (\ref{loopins})
it follows that
\beq
{\cal L}_0(z)=N^2{\cal L}(z)~~~~~,~~~~~\Winf_k^{(1)}=\frac{\partial}{\partial
g_k}
\label{n0ops}\eeq
To determine the operators $\Winf_k^{(m)}$ for $m\neq1$, we substitute
(\ref{hloopins})--(\ref{loopinsodd}) into (\ref{asloop1n}) and
(\ref{asloop2n}) and equate the coefficients of the $\frac{1}{z^{k+n}}$
terms. We then find
that the operators $\Winf_k^{(m)}$ obey the recurrence relations
\beq
\Winf_k^{(2n)}=\frac{1}{N^2}\sum_{m=0}^{k+n}\frac{\partial}{\partial g_m}
\Winf_{k-m}^{(2n+1)}-2\Winf_k^{(2n+1)}-\sum_{m\geq1}mg_m\Winf_{m+k}^{(2n+1)}
{}~~~,~~~k\geq-n
\label{wrec1}\eeq
\beq
\Winf_k^{(2n+1)}=\sum_{m\geq1}mg_m\Winf_{m+k}^{(2n-2)}-\frac{1}{N^2}
\sum_{m=0}^{k+n-1}\frac{\partial}{\partial g_m}\Winf_{k-m}^{(2n-2)}
{}~~~,~~~k\geq-n
\label{wrec2}\eeq

For example, the relations (\ref{n0ops})--(\ref{wrec2}) imply that the first
couple of $\Winf$-operators are
\beq
\Winf_k^{(0)}=\frac{1}{N^2}\sum_{m=0}^k\frac{\partial^2}{\partial g_m\partial
g_{k-m}}-\sum_{m\geq1}mg_m\frac{\partial}{\partial
g_{k+m}}-2\frac{\partial}{\partial g_k}
\label{wk0}\eeq
\beq\new{\begin{array}{ll}
\Winf_k^{(3)}=&\frac{1}{N^2}\sum_{m\geq1}mg_m\sum_{i+j=k+m}\frac{\partial^2}
{\partial g_i\partial g_j}-\sum_{m,n\geq1}mng_mg_n\frac{\partial}{\partial
g_{k+m+n}}\\&-2\sum_{m\geq1}mg_m\frac{\partial}{\partial
g_{k+m}}-\frac{1}{N^4}\sum_{i+j+m=k}\frac{\partial^3}{\partial g_i\partial
g_j\partial g_m}+\frac{2}{N^2}\sum_{i+j=k}\frac{\partial^2}{\partial
g_i\partial
g_j}\\&+\frac{1}{N^2}\sum_{m\geq1}mg_m\sum_{i+j=k+1}\frac{\partial^2}{\partial
g_i\partial g_{j+k-1}}+\frac{1}{N^2}\frac{k(k+1)}{2}\frac{\partial}{\partial
g_k}\end{array}}
\label{wk3}\eeq
Notice that the operators $\Winf_k^{(1)}$ generate a $U(1)$ Kac-Moody algebra
and the operators $\Winf_k^{(0)}$ coincide with generators $L_k$ of the
Virasoro algebra that we encountered in Subsection 3.5 above. The
$\Winf$-operators above therefore resemble the generators of the conventional
$W_k$-algebras. Again, these operators differ from the $\Winf$-operators that
appear in Hermitian multi-matrix models \cite{dijk1} by extra derivative
operators. In these latter models the algebra generated by the
$\Winf$-operators coincide in the double-scaling limit with the canonical
continuum $W$-algebras and lead to new symmetries of the underlying conformal
matrix models \cite{mir,mor}.

If $\tilde g_k=0$ in (\ref{poloop3}) for all $k\neq n$ and $\tilde g_n=
\frac{1}{n}$ for some $n\neq0$, then the $\frac{1}{z}$-expansion of
(\ref{poloop3}) can be written as
\beq
\frac{1}{Z_2}\sum_{k=1-n}^\infty\frac{1}{z^{k+n}}\left(\Winf_k^{(2n-2)}-
\Winf_{k+n}^{(1)}\right)Z_2=0
\eeq
and so the loop equations of the fermionic two-matrix model (\ref{part2}) are
equivalent to a set
of discrete $W$-constraints imposed on the partition function $Z_2$
\beq
\Winf_k^{(2n-2)}Z_2=\Winf_{k+n}^{(1)}Z_2~~~~~,~~~~~k\geq1-n
\label{wconstr}\eeq
The action of the $W$-algebra generators in (\ref{wconstr}) represent the Ward
identities of the two-matrix model (\ref{part2}) associated with the
infinitesimal fermionic variable changes
\beq\new{\begin{array}{c}
\chi\to\chi+\epsilon\chi(\ps2)^n~~~~~,~~~~~\bar\chi\to\bar\chi\\\psi\to\psi+
\epsilon\psi(\ps2)^n\left(V'(\ps2)-\ch2\right)~~~~~,~~~~~\bar\psi\to\bar\psi
\end{array}}
\label{2matrixinfchange}\eeq
under which the potential $V$ transforms as
\beq
V(\ps2)\to V(\ps2)-\epsilon V'(\ps2)^2(\ps2)^n+\epsilon(\ps2)^{n+1}
\eeq

These $W$-constraints are the two-matrix analogs
of the Virasoro constraints (\ref{virconstrexp})--(\ref{viralg}). In the
Hermitian case they are the basis of the integrable hierarchy structure in
multi-matrix models and generalized Kontsevich models
\cite{kharch1,marsh,mir,mor}.
Thus the fermionic multi-matrix models, in addition to providing some
novel generalization of the $(p,q)$ conformal models of string theory
\cite{daul,fgz,doug1}, also admit the conformal algebraic structure that
relates their integrability features to KdV and generalized Toda-chain
hierarchies
\cite{fgz,kharch1,mir,mor}. Since the discrete $W$-constraints here again
represent the
full set of Schwinger-Dyson equations of the model, the solutions of the
matrix model can be represented in more convenient forms relevant to conformal
string theory and the double-scaling continuum limit string theory can be
obtained from the Borel subalgebras of the continuum $W_{1+
\infty}$-algebras \cite{mir,mor}. These hierarchical structures should be quite
different though than in the conventional matrix models because of the
convergence properties of the fermionic partition functions. From the
fermionic matrix models some novel integrable hierarchies may therefore
emerge. Moreover, the $W$-constraints relate the Hermitian multi-matrix models
to more general types of topological field theories coupled to topological
gravity \cite{dijk2,wit}. This might make a more explicit connection between
matrix models
and the recent implementations of the ideas of representing topological
intersection indices in terms of matrix models using twisted $N=2$
superconformal quantum field theories \cite{dijk3,eguchi,li}.

These $W$-symmetries contain the usual Kac-Moody and Virasoro algebra
invariances which characterize conformal field theory and string theory
\cite{ginsparg}. This once again indicates the connection between the adjoint
fermion matrix models and string theory, and it would be interesting to use the
$W$-constraints to find other ``physical" (e.g. conformal field theoretic)
representations of the fermionic matrix models which makes this connection
complete. The appearence of the new $W$-algebra symmetries of the loop
equations (beyond the usual Kac-Moody and Virasoro ones) may have severe
implications for their extensions to higher-dimensional continuum string and
gauge theories where loop equations exist \cite{mak3,mig1}.

\section{Induced Gauge Theories and Mean Field Theory for Higher Dimensional
Matrix Models}

In this Section, we shall introduce a class of higher dimensional
matrix models related to the Kazakov-Migdal model of induced gauge
theory, which can be formulated with either bosonic or fermionic
adjoint matter fields.  We will concentrate on the fermionic case.  We
discuss the relation of this model to other lattice gauge theories, such
as the so-called adjoint model \cite{can,halliday,mp}. In this Section, we
review the mean field analysis of these models and postpone detailed
discussion of the loop equation approach to the next Section.  Using
loop equations, we shall see that the Gaussian model, and some others
in principle, are exactly solvable in the strong coupling regime.  In
this Section we shall present some speculations about the structure of
the theory in the weak coupling region.  Since no exact results are
available, we appeal to mean field theory techniques.  The main idea
is that there could be a phase transition in the Itzykson-Zuber
integral, which drastically modifies the solution of the model in the
weak coupling regime.  The motivation is to look for solutions of the
Kazakov-Migdal model which have a continuum limit resembling quantized
Yang-Mills theory.  It also can be used to say something about whether
the Kazakov-Migdal model can be obtained as a limit of more
conventional lattice gauge theory and whether the continuum theory
which describes Yang-Mills theory can be obtained in that limit.

In the present Section, we shall use mean field theory as an intuitive
introduction to the subject of induced gauge theory.  Although our
central purpose here is to discuss fermionic models, we devote some
space to comparison with similar results for Hermitian adjoint models.

\subsection{The Kazakov-Migdal Model with Adjoint Fermion Fields}

Consider a $D$-dimensional oriented hypercubic lattice $\lat
\subset\IR^D$ with sites $x$ and links $\ell=\langle x,y\rangle$
connecting nearest neighbour sites $x$ and $y$. At each site there are
fermion fields $\psi_j(x)$ and their conjugates $\bar\psi_j(x)$,
$j=1,\ldots,N_f$, which are independent $N\times N$ Grassmann-valued
matrices describing $N_f$ flavours of fermions with $N$ colours. These
fields also have an implicit spin index $\mu=1,\ldots,2^{[D/2]}$
(i.e. $\bar\psi$ and $\psi$ are actually $N\times(2^{[D/2]}N)$ and
$(2^{[D/2]}N)\times N$ matrix fields, respectively) which labels the
spinor representation of the Euclidean group in $D$-dimensions.  On
each link there is a gauge field $U(x,y)=U^\dagger(y,x)\in U(N)$. The
natural generalization of (\ref{part2}) to $D$-dimensions and $N_f$
fermion flavours is the lattice gauge theory
\beq
Z_D=\int\prod_{\langle x,y\rangle\in\lat}[dU(x,y)]~\int\prod_{x\in\lat}\prod
_{j=1}^{N_f}d\psi_j(x)~d\bar\psi_j(x)~\e^{S_F[\psi,\bar\psi;U]}
\label{partd}\eeq
where
\beq\new{\begin{array}{ll}
S_F[\psi,\bar\psi;U]=&\sum_{j=1}^{N_f}\sum_{x\in\lat}N^2\tr\biggm(V[\bar\psi_
j(x)\psi_j(x)]\\&-\sum_{\ell=1}^D\Bigm[\bar\psi_j(x)\proj_\ell^-U(x,x+\ell)
\psi_j(x+\ell)U^\dagger(x,x+\ell)\\&+\bar\psi_j(x+\ell)\proj_\ell^+U^\dagger
(x+\ell,x)\psi_j(x)U(x+\ell,x)\Bigm]\biggm)\end{array}}
\label{lataction}\eeq
is the gauge-invariant lattice fermion action \cite{ks,wilson}. It
includes the usual fermion kinetic term and fermion-gauge minimal
coupling in the last sum through the gauge-covariant lattice
derivative for Dirac fermions which transform in the adjoint
representation of the colour gauge group $U(N)$. Here
\beq
\proj_\ell^\pm=r\pm\gamma_\ell
\label{proj}\eeq
are the usual projection operators acting on the spin components in
(\ref{lataction}), so that $r=0$ for chiral fermions while
$r=1$ for Wilson fermions \cite{ks,wilson}, and $\gamma_\ell$ are the
gamma-matrices which generate the $D$-dimensional Euclidean Dirac algebra
\beq
\{\gamma_\ell,\gamma_{\ell'}\}=2\delta_{\ell\ell'}
\eeq
The potential is the usual Dirac potential
\beq
V(\ps2)=m\ps2+V_{\rm int}(\ps2)
\label{fermpot}\eeq
with $m$ the bare fermion mass.

The action (\ref{lataction}) differs from that of ordinary Wilson lattice
gauge theory \cite{wilson} in that the kinetic term for the gauge
field\footnote{\baselineskip=12pt Recall that in lattice gauge theory a gauge
field $U(x,x+\ell)\sim\e^{iA_\ell(x)}$ (with $A_\ell(x)$ the continuum gauge
field) is a function on lattice links, so that the curvature $F\sim dA+[A,A]$
is a function on plaquettes of the lattice. Consequently, (\ref{wilson})
coincides with the usual Yang-Mills action $F^2$ in the continuum limit where
the (unit) lattice spacing goes to zero, or equivalently $g\to0$.}
\beq
S_W[U]=\sum_{\Box\in\lat}\frac{N^2}{g^2}~{\rm Re}~W[U;\Box]
\label{wilson}\eeq
is absent. Here $\Box$ denotes the elementary plaquettes of the lattice
and for any oriented contour $C\in\lat$, $W[U;C]$ is the Wilson line
operator associated with $C$, i.e. the trace of the counterclockwise-oriented
path-ordered product of the link operators along $C$,
\beq
W[U;C]=\tr P\prod_{\langle x,y\rangle\in C}U(x,y)=W[U;-C]^*
\label{wilsonline}\eeq
where $-C$ denotes the contour $C$ with the opposite orientation.
However, because of the local $U(1)$ gauge-invariance (\ref{znsym})
of the matrix model (\ref{partd}) (applied to a {\it single} link),
the expectation value of any fundamental representation Wilson
loop\footnote{\baselineskip=12pt Note that for the fundamental representation
$N$ of $U(N)$, whose action on matrices $M$ is defined as $M\to U\cdot M$,
$U\in U(N)$, we have the group-theoretical decomposition
\bd
N\otimes\bar N=A\oplus{\bf1}
\ed
where $\bar N$ is the complex conjugate representation of $N$, and $A$ and
$\bf1$ are the adjoint and trivial (singlet) representations of $U(N)$,
respectively. This means that invariant traces in the adjoint representation
are related to (ordinary) invariant traces in the fundamental representation by
\bd
{\rm tr}_A~U=|\tr U|^2-1/N^2
\ed}
vanishes
and this leads to ultraconfinement of the theory. It is this feature that
makes the adjoint fermion model effectively solvable in contrast to the
conventional Wilson lattice gauge theory with fundamental representation
fermions \cite{kmnp,ks,wilson}.

The gauge-invariant matrix model (\ref{partd}) can be studied from several
different points of view. First of all, it can be viewed as an induced
gauge theory with partition function
\beq
Z_D=\int\prod_{\langle x,y\rangle\in\lat}[dU(x,y)]~\e^{S_{\rm ind}[U(x,y)]}
\label{partind}\eeq
and gauge field action $S_{\rm ind}[U(x,y)]$ defined by integrating out the
fermion fields in (\ref{partd})
\beq
\e^{S_{\rm ind}[U(x,y)]}=\int\prod_{x\in\lat}\prod_{j=1}^{N_f}d\psi_j(x)~
d\bar\psi_j(x)~\e^{S_F[\psi,\bar\psi;U]}
\label{indaction}\eeq
In the large-mass limit ($m\to\infty$ in (\ref{fermpot})), the Gaussian
integrations over the fermion fields give
\bd
S_{\rm ind}[U(x,y)]=N^2N_f~{\rm Tr}~\log\biggm[\delta_{x,y}-\frac{1}{m}
\sum_{\ell=1}^D\Bigm(\proj_\ell^-U(x,x+\ell)\otimes U^\dagger(x,x+\ell)
\ed
\beq
+\proj_\ell^+U^\dagger(x+\ell,x)\otimes U(x+\ell,x)\Bigm)\delta_{x+\ell,y}
\biggm]
\label{fermint}\eeq
The $\frac{1}{m}$-expansion of (\ref{fermint}) can be represented as a
sum over lattice loops $\Gamma$ of a spinor particle in an external gauge field
\cite{dz,wilson}
\beq
S_{\rm ind}[U]=N^2N_f\sum_{\Gamma\in\lat}\frac{|W[U;\Gamma]|^2}{l(\Gamma)
m^{l(\Gamma)}}~{\rm TR}~P\prod_{\ell\in\Gamma}\proj_\ell^\pm
\label{massexp}\eeq
where $l(\Gamma)$ is the lattice length of the loop $\Gamma$ (i.e. the
number of links in $\Gamma$) and TR denotes the trace over spinor indices
(the $\pm$ sign depends on the orientation of the link $\ell$).

If we consider in addition the large-$N_f$
limit correlated with the large-mass expansion so that $m/N_f^{1/4}\to1$
as $m,N_f\to\infty$, then all terms in (\ref{massexp}) except the
leading-order contributions in $1/m$ from single plaquettes are supressed
by factors of order $N_f^{-1/2}$. Then the induced action
$S_{\rm ind}[U]$ becomes the single-plaquette adjoint action
\beq
S_A[U]=N^2\frac{\beta_A}{2}\sum_{\Box\in\lat}|W[U;\Box]|^2
\label{adjaction}\eeq
with coupling constant
\beq
\beta_A=\frac{2^{\frac{D}{2}-1}N_f(1+2r^2-r^4)}{m^4}
\label{betaa}\eeq
Thus the adjoint fermion matrix model (\ref{partd}) in these limits induces
a single-plaquette lattice gauge theory built from traces in the adjoint
representation of the gauge group (as opposed to the more conventional
fundamental representation as in (\ref{wilson})).

Notice that for a matrix chain, associated with dimension $D\leq1$,
the gauge fields $U(x,y)$ in (\ref{partd}) can be absorbed by a local gauge
transformation (\ref{gaugetrans2}) and the model (\ref{partd}) reduces to
a fermionic multi-matrix model. In particular the case $D=1$ is associated with
an infinite matrix chain while $D=\frac{1}{2}$ corresponds to the two-matrix
model of the last Section. The generic $D>1$ model is the fermionic
analog of the Kazakov-Migdal model which is defined using heavy scalar fields
in the adjoint representation of the gauge group and which has recently
been proposed as a model for (induced) QCD \cite{km,kmsw,mak2}. If the
lattice gauge theory (\ref{partd}) is to indeed have a continuum limit
which reproduces the characteristic properties of QCD, then it must have
a phase transition into a phase where Wilson loop observables obey an
area law, the characteristic feature of quark confinement. This phase
structure can be studied from various points of view, such as a mean field
analysis of the induced gauge theory (\ref{massexp}) and loop equations
for the matrix model (\ref{partd}). These approaches, as well as the
possibility of using (\ref{partd}) as a model for strings in $D>1$
dimensions, will be discussed in this and the next Section.

\subsection{Phase Transitions in Unitary Matrix Integrals at Infinite $N$}

Before embarking on a detailed analysis of the critical behaviour of the matrix
model (\ref{partd}) when it is viewed as an induced gauge theory, we need some
preliminary results concerning the phase structure of unitary matrix models.

\subsubsection{The Fundamental Model}

Consider first the ``fundamental" unitary matrix model \cite{bg,gw} with
partition function
\beq
Z_F(\alpha)=\int[dU]~\e^{N^2\tr(\alpha^*U+\alpha U^\dagger)}
\label{zfund}\eeq
This integral can be computed explicitly in the large $N$ limit to get
the free energy \cite{gw},
\beq
F_F(\alpha)=-\lim_{N\rightarrow\infty}{1\over N^2}\log Z_F(\alpha)
=\left\{\matrix{- \vert\alpha\vert^2 & \vert\alpha\vert\leq 1/2 \cr
-2\vert\alpha\vert+{1\over2}\log2\vert\alpha\vert+3/4 &
\vert\alpha\vert\geq 1/2
\cr} \right.
\label{ffund}\eeq
Because of the invariance of the integration measure under a change
in phase of the unitary matrices, the free energy depends only on the
modulus of $\alpha$.  There is a third order phase transition at
$\vert\alpha\vert =1/2$.  We shall call the phase with
$\vert\alpha\vert\leq1/2$ the ``strong coupling phase'' and that with
$\vert\alpha\vert\geq1/2$ the ``weak coupling phase''.

{}From this expression, connected correlators of $\tr U$ and $\tr
U^{\dagger}$ can be obtained by taking derivatives with respect to
$\alpha$ and $\alpha^*$.  For example,
\beq
\langle\tr U\rangle_F\equiv-{\partial\over\partial\alpha^*} F_F(\alpha)
=\left\{
\matrix{ \alpha & \vert\alpha\vert\leq 1/2 \cr
{\alpha\over\vert\alpha\vert}\left(1-{1\over4\vert\alpha\vert}
\right) & \vert\alpha\vert\geq 1/2\cr } \right.
\label{3}\eeq
and
\beq
\langle\vert\tr U\vert^2\rangle_F -\vert\langle\tr U\rangle_F \vert^2 =0
\label{4}\eeq
The latter result, which is correct in both the weak and strong
coupling phases, is a result of factorization, i.e. the expectation value
of a product of any two functions of $U$ and $U^{\dagger}$ which are
separately invariant under unitary conjugation, $U\rightarrow
WUW^{\dagger}$ and $U^{\dagger}\rightarrow WU^{\dagger} W^{\dagger}$,
factorizes into the product of expectation values.

The result (\ref{ffund}) can be obtained explicitly \cite{gw} by using the
invariance of the integrand and measure in the partition function under
conjugation by unitary matrices to diagonalize $U$ and $U^{\dagger}$
and to write the partition function as an integral over the
eigenvalues.  The infinite $N$ limit is then obtained by saddle point
integration which can be done explicitly.
It is straightforward to check the small and large $\vert\alpha\vert$
limits.  First, consider small $\vert\alpha\vert$.  One can Taylor
expand the integrand in $\vert\alpha\vert$ to obtain
\beq\new{\begin{array}{c}
Z_F(\alpha)=\int[dU]~\left\{1+N^2\tr\left(\alpha^*U+\alpha
U^{\dagger}\right)+{1\over2!}\left( N^2\tr\left(\alpha^*U+\alpha
U^{\dagger}\right)\right)^2\right.\\
\left.+{1\over3!}\left( N^2\tr\left(\alpha^*U+\alpha
U^{\dagger}\right)\right)^3+\ldots\right\}\end{array}}
\label{zfexp}\eeq
Using the explicit integrals over the normalized Haar measure (see Subsection
1.1.1) we obtain
\beq
Z_F(\alpha)=1+N^2\vert\alpha\vert^2+{1\over2}N^4\vert\alpha\vert^4+\ldots
\label{9}\eeq
so that
\beq
F_F(\alpha)=-\vert\alpha\vert^2+\dots
\label{10}\eeq
which agrees with the strong coupling phase in (\ref{ffund}).

Also, when $\vert\alpha\vert$ is large, we perform a saddle point
integration.  Consider the variation $\delta U=iUH$ and $\delta
U^{\dagger}= -iHU^{\dagger}$ where $H$ is a Hermitian matrix.  The
exponent in the integrand of the partition function is maximized by
matrices which satisfy the equation
\beq
U_0= {\alpha\over\alpha^*}U_0^{\dagger}
\label{11}\eeq
or equivalently
\beq
U_0={\alpha\over\vert\alpha\vert}I
\label{12}\eeq
Then, defining $U=U_0 \e^{iH}$, $U^{\dagger}=\e^{-iH}U^{\dagger}_0$, the
partition function is approximated as\footnote{\baselineskip=12pt Here we use
the fact that a Hermitian Gaussian matrix integral is easily evaluated to be
\bd\new{\begin{array}{ll}
\int dH~\e^{-N^2\frac{m^2}{2}\tr H^2}&=\prod_{i=1}^N\int
dH_{ii}~\e^{-N\frac{m^2}{2}H_{ii}^2}\prod_{1\leq i<j\leq N}\int
d^2H_{ij}~\e^{-Nm^2|H_{ij}|^2}\\&=(2\sqrt{2})^N\left(\frac{\pi}{Nm^2}\right)^
{N^2/2}\end{array}}
\ed
for $m\in\IR$.}
\beq
Z_F(\alpha)=\e^{2N^2\vert\alpha\vert}\int dH~\e^{-N^2\vert\alpha
\vert\tr H^2}=\e^{2N^2\vert\alpha\vert-{N^2\over2}\log(2\vert\alpha\vert)+
\ldots}
\label{13}\eeq
which agrees with (\ref{ffund}) to the leading orders. It is interesting that,
up to normalization, it agrees with (\ref{zfund}) completely. This means
that, in the ``weak coupling'' phase, the large $N$ limit of the integral is
given exactly by its semiclassical (spherical) approximation. This is already
known to be the case for the Itzykson-Zuber integral (at finite-$N$) with
Hermitian matrices \cite{kmsw,mor}, and it would be
interesting to know whether this happens in other closely related cases.

It is possible to use a more sophisticated semiclassical approximation
to the above integral.  For this, we identify all extrema of the
action (maxima, minima and saddle points).  It is easy to see that
they are given by
\beq
U_0=\alpha\left(\matrix{ \pm1 & 0 & 0 & 0 &
\ldots \cr 0 & \pm1&0 & 0 & \ldots \cr 0 & 0 & \pm1 & 0 & \ldots \cr 0
& 0 & 0 & \pm1 & \ldots \cr ~&\ldots & ~ & \ldots ~ & \ldots\cr
}\right)
\label{14}\eeq
The semiclassical limit is given by
\beq\new{\begin{array}{ll}
Z_F(\alpha)&=\sum_{U_0}\e^{2N^2\vert\alpha\vert\tr U_0}
\int dH~\e^{-N^2\vert\alpha\vert\tr H^2U_0}
\\&=\e^{2N^2\vert\alpha\vert}(N\vert\alpha\vert)^{-N^2\over2} +N
\e^{2N(N-2)\vert\alpha\vert}(-N\vert\alpha\vert)^{-N^2\over2}+\ldots\\&~~~~~
+\left(\matrix{N\cr m\cr}\right)
\e^{2N(N-2m)\vert\alpha\vert}((-1)^mN\vert\alpha\vert)^{-N^2\over2}+
\ldots\\&=\e^{2N^2\vert\alpha\vert}(N\vert\alpha\vert)^{-N^2\over2}\cdot
\left( 1+(-1)^{-N^2\over2}\e^{-2N\vert\alpha\vert}\right)^N\end{array}}
\label{15}\eeq
This (exact) partition function is real when $N$ is even.  In the large $N$
limit, it reduces to (\ref{13}).

\subsubsection{The Adjoint Model}

We are primarily interested in the model with partition function
\beq
Z_A(\beta)=\int[dU]~\e^{N^2\beta\vert\tr U\vert^2}
\label{16}\eeq
This model has, besides the symmetry under conjugation of $U$ by
unitary matrices exhibited by (\ref{zfund}), a symmetry under redefining $U$ by
an element of the center of the unitary group, $U\rightarrow\e^{i\theta}U$.
This symmetry guarantees that a correlator vanishes unless it contains
the same number of $U$'s and $U^{\dagger}$'s. In particular, we have (c.f. eq.
(\ref{gaugecorr}))
\beq
\langle U_{jk}U_{\ell i}^\dagger\rangle_A=N\langle|\tr
U|^2\rangle_A\delta_{ij}\delta_{k\ell}
\label{uija0}\eeq

The partition function
$Z_A$ can be obtained as a Gaussian integral transform of $Z_F$ as
\beq
Z_A(\beta)\sim \int d\eta ~d\eta^* ~\e^{-N^2\vert\eta\vert^2/\beta}Z_F(\eta)
\label{17}\eeq
In the infinite $N$ limit, this integral can be performed by saddle
point integration.  It is necessary to find the value of $\eta$
which minimizes the normalized free energy function
\beq
F_A(\beta)=\min_\eta\left\{ \matrix{ (\beta^{-1}-1)\vert\eta\vert^2 &
\vert\eta\vert\leq 1/2 \cr
\beta^{-1}\vert\eta\vert^2-2\vert\eta\vert+{1\over2}\log 2
\vert\eta\vert+3/4 & \vert\eta\vert\geq 1/2\cr }\right.
\label{18}\eeq
When $\beta<1$, $\vert\eta\vert=0$.  When $\beta>1$,
$\vert\eta\vert = {1\over2}\beta\left(1+\sqrt{1-1/\beta}\right)$, so
that
\beq
F_A(\beta)=\left\{ \matrix{ 0 & \beta\leq 1\cr
{\beta\over4}(1-\sqrt{1-1/\beta})^2-\beta+3/4
+{1\over2}\log\left(\beta(1+\sqrt{1-1/\beta})\right) & \beta\geq 1\cr}\right.
\label{19}\eeq
This solution has the interesting property that
\beq
\left\langle\vert\tr U\vert^2\right\rangle_A
\equiv-{\partial\over\partial\beta}F_A(\beta) =\left\{\matrix{ 0 &
\beta<1 \cr {1\over4}\left(1+\sqrt{1-1/\beta}\right)^2 &\beta\geq1\cr}\right.
\label{20}\eeq
This expectation value is discontinuous at $\beta=1$ and the phase
transition there is of first order.

This behavior can be checked by perturbation theory in both the strong
and weak coupling regions.  In strong coupling, we expand the
exponents in the integrand in $\beta$,
\beq
\left\langle\vert\tr U\vert^2\right\rangle_A = {
\int [dU]~\vert\tr U\vert^2
\left(1+N^2\beta\tr(U+U^{\dagger})+\ldots\right)
\over \int [dU] ~\left(1+N^2\beta\tr(U+U^{\dagger})+\ldots\right)}
\label{21}\eeq
which, using the explicit integrals for the Haar measure correlation functions
given in Subsection 1.1.1, we can easily see is zero to
order $\beta^2$. In the large $\beta$ limit, we use the saddle point method.
The variation of the action gives the saddle point equation
\beq
U\tr U^{\dagger}~=~ U^{\dagger}\tr U
\label{22}\eeq
which has solution
\beq
U_0=\e^{i\theta}I
\label{23}\eeq
and
\beq
Z_A(\beta)\approx \e^{\beta N^2}\int dH ~\e^{-\beta(N^2\tr H^2-N^3(\tr H)^2)}=
\e^{N^2(\beta-{1\over2}\log\beta)}
\label{24}\eeq
which agrees with an asymptotic expansion of (\ref{19}).

\subsubsection{The Mixed Model}

Finally, we consider the model with a mixed action,
\beq
Z_M(\alpha,\beta)=\int [dU] ~\e^{N^2\beta\vert\tr U\vert^2+N^2
\tr(\alpha^* U+\alpha U^{\dagger})}
\label{25}\eeq
This action has a term which explicitly breaks the symmetry under
phase transformation of $U$ and $U^{\dagger}$. The mixed model (\ref{25}) is
invariant under the unitary transformations $U\to VUW$ where either $V$ is
taken as the matrix that multiplies the $n$-th row of $U$ by $-1$ and $W$
multiplies the $n$-th column by $-1$, or when $V$ interchanges the $m$-th and
$n$-th rows while $W$ interchanges the $m$-th and $n$-th columns. This symmetry
combined with the invariance of (\ref{25}) under conjugation by unitary
matrices leads to the identities
\beq
\langle U_{jk}U_{\ell
i}\rangle_M=\frac{N^2(N-1)-1}{N^2-1}\left(\left\langle|\tr
U|^2\right\rangle_M-1\right)\delta_{jk}\delta_{\ell i}+\frac{N^2}{N^2-1}\left(
1-\left\langle|\tr U|^2\right\rangle_M\right)\delta_{ij}\delta_{k\ell}
\label{uijm0}\eeq
\beq
\langle U_{ij}\rangle_M=\langle\tr U\rangle_M\delta_{ij}
\label{um0}\eeq

The partition function can be obtained by the offset Gaussian transform
\beq
Z_M(\alpha,\beta) \sim \int d\eta~
d\eta^*~\e^{-N^2\vert\eta\vert^2/\beta+N^2(\eta^*
\alpha+\eta\alpha^*)/\beta -|\alpha|^2 N^2 / \beta}~Z_F(\eta)
\label{26}\eeq
In the large $N$ limit, this can again be found by minimizing the free
energy function
\beq
F_M(\alpha,\beta)=\min_\eta\left\{\matrix{
(\beta^{-1}-1)\vert\eta\vert^2-\frac{(\eta^*\alpha+\alpha^*\eta)}{\beta}
+\frac{\vert\alpha\vert^2}{\beta}&
\vert\eta\vert\leq\frac{1}{2}\cr\beta^{-1}\vert\eta\vert^2-2\vert\eta\vert+
{1\over2}\log 2\vert\eta\vert-\frac{(\eta^*\alpha+\eta\alpha^*)}{\beta}
+\frac{\vert\alpha\vert^2}{\beta}+\frac{3}{4}&\vert\eta\vert\geq\frac{1}{2}\cr}
\right.
\label{27}\eeq
The minima occur at
$\eta=\alpha/(1-\beta)$ when $\vert\eta\vert\leq 1/2$,
which is the region $\vert\alpha|\leq (1-\beta)/2$.  In the other
region, $\vert\eta\vert={1\over2}(\beta+\vert\alpha\vert
+\sqrt{(\beta+\vert\alpha\vert)^2-\beta})$ which exists where
$\vert\alpha\vert\geq\sqrt{\beta}-\beta $ which is always satisfied in
the region $\vert\alpha\vert\geq {1\over2}(1-\beta)$.  Thus, the
solution for the free energy is
\beq\new{\begin{array}{ll}
{1\over N^2}F_M(\alpha,\beta)&=
{-\vert\alpha\vert^2\over(1-\beta)}~~{\rm for}~~ \vert\alpha\vert\leq
{1\over2}\vert 1-\beta\vert,0\leq\beta\leq1\\&
={1\over4\beta} \left( (\beta+\vert\alpha\vert) +
\sqrt{ (\beta+\vert\alpha\vert)^2-\beta}\right)^2+3/4 + \vert\alpha\vert ^2
/ \beta\\&~~~~-\left(  (\beta+\vert\alpha\vert)+\sqrt{(\beta + \vert
\alpha\vert)^2-\beta}\right) ( 1 + \vert\alpha\vert / \beta)\\&~~~~+
{1\over2}\log\left((\beta+\vert\alpha\vert)+\sqrt{(\beta + \vert
\alpha\vert)^2-\beta})\right)  ~~{\rm for}~~
\vert\alpha \vert \geq{1\over2}(1-\beta),\beta\geq0\end{array}}
\label{28}\eeq

In this case,
\beq
\langle\tr U\rangle_M=\left\{ \matrix{ {\alpha\over
1-\beta}& \vert\alpha\vert\leq {1\over2}(1-\beta),0\leq\beta\leq1 \cr
{-\alpha\over\vert\alpha\vert}{1\over2\beta}
\left(\vert\alpha\vert-\beta-\sqrt{(\beta+\vert\alpha\vert)^2-\beta}\right)
& \vert\alpha\vert\geq {1\over2}(1-\beta),\beta \geq 0
\cr }\right.
\label{29}\eeq
and
\beq
\left\langle\vert\tr U\vert^2\right\rangle_M=\left\{ \matrix{
\left({\vert\alpha\vert\over 1-\beta} \right)^2 & \vert\alpha\vert \leq
{1\over2} (1-\beta),0\leq\beta\leq1 \cr {1\over4\beta^2}\left(
\vert\alpha\vert -\beta-
\sqrt{(\beta+\vert\alpha\vert)^2-\beta}\right)^2 &
\vert\alpha\vert \geq {1\over2}(1-\beta),\beta\geq 0 \cr } \right.
\label{30}\eeq
This result exhibits factorization in all phases.  Furthermore, it
exhibits {\it spontaneous symmetry breaking} in the ``weak coupling
phase'' $\beta>1$,
\beq
\lim_{\vert\alpha\vert\rightarrow0}\langle\tr U\rangle_M= \left\{
\matrix{ 0 & \vert\alpha\vert\leq {1\over2}(1-\beta),0\leq\beta\leq1
\cr {\alpha\over2\vert\alpha\vert}\left(1 + \sqrt{1-1/\beta}\right) &
\vert\alpha\vert\geq{1\over2}(1-\beta),\beta\geq 0\cr}\right.
\label{31}\eeq
It is in such a symmetry-breaking scenario that the local $U(1)$ symmetry of
the adjoint lattice gauge model (\ref{adjaction}) could be viewed as being
eliminated and the resulting model possessing only the usual symmetries of the
standard Wilson lattice gauge theory \cite{ksw} (which gives the latticized
version of QCD). We shall examine this possibility further in Subsection 5.3.

The mixed unitary matrix model (\ref{25}) has been studied by Makeenko and
Polikarpov \cite{mp} (in the context of lattice gauge theory) and they showed
that at $N=\infty$ the mixed model is equivalent to the fundamental model
(\ref{zfund}). This follows from the factorization property of the correlators
above of the mixed model. It can be verified explicitly within the weak and
strong coupling expansions, and it can be proved exactly using loop equations
\cite{mp}. To show this, we assume for simplicity that the coupling constant
$\alpha$ of the fundamental term in (\ref{25}) is real and positive. At
$N=\infty$ we use the factorization property
\beq
\langle|\tr U|^2\rangle_M=\langle\tr U\rangle^2_M
\eeq
to write a differential equation for the free energy $F_M$,
\beq
2\frac{\partial F_M}{\partial\beta}=\left(\frac{\partial
F_M}{\partial\alpha}\right)^2
\label{32}\eeq
Differentiating (\ref{32}) with respect to $\alpha$ then yields to ${\cal
O}(1/N^2)$
\beq
\frac{\partial}{\partial\beta}\langle\tr U\rangle_M=\langle\tr
U\rangle_M\cdot\frac{\partial}{\partial\alpha}\langle\tr U\rangle_M
\label{33}\eeq
If we substitute the ansatz
\beq
\langle\tr U\rangle_M=\langle\tr U\rangle_{\bar F}
\label{34}\eeq
with the fundamental average $\langle\tr U\rangle_{\bar F}$ taken in the
fundamental model (\ref{zfund}) with the redefined coupling $\bar\alpha$
defined by the implicit function relation
\beq
\bar\alpha=\alpha+\beta\langle\tr U\rangle_{\bar F}~~~~~,
\label{35}\eeq
then from
\beq\new{\begin{array}{c}
\frac{\partial}{\partial\alpha}\langle\tr
U\rangle_M=\frac{d}{d\bar\alpha}\langle\tr U\rangle_{\bar
F}\left[1-\beta\frac{d}{d\bar\alpha}\langle\tr U\rangle_{\bar
F}\right]^{-1}\\\frac{\partial}{\partial\beta}\langle\tr U\rangle_M=\langle\tr
U\rangle_M\cdot\frac{d}{d\bar\alpha}\langle\tr U\rangle_{\bar
F}\left[1-\beta\frac{d}{d\bar\alpha}\langle\tr U\rangle_{\bar F}\right]^{-1}
\end{array}}
\label{36}\eeq
it is readily seen that (\ref{34}),(\ref{35}) satisfies (\ref{33}).

Thus in the mixed model (\ref{25}), the gauge field
correlators at large-$N$ depend not on the 2 variables $\alpha$ and $\beta$
separately, but are a function of their combination determined by the effective
charge $\bar\alpha$ in (\ref{35}). Using (\ref{3}) this effective charge is
\beq
\bar\alpha(\alpha,\beta)=\left\{\new{\begin{array}{l}\frac{\alpha}{1-\beta}
{}~~~,~~~\frac{\alpha}{1-\beta}\leq\frac{1}{2}~,~0\leq\beta\leq1\\\frac{1}{2}
\left(\alpha+\beta+\sqrt{(\beta+\alpha)^2-\beta}\right)~~~,~~~\frac{
\alpha}{1-\beta}\geq\frac{1}{2}~,~\beta\geq0\end{array}}\right.
\label{37}\eeq
The first order phase transition in the model occurs at the points where the
specific heat changes its sign (and becomes infinite). These are the points
where the right-hand sides of (\ref{36}) vanish, i.e.
\beq
\beta_c=\left(\frac{d}{d\bar\alpha}\langle\tr U\rangle_{\bar
F}\right)^{-1}~~~,~~~\alpha_c=\bar\alpha-\beta_c\langle\tr U\rangle_{\bar F}
\label{38}\eeq
Finally, the string tension $\sigma_M$, which is defined by the area law
behaviour of the Wilson loops
\beq
\langle\tr U\rangle_M\equiv\e^{-\sigma_MA}
\label{39}\eeq
can easily be found from (\ref{3}) and (\ref{35}) to be
\beq
\sigma_M(\alpha,\beta)=\left\{\new{\begin{array}{l}\frac{1}{A}\log\left
(\frac{1-\beta}{\alpha}\right)~~~,~~~\frac{\alpha}{1-\beta}\leq\frac{1}{2}~,~0
\leq\beta\leq1\\\frac{1}{A}\log\left(\frac{2\left(\alpha+\beta+\sqrt{(\beta+
\alpha)^2-\beta}\right)}{2\left(\alpha+\beta+\sqrt{(\beta+
\alpha)^2-\beta}\right
)-1}\right)~~~,~~~\frac{\alpha}{1-\beta}
\geq\frac{1}{2}~,~\beta\geq0\end{array}}\right.
\label{40}\eeq
Notice that for finite-$N$ one generally expects a perimeter law for the
adjoint Wilson loops $\langle|\tr U|^2\rangle_A$. However, the above analysis
shows that at $N=\infty$ they satisfy the same area law as the fundamental
Wilson loops $\langle\tr U\rangle_F$. The perimeter law at finite-$N$ thus
enters at ${\cal O}(1/N^2)$. We also note from (\ref{20}) that while the string
tension in the adjoint unitary matrix model in the weak coupling phase is
finite and corresponds to normal quark confinement, that in the strong coupling
phase is infinite and is associated with local confinement (i.e. the absence of
quarks in that phase).

\subsection{Mean Field Theory and Confinement at $N=\infty$}

We now carry out a detailed analysis of the phase structure at $N=\infty$ of
the gauge theory induced by (\ref{partd}). We first note that factorization
implies that the gauge theory
with adjoint action (\ref{adjaction}) is equivalent at $N=\infty$ to that
with the fundamental representation Wilson action $S_W[U]$ provided that the
coupling in (\ref{wilson}) obeys the self-consistency condition \cite{mp}
\beq
\bar\beta\equiv\frac{1}{g^2}=\beta_AW_F(\Box;\bar\beta)
\label{wfdef}\eeq
where $W_F(\Box;\bar\beta)$ denotes the fundamental plaquette average
\beq
W_F(\Box;\bar\beta)=\frac{\int\prod_{\langle x,y\rangle\in\lat}[dU(x,y)]~
\e^{S_W[U(x,y)]}W[U;\Box]}{\int\prod_{\langle x,y\rangle\in\lat}[dU(x,y)]~
\e^{S_W[U(x,y)]}}
\label{plaqavg}\eeq
(\ref{wfdef}) can be solved for $\bar\beta(\beta_A)$ at strong
coupling by substituting in the strong coupling expansion of (\ref{plaqavg})
\cite{dz,wilson}
\beq
W_F(\Box;\bar\beta)=\frac{\bar\beta}{2}+\frac{\bar\beta^5}{8}~~~{\rm for}~~~
\bar\beta\ll1
\label{avgstrong}\eeq
(\ref{wfdef}) in the strong coupling regime always possesses the
trivial solution $\bar\beta=0$. However, for $\beta_A\sim2$ there is the
nontrivial solution
\beq
\bar\beta\propto\left(\frac{1}{2}-\frac{1}{\beta_A}\right)^{1/4}
\label{betanontr}\eeq
which matches the weak coupling solution
\beq
\bar\beta\sim\beta_A-1/4~~~{\rm for}~~~\bar\beta\gg1
\label{betaweak}\eeq
as $\beta_A\to\infty$.

The basic observable of the induced gauge theory (\ref{adjaction}) is the
adjoint plaquette average
\beq
W_A(\Box;\beta_A)=\frac{\int\prod_{\langle x,y\rangle\in\lat}[dU(x,y)]~
\e^{S_A[U(x,y)]}\left(|W[U;\Box]|^2-\frac{1}{N^2}\right)}{\int\prod_{\langle
x,y\rangle\in\lat}[dU(x,y)]~\e^{S_A[U(x,y)]}}
\label{plaqadj}\eeq
Note that since the product of 2 fundamental representation link operators
contains both the adjoint and scalar representations, i.e. $(U_{ij}U^\dagger
_{k\ell})_A=U_{ij}U^\dagger_{k\ell}-\delta_{i\ell}\delta_{jk}$, we subtract off
the latter in (\ref{plaqadj}). Factorization implies that at $N=\infty$ it
is given by
\beq
W_A(\Box;\beta_A)=\left(W_F(\Box;\bar\beta)\right)^2=\left(\frac{\bar\beta}
{\beta_A}\right)^2
\label{plaqfact}\eeq
Since the sign of the second term in (\ref{avgstrong}) is positive, we find
that the slope of (\ref{plaqfact}) is negative for the solution
(\ref{betanontr}) near $\beta_A=2$. Consequently, a first order phase
transition in the induced gauge theory (\ref{adjaction}) must occur with
increasing $\beta_A$ at some critical coupling $\beta_A^c<2$ \cite{kmak1,mak2}.

Although the integrations in (\ref{partd}) cannot be carried out explicitly,
the phase diagram can be studied by employing a variational mean field
analysis \cite{dz,kmak1,kmak2} to the partition function (\ref{partd}) for
$N_f=1$ and $m\gg1$. The mean field approximation is based on a result known as
Jensen's inequality \cite{dz} which is easily derived as follows. Consider 2
partition functions $Z^{(1)}$ and $Z^{(0)}$ which are defined in terms of
actions $S_1$ and $S_0$, i.e. $Z^{(i)}=\int\e^{-S_i}$, with statistical
averages $\langle A\rangle_i=\int A\e^{-S_i}/Z^{(i)}$. For $t\in[0,1]$, we
introduce the one-parameter family of partition functions
$Z^{(t)}=\int\e^{-tS_0-(1-t)S_1}$. Then a simple calculation shows that
\beq
\frac{d^2\log
Z^{(t)}}{dt^2}=\left\langle(S_1-S_0)^2\right\rangle_t-\left\langle
S_1-S_0\right\rangle_t^2\geq0
\eeq
where we have used the fact that $\langle A^2\rangle\geq\langle A\rangle^2$ for
any statistical average. From this it follows that, as a function of $t$, $\log
Z^{(t)}$ is concave up everywhere, and so it lies above all of its tangent
lines. In particular,
\beq
\log Z^{(1)}\geq\log Z^{(0)}+\frac{d\log Z^{(t)}}{dt}\biggm|_{t=0}
\eeq
which leads to
\beq
Z^{(1)}\geq Z^{(0)}\e^{\langle S_0-S_1\rangle_0}
\label{jenseneq}\eeq
(\ref{jenseneq}) is a special case of what is known in functional analysis as
Jensen's inequality.

Going back to our analysis, we introduce the trial partition function
\beq
Z_D^{(A)}=\int\prod_{\langle x,y\rangle\in\lat}[dU(x,y)]~\e^{-\frac{N^2b_A}{2}
\sum_{\langle x,y\rangle\in\lat}|\tr U(x,y)|^2}
\label{trialpart}\eeq
which is similar in form to the unitary matrix integrations in (\ref{partd}).
The partition function (\ref{trialpart}) is a product of one-link
unitary matrix integrals which are the simplest ones with the local
$U(1)$-gauge invariance of (\ref{partd}). It can therefore be evaluated using
the results of Subsection 5.2.2 above as
\beq
Z_D^{(A)}=\e^{-N^2~{\rm vol}(\lat) DF_A(-b_A/2)}
\eeq
where $~{\rm vol}(\lat)$ is the volume of the lattice $\lat$ so that $~{\rm
vol}(\lat) D$ is the total number of links in the lattice (there are $2D$
nearest neighbours to each site). Then Jensen's inequality (\ref{jenseneq})
with
$Z^{(1)}=Z_D$, $Z^{(0)}=Z_D^{(A)}$
gives a bound on the partition function (\ref{partd})
\beq\new{\begin{array}{l}
Z_D\geq Z_D^{(A)}\int\prod_{x\in\lat}d\psi(x)~d\bar\psi(x)~\exp\biggm[
\sum_{x\in\lat}\Bigm\{N^2m\tr\bar\psi(x)\psi(x)\\-\left\langle\sum_{\ell=1}^D
\Bigm(N^2\tr\bigm(\bar\psi(x)\proj_\ell^-U(x,x+\ell)\psi(x+\ell)U^\dagger(x,x+
\ell)\right.\\\left.+\bar\psi(x+\ell)\proj_\ell^+U^\dagger(x+\ell,x)\psi(x)U(x+
\ell,x)\bigm)-\frac{N^2b_A}{2}|\tr U(x,x+\ell)|^2\Bigm)\right\rangle_0\Bigm\}
\biggm]\end{array}}
\label{jensen}\eeq
where the subscript 0 denotes the normalized average with respect
to the statistical ensemble (\ref{trialpart}). Since the argument of the
exponential function in (\ref{jensen}) is a sum of one-link averages, it
can be written in terms of the unitary one-matrix integral
\beq
u^2=\langle|\tr
U|^2\rangle_A(\beta=-b_A/2)\equiv\frac{\int[dU]~\e^{-\frac{N^2b_A}{2}|\tr
U|^2}|\tr
U|^2}{\int[dU]~\e^{-\frac{N^2b_A}{2}|\tr U|^2}}
\label{1matrixint}\eeq
using the identity
\beq
\langle\tr\bar\psi(x)\proj_\ell^\pm U(x,x+\ell)\psi(x+\ell)U^\dagger(x,x+\ell)
\rangle_0=u^2\tr\bar\psi(x)\proj_\ell^\pm\psi(x+\ell)
\label{1linkid}\eeq
where we have used (\ref{uija0}).

The idea behind the variational mean field method is to now fix the
coupling $b_A$ by the condition that the explicitly solvable trial partition
function
(\ref{trialpart}) gives the best approximation to (\ref{partd}) in this
class of models, from which it is hoped that (\ref{trialpart}) will have a very
close behaviour to $Z_D$. Taking the derivative with respect to $b_A$ using
(\ref{19}), (\ref{1matrixint}) and (\ref{1linkid}), we find after some algebra
that the maximum value of the right-hand side of (\ref{jensen}) is at the
coupling
\beq
b_A=\frac{\int\prod_{x\in\lat}d\psi(x)~d\bar\psi(x)~\e^{S_F[\psi,\bar\psi;u]}
\tr\left(\bar\psi(0)\proj_\ell^-\psi(0+\ell)+\bar\psi(0+\ell)\proj_\ell^+
\psi(0)\right)}{\int\prod_{x\in\lat}d\psi(x)~d\bar\psi(x)~\e^{S_F[\psi,\bar
\psi;u]}}
\label{ba}\eeq
where
\beq\new{\begin{array}{l}
S_F[\psi,\bar\psi;u]\\=\sum_{x\in\lat}N^2\tr\left(m\bar\psi(x)\psi(x)
-u^2\sum_{\ell=1}^D\left[\bar\psi(x)\proj_\ell^-\psi(x+\ell)+\bar\psi(x+
\ell)\proj_\ell^+\psi(x)\right]\right)\end{array}}
\eeq
(\ref{ba}) expresses the parameter $b_A$ as a function of the fermion
mass $m$. Thus the variational method above amounts to substituting in a mean
field value $[U(x,y)]_{ij}=u\delta_{ij}$ for every link operator except the
one on the fixed distinguished link $\langle0,0+\ell\rangle$, and then
(\ref{1matrixint}) gives a self-consistency condition at this link.

Notice that the large-mass expansion of (\ref{ba}) can be represented as
\beq
b_A=\frac{1}{m}\sum_{\Gamma_{x,\ell}\in\lat}\left(\frac{u^2}{m}\right)^{l(
\Gamma_{x,\ell})}~{\rm TR}~P\prod_{\ell'\in\Gamma_{x,\ell}}\proj_{\ell'}^\pm
\label{baexp}\eeq
where the sum is over all open contours $\Gamma_{x,\ell}$ with endpoints
at $x$ and $x+\ell$. It is only for chiral fermions ($r=0$ in (\ref{proj}))
that the mean field method gives the same large-mass expansion (\ref{massexp}),
since for Wilson fermions ($r=1$ in (\ref{proj})) backtracking paths never
contribute to the sum in (\ref{massexp}) because of the identity
\beq
\sigma\equiv\proj_\ell^\pm\proj_\ell^\mp=r^2-1
\label{bactrackpar}\eeq
and because of the unitarity of the gauge fields. In (\ref{baexp}) the
contribution from backtracking paths is explicitly taken into account.
The parameter $b_A$ can still, however, be considered as an upper bound to
the actual value which suffices to give the point of the first order
phase transition in the induced gauge theory \cite{dz,kmak1,kmak2}.

In the weak coupling region where $b_A>2$ and $\frac{1}{2}\leq
u\leq1$, the integral (\ref{1matrixint}) can be evaluated (see Subsection 5.2.2
above) and the self-consistency condition can be written as
\beq
2(u-u^2)=1/b_A
\label{selfcons}\eeq
In the strong-coupling phase the unitary matrix integral (\ref{1matrixint})
vanishes. The geometric criterion for the location of the
first order phase transition is that point where the $u\neq0$ solution
terminates with increasing $m$ \footnote{\baselineskip=12pt This requirement
is equivalent to the thermodynamic criterion that the free energies of
the $u=0$ and $u\neq0$ phases coincide.}. The large-$N$ phase transition is
then
associated with the freezing of the gauge field $U(x,y)$ at some mean
field value $u$ while $u=0$ in the strong coupling regime of the induced
gauge theory. Notice that the condition here for the corresponding unitary
one-matrix model to possess a weak coupling solution is $u>\frac{1}{2}$.

\subsubsection{Minimization of the Mean Field Coupling}

To explicitly calculate (\ref{ba}), we introduce the lattice Fourier transform
\beq
f(x)=\int_{-\pi}^{+\pi}\prod_{\ell=1}^D\frac{dp_\ell}{2\pi}~\e^{-ip\cdot
x}f(p)~~~~~,~~~~~x\in\lat
\label{latticefourier}\eeq
where the momentum space integral in (\ref{latticefourier}) is restricted to
the first Brillouin zone of the lattice $\lat$ and $p_\ell\in[-\pi,+\pi]$ are
the Bloch momenta. For example, for chiral fermions which in the naive
continuum limit ($m\to0$) are associated generally with $2^DN_f$ flavours, the
Gaussian integral in (\ref{ba}) can be evaluated by the usual lattice technique
\cite{dz,wilson} to give
\beq\new{\begin{array}{ll}
b_A&=\frac{2^D}{u^2}\int_{-\pi}^{+\pi}\prod_{\ell=1}^D\frac{dp_\ell}{2\pi}
{}~\frac{1-\frac{1}{D}\sum_{\ell=1}^D\cos p_\ell}{\frac{m^2}{2u^4}+D
-\sum_{\ell=1}^D\cos p_\ell}\\&=\frac{2^D}{Du^2}\left(1-\frac{m^2}{2u^4}
\int_0^\infty dy~\e^{-(\frac{m^2}{2u^4}+D)y}[I_0(y)]^D\right)\end{array}}
\label{bachiral}\eeq
where $I_0(y)$ is the modified Bessel function of order 0. Thus in this case
$u^2b_A$ is a monotonically increasing function of decreasing mass $m$ which
takes its maximal value $2^D/D$ at $m=0$. Moreover, the second integral in
(\ref{bachiral}) can be well-approximated by the leading order term in the
expansion in powers of $(\frac{m^2}{2u^4}+D)$ \cite{kmak2} and so we can write
\beq
\frac{1}{b_A}\sim2^{-D}\left(\frac{m^2}{2u^4}+D\right)u^2
\label{bapprox}\eeq
The critical point $m=m_c$ can now be found from (\ref{selfcons}) and
the geometric condition mentioned above which is
\beq
\frac{\partial u}{\partial m}\biggm|_{m=m_c}=\infty
\label{geomcrit}\eeq
For example, for $D=4$ we find $m_c\sim\frac{13}{2}$ and $u(m_c)\sim
\frac{2}{3}$. Notice that since $u^2b_A$ can always reach values larger than
$\frac{1}{2}$, the equation (\ref{selfcons}) always has a solution. Notice
also that the integral in (\ref{bachiral}) always converges for $0<m^2<\infty$
which is a consequence of the stability of the fermionic system.

The situation is the same for Kogut-Susskind fermions \cite{ks}, which are
associated
with $2^{D/2}N_f$ continuum fermion flavours and can be obtained from the
chiral fermion fields by a flavour-number reduction transformation as
\beq
\psi(x)=(\gamma_1)^{x_1}(\gamma_2)^{x_2}\cdots(\gamma_D)^{x_D}\chi(x)
\label{kogsusfer}\eeq
where $x_\ell$ are the components of the lattice vector $x$. In these new
spinor variables the chiral fermion action (\ref{lataction}) is
\bd\new{\begin{array}{l}
S_F[\chi,\bar\chi;U]\\=\sum_{x\in\lat}N^2\tr\biggm(m\bar\chi(x)\chi(x)
-\sum_{\ell=1}^D\eta_\ell(x)\Bigm[\bar\chi(x)U(x,x+\ell)\chi(x+\ell)U^\dagger
(x,x+\ell)\end{array}}
\ed
\beq
-\bar\chi(x+\ell)U^\dagger(x+\ell,x)\chi(x)U(x+\ell,x)\Bigm]\biggm)
\label{kogaction}\eeq
where
\beq
\eta_1(x)=1~~~~~;~~~~~\eta_\ell(x)=(-1)^{x_1+\dots+x_{\ell-1}}~~{\rm for}~~
\ell=2,\dots,D
\label{eta}\eeq
Since the action (\ref{kogaction}) is now diagonal in the spin indices, it
suffices to consider only one spin component of the field $\chi(x)$ so that
it can be considered as a spinless (scalar) Grassmann variable. The chiral
spinors are then reproduced in the continuum limit by combining $2^D$ of the
$\chi$ fields at nearest-neighbour sites of the lattice, so that this procedure
manifestly reduces the number of flavours in the continuum by a factor of
$2^{D/2}$. The mean field analysis for Kogut-Susskind fermions is now
identical to that of chiral fermions above, the only change being that
the factor of $2^D$ in (\ref{bachiral}) and (\ref{bapprox}) is now replaced
by $2^{D/2}$, so that the self-consistency condition (\ref{selfcons}) at
$D=4$ becomes
\beq
2\left(\frac{1}{u}-1\right)=\frac{m^2}{8u^4}+1
\label{bakog}\eeq
This equation determines the critical values as above associated with the
large-$N$ phase transition to be $m_c\sim\frac{7}{10}$ and $u(m_c)\sim
\frac{1}{2}$, and therefore the large-$N$ phase transition occurs as well
for the case of Kogut-Susskind fermions.

However, the situation is less clear for Wilson lattice fermions, which
correspond to $N_f$ light fermions in the naive continuum limit (while the
others whose masses are of the order of the inverse lattice spacing still
play a role in the lattice dynamics). Then the standard lattice calculation
\cite{dz,wilson} of the Gaussian fermionic integral (\ref{ba}) yields
\beq
b_A=\frac{2^{D/2}}{Du^2}\int_{-\pi}^{+\pi}\prod_{\ell=1}^D\frac{dp_\ell}
{2\pi}~\left(1-\frac{m}{2u^2}\frac{\frac{m}{2u^2}-\sum_{\ell=1}^D\cos p_\ell}
{\left(\frac{m}{2u^2}-\sum_{\ell=1}^D\cos p_\ell\right)^2+\sum_{\ell=1}^D
\sin^2p_\ell}\right)
\label{bawilson}\eeq
As mentioned above, the nice feature of (\ref{bawilson}) is that its
large-mass expansion coincides with that of the induced action (\ref{massexp})
where the backtracking paths never contribute due to the unitarity of the
gauge fields $U$. However, the price to pay for this nice property is the
very complicated integral which now appears in (\ref{bawilson}). Nonetheless,
since $b_A$ in
(\ref{bawilson}) varies from $b_A=0$ at $m=\infty$ to $b_A=1$ at $m=0$, we
expect that (\ref{ba}) and (\ref{geomcrit}) will possess a solution at some
critical value $m_c$. Indeed, an approximate numerical evaluation of the
integral in (\ref{bawilson}) \cite{kmak2} gives $m_c\sim
u(m_c)\sim\frac{1}{2}$, which resembles the situation for the case of
Kogut-Susskind fermions above.
However, the uncertainty in the mean field method is even larger now so
that a conclusion about the existence of the large-$N$ phase transition
is less certain for Wilson fermions. In any case, the large-$N$ phase
transition
definitely occurs in the case of Wilson fermions for $N_f\geq2$ \cite{kmak2}.
As mentioned in the introduction, for large enough flavours $N_f$ of the
fermion fields there is no asymptotic freedom for the adjoint fermion matrix
model (\ref{partd}). The mean field theory of this Subsection therefore
perfectly agrees with this, in that the phase transition occurs outside of the
asymptotically free region which is given by \cite{kmak2,mak2}
\beq
\frac{11}{3}-\frac{4N_f}{3}>0
\label{asymfree}\eeq

\subsubsection{Area Law}

{}From the analysis of the last Subsection we can conclude that the adjoint
fermion lattice gauge theory (\ref{partd}) has 2 phases
separated by a first order phase transition at some $m=m_c$ above which
the model is in the local confinement regime (see below) with dynamics
governed by the induced single-plaquette adjoint action (\ref{adjaction}).
For $m<m_c$ the model is in the perturbative area law phase and the local
$U(1)$ symmetry is broken (in some sense) \cite{mig5}. The phase
structure here is quite different from that of the Kazakov-Migdal model
\cite{mak2} in several respects. For instance, in the fermionic
model there is no Higgs phase or instability region due to the fermionic nature
of the inducing fields (but there can be a {\it composite} Higgs phase
associated with $\langle\ps2\rangle\neq0$). Furthermore, the first order phase
transition is present already for a single fermion flavour, whereas in the
Kazakov-Migdal model the large-$N$ first order phase transition appears only
for $N_f>N_f^c\sim30$ \cite{kmak2}. For the Kazakov-Migdal model the continuum
gauge theory is reached at the line of a second order phase transition which
separates
the area law and Higgs phases provided that one approaches it from the area
law phase \cite{mak2}. For the adjoint fermion matrix model (\ref{partd})
there is no Higgs phase and the continuum gauge theory is reached as $m\to0$.
Thus in this case the naive continuum limit is in fact the true continuum
limit of the lattice gauge theory.

Of course, these conclusions are only based on the mean field approximation
used above, but these arguments are quite reasonable because the phase
transition here occurs only for systems which are not asymptotically free.
At the point of this first order large-$N$ phase transition, the area law
behaviour of the adjoint Wilson loops which is associated with normal
confinement \cite{wilson} is restored just as it is for the single-plaquette
adjoint action (\ref{adjaction}). To see this, consider the adjoint Wilson loop
\beq
W_A(\Gamma)=\langle|W[U;\Gamma]|^2\rangle-1/N^2
\label{adjwilsonloop}\eeq
We can alternatively average this with respect to the induced action
(\ref{massexp}) which at $N_f=\infty$ becomes the single-plaquette adjoint
action (\ref{adjaction}). In this limit, the same factorization formula
(\ref{plaqfact}) holds at $N=\infty$ with the elementary plaquette $\Box$
replaced by the more general loop $\Gamma$ and $\bar\beta(\beta_A)$ given
by (\ref{wfdef}). Since $\bar\beta=0$ for $\beta_A<\beta_A^c$, from
(\ref{plaqfact}) it follows that the adjoint Wilson loop (\ref{adjwilsonloop})
vanishes there, unless the loop folds onto itself so that the area
$A_{\rm min}(\Gamma)$ of the minimal surface spanned by $\Gamma$ vanishes
\beq
W_A(\Gamma)=\delta_{0,A_{\rm min}(\Gamma)}+{\cal O}(1/N^2)
\label{amin0}\eeq
This means that in this phase all loops are contractable to a point, due
to unitarity, and formally this corresponds to local confinement with
an infinite string tension $\sigma_A(\beta_A)=\infty$. Thus in this phase
quarks aren't even asymptotically free and cannot propagate even within
hadrons. On the other hand, the area law \cite{wilson}
\beq
W_A(\Gamma)\sim\e^{-\sigma_A(\beta_A)A(\Gamma)}
\eeq
with string tension
\beq
\sigma_A(\beta_A)=2\sigma_F(\bar\beta(\beta_A))
\eeq
holds for $\beta_A>\beta_A^c$ when (\ref{wfdef}) has a non-trivial solution.
In this area law phase, the dynamics of extended objects are nontrivial.
Although the local confinement condition (\ref{amin0}) holds to all orders of
the large-mass expansion (\ref{massexp}) \cite{mak1}, it is not clear if the
area law behaviour is valid for finite $N_f$.

\subsection{Mean Field Theory of a Bosonic Lattice Gauge Theory at Large-$N$}

The above analysis has suggested that the Itzykson-Zuber integral undergoes a
large-$N$ first order phase transition when it is defined using
Grassmann-valued matrices which transform under the adjoint representation of
the Lie group $U(N)$.
It is instructive at this stage to demonstrate the breakdown of these results
in the scalar case. We shall therefore
in this Subsection briefly digress from our study of the adjoint fermion matrix
model (\ref{partd}) and discuss the analog of the phase structure in a
generalized bosonic lattice field theory which reduces to the Kazakov-Migdal
model (\ref{kazmigmodel}) as a limiting case. We present here a modified
version
\cite{van} of the mean field analysis of Khokhlachev and Makeenko \cite{kmak1}
who used the same adjoint unitary matrix integral (\ref{trialpart}) as the
trial partition function. We consider the $D$-dimensional Hermitian matrix
model (\ref{kazmigmodel}), except that we include the Wilson term
(\ref{wilson}) into the action to get a model more closely resemblant to QCD.
We are therefore interested in the adjoint scalar lattice gauge theory
\beq
Z_{KMW}=\int\prod_{x\in\lat}d\phi(x)~\e^{-N^2\sum_{x\in\lat}\tr
V(\phi(x))}\cdot
Z_U[\phi;\bar\beta]
\label{kazmigwil}\eeq
where
\bd
Z_U[\phi;\bar\beta]=\int\prod_{\langle
x,y\rangle\in\lat}[dU(x,y)]~\exp\left(N^2\sum_{\ell=-D}^D\tr\phi(x)U(x,x+\ell)
\phi(x+\ell)U^\dagger(x,x+\ell)\right.
\ed
\beq
\left.+\frac{N^2\bar\beta}{2}
\sum_{\Box\in\lat}\left(W[U;\Box)+W[U;\Box]^*\right)\right)
\label{kmwunitary}\eeq
The analysis of the matrix model (\ref{kazmigwil}) will therefore include the
phase structure of the exactly solvable Kazakov-Migdal model when we set
$\bar\beta=0$, and more generally the behaviour of the model for all values of
$\bar\beta$ for which (\ref{kazmigwil}) is a highly non-trivial field theory
which may have an interesting critical behaviour that is relevant to a
continuum limit for QCD.

To approximate (\ref{kazmigwil}) using mean field theory, we consider instead
of the adjoint model (\ref{trialpart}) the mixed trial partition function
\bd
Z_D^{(M)}=\int\prod_{\langle x,y\rangle\in\lat}[dU(x,y)]~\exp\left\{N^2\sum_{
\langle x,y\rangle\in\lat}\left[\tr\left(\alpha^*U(x,y)+\alpha
U^\dagger(x,y)\right)\right.\right.
\ed
\beq
\left.\left.+\beta|\tr U(x,y)|^2\right]\right\}
\label{trialpartkm}\eeq
which resembles (\ref{kmwunitary}). Again, we wish to find an optimized choice
of the parameters $\alpha$ and $\beta$ for which the behaviour of
(\ref{trialpartkm}) will closely resemble that of (\ref{kmwunitary}). Since
(\ref{trialpartkm}) is also a product of one-link unitary matrix integrals, it
can be evaluated using the results of Subsection 5.2.3 above from
\beq
Z_D^{(M)}=\e^{-N^2~{\rm vol}(\lat) DF_M(\alpha,\beta)}
\eeq
Now, however, the evaluation of the expectation values in Jensen's inequality
(\ref{jenseneq}) with $Z^{(1)}=Z_U$, $Z^{(0)}=Z_D^{(M)}$ and the maximization
over the parameters $\alpha$ and $\beta$ is a bit more involved and we shall
therefore go through the analysis more carefully than in the last Subsection.

First of all, the expectation value on the right-hand side of Jensen's
inequality (\ref{jenseneq}) for $\bar\beta=0$ is again a sum over one-link
averages and the normalized average $\langle S_0\rangle_0$ is easily computed
using (\ref{29}) and (\ref{30}).
To evaluate $\langle S_{KMW}\rangle_0$, with $S_{KMW}$ the action in
(\ref{kmwunitary}), we note first that (\ref{uijm0}) leads to the identity
\beq\new{\begin{array}{l}
\langle\tr\phi(x)U(x,x+\ell)\phi(x+\ell)U^\dagger(x,x+\ell)\rangle_0\\=\langle|
\tr U|^2\rangle_M\tr\phi(x)\phi(x+\ell)+(1-\langle|\tr
U|^2\rangle_M)\tr\phi(x)\tr\phi(x+\ell)\end{array}}
\label{km0id}\eeq
which holds at $N=\infty$.
Next, we need the normalized average $\langle S_W\rangle_0$ of the Wilson
action (\ref{wilson}). For this we note that the correlator $\langle
W[U;\Box]\rangle_0$ factorizes into a product of one-link averages involving
$\langle U_{ij}\rangle_M$ and $\langle U_{k\ell}^\dagger\rangle_M$. From
(\ref{um0}) it therefore follows that
\beq
\langle~{\rm Re}~W[U;\Box]\rangle_0=|\langle\tr U\rangle_M|^4
\label{wilson0id}\eeq
Notice that there is a total of $~{\rm vol}(\lat) D(D-1)/2$ plaquettes in the
lattice $\lat$. Substituting these identities into Jensen's inequality
(\ref{jenseneq})
then gives a lower bound on the partition function (\ref{kmwunitary})
\beq
Z_U[\phi;\bar\beta]\geq\e^{-N^2~{\rm vol}(\lat) D\Omega[\phi;\alpha,\beta]}
\label{zubound}\eeq
where
\bd
\Omega[\phi;\alpha,\beta]=-{\cal A}[\phi]+F_M(\alpha,\beta)+(\beta-{\cal
B}[\phi])\langle|\tr U|^2\rangle_M
+\alpha^*\langle\tr U\rangle_M+\alpha\langle\tr U^\dagger\rangle_M
\ed
\beq
-\frac{\bar\beta(D-1)}{2}\left|\langle\tr U\rangle_M\right|^4
\label{lambdadef}\eeq
and we have defined
\beq\new{\begin{array}{c}
{\cal A}[\phi]=\frac{1}{~{\rm vol}(\lat) D}\sum_{\langle
x,y\rangle\in\lat}\tr\phi(x)\tr\phi(y)\\
{\cal B}[\phi]=\frac{1}{~{\rm vol}(\lat) D}\sum_{\langle
x,y\rangle\in\lat}(\tr\phi(x)\phi(y)-\tr\phi(x)\tr\phi(y))
\end{array}}
\label{abdef}\eeq

\subsubsection{Minimization of the Mean Field Action}

We now want to minimize the action (\ref{lambdadef}) over all values of
$\alpha$ and $\beta$. Since $\Omega[\phi;\alpha,\beta]$ depends only
on $|\alpha|$, we may assume without loss of generality that the parameter
$\alpha$ is real and positive in the above. From (\ref{28})--(\ref{30}) we see
that $\Omega[\phi;\alpha,\beta]$ is given explicitly by
\beq
\Omega[\phi;\alpha,\beta]=\left\{\new{\begin{array}{ll}-{\cal A}+(1-{\cal
B})x-16qx^2~~,~&\alpha\leq(1-\beta)/2,0\leq\beta\leq1\\-{\cal A}+\frac{1}{4}
+\frac{1}{2}\log(2(1-y))-{\cal B}y^2-qy^4~~,~&\alpha\geq(1-\beta)/2,\beta\geq0
\end{array}}\right.
\label{lambdaexpl}\eeq
where we have introduced the constants
\beq
q=\bar\beta(D-1)/32~~~,~~~x=\alpha^2/(1-\beta)^2~~~,~~~y=(\beta-\alpha+
\sqrt{(\alpha+\beta)^2-\beta})/2\beta
\label{qxydef}\eeq
We see therefore that the action $\Omega$ can be simplified to a function of a
single variable in both regions where it is defined. In the first region, which
we shall call region 1, it is a function of $x\in[0,\frac{1}{4}]$ while in
region 2 it is function of $y\in[\frac{1}{2},1]$.

Consider first region 1. It is easy to see that there $\Omega$ is minimized
at either $x=0$ or $x=1/4$. However, $\Omega(x=0)\geq\Omega(x=1/4)$ is
 equivalent to ${\cal B}\leq1-4q$, and so over region 1 the
action (\ref{lambdaexpl}) is minimized at
\beq
\Omega_0[\phi;\bar\beta]=\left\{\new{\begin{array}{ll}-{\cal A}~~~,~~&{
\cal B}\leq1-4q\\-{\cal A}-q+(1-{\cal B})/4~~~,~~&{
\cal B}\geq1-4q\end{array}}\right.
\label{lambdamin1}\eeq
The behaviour of $\Omega$ in region 2 is somewhat more complicated. For
 the moment we ignore the constant $\cal A$ that appears in
(\ref{lambdaexpl}) above. In this region, for each fixed value
of $y\in[\frac{1}{2},1]$, as a function of $\cal B$ and $q$, $\Omega$
describes a plane. The minimum $\Omega_0$ of $\Omega$ over this region is then
the envelope function (i.e. lower bound) of the set of planes parametrized by
 $y$. This implies that $\Omega_0$ in this region is a concave down function
of $q$ and $\cal B$. Furthermore, since $\frac{1}{2}\leq y\leq1$, for each
 fixed $q$ the slope of $\Omega_0$ is bounded between $-\frac{1}{4}$ and
$-1$ and it approaches these limits as ${\cal B}\to\pm\infty$. Since the
 slope of $\Omega_0$ is 0 or $-1/4$ in region 1, it follows that there is
a unique point ${\cal B}_0(q)$ for each $q$ below which $\Omega$ takes
its minimum in region 1 and above ${\cal B}_0(q)$ in region 2.
 Since $\Omega_0({\cal B}=1)\geq-q$ for each fixed value of $q$, it is
greater than $\Omega_0$ in region 1 and hence ${\cal B}_0(q)\geq1$.

To find the explicit form of $\Omega_0$ in region 2, we
differentiate (\ref{lambdaexpl}) to get
\beq
\frac{\partial\Omega}{\partial y}=\frac{1-4{\cal B}y(1-y)-8qy^3(1-y)}{2(1-y)}
\label{lambdaderivy}\eeq
For $q$ and $\cal B$ positive, it follows from (\ref{lambdaderivy})
that $\partial_y\Omega$ can have at most one zero where it is increasing,
 and so $\Omega$ can have at most one minimum for $y\in[\frac{1}{2},
1]$. Furthermore, since $\partial_y\Omega\to\infty$ as $y\to1$
 and $\partial_y\Omega(y=1/2)=1-{\cal B}-q/2$, it follows
that $\partial_y\Omega$ has a unique zero for $y\in[\frac{1}{2},1]$ at
which point $\Omega$ attains its global minimum. Note that we need only
consider the region ${\cal B}>1$, since
as discussed above for ${\cal B}<1$ the minimum of $\Omega$ lies in
 region 1. Thus for ${\cal B}>1$, the minimum of $\Omega$
 is $\Omega_0[\phi;\bar\beta]=\Omega(y(q))$ where $y(q)$ is the unique
solution with $y\in[\frac{1}{2},1]$ of the quartic equation
\beq
1-4{\cal B}y(1-y)-8qy^3(1-y)=0
\label{quarteq}\eeq
Although an analytic solution for $y(q)$ is possible, it is rather
complicated, and it is more instructive to instead explore a small
$q$ (strong coupling) expansion of the minimum value $\Omega_0$.

First, we note that for $q=0$, the admissible solution of (\ref{quarteq})
is $y_0=(1+(1-1/{\cal B})^{1/2})/2$, so that over region 2 the minimum
of $\Omega$ for $q=0$ is
\beq
\Omega_0^{(0)}=\frac{1}{2}+\frac{1}{2}\log\left({\cal B}\left(1+\sqrt{1-1/
{\cal B}}\right)\right)-\frac{1}{2}{\cal B}\left(1+\sqrt{1-1/{\cal B}}\right)
\label{omminq0}\eeq
We now Taylor expand (\ref{lambdaexpl}) in region 2 about $y=y_0$
using (\ref{quarteq}) to compute the derivatives $\frac{d^ny(q)}{dq^n}|
_{q=0}$. After some algebra, we find to first order in $q$ that
\bd
\Omega_0=\frac{1}{2}+\frac{1}{2}\log\left({\cal B}\left(1+\sqrt{1-1/
{\cal B}}\right)\right)-\frac{1}{2}{\cal B}\left(1+\sqrt{1-1/{\cal B}}\right)
\ed
\beq
-\frac{q}{16}\left(1+\sqrt{1-1/{\cal B}}\right)^4+{\cal O}(q^2)
\label{om02strong}\eeq
Note that (\ref{om02strong}) approaches the value $-q+\frac{1}{2}\log(-2{
\cal B})-{\cal B}+3/4$ as $\cal B$ becomes large, which justifies the
assertion made earlier that (\ref{lambdaderivy}) as a function of $q$
tends to $-1$ as ${\cal B}\to\infty$ and that $\Omega$ eventually gets
 smaller than the minimum (\ref{lambdamin1}) from the previous region
at some value ${\cal B}_0(q)>1$. The actual value of ${\cal B}_0(q)$ is
 defined by the intersection of (\ref{om02strong}) with the
function $-q+(1-{\cal B})/4$, but notice that we can also determine it as
\beq
{\cal B}_0(q)=\min_{y\in[\frac{1}{2},1]}\frac{q(1-y^4)-\frac{1}{2}\log(2(1-y))}
{y^2-1/4}
\label{b0min}\eeq
which defines the intersection of the envelope function for the collection
of lines in region 2 in (\ref{lambdaexpl}) with the minimum value of $\Omega$
in region 1 in (\ref{lambdamin1}).

It can be shown by a numerical calculation \cite{van} that for small values
of $q$ a good approximation to ${\cal B}_0(q)$ is
\beq
{\cal B}_0(q)\approx 1+(1.32)\cdot y^{2/3}~~~{\rm for}~~~q\ll1
\label{boapprox}\eeq
For large values of $q$ (weak coupling), the minimizing value of $y$
 in (\ref{b0min}) is very close to 1, so that ${\cal B}_0(q)$
tends asymptotically to
\beq
{\cal B}_0(q)\approx2(1+\log4q)/3~~~{\rm for}~~~q\gg1
\label{b0qlarge}\eeq
Then we can write the minimum of the action $\Omega$ over both regions 1 and
 2 as
\beq
\Omega_0[\phi;\bar\beta]=\left\{\new{\begin{array}{ll}-{\cal A}~,&{
\cal B}\leq1-4q\\-{\cal A}-q+(1-{\cal B})/4~,&1-4q\leq{\cal B}\leq{
\cal B}_0(q)\\\new{\begin{array}{l}-{\cal A}+\frac{1}{2}+\frac{1}{2}\log(
{\cal B}(1+\sqrt{1-1/{\cal B}}))\\-\frac{1}{2}{\cal B}(1+\sqrt{1-1/{
\cal B}})-\frac{1}{16}q(1+\sqrt{1-1/{\cal B}})^4\\+{
\cal O}(q^2)\end{array}}~,&{\cal B}>{\cal B}_0(q)\end{array}}\right.
\label{omegamintot}\eeq

\subsubsection{Observables and Critical Behaviour}

We shall now use the action (\ref{omegamintot}) as an approximation in
the partition function (\ref{kmwunitary}) and examine the phase structure
of the induced bosonic field theory (\ref{kazmigwil}).
We further assume now that the field $\phi(x)$ is frozen at a mean value
$\phi$ at each site $x$ of the lattice $\lat$. This is justified if we
assume that the lattice is large enough so that the expectation values
$\langle f(\phi(x))\rangle$ in (\ref{kazmigwil}) are approximately the
same for each site $x$. For definiteness, we shall also now take $V$ to be
 the Gaussian potential
\beq
V(\phi)=M^2\phi^2
\eeq
for the scalar field. Thus we effectively want to study the properties of
the Hermitian one-matrix model
\beq
Z_{KMW}^{(1)}=\int d\phi~\e^{-N^2M^2\tr\phi^2-N^2D\Omega_0[\phi;\bar\beta]}
\label{kmwherm}\eeq
where from (\ref{abdef}) the functionals ${\cal A}[\phi]$ and ${\cal B}[
\phi]$ appearing in (\ref{omegamintot}) in the mean-field approximation are
\beq
{\cal A}[\phi]=(\tr\phi)^2~~~~~,~~~~~{\cal B}[\phi]=\tr\phi^2-(\tr\phi)^2
\label{meanab}\eeq
The Hermitian one-matrix model (\ref{kmwherm}) therefore possesses
the reflection symmetry $\phi\to-\phi$ so that $\langle\tr\phi\rangle=0$.
This suggests that we could approximate (\ref{meanab}) by ${\cal A}[\phi]=0$
 and ${\cal B}[\phi]=\tr\phi^2$ without significantly changing the behaviour
 of the model. This approximation (i.e.
 that $\langle(\tr\phi)^2\rangle\ll\langle\tr\phi^2\rangle$) is justified in
the large-$N$ limit of the matrix model (\ref{kmwherm}) because
of factorization. With this observation,  $\Omega_0$ is now a function of $
{\cal B}=\tr\phi^2$ with parameter $q$ which we will denote by $\Omega_q(
{\cal B})$.

Recall that for large $\cal B$, $\Omega_q$ is asymptotically equal to $3/4-
{\cal B}+\frac{1}{2}\log2{\cal B}-q$. It follows that for $M^2<D$ the
partition function (\ref{kmwherm}) diverges and the action there
becomes infinite as $\tr\phi^2\to\infty$. This corresponds to an
 unphysical situation where $\langle\tr\phi^2\rangle=\infty$. We must
  therefore require that $M^2\geq D$ in which case all correlators of
 the matrix model (\ref{kmwherm}) should be well-defined. To analyse
the behaviour of this matrix model, we consider the case $D=0$ first
where the partition function is a pure Gaussian matrix model. Rescaling
 the field variables in (\ref{kmwherm}) as $\phi\to M\phi$, we see that
the $M$-dependence of the partition function is $Z_{KMW}^{(1)}(D=0)
\sim M^{-N^2}$. From (\ref{kmwherm}) it follows that
\beq
\left\langle\tr\phi^2\right\rangle(D=0)=-\frac{1}{N^2}\frac{\partial
\log Z_{KMW}^{(1)}(D=0)}{\partial M^2}=\frac{1}{2M^2}
\label{D0sqcorr}\eeq
and that the root mean square fluctuation about this average value is
\beq
\left[\left\langle\left(\tr\phi^2-\left\langle\tr\phi^2\right\rangle(D=0)
\right)^2\right\rangle\right]^{1/2}=\left[\frac{1}{N^4}\frac{\partial^2
\log Z_{KMW}^{(1)}(D=0)}{\partial(M^2)^2}\right]^{1/2}=\frac{1}{NM^2\sqrt{2}}
\label{rms}\eeq
This shows that in the large-$N$ limit the configurations with $\tr\phi^2$
away from its mean value are largely suppressed.

In particular, the function
$f(\tr\phi^2)=M^2{\cal B}+D\Omega_q({\cal B})$ which is the action
in (\ref{kmwherm}) will be very close to its value at ${
\cal B}\equiv\langle\tr\phi^2\rangle$. Thus, at $N=\infty$ we can replace $f$
by its first order Taylor series expansion about $\cal B$, and the
 partition function (\ref{kmwherm}) can be written as
\beq
\lim_{N\to\infty}Z_{KMW}^{(1)}=\int d\phi~\e^{-N^2f'({\cal B})\tr\phi^2}
\label{kmwgauss}\eeq
up to an irrelevant constant. This is simply a Gaussian Hermitian
matrix integral, and so using (\ref{D0sqcorr}) we have
\beq
{\cal B}\equiv\left\langle\tr\phi^2\right\rangle=1/2f'({\cal B})
\label{selfcons1}\eeq
The correlator $\langle\tr\phi^2\rangle$, which displays the local behaviour
 of the scalar fields in the matrix model (\ref{kazmigwil}), is
 therefore determined as the solution of the self-consistency
condition (\ref{selfcons1}) which explicitly reads
\beq
D\Omega_q'({\cal B})=1/2{\cal B}-M^2
\label{selfconsexpl}\eeq
Thus for each value of $q$ we can determine the value of ${
\cal B}=\langle\tr\phi^2\rangle$ for each value of $M^2$ by finding the
point $\cal B$ where the 2 functions in (\ref{selfconsexpl}) intersect.

{}From (\ref{omegamintot}) we see that there are 3 different regions of
behaviour to consider. For ${\cal B}<1-4q$, we have $\Omega_q'({\cal
 B})=0$, while for $1-4q<{\cal B}<{\cal B}_0(q)$ we find $\Omega_q'
({\cal B})=-1/4$. In the final region, we recall that $\Omega_q({
\cal B})\equiv\Omega(y({\cal B},q),{\cal B},q)$ where $y({\cal B},q)$ is
the unique solution to $\partial_y\Omega(y,{\cal B},q)=0$
with $y\in[\frac{1}{2},1]$ (i.e. of (\ref{quarteq})). Thus in the third
region in (\ref{omegamintot}) we have
\beq
\Omega_q'({\cal B})=\frac{\partial}{\partial{\cal B}}\Omega(y({\cal B},q),
{\cal B},q)=-[y({\cal B},q)]^2
\label{omq3}\eeq
and so $-D/4\leq D\Omega_q'({\cal B})\leq-D$ in this region.
 Furthermore, $D\Omega_q'({\cal B})$ asymptotically approaches $D/2{\cal B
}-D$ for large $\cal B$, and, since $\Omega_q({\cal B})$ is concave
down everywhere, $\Omega_q'({\cal B})$ decreases from some value less
 than $-1/4$ at ${\cal B}={\cal B}_0(q)$ towards $-1$ as ${\cal B}\to
\infty$. The value of $\Omega_q({\cal B})$ in the third region can be
 calculated numerically for any $q$ using (\ref{omq3}) and
 (\ref{quarteq}) \cite{van}.

The function $1/2{\cal B}-M^2$ will intersect $D\Omega_q'({\cal B})$ in
the first region if the equation $1/2{\cal B}-M^2=0$ has a solution with
${\cal B}<1-4q$. This implies that $\langle\tr\phi^2\rangle=1/2M^2$
 for $q<\frac{1}{4}(1-1/2M^2)$. Next, (\ref{selfconsexpl}) has a solution in
the second region if the equation $1/2{\cal B}-M^2=-D/4$ has a solution
 with $1-4q<{\cal B}<{\cal B}_0(q)$. In the physical region, $1/2{
\cal B}-M^2<1/2-D<-D/4$ at ${\cal B}=1$ and for $D\geq1$, and so such a
 solution always exists and we find $\langle\tr\phi^2\rangle=1/2(M^2-D/4)$
for $q>\frac{1}{4}(1-1/(M^2-\frac{D}{4}))$. Finally, we note that the
third region never plays a role in determining the behaviour of $\cal B$
 with respect to $M^2$ and $q$, because the restriction $M^2>D$ ensures
that the functions in (\ref{selfconsexpl}) always intersect in one of the
first 2 regions. Recalling the definition of $q$ in (\ref{qxydef}), we
can summarize the behaviour of the scalar field correlator as
\beq
\left\langle\tr\phi^2\right\rangle=\left\{\new{\begin{array}{ll}
\frac{1}{2M^2}~~~,~~&\bar\beta<\frac{8}{D-1}\left(1-\frac{1}
{2M^2}\right)\\\frac{1}{2(M^2-D/4)}~~~,~~&\bar\beta>\frac{8}{D-1}\left(1-\frac
{1}{2(M^2-D/4)}\right)\end{array}}\right.
\label{scalarcorr}\eeq
for $M^2>D$. For $M^2\leq D$ we are in the forbidden region of the theory
where $\langle\tr\phi^2\rangle$ diverges. Notice that there is a small
region of ambiguity between the 2 regions in (\ref{scalarcorr}) where the
scalar correlator is not determined. This is most likely due to the fact
that the action $\Omega_q({\cal B})$ is not smooth at ${\cal B}=1-4q$, so
 that the assumption that it can be replaced by its Taylor series expansion
 to first order is questionable. The Hermitian one-matrix model
 (\ref{kmwherm}) may have a smooth transition between ${\cal B}=1/2M^2$
and ${\cal B}=1/2(M^2-D/4)$ in the vicinity of this ambiguous region,
 but there is a reasonable chance that the transition there is some sort
of singular phase transition, and that there is a critical line somewhere
 in this region.

Finally, we would like to determine the correlator $\langle~{
\rm Re}~W[U;\Box]\rangle$ which describes the local behaviour of the
 gauge fields in the model (\ref{kazmigwil}). The average of this quantity
over all plaquettes is given by
\beq
\langle~{\rm Re}~W[U;\Box]\rangle=\frac{2}{N^2~{\rm vol}(\lat) D(D-1)}\frac{
\partial Z_{KMW}^{(1)}}{\partial\bar\beta}
\label{gaugecorrdef}\eeq
{}From (\ref{omegamintot}), we easily find that
\beq
\langle~{\rm
Re}~W[U;\Box]\rangle=\left\{\new{\begin{array}{ll}0~~~~~,~~~~
&\bar\beta<\frac{8}
{D-1}\left(1-\frac{1}{2M^2}\right)\\1/8~~~~~,~~~~&\bar\beta>\frac{8}{D-1}
\left(1-\frac{1}{2(M^2-D/4)}\right)\end{array}}\right.
\label{gaugeexpl}\eeq

Thus, setting $\bar\beta=0$ in the above to recover the Kazakov-Migdal model,
 we see that the mean field theory results seem to suggest the existence of
 a phase transition in the Itzykson-Zuber integral $I[\phi(x),\phi(y)]
$. Although this seems to contradict the exact solutions of the
Gaussian Kazakov-Migdal model which are based on the Itzykson-Zuber formula,
 the latter formula is valid only with the {\it a priori} assumption that
there is no phase transition in this unitary matrix integral separating
 2 regions of different behaviour for different sizes of the components
 of $\phi$. Nevertheless, we do recover here the previous exact results
 of Gross \cite{gross} and the mean field theory results of Khokhlachev
 and Makeenko \cite{kmak1} which both showed that there is a barrier in
the Gaussian model at $M^2=D$ (below which lies an unstable Higgs phase
due to an unlimited Bose-Einstein condensation of the scalar fields) and
that there is no phase transition as $M^2$ is varied in the physical
 region $M^2>D$. Thus the Gaussian Kazakov-Migdal model has no continuum limit.

What is particularly interesting in the above extended mean field
theory analysis is the appearence of a possible phase transition between
2 regions as $\bar\beta$ is increased for any given value of $M^2$.
Furthermore, the critical curve actually intersects the $\bar\beta=0$ axis
at some $M^2>0$, although this intersection point lies in the
 unphysical $M^2<D$ region. If this intersection occured at some value
$M^2>D$ then the corresponding phase transition could be exactly what is
 needed to define the continuum limit of the Kazakov-Migdal model. There
 is therefore some hope that a similar model may have the correct behaviour
 to define a continuum limit. Of course, there is less of an accuracy in
the mean field theory results here than in the fermionic case \cite{mak2}.
 It could be that the non-trivial phase structure observed above is merely
an artifact of the phase transitions that occur in the trial partition
function of the mixed unitary matrix model. However, it has been argued
by Khokhlachev and Makeenko \cite{kmak1} that such mean field approaches
 are reasonably accurate for the description of phase transitions. Moreover,
the mean field value (\ref{scalarcorr}) for the scalar correlator
at $\bar\beta=0$ is very close to the exact result \cite{dms,gross} (see
Subsection 1.1.3 with $D=0$ there)
\beq
\left\langle\tr\phi^2\right\rangle_{KM}=\frac{1}{2\sqrt{M^4-1}}
\eeq
In addition, because of the local phase invariance $U\to\e^{i\theta}U$ of
 the Kazakov-Migdal model, the mean-field value of 0 for the Wilson
loop expectation value in the first phase in (\ref{gaugeexpl}) is correct.
 On the other hand, at weak-coupling $\bar\beta\to\infty$ the Wilson
 action (\ref{wilson}) is minimized when all the unitary matrices there are
set equal to the identity matrix, so one expects a value of 2 for the
Wilson loop expectation value in the second phase in (\ref{gaugeexpl}).
The discrepency in (\ref{gaugeexpl}) is due to the fact that the mean
field approximation is a bit too crude for the calculation of gauge
field correlators. To properly compute these correlators, one must turn to
the method of loop equations for the matrix model \cite{dms,dmsw}. This will
 be the topic of the next Section for the fermionic case.

\section{Loop Equations and Observables in Higher Dimensional Adjoint Fermion
Matrix Models}

We now turn to a more precise investigation of the characteristics of
 the adjoint fermion matrix model (\ref{partd}). To get a more
exact understanding of the behaviour of its observables, we must once
again resort to the Makeenko-Migdal method of loop equations.
The loop equations for the matrix model (\ref{partd}) are the only way here
to study the precise behaviour of observables of this model, and to obtain
the exact characteristics of the critical behaviour of the induced gauge
theory. Furthermore, by investigating the observables associated with extended
objects in the model (the open-loop averages) one can check to see if, while
the gravitational part of the system is continuous, the matter fields become
critical at a given fixed point \cite{mak5}.

\subsection{Loop Equations for Extended Averages}

To generate the observables of the adjoint fermion matrix model
(\ref{partd}), we must generate all the correlators of the two-matrix model of
Section 4 at each site and link. In this case the pair correlators of
the gauge fields can be evaluated from the one-link expectation values
(\ref{gaugecorr}), since at large-$N$ the expectation value of any quantity
such as $U_{ij}U^\dagger_{k\ell}$ factorizes into one-link averages. Because
of the higher-dimensionality of the model now there are, however, numerous
other
gauge correlators that can be formed along curves in the lattice. Gauge
invariance implies that any such observable must be some sort of combination
of the Wilson line operators (\ref{wilsonline}). However, the space of
gauge-invariant operators must be further reduced because of the local
$U(1)$-gauge symmetry (\ref{znsym}) at a given link. In particular, as
mentioned before, this extra symmetry excludes the conventional Wilson line
operator (\ref{wilsonline}) since then
\beq
\langle W[U;C]\rangle=0
\eeq
The non-vanishing observables at $N=\infty$ must contain the same number of $U$
and $U^\dagger$ operators at each link. The complete set of gauge correlators
of
the model (\ref{partd}) are therefore generated by the operator products
\beq
W[U;C_1,\ldots,C_k]=W[U;C_1]\cdots W[U;C_k]
\label{gaugeopgen}\eeq
where
\beq
C_1+\ldots+C_k=0
\eeq

The simplest examples of such gauge-invariant operators are the adjoint
Wilson loop
\beq
W_A[U;C]=W[U;C,-C]-1/N^2=|W[U;C]|^2-1/N^2
\label{adjwilson}\eeq
and for $D>1$ the filled Wilson loop
\beq
\Winf[U;\Sigma]=W[U;\partial\Sigma]\prod_{\Box\in\Sigma}W[U;\Box]^*
\label{filledwilson}\eeq
where $\Sigma$ is a 2-dimensional surface in the lattice $\lat$ with
boundary the loop $\partial\Sigma$. The filled Wilson loops are
particularly important if we wish to interpret the induced gauge theory as
QCD. If this is to be so then there must be an observable such as the Wilson
loop which for $N_f<\infty$ obeys an area law in the confining phase
\cite{wilson}. Since the adjoint Wilson loop generally only obeys a perimeter
law, the filled Wilson loop (\ref{filledwilson}) is the only reasonable
candidate for such an observable \cite{kmsw}. Furthermore, in the
continuum limit it depends only on $\partial\Sigma$ and reduces to the
conventional Wilson loop \cite{kmsw}. It is in this way that the adjoint
fermion model (\ref{partd}) may induce QCD in the continuum limit even though
in QCD the extra local $U(1)$-gauge invariance is absent. Moreover, the
filled Wilson loop observables in the Kazakov-Migdal model can be represented
as sums over fluctuating 2-dimensional surfaces in such a way that surfaces
with contractable boundaries play the role of the physical states (gluons)
while uncontractable boundaries describe world-sheet vortices \cite{kmsw}. The
filled Wilson loops are therefore important for constructing string theories in
dimension $D>1$ from the induced gauge theory model.
As discussed in the Section 4, however, there is no direct
way to evaluate these gauge field observables from the Schwinger-Dyson
equations for (\ref{partd}).

It is possible that the critical behaviours of observables in the induced gauge
theory could be seen from the criticality of the inducing matter fields.
To explore the possible criticality of the matter fields of the matrix model
(\ref{partd}), we investigate now the observables which are associated with
extended objects in the model. The quantum equations of motion for the
open-loop fermion observables (and in particular the lattice Dirac equation)
relate closed adjoint Wilson loops to open ones with fermions at the ends. As
usual, the loop equations are formally solved at $N=\infty$. However, for the
$D$-dimensional model (\ref{partd}) it turns out that in addition they are
only amenable to explicit solution in the strong coupling regime of the theory
where the local confinement condition (\ref{amin0}) holds. To see this,
consider as an example the open Wilson line correlator
\bd\new{\begin{array}{l}
\delta_{ij}\Hoo^{\mu\nu}(z;C_{x,y})\\=\left\langle\tr\psi^\mu_i(x)\left(P\prod
_{\langle x',y'\rangle\in C_{x,y}}U(x',y')\right)\frac{1}{z-\bar
\psi_j(y)\psi_j(y)}\bar\psi^\nu_j(y)\right.\end{array}}
\ed
\beq
\left.\times P\prod_{\langle x',y'\rangle\in C_{x,y}}U^\dagger(x',y')\right
\rangle
\label{opencorrgen}\eeq
where $\mu$ and $\nu$ are spin indices and $C_{x,y}$ is a contour
connecting the sites $x$ and $y$. The delta-function on the left-hand side
of (\ref{opencorrgen}) arises because the action (\ref{lataction}) is
diagonal in the flavour indices. It therefore suffices in the following to
consider only a single fermion flavour, $N_f=1$. The leading $1/z$ term in the
asymptotic expansion of (\ref{opencorrgen}) is the open Wilson line with
fermions at its ends
\bd
H^{\mu\nu}(C_{x,y})=\left\langle\tr\psi^\mu(x)\left(P\prod_{\langle x',y'
\rangle\in C_{x,y}}U(x',y')\right)\bar\psi^\nu(y)\right.
\ed
\beq
\left.\times P\prod_{\langle x',y'\rangle\in C_{x,y}}U^\dagger(x',y')\right
\rangle
\label{openloop}\eeq
At $N=\infty$ the multi-link correlator
(\ref{openloop}) can be factored into a product of one-link correlators
\beq
H^{\mu\nu}_\ell=\langle\tr\psi^\mu(x)U(x,x+\ell)\bar\psi^\nu(x+\ell)U^\dagger
(x,x+\ell)\rangle
\label{1linkfact}\eeq

Since the Wilson line in (\ref{openloop}) is in the same (adjoint)
representation as the fermion fields, we can use the standard large-mass
expansion of lattice gauge theory \cite{dz,wilson} to represent
(\ref{openloop}) as a sum over lattice loops
\beq
H^{\mu\nu}(C_{x,y})=\sum_{\Gamma_{y,x}\in\lat}\frac{W_A(C_{x,y}\circ
\Gamma_{y,x})}{m^{l(\Gamma_{y,x})+1}}\left(P\prod_{\ell\in\Gamma_{y,x}}
\proj_\ell^\pm\right)^{\mu\nu}
\label{amploops}
\eeq
where the sum is over all paths $\Gamma_{y,x}$ which when joined to $C_{x,y}$
result in a closed contour. In the strong coupling regime the paths
$\Gamma_{y,x}$ coincide with $C_{x,y}$ with opposite orientation modulo
backtrackings which form a 1-dimensional tree graph. Thus in this case the
problem of calculating the path amplitude (\ref{amploops}) is reduced to
the combinatorial problem of summing over 1-dimensional tree graphs
embedded in a $D$-dimensional space. In the phase with normal area law,
non-trivial loops contribute to the sum in (\ref{amploops}) and lead to
rather complicated unitary matrix integrals in what follows. It
seems that only in the local confinement phase, where the dynamics
of extended objects are trivial, can one write down a complete set of
Schwinger-Dyson equations for the local objects. In this Section we shall be
concerned with the contributions from the tree-like, or polymer, graphs in
(\ref{amploops}).

To explicitly see how the strong-coupling assumption formally simplifies the
evaluation of correlators, consider a general expectation value of the form
\beq
H_\ell^\mu({\cal O})=\langle\tr{\cal O}U(x,x+\ell)\psi^\mu(x+\ell)U^\dagger(
x,x+\ell)\rangle
\label{gencorrop}\eeq
where the operator $\cal O$ is independent of the operators $U(x,x+\ell)$,
$U^\dagger(x,x+\ell)$, $\bar\psi(x+\ell)$ and $\psi(x+\ell)$ (e.g. see
(\ref{1linkfact})). If we expand the exponent in (\ref{partd}) in a power
series in all of the coupling constants of the potential except the mass $m$
of the fermion field, then the calculation of (\ref{gencorrop}) reduces to the
evaluation of Wick contractions among $\bar\psi(y)$ and $\psi(y)$ and an
integration over all $U(y,y+\ell)$. The $U(N)$ integration can be carried
out independently on each link because there is no kinetic term for the
gauge fields. Thus to evaluate (\ref{gencorrop}) it suffices to consider the
one-link correlator
\beq\new{\begin{array}{l}
\langle\tr T^aU\chi^\mu U^\dagger\rangle_{1L}\\\equiv\frac{\int[dU]~\int d\chi
{}~d\bar\chi~\e^{N^2\tr(V(\ch2)-\bar\psi\proj_\ell^-U\chi U^\dagger-\bar\chi
\proj_\ell^+U^\dagger\psi U)}\tr T^aU\chi^\mu U^\dagger}{\int[dU]~\int d\chi~d
\bar\chi~\e^{N^2\tr(V(\ch2)-\bar\psi\proj_\ell^-U\chi U^\dagger-\bar\chi\proj_
\ell^+U^\dagger\psi U)}}\end{array}}
\label{1linkcorr}\eeq
since the remaining part of the large-mass expansion (see (\ref{amploops}))
decouples due to the properties of the unitary matrix integrals in the local
confinement phase and large-$N$ factorization. In (\ref{1linkcorr}) the
average is with respect to $U$ and $\chi$ only with $\psi$ playing the role
of an external field.

The integration in the large-mass expansion of (\ref{1linkcorr}) can be
carried out over $\bar\chi$ and $\chi$ using the Wick rules, after which the
unitary matrix integration is rather simplified, because all the $U$ matrices
cancel with their conjugates $U^\dagger$, and leaves a functional of
only $\tr T^a\psi^\mu(\ps2)^n$ \cite{mz}. This is because the potential
depends only on $\ps2$ so that terms such as $\tr T^a\bar\psi^n\psi^n$ do not
appear when evaluating (\ref{1linkcorr}) using the above rules. The one-link
correlator (\ref{1linkcorr}) at $N=\infty$ therefore has the form
\beq\new{\begin{array}{ll}
\langle\tr T^aU\chi^\mu U^\dagger\rangle_{1L}&=\sum_{\nu=1}^s(\proj_\ell^+)^{
\mu\nu}\tr T^a\psi^\nu F(\ps2)\\&=\sum_{\nu=1}^s(\proj_\ell^+)^{\mu\nu}\sum_
{n=1}^\infty F_n\tr T^a\psi^\nu(\ps2)^n\end{array}}
\label{1linkexp}\eeq
where $s=2^{[D/2]}$ is the number of spin components of the fermion fields and
\beq
F(z)=\sum_{n=1}^\infty F_nz^n
\label{F}\eeq
is some analytic function\footnote{\baselineskip=12pt In the Kazakov-Migdal
model, this function appears as the large-$N$ limit of the logarithmic
derivative of the Itzykson-Zuber integral in the saddle point equation
(\ref{saddlecor}),
\bd
F_i(\phi)\equiv\frac{1}{N}\sum_jC_{ij}(\phi)\phi_j
\ed}. Thus the general correlator (\ref{gencorrop}) has the form
\beq
H_\ell^\mu({\cal O})=\sum_{\nu=1}^s(\proj_\ell^+)^{\mu\nu}\left\langle\tr{
\cal O}F(\bar\psi(x)\psi(x))\psi^\nu(x)\right\rangle
\label{gencorrf}\eeq
The coefficients $F_n$ for (\ref{1linkcorr}) will be different than
those for the full lattice correlator (\ref{gencorrop}), but the analytic
function $F(z)$ in (\ref{gencorrf}) is universal in that it is independent
of the explicit form of the operator $\cal O$. The function (\ref{F})
depends on the potential (\ref{fermpot}) and in general it is not possible
to determine its exact form. As will be discussed later on, one
must substitute for it an appropriate ansatz.

The loop equation for the open-loop average (\ref{opencorrgen}) follows from
the invariance of the integration measure over $\bar\psi(x)$ in
(\ref{opencorrgen}) under the infinitesimal field shifts
\beq
\bar\psi(x)\to\bar\psi(x)+\epsilon^a(x)T^a
\label{dshift}\eeq
in analogy with (\ref{sdtransf}). In analogy with (\ref{sdc1}), we consider the
loop average
\beq
\left\langle\tr T^a\left(P\prod_{\langle x',y'\rangle\in
C_{x,y}}U(x',y')\right)
\frac{1}{z-\bar\psi(y)\psi(y)}\bar\psi(x)\left(P\prod_{\langle x',y'\rangle\in
C_{x,y}}U^\dagger(x',y')\right)\right\rangle\equiv0
\label{dvanavg}\eeq
which vanishes because of the gauge invariance of (\ref{partd}). Working out
the resulting correlators in the usual way by performing the shift
(\ref{dshift}) of $\bar\psi(x)$, using the invariance of the integration
measure and calculating the derivatives
$\frac{\partial}{\partial\bar\psi^b(x)}$, it is straightforward to show that
this leads to the Dirac open-loop equation for $\psi(x)$
\beq\new{\begin{array}{l}
\oint_{\cal C}\frac{d\lambda}{2\pi i}~\frac{V'(\lambda)}{z-\lambda}
\Hoo(\lambda;C_{x,y})-\sum_{\ell=1}^D\left(\proj_\ell^+\Hoo(z;C_{x+\ell,x}\circ
C_{x,y})+\proj_\ell^-\Hoo(z;C_{x-\ell,x}\circ C_{x,y})\right)\\
=\delta_{x,y}\left\langle\tr \left(P\prod_{\langle x',y'\rangle\in C_{x,y}}
U(x',y')\right)\frac{1}{z-\bar\psi(y)\psi(y)}\tr\frac{1}{z-\bar\psi(y)\psi(y)}
\right.\\\left.~~~~~~~~~~~~~~~~~~~~\times P\prod_{\langle x',y'\rangle\in
C_{x,y}}U^\dagger(x',y')\right\rangle\end{array}}
\label{diraceq}\eeq
where the path $C_{x\pm\ell,x}\circ C_{x,y}$ is obtained by attaching the link
$\langle x,x\pm\ell\rangle$ to the loop $C_{x,y}$ at the endpoint $x$, and an
implicit matrix multiplication over spin indices is assumed in (\ref{diraceq}).
The left-hand side of the loop equation (\ref{diraceq}) results from the
variation of the action in (\ref{partd}) while the right-hand side represents
the commutator term resulting from the variation of the integrand in
(\ref{dvanavg}).

The right-hand side of (\ref{diraceq}) is non-zero only for closed contours
and at large-$N$, when factorization holds, it becomes
\beq
\delta_{x,y}W_A(C_{x,x})\omega(z)^2+{\cal O}(1/N^2)=\delta_{x,y}
\delta_{0,A_{\rm min}(C_{x,x})}\omega(z)^2+{\cal O}(1/N^2)
\label{strfact}\eeq
in the strong coupling regime where all closed contours $C_{x,x}$ are
contractable. Here
\beq
\omega(z)=\left\langle\tr\frac{1}{z-\bar\psi(x)\psi(x)}\right\rangle
\label{omegad}\eeq
and thus at $N=\infty$ the right-hand side of the Dirac open-loop equation
involves only the generators of the $\ps2$-moments. This is the primary
simplification at strong coupling, since otherwise the Dirac equation
(\ref{diraceq}) involves non-trivial adjoint Wilson loops which are not
readily determined.

We should point out here that there is one more loop equation that is obeyed
by the open-loop averages. These are the Schwinger-Dyson equations (and in
particular the lattice Yang-Mills equation) which express solely the invariance
of the Haar measure over $U(x,x+\ell)$ under arbitrary changes of variables.
In particular, the Haar measure $[dU(x,x+\ell)]$ is invariant under
the infinitesimal shift
\beq
U(x,x+\ell)\to(1+i\epsilon(x,x+\ell))U(x,x+\ell)
\label{ushift}\eeq
of the gauge field $U(x,x+\ell)$ at the link $\langle x,x+\ell\rangle$, where
$\epsilon(x,x+\ell)$ is an infinitesimal Hermitian matrix. For example,
consider the open-loop correlator
\beq\new{\begin{array}{c}
\Gee(z,w;C_{x,y})=\left\langle\tr\frac{1}{z-\bar\psi(x)\psi(x)}\left(P\prod_{
\langle x',y'\rangle\in C_{x,y}}U(x',y')\right)\frac{1}{w-\bar\psi(y)\psi(y)}
\right.\\\left.\times P\prod_{\langle x',y'\rangle\in
C_{x,y}}U^\dagger(x',y')\right\rangle\end{array}}
\label{gextcorr}\eeq
The loop equation for (\ref{gextcorr}) then follows from the identity
\bd
0\equiv\left\langle\tr\frac{1}{z-\bar\psi(x)\psi(x)}\left(P\prod_{\langle x',y'
\rangle\in C_{x,z}}U(x',y')\right)T^a~P\prod_{\langle x',y'\rangle\in
C_{x,z}}U^\dagger(x',y')\right.
\ed
\beq
\times\left.\tr T^b\left(P\prod_{\langle x',y'\rangle\in C_{z,y}}U(x',y')
\right)\frac{1}{w-\bar\psi(y)\psi(y)}P\prod_{\langle x',y'\rangle\in C_{z,y}}U^
\dagger(x',y')\right\rangle
\label{gaugeinv0}\eeq
which vanishes due to the gauge invariance of (\ref{partd}). Performing the
shift (\ref{ushift}) in (\ref{gaugeinv0}) of $U(z,z+\ell)$ at some link
$\langle z,z+\ell\rangle\in C_{x,y}$ and using the invariance of the Haar
measure leads to a set of equations which is not closed because it introduces
higher-order correlators of the gauge fields (i.e. those of the form $\langle
UUU^\dagger U^\dagger\rangle$). Thus the Yang-Mills open-loop equation
expresses the one-link correlators of products such as $UUU^\dagger U^\dagger$
in terms of the pair correlators of the gauge fields $\frac{1}{N}C_{ij}=
\langle U_{ij}U^\dagger_{ji}\rangle$ \cite{mak5}. As mentioned before, however,
there is no direct way in the adjoint fermion matrix models to generate the
pair
correlators for the gauge fields from the equations of motion of the model.

\subsection{Loop Equations for One-link Correlators}

The extended loop averages such as (\ref{opencorrgen}) can be computed at
$N=\infty$ by solving the loop equations (\ref{diraceq}) directly (see
below) or by first computing the corresponding one-link correlators
and then convoluting them together using appropriate variations of the
pair correlators for the gauge fields. In the Kazakov-Migdal model this
convolution can be carried out straightforwardly \cite{dms} for the same
reasons discussed in Subsection 4.1 (see eq. (\ref{geeherm})). For example, the
adjoint Wilson loop in the Kazakov-Migdal model at $N=\infty$ can be computed
as
\beq
W_A^H(\Gamma)=\int\prod_{j=1}^{l(\Gamma)}d\alpha_j~\rho(\alpha_j)C(\alpha_1,
\alpha_2)C(\alpha_2,\alpha_3)\cdots
C(\alpha_{l(\Gamma)-1},\alpha_{l(\Gamma)})C(\alpha_{l(\Gamma)},\alpha_1)
\eeq
and similarly the analogs of the extended loop averages such as
(\ref{gextcorr}) can be explicitly calculated by substituting the gauge field
correlation function at each link $\ell$ by $C(\alpha_\ell,\alpha_{\ell+1})$
and integrating over the $\alpha_\ell$'s. Although it is not clear
how this latter method will carry through in the adjoint fermion model, we can
at least formally solve the one-link problem from a complete set of
Schwinger-Dyson equations in the hope of being able to study the critical
behaviour of the model in terms of the one-link averages.

The complete set of
observables of the matrix model (\ref{partd}) at $N=\infty$ and in the local
confinement phase are generated by the even-even one-link correlator
\beq
\Gee(z,w)=\left\langle\tr\frac{1}{z-\bar\psi(x)\psi(x)}U(x,x+\ell)\frac{1}{w-
\bar\psi(x+\ell)\psi(x+\ell)}U^\dagger(x,x+\ell)\right\rangle~~~~~,
\label{evend}\eeq
the odd-odd one-link correlator
\bd
\Hoo_\ell^{\mu\nu}(z,w)=\left\langle\tr\psi^\mu(x)\frac{1}
{z-\bar\psi(x)\psi(x)}
U(x,x+\ell)\frac{1}{w-\bar\psi(x+\ell)\psi(x+\ell)}\right.
\ed
\beq
\left.\times\bar\psi^\nu(x+\ell)U^\dagger(x,x+\ell)\right\rangle
\label{oddd}\eeq
and the generators (\ref{omegad}) of the $\bar\psi(x)\psi(x)$-moments at
the same site $x\in\lat$. From the symmetry $U\to U^\dagger$ of the Haar
measure and the fact that the potential $V$ in (\ref{lataction}) is the same
at each lattice site it follows that the functions
(\ref{omegad}), (\ref{evend}) and (\ref{oddd}) obey the usual symmetries
\beq
\Gee(z,w)=\Gee(w,z)~~~~~,~~~~~\Hoo_\ell^{\mu\nu}(z,w)=\Hoo_{-\ell}^{\mu\nu}
(w,z)
\label{gensymd}\eeq
for this symmetric case. The generating function $\Gee(z,w)$ therefore has
the same asymptotic expansions as in (\ref{gasymp}) but now with
$\Gee_n(z)=\tilde\Gee_n(z)$. Furthermore, these generators are all
independent of the lattice position $x$ because (\ref{partd}) is invariant
under lattice translations and rotations of the lattice by $\frac{\pi}{2}$.

The loop equations can be derived just as in Section 4 by considering
the identical identities there with $\psi=\psi^\mu(x)$,
$\chi=\psi^\mu(x+\ell)$,
etc. For instance, consider the identity
\beq\new{\begin{array}{ll}
0&\equiv\frac{N^{-2}}{Z_D}\sum_{i,j}\int\prod_{\langle
x,y\rangle\in\lat}[dU(x,y)]~\int\prod_{x\in\lat}d\psi(x)~
d\bar\psi(x)~\frac{\partial}{\partial\bar\psi^\mu(x)_{ij}}
\e^{S_F[\psi,\bar\psi;U]}\\&~~~~~\times
\left(\frac{1}{z-\bar\psi(x)\psi(x)}
U(x,y)\frac{1}{w-\bar\psi(y)\psi(y)}\bar\psi^\nu(y)U(y,x)\right)_{ij}\\
&=\left\langle\tr\frac{1}{z-\bar\psi(x)\psi(x)}\tr\psi^\mu(x)
\frac{1}{z-\bar\psi
(x)}U(x,y)\frac{1}{w-\bar\psi(y)\psi(y)}\bar\psi^\nu(y)U(y,x)\right\rangle\\&
{}~~~~~-\left\langle\tr\frac{V'[\bar\psi(x)\psi(x)]}
{z-\bar\psi(x)\psi(x)}\psi^\mu
(x)U(x,y)\frac{1}{w-\bar\psi(y)\psi(y)}\bar\psi^\nu(y)U(y,x)\right\rangle
\\&~~~~~-\sum_{u\in\lat}\sum_{\lambda=1}^s\left\{\left(\proj^-_{\langle
x,u\rangle}\right)^{\mu\lambda}\left\langle\tr U(x,u)\psi^\lambda(u)U(u,x)
\frac{1}{z-\bar\psi(x)\psi(x)}U(x,y)\right.\right.
\\&~~~~~\left.\times\frac{1}{w-\bar\psi(y)\psi(y)}\bar\psi^\nu(y)U(y,x)\right
\rangle+\left(\proj^+_{\langle u,x\rangle}\right)^{\mu\lambda}\left\langle\tr
U(u,x)\psi^\lambda(u)U(x,u)\right.\\&~~~~~\times\left.\left.
\frac{1}{z-\bar\psi(x)\psi(x)}U(x,y)\frac{1}{w-\bar\psi(y)\psi(y)}\bar\psi^
\nu(y)U(y,x)\right\rangle\right\}\end{array}}
\label{loopdderiv}\eeq
where $y=x+\ell$. The first correlation function in (\ref{loopdderiv}) can be
factored at large-$N$ into the 2 terms $\omega(z)\Hoo_\ell^{\mu\nu}(z,w)$. The
third term contains link operators which connect the site $x$ to all
neighbouring points. One of these connects $x$ to $y=x+\ell$ and thus
contributes a term proportional to $-\omega(z)+w\Gee(z,w)$. In this same term,
there are also $2D-1$ links which connect $x$ to other sites. For these we can
use the fact that the quantity $U(u,x)\psi^\lambda(u)U^\dagger(u,x)$ inside the
expectation value bracket can be replaced by the analytic function
$F(\bar\psi(x)\psi(x))$ at $N=\infty$ as discussed in the last Subsection (c.f.
eq. (\ref{gencorrf})).

With this input, we arrive at the analog of the loop
equation (\ref{loop23}) for the $D$-dimensional matrix model (\ref{partd}),
\beq
\omega(z)\Hoo_\ell^{\mu\nu}(z,w)-\frac{1}{s}(\proj_\ell^\mp)^{\mu\nu}\left(
w\Gee(z,w)-\omega(z)\right)-\oint_{\cal C}\frac{d\lambda}{2\pi i}~\frac
{{\cal V}'(\lambda)\lambda}{z-\lambda}\Hoo_\ell^{\mu\nu}(\lambda,w)=0
\label{loopd2}\eeq
Multiplying (\ref{loopd2}) by $(\proj_\ell^\pm)^{\mu\nu}$ and summing over
the spin indices, we can rewrite it as
\beq
\omega(z)\hat\Hoo(z,w)-\sigma\left(w\Gee(z,w)-\omega(z)\right)-\oint_{\cal C}
\frac{d\lambda}{2\pi i}~\frac{{\cal V}'(\lambda)\lambda}{z-\lambda}\hat
\Hoo(\lambda,w)=0
\label{loopd3}\eeq
where the effective potential ${\cal V}(z)$ is defined through
\beq
{\cal V}'(z)\equiv V'(z)-2\sigma(2D-1)F(z)
\label{poteff}\eeq
and we have introduced
\beq
\hat\Hoo(z,w)\equiv~{\rm TR}~\proj_\ell^\pm\Hoo_\ell(z,w)
\label{oddhat}\eeq
which has the asymptotic expansion
\beq
\hat\Hoo(z,w)=\sum_{n=0}^\infty\frac{\hat\Hoo_n(z)}{w^{n+1}}
\label{oddhatasymp}\eeq
and from the symmetries (\ref{gensymd}) and the definition (\ref{oddhat})
it follows that it has the symmetry property
\beq
\hat\Hoo(z,w)=\hat\Hoo(w,z)
\label{oddhatsym}\eeq

In a similar fashion, we can write the analog of the loop equation
(\ref{loop21}) for the $D$-dimensional matrix model (\ref{partd}) as
\beq
\left(s+1-z\omega(z)\right)\Gee(z,w)+\hat\Hoo(z,w)+\oint_{\cal C}
\frac{d\lambda}{2\pi i}~\frac{{\cal V}'(\lambda)\lambda}{z-\lambda}
\Gee(\lambda,w)=0
\label{loopd1}\eeq
The analogs here of the other 2 loop equations (\ref{loop22})
and (\ref{loop24}) are identical to the above loop equations because of the
symmetries (\ref{gensymd}) and (\ref{oddhatsym}). For $D=\frac{1}{2}$
($s=1$) and chiral fermions ($\sigma=-1$), the loop equations (\ref{loopd1})
and (\ref{loopd3}) coincide, respectively, with the loop equations
(\ref{loop21}) and (\ref{loop23}) of the fermionic two-matrix model
(\ref{part2}) with $V=\tilde V$. The analogs of the asymptotic equations
(\ref{asloop1})--(\ref{asloop3})
of the two-matrix model in the present case are respectively then
\beq
\hat\Hoo_n(z)=\left(z\omega(z)-(s+1)\right)\Gee_n(z)-\oint_{\cal C}\frac{d
\lambda}{2\pi i}~\frac{{\cal V}'(\lambda)\lambda}{z-\lambda}\Gee_n(\lambda)
\label{asloopd1}\eeq
\beq
\sigma\Gee_{n+1}(z)=\omega(z)\hat\Hoo_n(z)-\oint_{\cal C}\frac{d\lambda}
{2\pi i}~\frac{{\cal V}'(\lambda)\lambda}{z-\lambda}\hat\Hoo_n(\lambda)
\label{asloopd2}\eeq
and
\beq
\sigma z\omega(z)=\sigma-\oint_{\cal C}\frac{d\lambda}{2\pi i}~{\cal V}'
(\lambda)\lambda\hat\Hoo(z,\lambda)
\label{asloopd3}\eeq

\subsection{The Ansatz}

The equations (\ref{asloopd1})--(\ref{asloopd3}) formally determine the
one-loop correlator $\omega(z)$ in terms of the {\it effective} potential
(\ref{poteff}), rather than the given potential in (\ref{lataction}). In
general this effective potential is not known and must be assumed to take a
particular form at the onset. To determine ${\cal V}(z)$ formally from the
above equations, and hence find $F(z)$, we note that from (\ref{gencorrf}),
(\ref{oddd}), (\ref{oddhat}) and (\ref{oddhatasymp}) we have the equation
\beq
\hat\Hoo_0(z)=-\sigma\oint_{\cal C}\frac{d\lambda}{2\pi i}~\frac{F(\lambda)
\lambda}{z-\lambda}\omega(\lambda)
\label{fint}\eeq
(\ref{fint}) can be rewritten using (\ref{asloopd1}) as
\beq
(s+1-z\omega(z))\omega(z)+\oint_{\cal C}\frac{d\lambda}{2\pi i}~\frac
{{\cal V}'(\lambda)\lambda}{z-\lambda}\omega(\lambda)-\sigma\oint_{\cal C}
\frac{d\lambda}{2\pi i}~\frac{F(\lambda)\lambda}{z-\lambda}\omega(\lambda)=0
\label{omeqd}\eeq
which resembles the large-$N$ loop equation (\ref{loopcont}) of the
fermionic one-matrix model. From this we can conclude that the continuous
part of $\omega(z)$ across its singularities leads to the saddle point equation
\beq
2\left(\omega(\alpha+\epsilon_\perp)+\omega(\alpha-\epsilon_\perp)\right)
=(s+1)/\alpha-{\cal V}'(\alpha)+\sigma F(\alpha)
\label{contomegad}\eeq
where $\alpha$ lies on a branch cut of $\omega(z)$. Notice that for
Kogut-Susskind fermions where $s=1$ and $\sigma=-1$ \cite{kmak2,ks}, the
above equations coincide with those of the fermionic two-matrix model
(\ref{part2}) with $V\equiv{\cal V}$. Notice also that the above equations are
drastically simplified for Wilson
fermions which have $\sigma=0$. As mentioned before, this simplifying
feature for Wilson fermions is a consequence of the fact that if
(\ref{bactrackpar}) vanishes then backtracking paths never contribute to the
representations of the correlators as sums over lattice paths.

Although the equations above formally determine the effective potential
${\cal V}(z)$, they are still rather complicated because in general
${\cal V}(z)$ can be non-polynomial and have singularities in the complex plane
away from the branch cut and pole singularities of $\omega(z)$. The
problem simplifies, however, if we assume that ${\cal V}(z)$ is analytic
on the entire complex plane except possibly at infinity,
\beq
{\cal V}(z)=\sum_{n=1}^\infty\frac{g_n}{n}z^n
\label{ansatz}\eeq
so that the asymptotic equation (\ref{asloopd3}) becomes
\beq
\sigma z\omega(z)=\sigma-\sum_{n=1}^\infty g_n\hat\Hoo_{n-1}(z)
\label{asloopansatz}\eeq
and similarly the other contour integrals simplify to forms analogous to
those in Subsection 4.1. Then the equations (\ref{poteff}), (\ref{contomegad})
and (\ref{asloopansatz}) unambiguously fix the functions ${\cal V}(z)$,
$F(z)$ and $\omega(z)$ for a given potential $V(z)$. (\ref{ansatz}) is
considered as an {\it ansatz} for ${\cal V}(z)$ to be used in solving the
loop equations. Once ${\cal V}(z)$ and $F(z)$ are then explicitly determined,
the original potential $V(z)$ of the adjoint fermion matrix model
(\ref{partd}) can be determined from (\ref{poteff}). Because the terms
involving $F(z)$ disappear in certain instances (e.g. $D=\frac{1}{2}$) it is
natural to regard (\ref{ansatz}) as the potential of the proper
two-matrix model to which the $D$-dimensional matrix model has been
effectively simplified. As for the two-matrix model, however, the
equation determining $\omega(z)$ will be a $(2K)$-th order polynomial
equation when (\ref{ansatz}) is a polynomial of degree $K$, and moreover
these equations will all involve the set of unknown coefficients of
the ansatz (\ref{ansatz}).

Similarly, the loop equations (\ref{diraceq}) for the extended loop correlators
are solved by substituting in an appropriate ansatz for $\Hoo^{\mu\nu}
(z;C_{x,y})$. The proper ansatz that one should take can be deduced from
the large-mass expansion (\ref{amploops}). From the definitions
(\ref{opencorrgen}) and (\ref{gencorrop}) and the relation (\ref{gencorrf})
it follows that the Dirac equation (\ref{diraceq}) can be solved in the
strong coupling regime using the ansatz
\beq
\Hoo^{\mu\nu}(z;C_{x,y})=\left(\frac{F(z)}{z}\right)^{L(C_{x,y})}
\omega(z)\left(P\prod_{\ell\in C_{x,y}}\proj_\ell^\pm\right)
^{\mu\nu}
\label{extansatz}\eeq
where $L(C_{x,y})$ is the algebraic length of the path $C_{x,y}$ (i.e. the
lattice length after eliminating all backtrackings in $C_{x,y}$). The spin
factors in (\ref{extansatz}) are needed to cancel the projection operators
appearing in (\ref{diraceq}). One must now substitute an ansatz for $F(z)$,
and when this is substituted into
the loop equation (\ref{diraceq}) we will obtain an equation both
for $L(C_{x,y})\neq0$ when $x\neq y$ in (\ref{diraceq}) and also for
$L(C_{x,x})=0$ when $x=y$ and all closed contours are contractable. The
exact solutions of the loop equations for a given potential then represent
the solution to the combinatorial problem of summing over 1-dimensional
tree graphs embedded in a $D$-dimensional space.

\subsection{The Gaussian Model}

For the Gaussian potential $V(z)=mz$, we substitute into the loop equations
the ansatz
\beq
{\cal V}(z)=\Pi^{-1}z
\eeq
The asymptotic equation (\ref{asloopansatz}) then implies
\beq
\sigma z\omega(z)=\sigma-\Pi^{-1}\hat\Hoo_0(z)
\eeq
where from (\ref{asloopd1})
\beq
\hat\Hoo_0(z)=(z\omega(z)-(s+1))\omega(z)+\Pi^{-1}z\omega(z)-\Pi^{-1}
\eeq
These 2 equations combine to give as usual a quadratic equation for the
one-loop correlator $\omega(z)$, and the relations (\ref{poteff}) and
(\ref{contomegad}) imply that
\beq
\Pi=\frac{2}{\sqrt{m^2-4\sigma(2D-1)}+m}
\label{Lambda}\eeq
and
\beq
F(z)=\Pi z=\frac{2z}{\sqrt{m^2-4\sigma(2D-1)}+m}
\label{Fgauss}\eeq

The one-loop correlator is thus
\beq
\omega(z)=\frac{1}{2}\left(\frac{s+1}{z}-\mu+\frac{1}{z}\sqrt{\mu^2z^2
-2(s-1)\mu z+(s+1)^2}\right)
\label{omgaussd}\eeq
where
\beq
\mu=\Pi^{-1}+\sigma\Pi=\frac{(D-1)m+D\sqrt{m^2-4\sigma(2D-1)}}{2D-1}
\label{effmass}\eeq
Notice that only for Kogut-Susskind fermions, which are effectively spinless
Grassmann variables \cite{ks}, does the usual fermion chiral symmetry become an
invariance of the model. Finally, combining the loop equations
(\ref{loopd1}) and (\ref{loopd3}) determines the even-even and odd-odd
one-link correlators as
\beq
\Gee(z,w)=\frac{\sigma w(\omega(z)+\omega(w))-\sigma-\Pi^{-1}\omega(w)
(\omega(z)+\Pi^{-1}z)}{\omega(z)(2(s+1)-(\mu+\Pi^{-1})z-
\Pi^{-1}z^2)+\Pi^{-1}(s+1)z-\Pi^{-2}z^2-\sigma w+\mu}
\eeq
\beq
\hat\Hoo(z,w)=(\Pi^{-1}z+z\omega(z)-(s+1))\Gee(z,w)-\Pi^{-1}
\omega(w)
\eeq
It is easy to verify that these correlators are symmetric and non-singular
for any $z$ and $w$.

The constant $\Pi$ above and the one-loop correlator (\ref{omgaussd})
could also have been found by substituting the ansatz
\beq
\Hoo^{\mu\nu}(z;C_{x,y})=\Pi^{L(C_{x,y})}\omega(z)\left(P\prod_{\ell\in
C_{x,y}}\proj_\ell^\pm\right)^{\mu\nu}
\label{extgaussansatz}\eeq
into the Dirac equation (\ref{diraceq}) which in the case at hand reads
\bd
m\Hoo(z;C_{x,y})-\sum_{\ell=1}^D\left(\proj_\ell^+\Hoo(z;C_{x+\ell,x}\circ
C_{x,y})+\proj_\ell^-\Hoo(z;C_{x-\ell,x}\circ C_{x,y})\right)
\ed
\beq
=\delta_{x,y}\delta_{0,A_{\rm min}(C_{x,x})}\omega(z)^2
\label{diracgauss}\eeq
Setting $x\neq y$ in (\ref{diracgauss}) then yields (\ref{Lambda}) and the
$x=y$ equation for contractable loops yields the quadratic equation
above for $\omega(z)$. Notice that for Kogut-Susskind fermions,
(\ref{omgaussd}) coincides with the solution of the Gaussian model of the
Subsection 4.2. For Wilson fermions the above solutions simplify since
when the backtracking parameter $\sigma$ vanishes we have
\beq
\Pi^L=1/m^L
\eeq
and the solution therefore coincides with the leading-order of the
large-mass expansion. This is not surprising, since as discussed before
for Wilson fermions backtracking paths never contribute and thus the leading
term of the large-mass expansion yields the {\it exact} result for the
adjoint fermion matrix model (\ref{partd}) with Wilson fermions.

The $\frac{1}{z^2}$ coefficient in the asymptotic expansion of the one-loop
correlator (\ref{omgaussd}) is the expectation value
\beq
\xi=\langle\tr\bar\psi(x)\psi(x)\rangle=\frac{s}{\mu}=\frac{(2D-1)s}{(D-1)m+
D\sqrt{m^2-4\sigma(2D-1)}}
\label{2ptcorrd}\eeq
This correlator coincides with that obtained for lattice QCD with
fundamental representation fermions when the coefficient of the Wilson action
(\ref{wilson}) vanishes \cite{kmnp}. Indeed, there is an analogy based on the
Dirac equation between adjoint and fundamental fermions in the phase with
local confinement for the Gaussian potential \cite{kmak2}. The fact that
(\ref{2ptcorrd}) is the same for adjoint
and fundamental fermions follows from the lattice loop representation of
the propagators of the theory, i.e. in the case of the fundamental
representation the exact same combinatorial problem of summing over tree
diagrams emerges. In particular, as discussed in \cite{kmnp}, the non-vanishing
of (\ref{2ptcorrd}) implies that the adjoint fermion matrix model (\ref{partd})
in this way furnishes an interesting example of a quantum field theory with
spontaneous chiral symmetry breaking and associated non-trivial fermion chiral
condensate leading to a possible composite Higgs phase of the matrix model.
This coincidence is no longer valid, however,
for the higher order moments of these 2 fermion matrix models.

Notice that the above strong coupling solutions of the Gaussian model are
non-singular for any $m^2>0$, contrary to the Kazakov-Migdal model
\cite{dms,gross,kmak1,mak1,mak2} (see Subsection 5.4 above), and so the
solutions in the fermionic case
are stable
everywhere. However, the matrix model (\ref{partd}) undergoes a first order
large-$N$ phase transition beyond which the strong coupling solutions above
are no longer applicable. The critical point of this phase transition is
quite regular from the point of view of the large-mass expansion and the
strong coupling solution is insensitive to the large-$N$ phase transition
which occurs {\it inside} the region where the large-mass expansion converges.
In the phase with normal area law the ansatz of Subsection 6.3 above is no
longer valid and the solution of the loop equations in this regime will be
quite
different. The weak coupling phase where the dynamics of extended objects are
non-trivial is very similar to the standard Wilson lattice gauge theory
\cite{wilson}. We remark that one could also study the $\frac{1}{N}$-expansion
of the $D$-dimensional model (\ref{partd}) in the same way as in the previous
Sections. For instance, the next-to-leading order in $\frac{1}{N^2}$
contribution to the Dirac equation (\ref{diraceq}) for the Gaussian model
leads to the adjoint Wilson loop \cite{mig2}
\beq
W_A(\Gamma)=\delta_{0,L(\Gamma)}+[\Pi^{L(\Gamma)}
(1-\delta_{0,L(\Gamma)})-1]/N^2
\label{wan2}\eeq
The leading term here is the usual contribution from the backtracking loop
which corresponds to an infinite string tension. The $\frac{1}{N^2}$
correction describes the perimeter law for pointlike quarks propagating
along the adjoint double loop\footnote{\baselineskip=12pt A rectangular adjoint
Wilson loop, with space and time lengths $L$ and $T$, respectively, can be
interpreted in the limit $T\to\infty$
as the energy of a pair of mesons with separation $L$ \cite{wilson}.}.
 The Schr\"odinger wave equation is then effectively
obtained by summing over all of these closed contours. Thus the higher
order terms in the $\frac{1}{N}$-expansion of the adjoint fermion matrix
model (\ref{partd}) provide information relevant to the physical matter
content of the gauge theory.

\section{Supersymmetric Matrix Models}

In this final Section we shall present yet another application of the fermionic
matrix model formalism to the representation of the polymer phase of bosonic
string theory in target space dimensions $D>1$ \cite{makmean}. We shall be
ultimately interested in the next level of complexity of the matrix models we
have considered thus far, namely the combination of fermionic and bosonic
matrix  degrees of freedom into a supersymmetric matrix model. We will
primarily discuss models which are defined in terms of matrix superfields
defined in zero-dimensional target and supersurface spaces. These models are
therefore a version of the $D=1$ Marinari-Parisi superstring models
\cite{marinari} which were the first proposed supersymmetric extensions of the
matrix models describing strings and discretized random surfaces. These models
differ from the Gilbert-Perry Hermitian supermatrix models which are related to
the Penner matrix models \cite{alvarez,gilper}. As always, in these simple
$D=0$ cases a Hermitian fermionic matrix formalism is avoided because
nilpotency always implies that $\tr\psi^2=0$. The higher-dimensional
supermatrix models will be briefly described at the end of this Section.

{}From a combinatorical point of view, the supersymmetry of the matrix model
renders the partition function trivial. However, the matrix propagators in
$D=0$
correspond to a sum over polymer trees which describe a particular
combinatorical problem that we shall describe here in detail. Thus the
supersymmetric matrix model contains a dimensional reduction, because of the
cancellation of bosonic and fermionic matrix field loops due to supersymmetry,
so that for $D=0$ it is very similar to a vector model in that it counts random
polymers rather than random surfaces. Indeed, the Marinari-Parisi
supersymmetric matrix models in $D=1$ were originally introduced to describe
the same random surface theories as the $D=0$ Hermitian matrix models. Adding
supersymmetry to the purely fermionic (or bosonic) theory makes it more rigid
and reduces the number of degrees of freedom. As we will discuss, this is
observed in an extreme way, whereby only a very small subclass of graphs
corresponding to branched polymers survives when one examines the diagrammatic
expansions of non-supersymmetric observables. In the
$D=0$ cases, these models are motivated by the notion of `folding', an
important concept in polymer physics. We can consider a statistical model of
randomly branching polymer chains which are made up of $n$ independent
constituents, and which may therefore fold onto themselves. The entropy of such
a system is obtained by counting the number of inequivalent ways of folding the
chain. The combinatorial problem of enumerating all compact foldings is
equivalent to another geometrical problem, the meander problem
\cite{di1,lando}, i.e. the problem of enumerating the configurations of a
closed road crossing an infinite river through $n$ bridges.

\subsection{Meander Numbers and Random Matrix Models for the Meander Problem}

Consider an infinite straight line (river). A meander of order $n$ is defined
as a closed self-avoiding connected loop (road) which intersects the line
through $2n$ points (bridges). It can be viewed as a compact folding
configuration of a closed chain of $2n$ constituents (in one-to-one
correspondence with the $2n$ bridges) by putting a hinge on each section of the
road between 2 bridges. The principal meander number ${\cal M}_n$ is defined as
the number of topologically inequivalent meanders of order $n$. It also
therefore describes the number of different foldings of a closed strip of $2n$
stamps or of a closed polymer chain. One can also consider a generalized
version of this problem. The multi-component meander numbers ${\cal M}_n^{(k)}$
are defined as the number of topologically inequivalent meanders of order $n$
with $k$ connected components, i.e. made up of $k$ closed connected
non-intersecting but possibly interlocking bridges which cross the river
through a total of $2n$ bridges ($k$ loops of the road). Clearly ${\cal
M}_n\equiv{\cal M}_n^{(1)}$. The results of a computer enumeration of the
meander numbers up to $n=12$ have been presented in \cite{di1,lando}.

To obtain a matrix model representation of meanders \cite{di1}, we note that
the enumeration of (planar) meanders is very close to that of 4-valent (genus
$h=0$) fat-graphs constructed from 2 self-avoiding loops (1 black $B$, 1 white
$W$) intersecting each other at simple nodes. The $B$-loop corresponds to the
river (closed at infinity) while the $W$-loop corresponds to the road. Such a
fat-graph is called a black and white graph. Since the river now becomes a
loop, this replaces the order of the bridges by a cyclic order and
identifies the regions above and below the river. Thus ${\cal M}_n$ is now
given by the meander number for the inequivalent black and white graphs with
$2n$ intersections times $4n$ (2 for the up-down symmetry, $2n$ for the cyclic
symmetry) weighted by the symmetry factor  $|G({\cal F})|$ for each fat-graph
$\cal F$. The same connection holds between ${\cal M}_n^{(k)}$ and the black
and white graphs where the white loop has $k$ connected components.

The black and white graphs can be generated by the $N\times N$ Hermitian
multi-matrix integral
\beq
{\cal Z}(m,n,c;N)=\int\prod_{b=1}^ndB_b~\prod_{a=1}^mdW_a~\e^{-N^2\tr
P[B,W]}
\label{zmncN}\eeq
where the action is
\beq
P[B,W]=\sum_{b=1}^n\frac{B_b^2}{2}+\sum_{a=1}^m\frac{W_a^2}{2}-\frac{c}{2}
\sum_{a,b}B_bW_aB_bW_a
\label{PBW}\eeq
Here $a,b$ are ``colour" indices and $c$ is the coupling constant for the
quartic interactions between the fields $B_b$ and $W_a$. The perturbative
expansion of $\log{\cal Z}$ in powers of $c$ gives a series in which the term
of order $v$ can be readily evaluated as a Gaussian multi-matrix integral. The
Feynman rules lead to the propagators
\beq
\left\langle[B_b]_{ij}[B_{b'}]_{k\ell}\right\rangle_{c=0}=\frac{\delta_{i\ell}
\delta_{jk}}{N}\delta_{bb'}~~~,~~~\left\langle[W_a]_{ij}[W_{a'}]_{k\ell}
\right\rangle_{c=0}=\frac{\delta_{i\ell}\delta_{jk}}{N}\delta_{aa'}
\label{feynmean}\eeq
and the only vertices are 4-valent ones which have alternating black and white
edges and which therefore describe simple intersections of black and white
loops. The normalized averages in (\ref{feynmean}) are Gaussian ones obtained
from the partition function (\ref{zmncN}) at $c=0$. Thus the perturbative
expansion of $\log{\cal Z}$ can be obtained as the sum over 4-valent fat-graphs
whose $v$ vertices have to be connected by means of the 2 types of edges
(\ref{feynmean}) which have to alternate around each vertex. This gives an
exact realization of the black and white graphs except that any number of loops
in each colour is allowed.

With these Feynman rules, a graph of genus $h$ with $v$ vertices and $b$
(respectively $w$) black (white) loops receives a weight $N^{2-2h}c^vn^bm^w$.
We can reduce the number of black loops $b$ to 1 by the so-called `replica
trick' \cite{di1} which is similar to the replica symmetry of the $O(N)$ vector
model which picks out the ${\cal O}(N^0)$ component corresponding to a
self-avoiding random walk. Here we let the number $n$ of black matrices $B$
approach zero and retain only the contribution of order $n^0=1$ in $n$. This
leads to the generating function
\beq
{\cal E}_H(m,c;N)\equiv\lim_{n\to0}\frac{1}{n}\log{\cal Z}(m,n,c;N)=\sum_{\cal
F}\frac{N^{2-2h}c^vm^w}{|G({\cal F})|}
\label{black1}\eeq
where the sum is over all black and white connected graphs $\cal F$ with one
$B$ loop. The genus 0 contribution to (\ref{black1}) then gives a relation to
the meander numbers,
\beq
{\cal E}^{(0)}_H(m,c)\equiv\lim_{N\to\infty}\frac{1}{N^2}{\cal
E}(m,c;N)=\sum_{p=1}^\infty\frac{c^{2p}}{4p}\sum_{k=1}^p{\cal M}_p^{(k)}m^k
\label{genus0mean}\eeq
where we have used the aforementioned relation between the number of black and
white graphs and the multi-component meander numbers. Alternatively, the
generating function (\ref{black1}) can be obtained directly from the Hermitian
matrix model \cite{makmean,makpe}
\beq\new{\begin{array}{ll}
{\cal
E}_H(m,c;N)=\frac{2}{N^2}\int\prod_{a=1}^mdW_a&\e^{-\frac{N^2}{2}\sum_{a=1}^m
\tr W_a^2}\\&\times\log\left(\int
d\phi~\e^{-\frac{N^2}{2}\tr\phi^2+\frac{N^2c}{2}\sum_{a=1}^m\tr\phi W_a\phi
W_a}\right)\end{array}}
\label{emcN}\eeq
where the logarithm leaves only one closed loop of the Hermitian matrix field
$\phi$. The $\phi$-integral in (\ref{emcN}) is a version of the so-called
curvature matrix models considered in \cite{das} which provide explicit
examples of the polymer structures in $D>1$ string theory that we discussed at
the end of Subsection 1.1.3.

We can carry out the exact Gaussian integration over the $\phi$ matrices in
(\ref{emcN}) to get
\beq\new{\begin{array}{ll}
{\cal E}_H(m,c;N)&=-\frac{N}{2}\int\prod_{a=1}^mdW_a~\tr\log\left(I\otimes
I-c\sum_{a=1}^mW_a^T\otimes W_a\right)\e^{-\frac{N^2}{2}\sum_a\tr
W_a^2}\\&=N\sum_{p=1}^\infty\frac{c^p}{2p}\left\langle\!\!
\left\langle\sum_{a=1}^m\tr\left(W_a^T\otimes
W_a\right)^p\right\rangle\!\!\right\rangle_H\\&=N^2\sum_{p=1}^\infty\frac
{c^p}{2p}\sum_{1\leq a_1,\dots,a_p\leq m}\left\langle\!\!\left\langle\left(\tr
W_{a_1}\cdots W_{a_p}\right)^2\right\rangle\!\!\right\rangle_H\end{array}}
\label{Eexpl}\eeq
where we have used the Hermiticity of the $W$ matrices, $W^T=W^*$, and the
normalized average $\langle\!\langle\cdot\rangle\!\rangle_H$ over the $W$
matrices in (\ref{Eexpl}) is with respect to
the (free field) Gaussian $W$ measure in (\ref{emcN}). At $N=\infty$,
factorization implies
$\langle\!\langle(\tr\prod_iW_{a_i})^2\rangle\!\rangle_H=\langle\!\langle
\tr\prod_iW_{a_i}\rangle\!\rangle_H^2$, and reflection
symmetry $W\to-W$ of the Gaussian measure in (\ref{emcN}) implies that the only
nonvanishing contributions in (\ref{Eexpl}) are for $p=2n$ even. Comparing this
with (\ref{genus0mean}) leads to a representation of the meander numbers of
order $n$ with $k$ connected components in terms of Gaussian averages of the
$W$-fields,
\beq
\sum_{k=1}^n{\cal M}_n^{(k)}m^k=\sum_{1\leq a_1,\dots,a_{2n}\leq
m}\left(\lim_{N\to\infty}\left\langle\!\!\left\langle
\tr\prod_{i=1}^{2n}W_{a_i}\right\rangle\!\!\right\rangle_H\right)^2
\label{meangaussW}\eeq
The left-hand side of (\ref{meangaussW}) is a polynomial of degree $n$ in $m$
with vanishing constant coefficient, and it is therefore completely determined
by the first $n$ values $m=1,\dots,n$. Thus the right-hand side of
(\ref{meangaussW}) evaluated for $m=1,\dots,n$ determines completely the
coefficients ${\cal M}_n^{(k)}$. Notice that the Gaussian moments there are
given explicitly by making Wick pairing contractions to get
\beq
\langle\!\langle\tr W_{a_1}W_{a_2}\cdots
W_{a_{2n-1}}W_{a_{2n}}\rangle\!\rangle_H=
\delta_{a_1a_2}\delta_{a_3a_4}\cdots\delta_{a_{2n-1}a_{2n}}+{\cal
P}[a_1,a_2,\dots,a_{2n}]
\label{hermgaussavgsexpl}\eeq
where $\cal P$ contains the sum of delta-functions over all permutations of the
indices $a_i$. Thus a Hermitian random matrix model can be used to generate a
convenient and practical way of calculating meander numbers.

The ordered but cyclic-symmetric sequence of indices
$(a_1,a_2,\dots,a_{2n-1},a_{2n})$ in (\ref{meangaussW}) is called a `word'
constructed of $m$ letters. The average on the right-hand side of
(\ref{meangaussW}) is called the `meaning' of the word. The meander problem is
therefore equivalent to the problem of summing over all words with a Gaussian
meaning. The principal meander numbers ${\cal M}_n$ can be obtained from
(\ref{meangaussW}) using an analog of the replica trick which suppresses higher
loops of the field $W$, i.e. they are determined as the linear terms of the
expansion of (\ref{meangaussW}) in $m$,
\beq
{\cal M}_n=\lim_{m\to0}\frac{1}{m}\sum_{1\leq a_1,\dots,a_{2n}\leq
m}\left(\lim_{N\to\infty}\left\langle\!\!\left
\langle\tr\prod_{i=1}^{2n}W_{a_i}\right\rangle\!\!\right\rangle_H\right)^2
\label{princmeangaussW}\eeq
For $m=1$, the Gaussian moments are (see (\ref{wignerherm}))
\beq
\lim_{N\to\infty}\left\langle\!\!\left\langle\tr
W^{2n}\right\rangle\!\!\right\rangle_H=\frac{(2n)!}{(n+1)!n!}\equiv C_n
\label{catalan}\eeq
which is known as the Catalan number of order $n$. The quantity $C_n$ is the
number of paranthesings of words of $n+1$ letters with $n$ opening and $n$
closing parantheses. From (\ref{meangaussW}) it then follows that
\beq
\sum_{k=1}^n{\cal M}_n^{(k)}=C_n^2
\label{sumrule1}\eeq
which is just the first Di Francesco-Golinelli-Guitter meander sum rule
\cite{di1}.

It is also possible to define higher-genus meander numbers ${\cal
M}_p^{(k)}(h)$ which correspond to the higher-genus fat-graphs of the matrix
model. Here the indexation is now by the number of intersections (or bridges),
i.e. ${\cal M}_{2n}^{(k)}(0)={\cal M}_n^{(k)}$. The generating function for
these meanders is then given as the genus expansion
\beq
\sum_{h=0}^\infty\sum_{k=1}^\infty{\cal M}_p^{(k)}(h)m^kN^{-2h}=\sum_{1\leq
a_1,\dots,a_p\leq
m}\left\langle\!\!\left\langle
\left(\tr\prod_{i=1}^pW_{a_i}\right)^2\right\rangle\!\!\right\rangle_H
\label{meanh}\eeq
and the sum rule (\ref{sumrule1}) becomes \cite{di1}
\beq
\sum_{h=0}^\infty\sum_{k=1}^\infty{\cal M}_p^{(k)}(h)m^k=m(m+2)(m+4)\cdots
(m+2p-2)
\label{sumruleh}\eeq
For example, a genus 1 meander is made of a collection of $p$ loops
intersecting the river only once (since the right-hand side of (\ref{sumruleh})
is a polynomial of degree $p$ with leading coefficient 1).

The meander numbers can alternatively be represented as a Gaussian average over
complex matrices \cite{makmean,makpe}. The generating function is
\beq
{\cal E}_C(m,c;N)=\frac{1}{N^2}\left\langle\!\!\left\langle\log\left(\int
d\phi_1~d\phi_2~\e^{-\Sigma_C[\phi,W]}\right)\right\rangle\!\!\right\rangle_C
\label{ECmcN}\eeq
where
\beq
\left\langle\!\left\langle
Q(W)\right\rangle\!\right\rangle_C\equiv\frac{1}{\cal
N}\int\prod_{a=1}^m\prod_{i,j}
d[W_a]_{ij}~d[W_a^\dagger]_{ij}~Q(W)\e^{-N^2\sum_a\tr W_a^\dagger W_a}
\label{Cgauss}\eeq
with $\cal N$ a normalization constant, and the action is
\beq
\Sigma_C[\phi,W]=\frac{N^2}{2}\tr\phi_1^2+\frac{N^2}{2}\tr\phi_2^2-cN^2
\sum_{a=1}^m\tr\phi_1W_a^\dagger\phi_2W_a
\label{complexactionW}\eeq
Here $\phi_1$ and $\phi_2$ are $N\times N$ Hermitian matrices and $W_a$ are
general $N\times N$ complex-valued matrices. Notice that in the generating
function (\ref{emcN}) the Hermitian matrix $\phi$ can be represented in
diagonal form by absorbing the unitary matrices $U$ of the diagonalization
transformation into the Hermitian matrices by the adjoint action $W_a\to
U^\dagger W_aU$ of $U(N)$ which leaves the integration measure over Hermitian
matrices $W_a$ invariant. The same feature is true of the multi-matrix integral
(\ref{ECmcN}) -- diagonalizing the Hermitian matrices $\phi_1$ and $\phi_2$ by
unitary transformations generated by $U_1,U_2\in U(N)$, respectively, the
unitary matrices $U_1$ and $U_2$ can be absorbed by the $U(N)\otimes U(N)$
adjoint transformation $W_a\to U_2^\dagger W_aU_1$, $W_a^\dagger\to U_1^\dagger
W_a^\dagger U_2$ which leaves the integration measure in (\ref{ECmcN})
unchanged because it is over general complex matrices $W_a$.

Consider in addition the generating function
\beq
{\cal M}_C(m,c;N)=c\left\langle\!\!\left\langle\frac{\int
d\phi_1~d\phi_2~\e^{-\Sigma_C[\phi,W]}\tr\phi_1W_{a_0}^
\dagger\phi_2W_{a_0}}{\int
d\phi_1~d\phi_2~\e^{-\Sigma_C[\phi,W]}}\right\rangle\!\!\right\rangle_C
\label{Mc}\eeq
which is averaged in terms of a single component $W_{a_0}$. Differentiating
(\ref{ECmcN}) with respect to $c$ and noting that all $W_a$ in the resulting
expression are weighted the same way, we get a relation
\beq
c\frac{\partial}{\partial c}{\cal E}_C(m,c;N)=m{\cal M}_C(m,c;N)
\label{genfnrel}\eeq
between these 2 generating functions.

To see how the complex matrix model (\ref{ECmcN}) generates meander numbers, we
explicitly calculate the Gaussian integrals over $\phi_1$ and $\phi_2$ in
(\ref{ECmcN}) as before using
\beq\new{\begin{array}{ll}
\int d\phi_1~d\phi_2~\e^{-\Sigma_C[\phi,W]}&=\int
d\phi_2~\e^{-\frac{N^2}{2}\tr\phi_2^2+\frac{c^2N^2}{2}\sum_{a,b=1}^m\tr(\phi_2
W_aW_b^\dagger\phi_2W_bW_a^\dagger)}\\&={\det}^{-1/2}\left[I\otimes
I-c^2\sum_{a,b=1}^mW_aW_b^\dagger\otimes(W_bW_a^\dagger)^T\right]\end{array}}
\label{complexgaussint}\eeq
Using (\ref{genfnrel}) we get
\beq
\lim_{N\to\infty}{\cal M}_C(m,c;N)=\sum_{n=1}^\infty c^{2n}\sum_{k=1}^n{\cal
M}_n^{(k)}m^{k-1}
\label{MCmean}\eeq
with
\beq
\sum_{k=1}^n{\cal M}_n^{(k)}m^{k-1}=\sum_{1\leq a_2,\dots,a_{2n}\leq
m}\left(\lim_{N\to\infty}\left\langle\!\!\left\langle\tr
W_{a_0}W_{a_2}^\dagger\cdots
W_{a_{2n-1}}W_{a_{2n}}^\dagger\right\rangle\!\!\right\rangle_C\right)^2
\label{meanCgaussW}\eeq
The planar limit of the Gaussian average in (\ref{meanCgaussW}) coincides with
that of (\ref{hermgaussavgsexpl}) where the former is obtained from Wick
contractions among $W$ and $W^\dagger$. In particular, the analog of the $m=1$
formula (\ref{catalan}) for complex matrices is (see Subsection 1.2.2)
\beq
\lim_{N\to\infty}\left\langle\!\!\left\langle\tr(W^\dagger
W)^n\right\rangle\!\!\right\rangle_C=C_n
\label{m=1Cgaussavg}\eeq
which leads again to the sum rule (\ref{sumrule1}).

\subsection{Arch Configurations and Fermionic Matrix Models}

A general meander of order $n$ with an arbitrary number of connected components
is uniquely specified by its upper half (above the river) and its lower half
(below the river). Both halves form systems of $n$ non-intersecting arches
connecting $2n$ bridges by pairs. Each of these systems are called arch
configurations. Thus any meander is a superposition of 2 arch configurations.
Any 2 arches are either disjoint or included, one into the other. The number of
arch configurations of order $n$, linking $n$ bridges, is the Catalan number
$C_n$. Any arch configuration is completed by reflecting with respect to the
river. This gives a one-to-one correspondence between the $n$-component
meanders of order $n$ and the arch configurations of order $n$, so that
\beq
{\cal M}_n^{(n)}=C_n
\label{Mnn}\eeq
More generally, any multi-component meander of order $n$ is obtained by
superimposing any 2 arch configurations of order $n$, one above the river and
one below, and connecting them through the $2n$ bridges. This fact implies
immediately the sum rule (\ref{sumrule1}), which therefore expresses
combinatorically the total number of multi-component meanders of order $n$ as
the total number of pairs of (top and bottom) arch configurations of order $n$.

The construction of meanders is equivalent to 2 types of moves on arch
configurations \cite{di1}. The first move, which we shall denote by an
operator ${\cal I}_1$, is to pick any exterior arch configuration of a meander
and cut it, and then pull the 2 edges of the cut across the river (left part of
exterior arch to the left, right part to the right). Then paste them around the
lower `rainbow'. This increases the rainbow configuration by 1 arch (by
definition a rainbow configuration of order $n$ has 1 arch of each depth
between 1 and $n$) and the number of bridges by 2. This yields a meander of
order $n+1$ with the same number $k$ of connected components, and so from each
meander $\bf M$ of order $n$ with $k$ components we can construct $E({\bf
M})$ distinct meanders of order $n+1$ with $k$ components, where $E({\bf M})$
is the number of exterior arches of $\bf M$. The second move ${\cal I}_2$ is to
take a meander of order $n$ with $k-1$ connected components, add an extra
circular loop around it (which increases the lower rainbow configuration of
order $n$ by 1 arch) and then add 2 bridges. This yields an order $n+1$ meander
with $k$ connected components which cannot be obtained by the move ${\cal I}_1$
above (since it has only 1 exterior arch, whereas ${\cal I}_1$ yields at least
2 exterior arches). Conversely, it is possible to show that any meander of
order $n$ with $k$ connected components can be obtained in this way \cite{di1}.

An important idea for the combinatorics of the meander problem is the notion of
the signature of arch configurations \cite{di1}. This is defined recursively
for an arch configuration ${\bf A}$ with respect to the moves ${\cal I}_1$ and
${\cal I}_2$ above, starting from the empty arch,
\beq
{\rm sig}(\emptyset)=1~~~~~;~~~~~{\rm sig}({\cal I}_1{\bf A})=(-1)^{|{\bf
A}|}~{\rm sig}({\bf A})~~~,~~~{\rm sig}({\cal I}_2{\bf A})=(-1)^{|{\bf
A}|+1}~{\rm sig}({\bf A})
\label{sigarch}\eeq
where $|{\bf A}|$ is the number of arches in $\bf A$. The relevant
combinatorial quantity for the meander problem is the number
\beq
s(n)\equiv\sum_{{\bf A}_n}{\rm sig}({\bf A}_n)
\label{sn}\eeq
where the sum is over all arch configurations ${\bf A}_n$ of order $n$. Given a
meander ${\bf M}$ with upper and lower arch configurations ${\bf A}^{(u)}$ and
${\bf A}^{(d)}$, respectively, we define its signature as
\beq
{\rm sig}({\bf M})=~{\rm sig}({\bf A}^{(u)})\cdot~{\rm sig}({\bf A}^{(d)})
\label{sigmean}\eeq
For a meander ${\bf M}_n^{(k)}$ of order $n$ with $k$ connected components, it
follows from this definition that \cite{di1}
\beq
{\rm sig}({\bf M}_n^{(k)})=(-1)^{k+n}
\label{sigmnk}\eeq

We now turn to a matrix model representation of these notions. We can modify
the complex matrix model (\ref{ECmcN}) by considering Grassmann-valued matrices
instead of complex-valued ones. The generating function is
\beq
{\cal E}_F(m,c;N)=\frac{1}{N^2}\left\langle\!\!\left\langle\log\left(\int
d\phi_1~d\phi_2~\e^{-\Sigma_F[\phi,\psi,\bar\psi]}\right)\right\rangle\!\!
\right\rangle_F
\label{EFmcN}\eeq
with fermionic Gaussian average
\beq
\left\langle\!\!\left\langle
Q[\psi,\bar\psi]\right\rangle\!\!\right\rangle_F=\frac{1}{\cal
N}\int\prod_{a=1}^md\psi_a~d\bar\psi_a~Q[\psi,\bar\psi]\e^{N^2\sum_a
\tr\bar\psi_a\psi_a}
\label{fermgaussavg}\eeq
and action
\beq
\Sigma_F[\phi,\psi,\bar\psi]=\frac{N^2}{2}\tr\phi_1^2+\frac{N^2}{2}\tr\phi_2^2
-cN^2\sum_{a=1}^m\tr\phi_1\bar\psi_a\phi_2\psi_a
\label{sigmaF}\eeq
where $\phi_1$ and $\phi_2$ are as before $N\times N$ Hermitian matrices, and
$\bar\psi_a$ and $\psi_a$ are independent $N\times N$ fermionic
Grassmann-valued matrices. Most of the formalism which identifies the complex
matrix model representation of meanders that we discussed in Subsection 7.1
above now carries through in exactly the same way for the generating function
(\ref{EFmcN}) with the replacements $W_a^\dagger\to\bar\psi_a$, $W_a\to\psi_a$
there. There are, however, 2 important differences in the fermionic case. First
of all, we recall that the Feynman rules associate a factor of $-1$ to each
loop of the fermion fields. This modifies the meander generating function
(\ref{MCmean}) to
\beq\new{\begin{array}{ll}
\lim_{N\to\infty}{\cal M}_F(m,c;N)&\equiv\lim_{N\to\infty}
c\left\langle\!\!\left\langle\frac{\int
d\phi_1~d\phi_2~\e^{-\Sigma_F[\phi,\psi,\bar\psi]}\tr\phi_1\bar\psi_{a_0}
\phi_2\psi_{a_0}}{\int
d\phi_1~d\phi_2~\e^{-\Sigma_F[\phi,\psi,\bar\psi]}}\right\rangle\!\!
\right\rangle
_F\\&=\sum_{n=1}^\infty c^{2n}\sum_{k=1}^n{\cal
M}_n^{(k)}(-m)^{k-1}\end{array}}
\label{MFmcN}\eeq
with
\beq\new{\begin{array}{l}
\sum_{k=1}^n{\cal M}_n^{(k)}(-m)^{k-1}\\=\sum_{1\leq a_2,\dots,a_{2n}\leq
m}\lim_{N\to\infty}\left\langle\!\!\left\langle\tr\psi_{a_0}\bar\psi_{a_2}\cdots
\psi_{a_{2n-1}}\bar\psi_{a_{2n}}\right\rangle\!\!\right\rangle_F\left\langle\!
\!\left\langle\tr\bar\psi_{a_{2n}}\psi_{a_{2n-1}}\cdots\bar\psi_{a_2}\psi_{a_0}
\right\rangle\!\!\right\rangle_F\end{array}}
\label{meanfermgauss}\eeq
where we have kept track of the order of the matrices as they appear from the
generating function (\ref{MFmcN}) because the signs are essential for
Grassmann-valued matrices. As we shall see below, the factors of $(-1)^{k-1}$
are associated with the signatures of the arch configurations.

The second distinguishing difference from the bosonic case is the $m=1$,
$N=\infty$ Gaussian average, which from the
generating function (\ref{omgauss}) is easily found to be
\beq
\lim_{N\to\infty}\left\langle\!\!\left\langle
\tr(\ps2)^n\right\rangle\!\!\right\rangle_F=\left\{\new{\begin{array}{ll}
0&~~~,~~~{\rm for}~~n=2p~~{\rm
even}\\C_p&~~~,~~~{\rm for}~~n=2p+1~~{\rm odd}\end{array}}\right.
\label{largeNm=1avg}\eeq
Using (\ref{meanfermgauss}) this implies that the $m=1$ sum rule
(\ref{sumrule1}) is now modified to
\beq
\sum_{k=1}^n(-1)^{k-1}{\cal
M}_n^{(k)}=\left\{\new{\begin{array}{ll}0&~~~,~~~{\rm for}~~n=2p~~{\rm
even}\\C_p^2&~~~,~~~{\rm for}~~n=2p+1~~{\rm odd}\end{array}}\right.
\label{sumrule2}\eeq
which is just the second Di Francesco-Golinelli-Guitter meander sum rule
\cite{di1}. Notice, in particular, that the left-hand side of (\ref{sumrule2})
is given by the combinatorical quantity
\beq
\sum_{k=1}^n(-1)^{k-1}{\cal M}_n^{(k)}=(-1)^{n-1}\sum_{{\bf
A}_n^{(u)}}\sum_{{\bf A}_n^{(d)}}{\rm sig}({\bf A}_n^{(u)})~{\rm sig}({\bf
A}_n^{(d)})=(-1)^{n-1}s(n)^2
\label{sumrule2comb}\eeq
which combined with (\ref{sumrule2}) yields $s(2p)=0,s(2p+1)=(-1)^{p+1}C_p$ for
$p=1,2,\dots$ (with the initial values $s(0)=1,s(1)=-1$). Thus the adjoint
fermion matrix models provide a natural representation of the notion of the
signature of arch configurations that we discussed above.

\subsection{Principal Meander Numbers and Supersymmetric Matrix Models}

As a final example of a matrix model representation of the meander problem, we
shall now combine the models discussed in the previous 2 Subsections and
consider a general complex matrix model with both bosonic and fermionic
matrices, i.e. we combine the bosonic representation
(\ref{MCmean}),(\ref{meanCgaussW}) of meanders with the fermionic one
(\ref{MFmcN}),(\ref{meanfermgauss}). To this end, we consider a general
$m=m_b+m_f$ component matrix field $W_a$ with $m_b$ bosonic (complex)
components $\tilde\phi_a$ and $m_f$ fermionic (Grassmann) components $\psi_a$,
\beq
W_a=(\tilde\phi_1,\dots,\tilde\phi_{m_b},\psi_1,\dots,\psi_{m_f})~~~,~~~\bar
W_a=(\tilde\phi_1^\dagger,\dots,\tilde\phi_{m_b}^\dagger,\bar\psi_1,\dots,
\bar\psi_{m_f})
\label{Wagenbosferm}\eeq
The generating function is defined by
\beq
{\cal
E}_S^{(m_b,m_f)}(c,N)=\frac{1}{N^2}\left\langle\!\!\left\langle\log\left(\int
d\phi_1~d\phi_2~\e^{-\Sigma_S^{(m_b,m_f)}[\phi,W,\bar
W]}\right)\right\rangle\!\!\right\rangle_S
\label{EScN}\eeq
with Gaussian average
\beq
\left\langle\!\!\left\langle Q[W,\bar
W]\right\rangle\!\!\right\rangle_S=\frac{1}{\cal N}\int\prod_{a=1}^mdW_a~d\bar
W_a~Q[W,\bar W]\e^{-N^2\sum_{a=1}^m\tr\bar
W_aW_a}
\label{susygaussavg}\eeq
and action
\beq
\Sigma_S^{(m_b,m_f)}[\phi,W,\bar
W]=\frac{N^2}{2}\tr\phi_1^2+\frac{N^2}{2}\tr\phi_2^2-cN^2\sum_{a=1}^m\tr\phi_1
\bar W_a\phi_2W_a
\label{sigsusygen}\eeq
where the integration measure is
\beq
dW_a~d\bar W_a\equiv
d\psi_a~d\bar\psi_a~\prod_{i,j}d[\tilde\phi_a]_{ij}~
d[\tilde\phi_a^\dagger]_{ij}
\label{susymeas}\eeq
Since fermion loops are always accompanied by a minus sign, it follows that the
generating function (\ref{EScN}) is related to the meander numbers by
\beq
\lim_{N\to\infty}{\cal
E}_S^{(m_b,m_f)}(c,N)=\sum_{n=1}^\infty\frac{c^{2n}}{2n}\sum_{k=1}^n{\cal
M}_n^{(k)}(m_b-m_f)^k
\label{meansusygen}\eeq

To generalize the relations of the previous Subsections between the meander
problem and the problem of summing over all words with Gaussian meaning, we
again introduce another generating function
\beq
{\cal M}_S(m_b,m_f,c;N)=c\left\langle\!\!\left\langle\frac{\int
d\phi_1~d\phi_2~\e^{-\Sigma_S^{(m_b,m_f)}[\phi,W,\bar W]}\tr\phi_1\bar
W_{a_0}\phi_2W_{a_0}}{\int
d\phi_1~d\phi_2~\e^{-\Sigma_S^{(m_b,m_f)}[\phi,W,\bar
W]}}\right\rangle\!\!\right\rangle_S
\label{MSgensusy}\eeq
which satisfies
\beq
c\frac{\partial}{\partial c}{\cal E}_S^{(m_b,m_f)}(c,N)=(m_b-m_f){\cal
M}_S(m_b,m_f,c;N)
\label{diffeqsusyME}\eeq
{}From (\ref{meansusygen}) and (\ref{diffeqsusyME}) it follows that
(\ref{MSgensusy}) is related to the meander numbers by
\beq
\lim_{N\to\infty}{\cal M}_S(m_b,m_f,c;N)=\sum_{n=1}^\infty
c^{2n}\sum_{k=1}^n{\cal
M}_n^{(k)}(m_b-m_f)^{k-1}
\label{MSsusymean}\eeq
The Gaussian integrations above can be computed as before to get
\beq\new{\begin{array}{ll}
\int d\phi_1~d\phi_2~\e^{-\Sigma_S^{(m_b,m_f)}[\phi,W,\bar W]}&=\int
d\phi_2~\e^{-\frac{N^2}{2}\tr\phi_2^2+\frac{c^2N^2}{2}\sum_{a,b=1}^m\tr(\bar
W_a\phi_2W_a\bar W_b\phi_2W_b)}\\&=\int
d\phi_2~\e^{-\frac{N^2}{2}\tr\phi_2^2+\frac{c^2N^2}{2}\sum_{a,b=1}^m\sigma_a
\tr(\phi_2W_a\bar W_b\phi_2W_b\bar W_a)}\\&={\det}^{-1/2}\left[I\otimes
I-c^2\sum_{a,b=1}^m\sigma_aW_a\bar W_b\otimes(W_b\bar W_a)^T\right]\end{array}}
\label{gaussintsusy}\eeq
where $\sigma_a$ is the signature factor (or Klein number) of the component
$W_a$, defined as $+1$ for the bosonic components and $-1$ for the fermionic
ones.

Expanding the determinant in (\ref{gaussintsusy}) as before on the right-hand
side of (\ref{EScN}) in powers of $c^2$, we arrive at the representation
\beq\new{\begin{array}{l}
\lim_{N\to\infty}{\cal
E}_S^{(m_b,m_f)}(c,N)\\=\sum_{n=1}^\infty\frac{c^{2n}}{2n}\sum_{a_1,\dots,
a_{2n-1},a_{2n}=1}^m\sigma_{a_1}\sigma_{a_3}\cdots\sigma_{a_{2n-1}}\\~~~~~
\times\lim_{N\to\infty}\left\langle\!\!\left\langle\tr W_{a_1}\bar
W_{a_2}\cdots
W_{a_{2n-1}}\bar
W_{a_{2n}}\right\rangle\!\!\right\rangle_S\left\langle\!\!\left\langle\tr
W_{a_{2n}}\bar W_{a_{2n-1}}\cdots W_{a_2}\bar
W_{a_1}\right\rangle\!\!\right\rangle
_S\\=\sum_{n=1}^\infty\frac{c^{2n}}{2n}\sum_{a_1,\dots,a_{2n-1},a_{2n}=1}
^m\lim_{N\to\infty}\left\langle\!\!\left\langle\tr W_{a_1}\bar W_{a_2}\cdots
W_{a_{2n-1}}\bar W_{a_{2n}}\right\rangle\!\!\right\rangle_S\\~~~~~~~~~~~~~~~
\times\left\langle\!\!\left\langle\tr\bar
W_{a_{2n}}W_{a_{2n-1}}\cdots\bar
W_{a_2}W_{a_1}\right\rangle\!\!\right\rangle_S\end{array}}
\label{largeNsusyE}\eeq
where the signs of the fermionic components in the first equality in
(\ref{largeNsusyE}) have been absorbed by reordering the components in the
second Gaussian correlator there. Comparing this result with the above
expressions for the meander generating functions we find
\beq\new{\begin{array}{l}
\sum_{k=1}^n{\cal
M}_n^{(k)}(m_b-m_f)^{k-1}\\=\sum_{a_2,\dots,a_{2n}=1}^m\lim_{N\to\infty}
\left\langle\!\!\left\langle\tr W_{a_0}\bar W_{a_2}\cdots W_{a_{2n-1}}\bar
W_{a_{2n}}\right\rangle\!\!\right\rangle_S\\~~~~~~~~~~~~~~~\times\left\langle
\!\!\left\langle\tr\bar
W_{a_{2n}}W_{a_{2n-1}}\cdots\bar W_{a_2}W_{a_0}\right\rangle
\!\!\right\rangle_S\end{array}}
\label{meansumsusyexpl}\eeq
Notice that the order of the matrices in (\ref{meansumsusyexpl}) is chosen to
absorb the signature factors for the fermionic components.

We are especially interested in the situation above where the number of bosonic
and fermionic matrix fields coincide, $m_b=m_f$. In that case, in addition to
the usual gauge and charge-conjugation invariances of the complex and fermionic
matrix models, the model (\ref{EScN}) is invariant under the supersymmetry
(boson-fermion) transformation $\tilde\phi_a\leftrightarrow\psi_a$,
$\tilde\phi_a^\dagger\leftrightarrow\bar\psi_a$. We can use this supersymmetry
to kill the loops of the $W$-fields which represents an alternative to the
replica trick \cite{makmean,makpe}. For brevity of notation, we restrict our
attention to the case $m=2$ ($m_b=m_f=1$), i.e. we consider a 2-component field
$W_a$ whose first component is a bosonic, complex-valued matrix and whose
second component is a fermionic matrix,
\beq
W_a=(\tilde\phi,\psi)~~~~~,~~~~~\bar W_a=(\tilde\phi^\dagger,\bar\psi)
\label{Wsusy}\eeq
Since the propagators for the bosonic and fermionic matrices in (\ref{Wsusy})
coincide (see Section 1), the simple $D=0$ supersymmetry here reduces to just
rotations between the bosonic and fermionic components. The proper
infinitesimal supersymmetry transformation is
\beq
\delta_{\bar\epsilon}\tilde\phi=\bar\epsilon\psi~~~,~~~\delta_\epsilon
\psi=-\epsilon\tilde\phi~~~~~;{}~~~~~\delta_\epsilon
\tilde\phi^\dagger=\bar\psi\epsilon~~~,~~~\delta_{\bar\epsilon}\bar\psi=-\tilde
\phi^\dagger\bar\epsilon
\label{m2susytransf}\eeq
where $\epsilon$ and $\bar\epsilon$ are infinitesimal anticommuting parameters.
We shall briefly discuss more general supersymmetry transformations at the end
of this Section.

The key property of the supersymmetry is that the contributions from the
bosonic and fermionic loops are mutually cancelled for any potential which is
constructed symmetrically from the superfields (\ref{Wsusy}), and which is
therefore supersymmetric. In the generating function ${\cal E}_S^{(1,1)}(c,N)$
defined as the supersymmetric one-matrix integral in (\ref{EScN}), the
potential there is the simplest Gaussian superpotential
\beq
{\cal W}_{\rm Gauss}[\tilde\phi,\psi,\bar\psi]=\bar WW\equiv\sum_{a=1,2}\bar
W_aW_a=\tilde\phi^\dagger\tilde\phi+\ps2
\label{gausssuperpot}\eeq
which reproduces the propagators given in Section 1. Notice that it is even
invariant under the rotation (\ref{m2susytransf}) when the parameters
$\epsilon$ and $\bar\epsilon$ are fermionic $N\times N$ matrices. This large
degree of (super)symmetry means that the supersymmetric generating function
${\cal E}_S^{(1,1)}(c,N)$ is identically zero, due to the mutual cancellation
between the loops of the bosonic and fermionic matrix fields. Instead, one
should use the generating function (\ref{MSgensusy}) which for
the supersymmetric matrix model above can be represented as
\beq
{\cal M}_S(c;N)\equiv{\cal
M}_S(1,1,c;N)=\left\langle\!\!\left\langle
\tr\tilde\phi^\dagger\tilde\phi~\log\left(\int
d\phi_1~d\phi_2~\e^{-\Sigma_S^{(1,1)}[\phi,W,\bar
W]}\right)\right\rangle\!\!\right\rangle_S
\label{MScN}\eeq
where the action is explicitly given by
\beq
\Sigma_S^{(1,1)}[\phi,W,\bar
W]=\frac{N^2}{2}\tr\phi_1^2+\frac{N^2}{2}\tr\phi_2^2-cN^2\tr\phi_1\tilde
\phi^\dagger\phi_2\tilde\phi-cN^2\tr\phi_1\bar\psi\phi_2\psi
\label{susyaction}\eeq
That (\ref{MScN}) is equivalent to (\ref{MSgensusy}) (with $a_0=1$) in this
supersymmetric case follows from an integration by parts in the identity
\beq
0\equiv\left\langle\!\!\left\langle
N^{-1}\frac{\partial}{\partial\tilde\phi_{ij}}\left\{\tilde\phi_{k\ell}
\log\left(\int d\phi_1~d\phi_2~\e^{-\Sigma_S^{(1,1)}[\phi,W,\bar
W]}\right)\right\}\right\rangle\!\!\right\rangle_S~~~~~,
\label{susygaussid}\eeq
summing over all $i=k,j=\ell$, and using the fact that ${\cal
E}_S^{(1,1)}(c,N)\equiv0$.

Because the bosonic and fermionic generating functions
(\ref{MCmean}),(\ref{meanCgaussW}) and (\ref{MFmcN}),(\ref{meanfermgauss})
alternate in sign relative to one another, only the $k=1$ terms there survive
for the supersymmetric model. Thus the multi-component meanders vanish and from
(\ref{MSsusymean}),(\ref{meansumsusyexpl}) we get the representation
\beq\new{\begin{array}{l}
{\cal
M}_n\\=\sum_{a_2,\dots,a_{2n}=1,2}\lim_{N\to\infty}\left\langle\!\!\left\langle
\tr\tilde\phi\bar W_{a_2}\cdots W_{a_{2n-1}}\bar
W_{a_{2n}}\right\rangle\!\!\right\rangle_S\left\langle\!\!\left\langle\tr\bar
W_{a_{2n}}W_{a_{2n-1}}\cdots\bar
W_{a_2}\tilde\phi\right\rangle\!\!\right\rangle_S\end{array}}
\label{meansusy}\eeq
for the principal meander numbers. Again we keep care here of the order of
matrices as the signs are crucial for the fermionic components of the
$W$-field. Alternatively, replacing $\tilde\phi$ by $\psi$ and
$\tilde\phi^\dagger$ by $\bar\psi$ in (\ref{MScN}),(\ref{susygaussid}) ($a_0=2$
in (\ref{MSgensusy})), we also have
\beq\new{\begin{array}{l}
-{\cal
M}_n\\=\sum_{a_2,\dots,a_{2n}=1,2}\lim_{N\to\infty}\left\langle\!\!\left\langle
\tr\psi\bar W_{a_2}\cdots W_{a_{2n-1}}\bar
W_{a_{2n}}\right\rangle\!\!\right\rangle_S\left\langle\!\!\left\langle\tr\bar
W_{a_{2n}}W_{a_{2n-1}}\cdots\bar
W_{a_2}\psi\right\rangle\!\!\right\rangle_S\end{array}}
\label{meansusyalt}\eeq
Thus the supersymmetric matrix model provides a representation of the principal
meander numbers which looks much more natural than the one before based on the
replica trick. It is hoped that the large supersymmetry of the problem will
make it simpler to solve the $m=2$ supersymmetric matrix model than a pure
bosonic or fermionic one at arbitrary $m$, for example by solving it using Ward
identities associated with the supersymmetry (i.e. loop equations associated
with the transformations (\ref{m2susytransf})). For explicit results of the
calculations of the principal meander numbers up to $n=4$ based on the
supersymmetric matrix model equations (\ref{meansusy}) and (\ref{meansusyalt}),
as well as a comparison with those based on the purely bosonic matrix model
equations (\ref{meanCgaussW}), see \cite{makpe}.

In \cite{makmean,makpe} it is also shown how to represent the meander problem
in terms of unitary matrix models, and ultimately its connection with the
Kazakov-Migdal model on a $D$-dimensional lattice. The words are the same for
both the meander problem and the Kazakov-Migdal model, the only difference
residing in the meaning of non-vanishing words which is unity in the case of
averages over unitary matrices in Haar measure. This relation could give a hint
on how to solve the meander problem. Its connection with higher-dimensional
matrix models, and in particular the combinatorial problem of summing over all
closed loops of a given length in a $D$-dimensional embedding space with all
possible backtrackings (or foldings) included of which we found the Gaussian
fermionic representation in Section 6 above, is an interesting representation
of the polymer phase of higher-dimensional string theories.

\subsection{Loop Equations}

We shall now proceed to discuss the evaluation of observables in supersymmetric
matrix models. Again, we shall be primarily interested with the novel model of
the last Subsection describing the meander problem, and we shall start first
from the generic complex matrix model (\ref{EScN}) and then later on impose the
restricition to supersymmetry. However, the method for solving the
corresponding loop equations in these cases, developed recently in
\cite{makpe}, is an elaborate technique for dealing with general supersymmetric
matrix models, not just the one that we are dealing with thus far. The
observables in the cases at hand are words. To deal with these appropriately,
we must modify somewhat extensively the generating functions considered in our
earlier matrix theories. The observables of the matrix model are now generated
by introducing non-commutative sources which reduce the calculation to
computing averages in a Boltzmannian Fock space. This method has been applied
recently to the loop equations of bosonic matrix models in
\cite{carroll,doug3,dougli,gopgross}. To this end, we introduce a set of
non-commuting variables $\hu_a,\hu_a^\dagger$, $a=1,\dots,m$, which generate
the Cuntz algebra
\beq
\hu_a\hu_b^\dagger=\delta_{ab}
\label{cuntzalg}\eeq
and regard them as creation and annihilation operators on some representation
Hilbert space. The normalized vacuum state $|\Omega\rangle$,
$\langle\Omega|\Omega\rangle=1$, satisfies
\beq
\hu_a|\Omega\rangle=\langle\Omega|\hu_a^\dagger=0
\label{normvac}\eeq
and the completeness relation is
\beq
\sum_{a=1}^m\hu_a^\dagger\hu_a={\bf1}-|\Omega\rangle\langle\Omega|
\label{complrel}\eeq
where $\bf1$ is the identity operator on the given Hilbert space.

The non-commutative variables are introduced above to define the
generating functions for words,
\beq\new{\begin{array}{l}
\Gee(z;\hu)\\=\left\langle\!\!\left\langle
\tr\frac{1}{z-\sum_{a,b=1}^m\hu_a\hu_bW_a\bar
W_b}\right\rangle\!\!\right\rangle_S\\=
\frac{1}{z}+\sum_{n=1}^\infty\frac{1}{z^{n+1}}\sum_{a_1,\dots,a_{2n}
=1}^m\hu_{a_1}\hu_{a_2}\cdots\hu_{a_{2n-1}}\hu_{a_{2n}}\left\langle\!\!
\left\langle\tr W_{a_1}\bar W_{a_2}\cdots W_{a_{2n-1}}\bar
W_{a_{2n}}\right\rangle\!\!\right\rangle_S\end{array}}
\label{wordgenfn}\eeq
\beq\new{\begin{array}{l}
\bar\Gee(z;\hu)\\=\left\langle\!\!
\left\langle\tr\frac{1}{z-\sum_{a,b=1}^m\hu_a\hu_b\bar W_a
W_b}\right\rangle\!\!\right\rangle
_S\\=\frac{1}{z}+\sum_{n=1}^\infty\frac{1}{z^{n+1}}\sum_{a_1,\dots,a_{2n}
=1}^m\hu_{a_1}\hu_{a_2}\cdots\hu_{a_{2n-1}}\hu_{a_{2n}}\left\langle\!\!
\left\langle\tr\bar W_{a_1}W_{a_2}\cdots\bar
W_{a_{2n-1}}W_{a_{2n}}\right\rangle\!\!\right\rangle_S\end{array}}
\label{wordgenfnbar}\eeq
where the general complex matrices $W_a,\bar W_a$ are defined in
(\ref{Wagenbosferm}). These 2 generating functions are not independent
because of the cyclic symmetry of the traces. Rearranging the
components appropriately, we have
\beq
\bar\Gee(z;\hu_a)=\Gee(z;\sqrt{\sigma_a}\hu_a)
\label{wordgenfnrel}\eeq
so that the components of $\hu_a$ associated with bosonic components
are unchanged while those associated with fermionic components are
multiplied by a factor of $i$.

The Schwinger-Dyson equations for these generating functions follow in
the usual way by shifting the matrix integration variables
\beq
\bar W_{2n}\to\bar W_{2n}+\epsilon\bar W_{2n}^n~~~~~,~~~~~W_1\to W_1+\epsilon
W_1^n
\label{wordshift}\eeq
in (\ref{susygaussavg}). The shifts (\ref{wordshift}) lead,
respectively, to the standard set of recurrence relations at $N=\infty$
\beq\new{\begin{array}{l}
\left\langle\!\!\left\langle\tr W_{a_1}\bar W_{a_2}\cdots W_{a_{2n-1}}\bar
W_{a_{2n}}\right\rangle\!\!\right\rangle
_S\\=\sum_{k=0}^{n-1}\delta_{a_{2n},a_{2k+1}}\left\langle\!\!\left\langle\tr
W_{a_1}\bar W_{a_2}\cdots\bar
W_{a_{2k}}\right\rangle\!\!\right\rangle_S\left\langle\!\!\left\langle\tr\bar
W_{a_{2k+2}}\cdots
W_{a_{2n-1}}\right\rangle\!\!\right\rangle_S\end{array}}
\label{wordrecrel1}\eeq
\beq\new{\begin{array}{l}
\left\langle\!\!\left\langle\tr W_{a_1}\bar W_{a_2}\cdots W_{a_{2n-1}}\bar
W_{a_{2n}}\right\rangle\!\!\right\rangle_S\\=\sum_{k=1}^n\delta_{a_1,a_{2k}}
\left\langle\!\!\left\langle\tr\bar W_{a_2}
W_{a_3}\cdots W_{a_{2k-1}}\right\rangle\!\!\right\rangle_S
\left\langle\!\!\left\langle\tr
W_{a_{2k+1}}\cdots\bar W_{a_{2n}}\right\rangle\!\!\right\rangle_S\end{array}}
\label{wordrecrel2}\eeq
for the words with Gaussian meaning. Multiplying these equations by
$1/z^{n+1}$ and summing over all $n\in\IZ^+$ using
(\ref{wordgenfn}),(\ref{wordgenfnbar}) and the cyclic symmetry of the
trace results in the loop equations
\beq
1-z\Gee(z;\hu)=z\sum_{a=1}^m\hu_a\bar\Gee(z;\hu)\hu_a\Gee(z;\hu)=z\sum_{a=1}^m
\Gee(z;\hu)\hu_a\bar\Gee(z;\hu)\hu_a
\label{wordloop1}\eeq
\beq
1-z\bar\Gee(z;\hu)=z\sum_{a=1}^m\sigma_a\bar\Gee(z;\hu)\hu_a\Gee(z;\hu)\hu_a
=z\sum_{a=1}^m\sigma_a\hu_a\Gee(z;\hu)\hu_a\bar\Gee(z;\hu)
\label{wordloop2}\eeq
Notice that, because of the relation (\ref{wordgenfnrel}), these 2
loop equations are not independent and can be obtained from one
another by the substitution $\hu_a\to\sqrt{\sigma_a}\hu_a$. For a
combinatorial interpretation of these loop equations which ties in
directly with the ideas of the meander problem \cite{di1}, see
\cite{makpe}.

The loop equations (\ref{wordloop1}) and (\ref{wordloop2}) generalize
those of the Gaussian adjoint fermion one-matrix model. They
are easily solved for the pure fermionic case ($m_b=0,m_f=m$) when the
variables $\hu_a$ above are commutative sources (i.e. ordinary
Euclidean vectors $\vec u=(u_1,\dots,u_m)\in\IR^m$). In that case, we
can add (\ref{wordloop1}) and (\ref{wordloop2}) together to get the
pair of simple equations
\beq
\Gee(z;\vec u)=2/z-\bar\Gee(z;\vec u)~~~,~~~1-z\bar\Gee(z;\vec u)=\vec
u^2z\bar\Gee(z;\vec u)^2-2\vec u^2\bar\Gee(z;\vec u)
\label{purefermloop}\eeq
which have solutions
\beq
\bar\Gee(z;\vec u)=\frac{1}{z}-\frac{1}{2\vec u^2}+\frac{1}{2\vec
u^2z}\sqrt{z^2+4\vec u^4}~~~,~~~\Gee(z;\vec u)=\frac{1}{z}+\frac{1}{2\vec
u^2}-\frac{1}{2\vec u^2z}\sqrt{z^2+4\vec u^4}
\label{purefermsolns}\eeq
The solution for $\bar\Gee$ coincides with the one-loop
correlator (\ref{omgauss}) of the Gaussian fermionic one-matrix model with
$t\equiv1/\vec u^2$, while the large-$z$ expansion of $\Gee$ in
(\ref{purefermsolns}) yields the moments (\ref{largeNm=1avg}). Similarly, for a
purely bosonic model ($m_b=m,m_f=0$) it is possible to show that the above
results reproduce the Wigner semi-circle distribution for the Gaussian
Hermitian (or complex) one-matrix model \cite{makpe}.

The formal solution to the loop equations in the general case can be
found by rewriting (\ref{wordloop1}) and (\ref{wordloop2}) as
\beq
\Gee(z;\hu)=\frac{1}{z+z\sum_a\hu_a\bar\Gee(z;\hu)\hu_a}~~~,~~~
\bar\Gee(z;\hu)=\frac{1}{z+z\sum_a\sqrt{\sigma_a}\hu_a\Gee(z;\hu)
\sqrt{\sigma_a}\hu_a}
\label{contfracsoln}\eeq
As shown in \cite{makpe}, iterations of these equations lead to a
formal solution for the generating functions of words as continued
fractions which represents another form of the solutions for Gaussian
matrix models.  Alternatively, using the Cuntz algebra
(\ref{cuntzalg}) we can write the loop equations (\ref{wordloop1}) and
(\ref{wordloop2}) as
\beq
(1-z\Gee(z;\hu))\hu_a^\dagger=z\Gee(z;\hu)\hu_a\bar\Gee(z;\hu)~~~,~~~
\hu_a(1-z\Gee(z;\hu^\dagger))=z\bar\Gee(z;\hu^\dagger)\hu_a^\dagger
\Gee(z;\hu^\dagger)
\label{wordloopshort1}\eeq
\beq
(1-z\bar\Gee(z;\hu))\hu_a^\dagger=z\sigma_a\bar\Gee(z;\hu)\hu_a\Gee(z;\hu)~~~,
{}~~~\hu_a(1-z\bar\Gee(z;\hu^\dagger))=z\sigma_a\Gee(z;\hu^\dagger)
\hu_a^\dagger\bar\Gee(z;\hu^\dagger)
\label{wordloopshort2}\eeq
where the left-hand sides of (\ref{wordloopshort1}) and
(\ref{wordloopshort2}) are explicitly
\beq\new{\begin{array}{l}
\hu_a(z\Gee(z;\hu^\dagger)-1)\\=\sum_{n=1}^\infty\frac{1}{z^n}\sum_{a_2,\dots,
a_{2n}=1}^m\hu_{a_2}^\dagger\cdots\hu_{a_{2n-1}}^\dagger\hu_{a_{2n}}^\dagger
\left\langle\!\!\left\langle\tr W_a\bar W_{a_2}\cdots W_{a_{2n-1}}\bar
W_{a_{2n}}\right\rangle\!\!\right\rangle_S\end{array}}
\label{lhswordexpl1}\eeq
\beq\new{\begin{array}{l}
(z\bar\Gee(z;\hu)-1)\hu_a^\dagger\\=\sum_{n=1}^\infty\frac{1}{z^n}\sum_{a_1,
a_2,\dots,a_{2n-1}=1}^m\hu_{a_1}\hu_{a_2}\cdots\hu_{a_{2n}}\left\langle
\!\!\left\langle\tr\bar W_{a_1}W_{a_2}\cdots\bar
W_{a_{2n-1}}W_a\right\rangle\!\!\right\rangle_S\end{array}}
\label{lhswordexpl2}\eeq
The generating function (\ref{MSgensusy}) is determined as
\beq\new{\begin{array}{ll}
\langle\Omega|\Gee(a;\hu)\bar\Gee(z;\hu^\dagger)|\Omega\rangle&=\langle\Omega|
\bar\Gee(z;\hu)\Gee(z;\hu^\dagger)|\Omega\rangle\\&=c^2+c^2(m_b-m_f)
\lim_{N\to\infty}{\cal M}_S(m_b,m_f,c;N)\end{array}}
\label{genfnnoncomm}\eeq
with $z=1/c$. The equation (\ref{genfnnoncomm}) follows from (\ref{cuntzalg})
and (\ref{normvac}) which imply that the left-hand side of it correctly
reproduces the contraction of indices in (\ref{meansumsusyexpl}).

We now specialize to the supersymmetric case $m_b=m_f=1$. Then
(\ref{genfnnoncomm}) does not determine the meander numbers and one
should instead use (\ref{meansusy}) or (\ref{meansusyalt}), where
there is no summation over one of the indices $a_1$, to get the
principal meander numbers.  In this case we denote the components of
$\hu_a$ as $\hu_a=(u,v)$. Using the completeness relation
(\ref{complrel}) we have
\beq
z^2\Gee(z;\hu)u^\dagger u\bar\Gee(z;\hu^\dagger)=-z^2\Gee(z;\hu)v^\dagger
v\bar\Gee(z;\hu^\dagger)+z^2\Gee(z;\hu)\bar\Gee(z;\hu^\dagger)-|\Omega\rangle
\langle\Omega|
\label{complGid}\eeq
and because of the supersymmetry we further have from (\ref{genfnnoncomm})
\beq
z^2\langle\Omega|\Gee(z;\hu)\bar\Gee(z;\hu^\dagger)|\Omega\rangle=1
\label{susyexpvalue}\eeq
Thus, taking the vacuum expectation value of the expression (\ref{complGid}),
the supersymmetry of the matrix model implies that (\ref{lhswordexpl1}) and
(\ref{lhswordexpl2}) yield
\beq
\lim_{N\to\infty}{\cal M}_S(1/z;N)=z^2\langle\Omega|\bar\Gee(z;\hu)u^\dagger
u\Gee(z;\hu^\dagger)|\Omega\rangle=-z^2\langle\Omega|\bar\Gee(z;\hu)v^\dagger
v\Gee(z;\hu^\dagger)|\Omega\rangle
\label{susylhsword}\eeq
or alternatively
\beq
\lim_{N\to\infty}{\cal M}_S(1/z;N)=z^2\langle\Omega|\Gee(z;\hu)u^\dagger
u\bar\Gee(z;\hu^\dagger)|\Omega\rangle=-z^2\langle\Omega|\Gee(z;\hu)v^\dagger
v\bar\Gee(z;\hu^\dagger)|\Omega\rangle
\label{susylhswordalt}\eeq
Using (\ref{wordloopshort1}) and (\ref{wordloopshort2}) these equations can
then be written, respectively, as
\beq\new{\begin{array}{ll}
\lim_{N\to\infty}{\cal
M}_S(1/z;N)&=\langle\Omega|\Gee(z;\hu)u\bar\Gee(z;\hu)\Gee(z;\hu^
\dagger)u^\dagger\bar\Gee(z;\hu^\dagger)|\Omega\rangle\\&=\langle\Omega|
\Gee(z;\hu)v\bar\Gee(z;\hu)\Gee(z;\hu^
\dagger)v^\dagger\bar\Gee(z;\hu^\dagger)|\Omega\rangle\end{array}}
\label{meansusyvev1}\eeq
\beq\new{\begin{array}{ll}
\lim_{N\to\infty}{\cal
M}_S(1/z;N)&=\langle\Omega|\bar\Gee(z;\hu)u\Gee(z;\hu)\bar\Gee(z;\hu^
\dagger)u^\dagger\Gee(z;\hu^\dagger)|\Omega\rangle\\&=\langle\Omega|\bar
\Gee(z;\hu)v\Gee(z;\hu)\bar\Gee(z;\hu^
\dagger)v^\dagger\Gee(z;\hu^\dagger)|\Omega\rangle\end{array}}
\label{meansusyvev2}\eeq

These latter 2 representations of the generating function for the
principal meander numbers are convenient for an iterative evaluation.
The standard trick
\cite{carroll,dougli} for dealing with 2 non-commutative variables is to expand
the quantities in $v$. This will lead to the calculation of the prinicipal
meander numbers order by order in $c=1/z$. We therefore introduce the
expansions
\beq
\Gee(z;\hu)=\sum_{n=0}^\infty\Gee_n(z;\hu)~~~~~,~~~~~
\bar\Gee(z;\hu)=\sum_{n=0}^\infty(-1)^n\Gee_n(z;\hu)
\label{Geesusyexp}\eeq
where the $v$-expansion of $\bar\Gee$ is an alternating series because of the
relation (\ref{wordgenfnrel}). The coefficients $\Gee_n(z;\hu)$ in
(\ref{Geesusyexp}) are those terms in (\ref{wordgenfn}) that involve exactly
$2n$ factors of $v$. From the definition (\ref{Wsusy}) and (\ref{m=1Cgaussavg})
it follows that
\beq
\Gee_0(z;\hu)=\frac{1-\sqrt{1-4u^2/z}}{2u^2}
\label{wignerlaw}\eeq
which we note is just the Wigner semi-circle law for the distribution
of the complex Gaussian moments
$\langle\!\langle\tr(\tilde\phi^\dagger\tilde\phi)^n\rangle\!\rangle_C=C_n$.
The
functions $\Gee_n(z;\hu)$ for $n\geq1$ can be found recursively from
(\ref{contfracsoln}) which now takes the form
\beq\new{\begin{array}{l}
\sum_{n=0}^\infty\Gee_n(z;\hat
u)\\=\frac{\Gee_0(z;\hu)}{z}\frac{1}{1+\sum_{n=0}^\infty
(-1)^nu\Gee_n(z;\hu)u\Gee_0(z;\hu)+\sum_{n=0}^\infty(-1)^nv\Gee_n(z;\hu)v
\Gee_0(z;\hu)}\end{array}}
\label{contfracsusy}\eeq

At each order of the $v$-expansion in (\ref{contfracsusy}) we have to solve the
equation
\beq
\Gee_n(z;\hu)=-\frac{1}{z}\left((-1)^n\Gee_0(z;\hu)u\Gee_n(z;\hu)u
\Gee_0(z;\hu)+A_n(z;\hu)\right)
\label{eqnvorder}\eeq
for some functions $A_n(z;\hu)$ which will be determined recursively below. The
solution of (\ref{eqnvorder}) is found by iterating it to get
\beq\new{\begin{array}{c}
\Gee_{2p-1}(z;\hu)=-\frac{1}{z}\sum_{\ell=0}^\infty
\left(\Gee_0(z;\hu)u\right)^\ell
\frac{A_{2p-1}(z;\hu)}{z^\ell}\left(\Gee_0(z;\hu)u\right)^\ell\\
\Gee_{2p}(z;\hu)=-\frac{1}{z}\sum_{\ell=0}^\infty(-1)^\ell
\left(\Gee_0(z;\hu)u\right)^\ell
\frac{A_{2p}(z;\hu)}{z^\ell}\left(\Gee_0(z;\hu)u\right)^\ell\end{array}}
\label{solnvorder}\eeq
where $p\geq1$. The 2 solutions in (\ref{solnvorder}) can be written nicely as
the contour integral
\beq
\Gee_n(z;\hu)=-\frac{1}{z}\oint_{\cal C}\frac{d\lambda}{2\pi
i}~\frac{1}{\lambda-(-1)^{n+1}\Gee_0(z;\hu)u}~A_n(z;\hu)
{}~\frac{1}{1-\lambda\Gee_0(z;\hu)u/z}\equiv\left\{A_n\right\}
\label{contintGee0}\eeq
where the contour $\cal C$ encircles the origin of the complex $\lambda$-plane,
and we have introduced a short-hand bracket notation for the quantities on
the right-hand sides of (\ref{solnvorder}) or (\ref{contintGee0}) to simplify
some of the cumbersome formulas which follow.

The functions $A_n(z;\hu)$ above are now computed recursively using
(\ref{contfracsusy}) and (\ref{eqnvorder}). The first few lower order ones are
explicitly
\beq\new{\begin{array}{c}
A_1=\Gee_0v\Gee_0v\Gee_0~~~,~~~A_2=-\Gee_0v\Gee_1
v\Gee_0+z^2\Gee_1\frac{1}{\Gee_0}\Gee_1\\A_3=\Gee_0v\Gee_2v\Gee_0+z^2
\Gee_2\frac{1}{\Gee_0}\Gee_1+z^2\Gee_1\frac{1}{\Gee_0}\Gee_2+z^3\Gee_1\frac{1}
{\Gee_0}\Gee_1\frac{1}{\Gee_0}\Gee_1\end{array}}
\label{loworderA}\eeq
Generally, substituting the expansion (\ref{Geesusyexp}) into
(\ref{wordloopshort1}) (or (\ref{wordloopshort2})), equating the orders of the
$v$-expansions and using the fact that
\beq
\Gee_n(z;\hu)v^\dagger=-A_n(z;\hu)v^\dagger/z
\label{GeeAvid}\eeq
which follows from (\ref{solnvorder}) and (\ref{cuntzalg}), we see that the
$A_n$ satisfy the identities
\beq\new{\begin{array}{c}
A_n(z;\hu)v^\dagger=z\sum_{k=0}^{n-1}(-1)^k\Gee_{n-1-k}(z;\hu)v\Gee_k(z;\hu)
\\vA_n(z;\hu^\dagger)=z\sum_{k=0}^{n-1}(-1)^k\Gee_k(z;\hu^\dagger)v^\dagger
\Gee_{n-1-k}(z;\hu^\dagger)\end{array}}
\label{Anids}\eeq
Then, using (\ref{loworderA}) and the fact that $\Gee_0$ commutes with
the brackets defined by (\ref{contintGee0}), the first few $\Gee_n$'s
are given by
\bd
\Gee_1=\Gee_0\{v\Gee_0v\}\Gee_0~~~,~~~\Gee_2=-\Gee_0\{v\Gee_0\{v\Gee_0v\}
\Gee_0v\}\Gee_0+z^2\Gee_0\{\{v\Gee_0v\}\Gee_0\{v\Gee_0v\}\}\Gee_0
\ed
\beq\new{\begin{array}{ll}
\Gee_3&=-\Gee_0\{v\Gee_0\{
v\Gee_0\{v\Gee_0v\}\Gee_0v\}\Gee_0v\}\Gee_0+z^2\Gee_0\{v\Gee_0\{v\Gee_0v
\}\Gee_0\{v\Gee_0v\}\Gee_0v\}\Gee_0\\&~~~~~-z^2\Gee_0\{\{v\Gee_0v\}\Gee_0
\{v\Gee_0\{v\Gee_0v\}\Gee_0v\}\}\Gee_0-z^2\Gee_0\{\{v\Gee_0\{v\Gee_0v\}\Gee_0v\}
\Gee_0\{v\Gee_0v\}\}\Gee_0\\&~~~~~+z^3\Gee_0\{\{v\Gee_0v\}\Gee_0\{v\Gee_0v\}
\Gee_0\{v\Gee_0v\}\}\Gee_0+z^4\Gee_0\{\{\{v\Gee_0v\}\Gee_0\{v\Gee_0v\}\}
\Gee_0\{v\Gee_0\}\}\Gee_0\\&~~~~~+z^4\{\{v\Gee_0v\}\Gee_0\{\{v\Gee_0v\}
\Gee_0\{v\Gee_0v\}\}\}\Gee_0\end{array}}
\label{loworderG}\eeq
The brackets in (\ref{loworderG}) must in general always pair equal numbers of
open and closed parantheses, and an inner pair of brackets must always be
embedded in an overall pair of outer brackets. For general $n$, $\Gee_n(z;\hu)$
contains $C_n$ terms of the types in (\ref{loworderG}) with alternating signs
up to order $z^{2^{n-1}}$. Some rules for representing a general
term can be formulated which resemble the Wick pairing of bilinear combinations
of the $v$'s \cite{makpe}.

Most of the relations above are similar to those that one would obtain
for 2 bosonic matrix fields ($m_b=m=2,m_f=0$), except that there are
no alternating signs in the pure bosonic case. The occurence of these
minus signs at appropriate places makes the supersymmetric matrix
model somewhat simpler in structure than the pure bosonic $m=2$ case.
For instance, the leading order $v^4$ terms could be cancelled for
$A_2$ in (\ref{loworderA}) and for $\Gee_2$ in (\ref{loworderG}),
which follows from the general property (\ref{largeNm=1avg}) of
fermionic Gaussian averages. For this same reason the asymptotic
behaviours of the functions $\Gee_n(z;\hu)$ are
\beq
\lim_{|z|\to\infty}\Gee_n(z;\hu)=\left\{\new{\begin{array}{ll}
1/z^{2n+2}&~~~{\rm for}~~n~~{\rm odd}\\1/z^{2n+4}&~~~{\rm for}~~n\geq2~~
{\rm even}\end{array}}\right.
\label{Gnasymptz}\eeq
Some further relations are imposed by the supersymmetry of the model. The
simplest one follows from substituting the expansion (\ref{Geesusyexp}) into
(\ref{susyexpvalue}) which leads to
\beq
z^2\sum_{n=0}^\infty(-1)^n\langle\Omega|\Gee_n(z;\hu)\Gee_n(z;\hu^\dagger)|
\Omega\rangle=1
\label{simplesusyreln}\eeq
where only diagonal terms contribute in (\ref{simplesusyreln}) because
of the definitions (\ref{cuntzalg}) and (\ref{normvac}).

Finally, we evaluate the prinicipal meander numbers ${\cal M}_n$ using the
above relations. Notice that (\ref{Gnasymptz}) implies that a contribution to
${\cal M}_n$ in (\ref{susylhswordalt}) (the coefficient of $1/z^{2n}$) can come
at most from $\Gee_n$ with $n$ odd and from $\Gee_{n-1}$ with $n$ even.
Substituting the expansion (\ref{Geesusyexp}) into (\ref{susylhswordalt}) we
find
\beq\new{\begin{array}{ll}
\lim_{N\to\infty}{\cal
M}_S(1/z;N)&=z^2\sum_{n=1}^\infty(-1)^{n-1}\langle\Omega|\Gee_n(z;\hu)v^\dagger
v\Gee_n(z;\hu^\dagger)|\Omega\rangle\\&=\sum_{n=1}^\infty(-1)^{n-1}\langle
\Omega|A_n(z;\hu)v^\dagger vA_n(z;\hu^\dagger)|\Omega\rangle\end{array}}
\label{Mnexpvalue}\eeq
where in the second equality in (\ref{Mnexpvalue}) we have used
(\ref{GeeAvid}). From this expression, we can immediately calculate the first
and second principal meander numbers ${\cal M}_1$ and ${\cal M}_2$. From
(\ref{loworderG}) and (\ref{contintGee0}) we have
\beq
\Gee_1(z;\hu)v^\dagger=-\Gee_0(z;\hu)v\Gee_0(z;\hu)/z^2~~~,~~~v\Gee_1(z;\hu
^\dagger)=-\Gee_0(z;\hu^\dagger)v^\dagger\Gee_0(z;\hu^\dagger)/z^2
\label{G1v}\eeq
so that (\ref{cuntzalg}), (\ref{normvac}) and (\ref{wignerlaw}) give
\beq
\lim_{N\to\infty}{\cal
M}_S(1/z;N)=\frac{1}{z^2}\left(\sum_{k=0}^\infty\frac{C_k^2}{z^{2k}}\right)^2
+z^2\sum_{n=2}^\infty(-1)^{n-1}\langle\Omega|\Gee_n(z;\hu)v^\dagger
v\Gee_n(z;\hu^\dagger)|\Omega\rangle
\label{MS12}\eeq
The first term on the right-hand side of (\ref{MS12}) (the $n=1$ term in
(\ref{Mnexpvalue})) yields the anticipated values ${\cal M}_1=C_0^2=1$, ${\cal
M}_2=2C_0^2C_1^2=2$ \cite{di1,makpe}. The contributions of the next orders are
controlled by the asymptotic behaviours (\ref{Gnasymptz}) above.

For the terms in (\ref{Mnexpvalue}) up to $n=3$, at most the expansion
coefficients $\Gee_3$ are essential, and (\ref{loworderG}) along with
(\ref{wignerlaw}) give
\beq\new{\begin{array}{c}
\Gee_2(z;\hu)v^\dagger=(vuv^2u-uv^2uv)/z^8+{\cal O}(1/z^{10})\\
v\Gee_2(z;\hu^\dagger)=(-v^\dagger
u^\dagger(v^\dagger)^2u^\dagger+u^\dagger(v^\dagger)^2u^\dagger
v^\dagger)/z^8+{\cal O}(1/z^{10})\\\Gee_3(z;\hu)v^\dagger=v^6/z^8+{\cal
O}(1/z^{10})~~~,~~~v\Gee(z;\hu^\dagger)=(v^\dagger)^6/z^8+{\cal
O}(1/z^{10})\end{array}}
\label{G2G3}\eeq
Substituting (\ref{G2G3}) into (\ref{MS12}) yields the expected third
principal meander number ${\cal M}_3=8$ \cite{di1,makpe}. In general,
however, the iterative procedure described above becomes rather
cumbersome when trying to evaluate the vacuum expectation values in
(\ref{MS12}) which involve $C_n^2$ terms for each $n>1$. While each
individual term is in principle calculable, it is not as such clear if
this iterative procedure could lead to recurrence relations for the
meander numbers. The main difference between the non-commutative loop
equations for the supersymmetric model above and those for the
two-matrix model with polynomial potential \cite{dougli} is that the
$v$-expansion in the supersymmetric case does not lead to an algebraic
equation determining the coefficients $\Gee_n$. Thus, just as in the pure
fermionic cases, the supersymmetric models aren't as amenable to explicit
solution as the pure bosonic cases because of the complexity of the loop
equations involving fermionic degrees of freedom. As always, this complexity
can lead to novel critical phenomena so that the existence of fermionic
inducing fields leads to exotic types of random surface models.

\subsection{$D=0$ Supersymmetric Matrix Models and Branched Polymers}

In this Subsection we shall elaborate on the uses of $D=0$ supersymmetric
matrix models to represent the branched polymer phase of string theory,
following the approach of Ambj\o rn, Makeenko and Zarembo \cite{ammakz}. We
consider a more general interaction potential defined by
\beq
{\cal W}(\bar WW)=\sum_{k\geq1}\frac{g_k}k(\bar WW)^k
\label{gensusypot}\eeq
The invariance of (\ref{gensusypot}) under the matrix supersymmetry
transformations (\ref{m2susytransf}) follows from
\beq
\delta_\epsilon(\bar
WW)=\delta_\epsilon(\tilde\phi^\dagger\tilde\phi+\ps2)=\bar\psi\epsilon
\tilde\phi-\bar\psi\epsilon\tilde\phi=0
\eeq
and similarly for the action of $\delta_{\bar\epsilon}$. The supersymmetry
transformations can be generated using the matrix supercharges
\beq
Q_{ij}=N\sum_{k=1}^N\left(\psi_{ik}\frac\partial{\partial\tilde\phi_{jk}}-
\frac{\partial}{\partial\bar\psi_{ki}}\tilde\phi^\dagger_{kj}\right)~~~,~~~\bar
Q_{ij}=N\sum_{k=1}^N\left(\frac\partial{\partial\tilde\phi^\dagger_{ki}}
\bar\psi_{kj}-\tilde\phi_{ik}\frac\partial{\partial\psi_{jk}}\right)
\label{matrixsuperch}\eeq
whose action on matrix fields $\Psi$ of the model give the infinitesimal
supersymmetry variations,
\beq
\delta_\epsilon\Psi=[\tr\bar
Q\epsilon,\Psi]~~~,~~~\delta_{\bar\epsilon}\Psi=[\tr\bar\epsilon Q,\Psi]
\label{chargesusytransf}\eeq
The anti-commutation relations between these supercharges are
\beq
\{Q_{ij},Q_{mn}\}=\{\bar Q_{ij},\bar Q_{mn}\}=0
\label{supercomm1}\eeq
\beq
\left\{Q_{ij},\bar
Q_{mn}\right\}=-N\delta_{in}\sum_{k=1}^N\left(\tilde\phi_{mk}\frac\partial
{\partial\tilde\phi_{jk}}+\frac{\partial}{\partial\tilde\phi^\dagger_{km}}
\tilde\phi^\dagger_{kj}\right)-N\sum_{k=1}^N\left(\psi_{ik}
\frac\partial{\partial\psi_{nk}}-\frac\partial{\partial\bar\psi_{ki}}
\bar\psi_{kn}\right)\delta_{mj}
\label{supercomm2}\eeq

We now set $g_1=-1$ in (\ref{gensusypot}) and consider the supersymmetric
matrix model with partition function
\beq
Z_S[g]=\int dW~d\bar W~\e^{N^2\tr{\cal W}(\bar WW)}
\label{gensusypart}\eeq
We shall be interested in the generating function for the correlators of the
complex, bosonic matrices,
\beq
\Xi(z)=\left\langle\tr\frac1{z-\tilde\phi^\dagger\tilde\phi}\right\rangle
=\frac1z+\sum_{n=1}^\infty\frac1{z^{n+1}}\xi_n
\label{Xigenbos}\eeq
where the bosonic moments are
\beq
\xi_n=\left\langle\tr(\tilde\phi^\dagger\tilde\phi)^n\right\rangle
\label{xindef}\eeq
The Schwinger-Dyson equation for this generating function follows from the
identity
\beq
\frac1{N^2}\sum_{i,j}\int dW~d\bar
W~\frac\partial{\partial\tilde\phi_{ij}}\left(\e^{N^2\tr{\cal W}(\bar
WW)}~\tilde\phi\frac1{z-\tilde\phi^\dagger\phi}\right)_{ij}=0
\label{susysdint}\eeq
which can be expanded out into averages to give
\beq
\left\langle\tr{\cal W}'(\bar
WW)\frac{\tilde\phi^\dagger\tilde\phi}{z-\tilde\phi^\dagger\tilde\phi}
\right\rangle=z\left\langle\left(\tr\frac1{z-\tilde\phi^\dagger\tilde\phi}
\right)^2\right\rangle
\label{susysdavg}\eeq
In contrast to the loop equations derived previously, the equation
(\ref{susysdavg}) can actually be expressed in a closed form by exploiting the
invariance of the action and integration measure in (\ref{gensusypart}) under
the supersymmetry transformations (\ref{m2susytransf}). This leads to the Ward
identity
\beq
\left\langle\delta_\epsilon\left(\psi\frac1{w-\bar WW}\frac1{z-\tilde\phi^
\dagger\tilde\phi}\tilde\phi^\dagger\right)\right\rangle=0
\label{wardid}\eeq
which, after the proper contraction of matrix indices, can be expanded to give
\beq
\left\langle\tr\frac{\bar WW}{w-\bar
WW}\frac1{z-\tilde\phi^\dagger\tilde\phi}\left(1+\tr\frac{\tilde\phi^\dagger
\tilde\phi}
{z-\tilde\phi^\dagger\tilde\phi}\right)\right\rangle=\left\langle\tr\frac1
{w-\bar WW}\frac{\tilde\phi^\dagger\tilde\phi}{z-\tilde\phi^\dagger\tilde\phi}
\tr\frac{\tilde\phi^\dagger\tilde\phi}{z-\tilde\phi^\dagger\tilde\phi}
\right\rangle
\label{wardidavg}\eeq

As mentioned before, the supersymmetry leads to a cancellation of bosonic and
fermionic loops, so that the partition function is unity and all supersymmetric
correlators vanish. This can be formally proved by taking the limit
$z\to\infty$ in (\ref{wardidavg}) and comparing the ${\cal O}(1/z)$
coefficients on both sides to give
\beq
\left\langle\tr(\bar WW)^n\right\rangle=0
\eeq
for all $n\geq1$. Since $\langle\tr(\bar
WW)^n\rangle=\frac1{N^2}\frac\partial{\partial g_n}Z_S[g]$, the partition
function is independent of the potential $\cal W$ and is formally equal to 1.
However, non-trivial physical characteristics of the model reside in
non-supersymmetric correlators, such as those of the bosonic matrices
(\ref{xindef}).

In the large-$N$ limit, when factorization holds, the equations of motion
(\ref{susysdavg}) and (\ref{wardidavg}) of the supersymmetric matrix model can
be written succinctly as the respective integral equations
\beq
\oint_{\cal C}\frac{d\lambda}{2\pi i}~\oint_{\cal C}\frac{d\mu}{2\pi
i}~\frac{\lambda{\cal W}'(\mu){\cal Q}(\mu,\lambda)}{z-\lambda}=z\Xi(z)^2
\label{susysdcont}\eeq
\beq
\oint_{\cal C}\frac{d\mu}{2\pi i}~\frac{\mu{\cal
Q}(\mu,z)}{w-\mu}\left(1+\oint_{\cal C}\frac{d\lambda}{2\pi
i}~\frac{\lambda\Xi(\lambda)}{z-\lambda}\right)=\oint_{\cal C}\frac{d\mu}{2\pi
i}~\frac{\mu{\cal Q}(w,\mu)}{z-\mu}\oint_{\cal C}\frac{d\lambda}{2\pi
i}~\frac{\lambda\Xi(\lambda)}{z-\lambda}
\label{wardidcont}\eeq
where we have introduced the additional generating function
\beq
{\cal Q}(w,z)=\left\langle\tr\frac1{w-\bar
WW}\frac1{z-\tilde\phi^\dagger\tilde\phi}\right\rangle
\label{Qdef}\eeq
and the closed contour $\cal C$ above encircles all singularities of $\cal Q$
and $\Xi$, but not the points $z$, $w$ or $\infty$, with counterclockwise
orientation. Using the usual asymptotic behaviours
\beq
{\cal Q}(w,z)=1/wz+{\cal O}(1/z^2)=\Xi(z)/w+{\cal
O}(1/w^2)~~~,~~~\Xi(z)=1/z+{\cal O}(1/z^2)
\label{QXiasympt}\eeq
the contour integrations in (\ref{wardidcont}) can be evaluated by computing
the residues at $z$, $w$ and $\infty$ to yield, after some algebra, an equation
determining the generating function ${\cal Q}$ in terms of $\Xi$,
\beq
{\cal Q}(w,z)=\frac{wz\Xi(z)^2-z\Xi(z)+1}{wz(w\Xi(z)-z\Xi(z)+1)}
\label{QexplXi}\eeq

We now substitute (\ref{QexplXi}) into the other integral equation
(\ref{susysdcont}) to generate an equation for the generating function $\Xi$.
The contour integrals over $\mu$ and $\lambda$ can be carried out by first
computing the residues of the poles of ${\cal Q}(\mu,\lambda)$ in
(\ref{QexplXi}) as a function of $\mu$, and then computing the residues at
$\lambda=z$ and $\lambda=\infty$. After some algebra, we arrive finally at
\beq
(z\Xi(z)-1){\cal W}'(z-1/\Xi(z))=z\Xi(z)^2
\label{Xieqfinal}\eeq
Using the asymptotic expansion (\ref{Xigenbos}), the expression
(\ref{Xieqfinal}) can be equated order by order on both sides in $1/z$, leading
to various mixed equations for the bosonic correlators (\ref{xindef}). In
particular, equating the leading-order $1/z$ coefficients of (\ref{Xieqfinal})
leads to the expression
\beq
\xi_1{\cal W}'(\xi_1)=1
\label{xi1eq}\eeq
for the correlator $\xi_1=\langle\tr\tilde\phi^\dagger\tilde\phi\rangle$. We
immediately identify (\ref{xi1eq}) with the saddle-point equation
(\ref{ONstatgen}) (see also (\ref{M1algeq})) describing the continuum limit of
a branched polymer model, that we derived from the vector model representation.
The fact that we arrive at a closed equation for the bosonic propagator $\xi_1$
is a consequence of the cancellations between the bosonic and fermionic loops.
The Feynman graphs which survive the supersymmetric cancellation in the
diagrammatic expansion of the propagator are the so-called ``cactus diagrams"
\cite{ammakz} which consist of a closed loop of links with one marked vertex.
These Feynman diagrams have an orientation, since the cactus loops can only be
attached to the exterior of already existing loops, and thus the bosonic
correlators of the supersymmetric matrix model here describe ``chiral" branched
polymers, which branch out only at one side of an open line. This is one of the
fundamental differences between the supersymmetric matrix models and the
representation of branched polymers via cactus graphs of the vector models that
were studied in Section 2.

Thus in this way, the bosonic propagator generates the random polymer model
\beq
\xi_1=\sum_{\vec{\cal P}}w(\vec{\cal P})\e^{-\Lambda L(\vec{\cal P})}
\eeq
where the sum is over all chiral branched polymers $\vec{\cal P}$ and
\beq
w(\vec{\cal P})=\prod_{v\in\vec{\cal P}}g(v)=\prod_{k\geq2}(-g_k)
\eeq
are the branching weights, with the local weight factors $g(v)=-g_k$ depending
only on the order $k$ (number of nearest neighbours) of the vertex $v$. The
supersymmetric matrix model representation of the branched polymer theory
allows a detailed study of its spectral properties \cite{ammakz}, specifically
the determination of the universality classes of pure branched polymers as a
function of the weight attributed to the branching at the individual nodes. For
instance, the discontinuity of the generating function $\Xi(z)$ above
determines the eigenvalue distribution of the Hermitian matrix
$\tilde\phi^\dagger\tilde\phi$, and therefore the spectrum of the proper
statistical model. It is also possible to study multi-matrix versions of the
above supersymmetric model involving more than one matrix superfield $W$. For
example, a 2-matrix version of the above model was solved in \cite{ammakz} and
shown to reproduce the well-known features of the Ising model on a branched
polymer \cite{ambthor}.

\subsection{$D=-2$ Supersymmetric Matrix Models and Superstrings}

We have considered thus far in this Section relatively simple
supersymmetric matrix models. Nonetheless, these models result in more
complicated structures than the general $D$-dimensional pure fermionic
or pure bosonic matrix models.  The combinatorical problem of
enumerating meander numbers that they describe belongs to the same
generic class of problems of words as large-$N$ multi-colour QCD but
is presumably simpler. It would be interesting to study these novel
supersymmetric matrix models in connection with other physical
applications to quantum gravity, other than those represented by a
random polymer phase.  For instance, the connection between
supersymmetric matrix models and the meander problem could be useful
for analysing statistical theories of discretized super-Riemann
surfaces and superstrings. We conclude this Section with a brief
discussion about the possibility of such applications.

In the above models we considered supersymmetric matrix fields on a $D=0$
dimensional embedding space, whereby the bosonic and fermionic matrix
propagators coincide. This means essentially that we don't need introduce the
usual (dimensionless) superspace coordinates $\theta$ which are canonically
associated with the superspace formulation of supersymmetric field theories. As
we have mentioned before, the first attempt at incorporating fermionic degrees
of freedom into bosonic matrix theories was the Gilbert-Perry Hermitian
supermatrix model \cite{alvarez,gilper,yost}. However, the (Hermitian) fermion
matrix fields can be integrated out and the model in this case reduces to an
ordinary bosonic Hermitian one-matrix model (which is trivial when the number
of bosonic and fermionic degrees of freedom are the same). Here we shall
discuss a slight modification of the above supersymmetric matrix models. We
assume that our matrices are superfields defined on a supersurface associated
with an $N=1$ supersymmetry (the generalization to higher-component
supersymmetries is immediate). This means that we augment the zero-dimensional
space of the matrix model by 2 anticommuting superspace coordinates. Using the
usual interpretation of Grassmann coordinates as negative dimensions, we can
think of such a model as a $D=-2$ dimensional matrix model. The planar
triangulations of closed surfaces in $D=-2$ dimensions was studied by Kazakov,
Kostov and Migdal in \cite{kazkost}, and the matrix model approach was
developed in \cite{david0,klebwil,kosmet}.

The supersymmetric matrix model we consider here is thus defined by the
partition function
\beq
Z_{MP}=\int d\Phi~\e^{-N^2\Sigma_{WZ}[\Phi]}
\label{marparpart}\eeq
where the ``action" in (\ref{marparpart}) is that of the standard Wess-Zumino
model \cite{wess} in the superspace formulation
\beq
\Sigma_{WZ}[\Phi]=\int d\theta~d\bar\theta~\tr\left(\bar{\cal D}\Phi{\cal
D}\Phi+{\cal
W}(\Phi)\right)
\label{wesszumaction}\eeq
with ${\cal W}(\Phi)$ some superpotential. Here
\beq
\Phi(\theta,\bar\theta)=\phi+\psi\bar\theta+\bar\psi\theta+F\bar\theta\theta
\label{supermatrix}\eeq
are the usual supermatrix fields defined on a purely Grassmannian
supersurface, with $\phi$ a bosonic $N\times N$ Hermitian matrix,
and $\psi$ and $\bar\psi$ independent fermionic $N\times N$ Grassmann
matrices. The bosonic $N\times N$ Hermitian matrix $F$ is an external
field, $\theta$ and $\bar\theta$ are independent, anti-commuting
Grassmann coordinates and the covariant derivatives on the
supersurface are
\beq
{\cal D}=\frac{\partial}{\partial\theta}~~~~~,~~~~~\bar{\cal
D}=\frac{\partial}{\partial\bar\theta}
\label{covsuperderivs}\eeq
which generate the infinitesimal supersymmetry transformations.
The integration measure in (\ref{marparpart}) is
\beq
d\Phi=d\phi~d\psi~d\bar\psi~dF
\label{superintmeas}\eeq
and, as always, the complex-conjugation convention for the Grassmann numbers
implies that $\bar{\cal D}^*=-{\cal D}$.

The statistical model above is the $D=0$ direct simplification of the
Marinari-Parisi model \cite{marinari} which is the standard form for a
quantum field theory with Poincar\'e supersymmetry in a superspace
formulation. It describes super-triangulations of super-surfaces with
weights that are reparametrization invariant in parameter space and
supersymmetry invariant in target space. More precisely, the weights
here associate to each triangulation a factor proportional to the free
propagator of the superfield (\ref{supermatrix}) (the product of
functions of the relative super-distances of all contiguous triangles
which are therefore invariant under super-rotations) and the
supersymmetric string theory (in a non-existent embedding space) is
then described by summing over all possible topologies of these
triangulations and integrating the corresponding weights in the
superspace. For non-Gaussian super-potentials in
(\ref{wesszumaction}), it can be shown \cite{marinari} that there are
critical points where the correlation length becomes much larger than
the fundamental length of the discretization of the superstring
theory. This can be worked out by first performing the Grassmann integrals over
$\theta$ and $\bar\theta$ in (\ref{wesszumaction}) to get
\beq
\Sigma_{WZ}[\Phi]=\tr\left(-F^2+{\cal W}'(\phi)F+\bar\psi{\cal
W}''(\phi)\psi+\ps2{\cal W}''(\phi)\right)
\label{WZactionint}\eeq
The integration over the auxilliary field $F$ is thus Gaussian and can be
carried out explicitly. Then the partition function (\ref{marparpart}) becomes
\cite{marinari}
\beq
Z_{MP}=\int
d\phi~d\psi~d\bar\psi~\exp\left\{\frac{-N^2}{2}\tr\left(\frac{\partial{\cal
W}(\phi)}{\partial\phi^*}\frac{\partial{\cal
W}(\phi)}{\partial\phi}-\bar\psi\frac{\partial^2{\cal
W}(\phi)}{\partial\phi^2}\psi-\ps2\frac{\partial^2{\cal
W}(\phi)}{\partial\phi^2}\right)\right\}
\label{intsusypart}\eeq

Notice that when the superpotential ${\cal W}(\Phi)$ is quadratic in $\Phi$,
the
action in (\ref{intsusypart}) reduces to that considered for the
meander problem above. More complicated, non-Gaussian superpotentials
lead to a more complicated supersymmetry among the bosonic and
fermionic components which results in a critical behaviour for the
statistical system described above. In the above superspace
formulation in terms of superfields, the model is readily generalized
to higher-dimensional matrix fields. The matrix model (\ref{intsusypart}) is
invariant under the infinitesimal supersymmetry transformations
\beq
\delta\phi=\bar\epsilon\psi+\bar\psi\epsilon~~~,~~~\delta\psi=-\frac12\epsilon
{\cal W}'(\phi)~~~,~~~\delta\bar\psi=-\frac12\bar\epsilon{\cal W}'(\phi)
\label{-2susytransf}\eeq

Again, the supersymmetric matrix model is trivial, in that the partition
function is independent of the form of the superpotential $\cal W$. In the
present case this follows from using the invariance of the model
(\ref{intsusypart}) under the simultaneous $U(N)$ transformations $\phi\to
U\phi U^\dagger$, $\psi\to U\psi U^\dagger$ and $\bar\psi\to U\bar\psi
U^\dagger$ to write the partition function in terms of the eigenvalues of the
Hermitian matrix $\phi$, and then integrating out the fermion fields
\cite{klebwil}. From this it follows from differentiating the partition
function with respect to a set of couplings $g_k$ in the usual way that any
integrated supersymmetric correlator vanishes,
\beq
\left\langle\int d\theta~d\bar\theta~\tr\Phi^n\right\rangle=0
\eeq
Thus the supersymmetric matrix model free energy itself exhibits no critical
behaviour. But a critical model is obtained as before by examing the
correlators of the Hermitian matrix field $\phi$. The result of integrating out
the fermions above shows that these correlators are determined as the Hermitian
Gaussian correlators of the inverse of the map $\phi\to{\cal W}'(\phi)$ (i.e.
the Nicolai map),
\beq
\left\langle\tr\phi^n\right\rangle=\left\langle\!\!\left\langle\tr\left[{\cal
W}'^{-1}(\phi)\right]^k\right\rangle\!\!\right\rangle
\label{boscorrssusy}\eeq
Despite the triviality of the supersymmetric partition function, such
correlation function do display non-trivial critical behaviour. The
diagrammatics of the matrix theory for such correlation functions can be
written down and shown to coincide the sort of super-discretizations mentioned
earlier. Furthermore, the (planar) scaling limit reproduces the correct string
susceptibility exponent of the Liouville theory prediction for (non-unitary)
$D=-2$ matter fields coupled to two-dimensional quantum gravity
\cite{klebwil,kosmet}.

As mentioned before, the supersymmetry of a matrix model always leads to a
dimensional reduction, so that the supersymmetry of the matrix theory does not
coincide with the supersymmetry of the supergravity or superstring theory
\cite{alvarez,dadda}. This fact was originally exploited by Marinari
and Parisi \cite{marinari} who considered time-dependent matrix fields
above (so that the partition function describes a supersymmetric
matrix quantum mechanics). The Hamiltonian which follows from
quantizing this one-dimensional system then has the standard form of a
Witten-type supersymmetric quantum mechanics \cite{wittensusy} (in our $D=0$
case above, the Hamiltonian vanishes and the supersymmetry charges generating
the infinitesimal supersymmetry transformations of the theory coincide
with the covariant derivatives (\ref{covsuperderivs})). The
dimensional reduction mechanism in this $D=1$ model comes from the
stochastic quantization method associated with regarding the bosonic
sector Hamiltonian as the forward Fokker-Planck Hamiltonian in the
Langevin time-evolution equation \cite{fgz}. However, as for the pure
adjoint fermion matrix models, the supersymmetric theory is always
well-defined and no ambiguities occur, in contrast to the standard
$D=0$ Hermitian matrix models which exhibit non-perturbative
ambiguities, instabilities and violations of the Schwinger-Dyson
equations \cite{david3,nish} (as discussed previously). For $N\to\infty$, the
expectation values in the supersymmetric matrix model coincide with those of
the
ill-defined $D=0$, $m=2$ bosonic theory, and moreover the genus
expansion is completely reproduced. Thus the dimensionally reduced
Fokker-Planck supersymmetry can be taken as an alternative definition
of the $D=0$ bosonic matrix theory.

Thus, as with the adjoint fermion models that we have extensively studied
throughout this Review, supersymmetric matrix models define better
behaved random surface theories than the conventional bosonic matrix
models and reproduce features similar to these cases. Besides this
feature, it would be interesting to develop these supersymmetric models to
describe non-critical superstring theories and models of dynamical
supersymmetry breaking. The $D=1$ Marinari-Parisi model provides
a good example of the latter mechanism which is associated with the
phase transition and the double-scaling limit, and which arises from
the dimensional reduction mechanism discussed above, i.e. the supersymmetry of
the matrix model is broken leaving an effectively non-supersymmetric model.

However, the main obstacle in arriving at a matrix model representation of
supergravity and superstrings is precisely this dimensional reduction. The
critical string susceptibility index for a superstring embedded in a
$D$-dimensional space as calculated from super-Liouville theory \cite{polzam}
is
\beq
\gamma_{\rm str}^{(S)}=\frac14\left(D-1-\sqrt{(1-D)(9-D)}\right)
\eeq
It is not known at this stage how to relate superstring theories to
supersymmetric matrix models and therefore to geometrical descriptions of
discretized super-surfaces. The only succesful discrete approach to date to
2-dimensional supergravity coupled to minimal superconformal models are the
super-eigenvalue models introduced in \cite{aimz}. These models reproduce the
super-Virasoro algebra associated with the Neveu-Schwarz sector of the
superstring theory. The idea is to use the prominent role played by the
Virasoro constraints of the usual Hermitian matrix models to construct a
partition function for $N$ Grassmann even and odd variables (the
``super-eigenvalues") which obeys a set of super-Virasoro constraints. To
formulate these constraints, one must introduce, in addition to the usual
coupling constants $g_k$, a set of Grassmann-valued couplings $\xi_{k+1/2}$.
Many of the well-known features of matrix models, such as the genus expansion,
loop equations and loop insertion operators, have supersymmetric counterparts
in this formalism. The problem of solving a super-eigenvalue model can be
reformulated as a set of superloop equations obeyed by superloop correlators.
The double-scaling limit was further studied in \cite{abbem}, and the complete
iterative solution based on the moment technique of \cite{ackm} (see Subsection
3.4) was carried out by Plefka in \cite{plefka1,plefka2}. For the connections
between $N=1$ super-Liouville amplitudes and super-eigenvalue correlators, see
\cite{zaab}.

There is, unfortunately, no known connection between the super-eigenvalue model
and any type of super-matrix model. The idea to determine whether or not a
super-Virasoro algebra can be realized in a super-matrix model, in terms of a
set of differential operators in the coupling constants of a general matrix
potential, is to construct matrix generators analogous to the matrix
super-charges (\ref{matrixsuperch}) which generate, as $N\to\infty$, the
super-Virasoro algebra associated with the Ramond sector of superstring theory
in a $D=0$ dimensional target space. This problem has been discussed somewhat
by Makeenko in \cite{maksusy}. In order for such constructions to work, though,
the symmetry of the matrix model should be reduced by modifying the potential
so that a larger class of Feynman graphs (beyond the tree-like, cactus
diagrams) survives the perturbative expansion.

\section{Conclusions}

We have shown that a novel class of matrix models with fermionic
degrees of freedom are formally solvable by the method of loop
equations, even though they do not admit the standard Riemann-Hilbert
problem \cite{bipz,mig2,mig4} which arises for the statistical distribution of
eigenvalues in the more conventional matrix models. The solutions of the loop
equations for fermionic matrix models are completely analogous to those of
Hermitian and complex matrix models \cite{ackm}, and they coincide at
each order of the $\frac{1}{N}$-expansion with those of a Hermitian
matrix model with a generalized Penner interaction. However, the
observables in the fermionic case are always well-defined and
convergent quantities, contrary to Hermitian matrix models, and they
may have an interesting interpretation as dynamically triangulated
theories of random surfaces. A price to pay for this good convergence
is the larger degree of complexity of the equations which
completely determine the set of correlators of these models.  This
complexity leads to a more complicated phase structure of the models
and results in a critical behaviour which is non-characteristic of the
usual matrix models. It would be interesting to give the fermionic
nature of these matrix models an interpretation in terms of worldsheet
discretizations in string theory. One step along these lines could
involve examining the continuum limit of the fermionic models in
relation to their novel Virasoro and $W$-algebra constraints to
determine precisely what the double-scaling limit continuum theory is
\cite{mak3}.

The $D>0$ dimensional models can only be explicitly solved for Gaussian
potentials, since higher degree potentials lead to large degree algebraic
equations and even when their solutions are explicitly known (e.g. for
a Penner-type potential) they are rather obscure and are not at all
informative. It would be interesting to develop other methods to solve
these models, such as an explict formula for the Itzykson-Zuber integral
(\ref{iz}) with Grassmann-valued matrices, in order to explore the
critical behaviour associated with higher order potentials in these cases.
This critical behaviour might provide new insights into quantum gravity
in 2 and higher dimensions, and especially for the $D>1$ dimensional
models which might induce QCD. In these latter models the loop equations
are even more complex since one doesn't know right away what potential
the ansatz (\ref{ansatz}) is associated with. Moreover, the extended loop
correlators such as (\ref{opencorrgen}) can only be found from a separate
set of Schwinger-Dyson equations for the extended objects. Contrary to the
Kazakov-Migdal model, however, these higher-dimensional models {\it do}
have a first order phase transition in the Gaussian case which is associated
with the restoration of the area law. Furthermore, the fermionic nature of the
inducing fields in these cases makes the adjoint fermion matrix model
resemble ordinary QCD in many respects. It would be interesting though to
develop some other formalism, such as a master field formulation
\cite{mak1,mak2,mig2,mig4}, for solving these models in the weak coupling phase
where the dynamics of extended objects are non-trivial.

Insights into the nature of the solutions of the loop equations for these
higher dimensional fermionic matrix models could also come from further
investigations of the supersymmetric theories. It would be interesting to
generalize these reduced matrix theories to other combinatorial problems,
especially those relevant to superstring theory. As always the problem is the
dimensional reduction which appears in these models, so that the supersymmetry
of the matrix model does not represent the same supersymmetry of the
superstring theory \cite{alvarez,dadda}. However, a supersymmetric model in
$D>0$ dimensions could be relevant to a statistical theory of discretized
super-Riemann surfaces. As in the case of the pure matrix theories, modified
supersymmetric vector models could provide huge insights into these problems.

\section*{Acknowledgements}

We wish to thank Yu. Makeenko, N. Marshall, L. Paniak, J. Plefka and M. Van
Raamsdonk for valuable discussions and comments on the manuscript. This work
was supported in part by the Natural Sciences and Engineering Research Council
of Canada.

\end{document}